\documentclass[aps,prx,notitlepage,superscriptaddress,longbibliography,twocolumn,nofootinbib,floatfix]{revtex4-2}

\usepackage[utf8]{inputenc}
\usepackage{graphicx}
\usepackage{hyperref}
\usepackage{xcolor,graphicx}

\usepackage{lipsum}

\usepackage{amsmath,amsthm,amssymb,mathtools}
\usepackage{multirow}
\usepackage{braket}
\usepackage{bbm,bm}
\usepackage[bb=boondox]{mathalfa}

\usepackage{makecell}

\newcommand{\ie}{{\it i.e.},\ }
\newcommand{\eg}{{\it e.g.},\ }
\newcommand{\id}{\mathbb{1}}
\newcommand{\Tr}{\operatorname{Tr}}
\newcommand{\ii}{\operatorname{i}}
\newcommand{\ee}{\operatorname{e}}

\begin{document}
\title{Volume-law entanglement entropy of typical pure quantum states}
	
\author{Eugenio Bianchi}
\email{ebianchi@psu.edu}
\affiliation{Department of Physics, The Pennsylvania State University, University Park, PA 16802, USA}
\affiliation{Institute for Gravitation and the Cosmos, The Pennsylvania State University, University Park, PA 16802, USA}
	
\author{Lucas Hackl}
\email{lucas.hackl@unimelb.edu.au}
\affiliation{School of Mathematics and Statistics, The University of Melbourne, Parkville, VIC 3010, Australia}
\affiliation{School of Physics, The University of Melbourne, Parkville, VIC 3010, Australia}
	
\author{Mario Kieburg}
\email{m.kieburg@unimelb.edu.au}
\affiliation{School of Mathematics and Statistics, The University of Melbourne, Parkville, VIC 3010, Australia}
	
\author{Marcos Rigol}
\email{mrigol@phys.psu.edu}
\affiliation{Department of Physics, The Pennsylvania State University, University Park, PA 16802, USA}
	
\author{Lev Vidmar}
\email{lev.vidmar@ijs.si}
\affiliation{Department of Theoretical Physics, J. Stefan Institute, SI-1000 Ljubljana, Slovenia}
\affiliation{Department of Physics, Faculty of Mathematics and Physics, University of Ljubljana, SI-1000 Ljubljana, Slovenia}

\begin{abstract}
The entanglement entropy of subsystems of typical eigenstates of quantum many-body Hamiltonians has been recently conjectured to be a diagnostic of quantum chaos and integrability. In quantum chaotic systems it has been found to behave as in typical pure states, while in integrable systems it has been found to behave as in typical pure Gaussian states. In this tutorial, we provide a pedagogical introduction to known results about the entanglement entropy of subsystems of typical pure states and of typical pure Gaussian states. They both exhibit a leading term that scales with the volume of the subsystem, when smaller than one half of the volume of the system, but the prefactor of the volume law is fundamentally different. It is constant (and maximal) for typical pure states, and it depends on the ratio between the volume of the subsystem and of the entire system for typical pure Gaussian states. Since particle-number conservation plays an important role in many physical Hamiltonians, we discuss its effect on typical pure states and on typical pure Gaussian states. We prove that while the behavior of the leading volume-law terms does not change qualitatively, the nature of the subleading terms can change. In particular, subleading corrections can appear that depend on the square root of the volume of the subsystem. We unveil the origin of those corrections. Finally, we discuss the connection between the entanglement entropy of typical pure states and analytical results obtained in the context of random matrix theory, as well as numerical results obtained for physical Hamiltonians.
\end{abstract}
	
\maketitle
	
\tableofcontents

\section{INTRODUCTION}
Entanglement is a defining property of quantum theory, and plays a crucial role in a broad range of problems in physics, ranging from the black hole information paradox~\cite{page1993information} to the characterization of phases in condensed matter systems~\cite{eisert2010colloquium}. Put simply, entanglement refers to quantum correlations between different parts of a physical system that cannot be explained classically~\cite{bell1964einstein, horodeckiRMP2009}. Over the years, a wide range of \emph{entanglement measures} have been devised to quantify entanglement~\cite{pleniomeasures2007}. Prominent among those are the \emph{bipartite} entanglement measures, which involve splitting the system in two parts.

For the special case of globally pure quantum states $\ket{\psi}$ (our interest here) and a bipartition, the von Neumann entanglement entropy, also known as the entropy of entanglement or just the \emph{entanglement entropy}, is one of the simplest measures of quantum entanglement. It vanishes if and only if there is no quantum entanglement between the two parts, in which case the state must be a product state. We study the entanglement entropy in Hilbert spaces with a tensor product structure $\mathcal{H}=\mathcal{H}_A\otimes\mathcal{H}_B$\footnote{For fermionic systems, as considered later, one needs to work with a fermionic generalization of the tensor product, which also gives rise to a fermionic notion of the partial trace~\cite{szalay2021fermionic}.}. To compute the entanglement entropy of subsystem $A$ (with volume $V_A$) of $\ket{\psi}$, one traces out the complement subsystem $B$ (with volume $V-V_A$, where $V$ is the total volume) to obtain the mixed density matrix $\hat \rho_A=\mathrm{Tr}_{\mathcal{H}_B}\ket{\psi}\bra{\psi}$. The entanglement entropy $S_A$ of subsystem $A$ is then
\begin{equation}\label{Neumann.entropy}
    S_A=-\mathrm{Tr}(\hat \rho_A\ln\hat \rho_A),
\end{equation}
while the $n$th R\'enyi entropy is defined as
\begin{equation}
    S_A^{(n)} = - \ln[\mathrm{Tr}(\hat \rho_A^n)]\,.
\end{equation}
The second-order Rényi entropy $S_A^{(2)}$ has already been measured in experiments with ultracold atoms in optical lattices~\cite{islam2015measuring, kaufman2016quantum}.

We stress that the focus of this tutorial is in pure quantum states. Quantifying entanglement in globally mixed states is more challenging. In particular, the von Neumann and R\'enyi entanglement entropies are not entanglement measures for globally mixed states. Several of the bipartite entanglement measures defined for mixed states (\eg distillable entanglement, entanglement cost, entanglement of formation, relative entropy of entanglement, and squashed entanglement) reduce to the entanglement entropy when evaluated on pure states~\cite{pleniomeasures2007}.
	
\subsection{Ground-state entanglement}
	
In general one is interested in understanding the behavior of measures of entanglement in physical systems, and in determining what such a behavior can tell us about the physical properties of the system. Much progress in this direction has been achieved in the context of many-body ground states of local Hamiltonians, for which a wide range of theoretical approaches are available~\cite{amico_fazio_08, Peschel2009, calabrese_cardy_09, eisert2010colloquium}. Such ground states usually exhibit a leading term of the entanglement entropy that scales with the area, or with the logarithm of the volume, of the subsystem. Identifying and understanding universal properties of the entanglement entropy in ground states of local Hamiltonians has been a central goal~\cite{audenaert_eisert_02, osterloh_amico_2002, osborne_nielsen_02, vidal_latorre_03}. 
	
In one-dimensional systems of spinless fermions or $\tfrac{1}{2}$ spins, the leading (in the volume $V_A$) term in the entanglement entropy has been found to distinguish ground states of critical systems from those of noncritical ones~\cite{vidal_latorre_03, latorre_rico_04, hastings_07}. In the former the leading term exhibits a logarithmic scaling with the volume (when described by conformal field theory, the central charge is the prefactor of the logarithm~\cite{vidal_latorre_03, latorre_rico_04, calabrese_cardy_04}), while in noncritical ground states the leading term is a constant (which, in one dimension, reflects an area-law scaling). Subleading terms have also been studied, specially in the context of states that are physically distinct but exhibit the same leading entanglement entropy scaling. An example in the context of quadratic Hamiltonians in two dimensions are ground states that are critical with a pointlike Fermi surface versus noncritical, which both exhibit a leading area-law entanglement entropy~\cite{wolf_06, gioev_klich_06, barthel_chung_06, li_ding_06, cramer_eisert_07}. Remarkably, the subleading term in the former scales logarithmically with $V_A$ while it is constant for noncritical ground states~\cite{ding_brayali_08}. Also, in two-dimensional systems, critical states described by conformal field theory~\cite{fradkin_moore_06} and states with a spontaneously broken continuous symmetry~\cite{kallin_hastings_11, metlitski_grover_15} have been found to exhibit a universal subleading logarithmic term.
	
\subsection{Excited-state entanglement}
	
In recent years, interest in understanding the far-from-equilibrium dynamics of (nearly) isolated quantum systems and the description of observables after equilibration~\cite{polkovnikov2011colloquium, d2016quantum,  gogolin2016equilibration} have motivated many studies of the entanglement properties of highly excited eigenstates of quantum many-body systems (mostly in the context of lattice systems)~\cite{mejia_05, alba09, Deutsch_2010, santos_12, deutsch_li_13, storms_singh_14, moelter_barthel_14, lai_yang_15, beugeling_andreanov_15, yang_chamon_15, nandy_sen_16, vidmar2017entanglement, vidmar2017entanglement2, zhang_vidmar_18, dymarsky2018subsystem, garrisson_grover_18, nakagawa_watanabe_18, vidmar2018volume, huang_19, hackl2019average, lu_grover_19, murthy_19, jafarizadeh_rajabpour_19, wilming_goihl_19, leblond_mallayya_19, faiez_20a, modak_nag_20, kaneko_iyoda_20, bhakuni_sharma_20, faiez_20b, lydzba2020entanglement, lydzba2021entanglement, haque_mcclarty_20, miao_barthel_20}. Because of the limited suit of tools available to study entanglement properties of highly excited eigenstates of model Hamiltonians, most of the results reported in those works were obtained using exact diagonalization techniques, which are limited to relatively small system sizes.
	
In contrast to the ground states, typical highly excited many-body eigenstates of local Hamiltonians have a leading term of the entanglement entropy that scales with the volume of the subsystem. Also, in contrast to the ground states, the leading volume-law term exhibits a fundamentally different behavior depending on whether the Hamiltonian is nonintegrable (the generic case for physical Hamiltonians) or integrable. In the former case the coefficient has been found to be constant, while in the latter case it depends on the ratio between the volume of the subsystem and the volume of the entire system.
	
Many-body systems that are integrable are special as they have an extensive number of local conserved quantities~\cite{sutherland_book_04}. As a result, their equilibrium properties can in many instances be calculated analytically, and their near-equilibrium properties can be ``anomalous,'' e.g., they can exhibit transport without dissipation (ballistic transport). Also, isolated integrable systems fail to thermalize if taken far from equilibrium. Interested readers can learn about the effects of quantum integrability in the collection of reviews in Ref.~\cite{calabrese_essler_review_16}. 
	
There is a wide range of quadratic Hamiltonians in arbitrary dimensions (which include a wide range of noninteracting models), e.g., translationally invariant quadratic Hamiltonians, that can be seen as a special class of integrable models. A class in which the nondegenerate many-body eigenstates are Gaussian states, while their degenerate eigenstates can always be written as Gaussian states. This means that those many-body eigenstates are fully characterized by their one-body density matrix or their covariance matrix. The entanglement entropy of highly excited eigenstates of some of those ``integrable'' quadratic Hamiltonians was studied in Refs.~\cite{storms_singh_14, vidmar2017entanglement, zhang_vidmar_18, hackl2019average, jafarizadeh_rajabpour_19}. Other quadratic Hamiltonians in arbitrary dimensions that will be of interest to us here are quadratic Hamiltonians in which the single-particle sector exhibits quantum chaos (to be defined in the next subsections). We refer to such Hamiltonians as quantum-chaotic quadratic Hamiltonians. The entanglement entropy of highly excited eigenstates of quantum-chaotic quadratic Hamiltonians (on a lattice) was studied in Refs.~\cite{lydzba2020entanglement, lydzba2021entanglement}. It was found to exhibit a typical leading volume-law term that is qualitatively similar to that found in eigenstates of integrable quadratic Hamiltonians (in which the single-particle sector does not display quantum chaos), such as translationally invariant quadratic Hamiltonians (on a lattice)~\cite{vidmar2017entanglement, hackl2019average}. 
	
In the presence of interactions, many-body integrable systems mostly exist in one dimension~\cite{cazalilla_citro_review_11, guan2013fermi}. They come in two ``flavors,'' Hamiltonians that can be mapped onto noninteracting ones (a smaller class), and Hamiltonians that cannot be mapped onto noninteracting ones. Remarkably, both ``flavors'' have been found to describe pioneering experiments with ultracold quantum gases in one dimension~\cite{moritz_stoferle_03, kinoshita_wenger_04, paredes_widera_04, kinoshita_wenger_05, kinoshita_wenger_06, amerongen_es_08, gring_kuhnert_12, fukuhara2013microscopic, pagano2014one, langen_erne_15, Bloch2016, tang_kao_18, schemmer2019generalized, wilson_malvania_20, jepsen2020spin, lev2020, malvania_zhang_21}. The entanglement entropy of highly excited eigenstates of lattice Hamiltonians that can be mapped onto noninteracting ones (which exhibit the same leading volume-law terms as their noninteracting counterparts) was studied in Refs.~\cite{vidmar2018volume, hackl2019average}, while the entanglement entropy of highly excited eigenstates of a Hamiltonian (the spin-$\frac{1}{2}$ XXZ chain) that cannot be mapped onto a noninteracting one was studied in Ref.~\cite{leblond_mallayya_19}. Remarkably, in all the quadratic and integrable systems studied so far, the coefficient of the leading volume-law term of typical eigenstates has been found to depend on the ratio between the volume of the subsystem and the volume of the entire system.
	
Analytical progress understanding the previously mentioned numerical results has been achieved in some special cases. One such case is translationally invariant quadratic Hamiltonians, or models that can be mapped onto them in one dimension~\cite{cazalilla_citro_review_11}, for which tight bounds were obtained for the leading (volume-law) term in the average entanglement entropy~\cite{vidmar2017entanglement, hackl2019average}, and some understanding was gained about subleading corrections~\cite{vidmar2018volume}. This was possible thanks to the Gaussian nature of the eigenstates. Another case is nonintegrable models under the assumption that their eigenstates exhibit eigenstate thermalization~\cite{Deutsch_2010, dymarsky2018subsystem, garrisson_grover_18, murthy_19}.
	
\subsection{Random matrix theory in physics}
	
Random matrix theory has provided a more systematic approach to gaining an analytical understanding of the entanglement properties of many-body eigenstates in nonintegrable models~\cite{yang_chamon_15, vidmar2017entanglement2, liu_chen_18, huang_gu_19, pengfei_chunxiao_20, morampudi_chandran_20, haque_mcclarty_20}. Such an approach is justified by the fact that many studies (see, \eg Ref.~\cite{d2016quantum} for a review) have shown that nonintegrable models exhibit ``quantum chaos.'' By quantum chaos what is meant is that statistical properties of highly excited eigenstates of such models, \eg level spacing distributions, are described by the Wigner surmise~\cite{d2016quantum}. This was conjectured by Bohigas, Giannoni, and Schmit (BGS)~\cite{bohigas_giannoni_84} for quantum systems with a classical counterpart, in which case ``quantum chaos'' usually occurs when the classical counterparts are $K$-chaotic, where $K$ stands for Kolmogorov, and it is the class of systems that exhibit the highest degree of chaos. Remarkably, even statistical properties of eigenvectors such as the ratio between the variance of the diagonal and the off-diagonal matrix elements of Hermitian operators have been shown to agree with random matrix theory predictions~\cite{mondaini_rigol_17, jansen_stolpp_19, richter_dymarsky_20, schoenle_jansen_21}. Recently, two of us (M.R. and L.V., in collaboration with P. \L yd\.{z}ba) used random matrix theory in the context of quantum-chaotic quadratic Hamiltonians to obtain a closed-form expression that describes the average entanglement entropy of highly excited eigenstates of quadratic models whose single-particle spectrum exhibits quantum chaos, such as the three-dimensional Anderson model~\cite{lydzba2020entanglement, lydzba2021entanglement}.
	
The application of random matrix theory to many-body systems goes back to works by Wigner~\cite{wigner_55, wigner_57, Wigner-surmise, wigner_58} as well as Landau and Smorodmsky~\cite{landau1955} in the 1950s, who aimed at finding a statistical theory that described the excitation spectra in nuclei for elastic scattering processes. Their novel idea was that a sufficiently complicated operator like the Hamilton, or the lattice Dirac operator, can be replaced by a random matrix (whose entries are, preferably, Gaussian distributed as those are easier to deal with analytically) with the appropriate symmetries. For this to hold, it is not important that the physical operator has matrix entries that are all occupied with nonzero entries. In condensed matter models~\cite{d2016quantum}, as well as in lattice QCD~\cite{Berbenni-Bitsch-1998, damgaard-2000, Farchioni-2000, Deuzeman-2011, Kieburg:2017rrk}, numerical evidence has shown that very sparse matrices can also exhibit spectral characteristics of a random matrix with Gaussian distributed entries. It is the concept of universality that has made random matrices so versatile. Like in the central limit theorem, in which an infinite sum of independently and identically distributed random variables leads to a Gaussian random variable under very mild conditions, it happens that, for many spectral quantities, it does not matter how the random matrix is actually distributed. 
	
Over the years, random matrix theory has found many more applications in physics; for example, the local level density about Dirac points (also known as hard edges in random matrix theory) has been used to classify operators such as Hamiltonians and Dirac operators, and to discern global symmetries of a system. By global symmetries, it is meant those that are described by a linear involution (operators that square to unity) in terms of unitary and antiunitary operators. Well-known examples in physics are, time reversal, parity, charge conjugation, and chirality. Global symmetries play a central role when classifying systems in the context of quantum chaos~\cite{Dyson1962}, in superconductors and topological insulators~\cite{1997PhRvB..55.1142A, 2008PhRvB..78s5125S}, in quantum-chromodynamics-like theories in the continuum and on a lattice~\cite{Verbaarschot:1994qf, Kieburg:2017rrk}, and in Sachdev-Ye-Kitaev-models (SYK)~\cite{Garcia21, Kanazawa:2017dpd}.
	
\subsection{Local spectral statistics}\label{sec:localspec}
	
There are two spectral scales that are usually discussed in the context of random matrix theory, and to which different kinds of universalities apply. Those are the local and the global spectral scales.
	
The microscopic or local spectral scale is given by the local mean level spacing where the fluctuations of the individual eigenvalues are resolved. This scale is often of more physical interest as it analyses the level repulsion of eigenvalues that are very close to each other. Such a level repulsion is usually algebraic for very small distances $s$. Namely, the level spacing distribution $p(s)$, which is the distribution of the distance of two consecutive eigenvalues, is of the form $s^\beta$ (where $\beta$ is the Dyson index) for small distances. 
	
\begin{figure*}[!t]
	\centering
	\includegraphics[width=\linewidth]{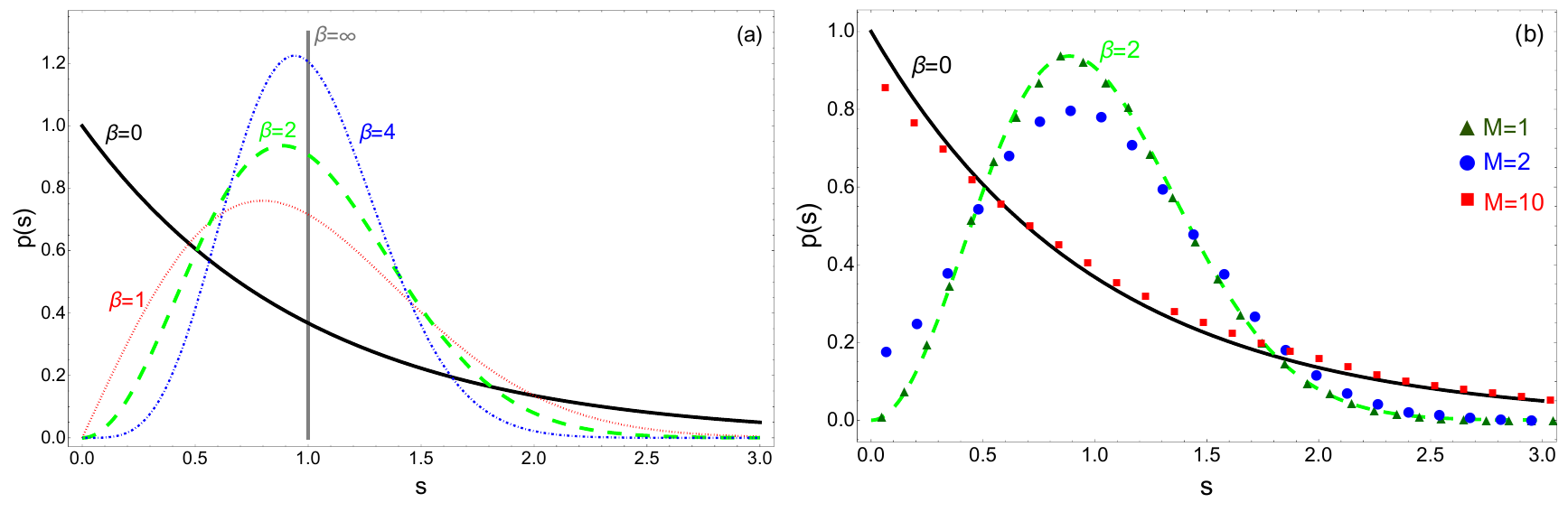}
	\caption{(a) The level spacing distributions of the Poisson distribution (solid line; $\beta=0$), the Wigner surmise of the GOE (dotted line; $\beta=1$), of the GUE (dashed line; $\beta=2$), of the GSE (dash-dot line; $\beta=4$), and the picket fence statistics (vertical line; $\beta=\infty$). (b) Three Monte Carlo simulations (symbols) of the spacing between eigenvalues $(50\cdot M)$ and $(50\cdot M+1)$ of the direct sum of $M$ GUEs with a matrix dimension $N=100$ (in total, the matrix dimension is $100^M\times 100^M$), compared to the Poisson distribution (solid line), and the Wigner surmise of the GUE (dashed line). The ensemble size is $10^5$ such that the statistical error is about $1\%$. The bin size is about $0.1$, but varies as the unfolding slightly changes their actual value.}
	\label{fig:level-spacing}
\end{figure*}

While the symmetry of a Hamiltonian, such as time reversal, chirality, or charge conjugation, is not very important for the global spectral scale, it is very important for the local spectral statistics as it influences the value of $\beta$. Wigner~\cite{Wigner-surmise} derived the distribution for two-level Gaussian random matrices with Dyson index $\beta=1$, which was soon generalized to $\beta=2,4$,
\begin{equation}
	p(s)=2\frac{(\Gamma[(\beta+2)/2])^{\beta+1}}{(\Gamma[(\beta+1)/2])^{\beta+2}}s^\beta\exp\left[-\left(\frac{\Gamma[(\beta+2)/2]}{\Gamma[(\beta+1)/2]}\right)^2s^2\right]
\end{equation}
with the gamma function $\Gamma[x]$. This distribution is nowadays called Wigner's surmise. The corresponding random matrices are known as the Gaussian orthogonal ensemble (GOE; $\beta=1$), the Gaussian unitary ensemble (GUE; $\beta=2$), and the Gaussian symplectic ensemble (GSE; $\beta=4$). Those are usually compared with the level spacing distribution of independently distributed eigenvalues ($\beta=0$), which gives the Poisson distribution
\begin{eqnarray}
	p(s)=e^{-s},
\end{eqnarray}
and with the level spacing distribution of the one-dimensional quantum harmonic oscillator (also known as the picket fence statistics), which is a simple Dirac delta function
\begin{eqnarray}
	p(s)=\delta(1-s).
\end{eqnarray}
All five benchmark distributions are shown in Fig.~\ref{fig:level-spacing}(a).
	
The use of the Wigner surmise as a diagnostic of quantum chaos and integrability followed fundamental conjectures by BGS~\cite{bohigas_giannoni_84} (mentioned before) and Berry and Tabor~\cite{berry_tabor_77}, respectively. The latter states that, for an integrable bounded system with more than two dimensions and incommensurable frequencies of the corresponding tori, the spectrum should follow the Poisson statistics. However, both conjectures have to be understood with the following care as the eigenvalue spectrum must be prepared appropriately.
\begin{itemize}
	\item[(i)] The spectrum must be split into subspectra with fixed ``good'' quantum numbers such as the spin, parity, and conserved charges. This requires knowledge of all the symmetries of the model. This step must be taken since a direct sum of independent GUE matrices can yield a level spacing distribution that resembles the Poisson statistics; see Fig.~\ref{fig:level-spacing}(b). 
	\item[(ii)] One needs to unfold the spectra, meaning, that the distance between consecutive eigenvalues must be in average equal to one. This second step is crucial as only then the level spacing distributions are comparable and universal statistics can be revealed. The eigenvalue spectrum of an irregularly shaped drum, a complex molecule, and that of a heavy nucleus have completely different energy scales. After the unfolding of their spectra these scales are removed and show common behavior. Yet, the procedure of unfolding is far from trivial for empirical spectra. There are other means such as the study of the ratio between the two spacings of three consecutive eigenvalues~\cite{Oganesyan-2007}. But this observable also has its limitations as this kind of ``automatic unfolding'' only works in the bulk of the spectrum. It fails at spectral edges and other critical points in the spectrum.
\end{itemize}
	
In the context of the Wigner surmise, we should stress that even though the statistics of the spectral fluctuations are well described at the level of the mean level spacing~\cite{PhysRev.120.1698, FRENCH19715, BOHIGAS1971383} (even beyond the context of many-body systems; see, \eg the reviews and books~\cite{Guhr1998, mehta2004, akemann2011, haake2019} and the references therein), it was soon realized that there are statistical properties of the spectral fluctuations of many-body Hamiltonians that cannot be described using full random matrices; see Refs.~\cite{BOHIGAS1971261, FRENCH1970449, monfrench1975, Benet:2000cy}. This is due to the fact that usually only one-, two- and maybe up to four-body interactions represent the actual physical situation. Random matrices that reflect these sparse interactions are called embedded random matrix ensembles~\cite{monfrench1975, RevModPhys.53.385, Guhr1998, Kota2001, Kota2014}. In the past decades, they have experienced a revival due to studies of the SYK model~\cite{1993PhRvL..70.3339S, 2016PhRvD..94j6002M, Garcia-Garcia:2016mno, Garcia-Garcia:2017pzl, Garcia-Garcia:2018fns, 2014MPAG...17..441E}, and two-body interactions~\cite{Vyas:2018aal, 2017AIPC.1912b0003B, 2018tqrf.book..457S}. A full understanding of how these additional tensor structures, which arise naturally in quantum many-body systems, impact the entanglement of the energy eigenstates is currently missing.
	
\subsection{Global spectral statistics and eigenvector statistics}
	
The second scale is the macroscopic or global spectral scale, which is usually defined as the average distance between the largest and the smallest eigenvalues. For this scale, Wigner~\cite{wigner_55, wigner_58} derived the famous Wigner semicircle, which describes the level density of a Gaussian distributed real symmetric matrix. He was also the first to show, again under mild conditions, that the Gaussian distribution of the independent matrix entries can be replaced by an arbitrary distribution, and nevertheless one still obtains the Wigner semicircle. One important feature of this kind of universality is that it does not depend on the symmetries of the operators. For instance, whether the matrix is real symmetric, Hermitian, or Hermitian self-dual has no impact on the level density, which is in all those cases a Wigner semicircle~\cite{Forrester_2010}. The global spectral scale also plays a crucial role in time series analysis~\cite{Giraud2015} and telecommunications~\cite{Couillet2011}, where instead of the Wigner semicircle the Mar\v{c}enko-Pastur distribution~\cite{marcenko} describes the level density. 
	
The global scale is always important when considering the so-called linear spectral statistics, meaning an observable that is of the form $\sum_{j=1}^Nf(\lambda_j)$, where the $\lambda_j$ are the eigenvalues of the random matrix. This is the situation that we encounter when computing the entanglement entropy, where the $\lambda_j$ are the eigenvalues of the density matrix; cf. Eq.~\eqref{Neumann.entropy}. Therefore, we expect that the leading terms in the entanglement entropy are insensitive to the Dyson index $\beta$, so that the entanglement entropy can serve as an excellent diagnostic for integrable or chaotic behavior. 
	
A related diagnostic for the amplitude $A$ of vector components of eigenstates is the Porter-Thomas distribution~\cite{PhysRev.104.483}, which is used to decide whether a state is localized or delocalized. The Porter-Thomas distribution is a $\chi^2$ distribution,
\begin{equation}
	\mathcal{I}(A)= \left(\frac{\beta N}{2}\right)^{\beta/2}\frac{A^{\beta/2-1}}{\Gamma[\beta/2]}\exp\left[-\frac{\beta N}{2}A\right]   ,
\end{equation}
where the normalization of the first moment is chosen to be equal to $1/N$. Note that in the quaternion case one defines the amplitude as the squared modulus of a quaternion number. Hence, as a sum of four squared real components, similar to the squared modulus of a complex number (which is the sum of the square of the real and imaginary parts). Actually, the application of random matrices for computing the entanglement entropy is based on this idea. We can only replace a generic eigenstate by a Haar-distributed vector on a sphere after assuming that the state is delocalized. Unlike the Porter-Thomas distribution, as previously mentioned, the leading terms in the entanglement entropy are expected to be independent of the Dyson index $\beta$ (which has yet to be proved).
	
The relation between certain quantum informational questions and random matrix theory also has a long history, and the techniques developed are diverse (see, e.g., the review~\cite{2016JMP....57a5215C} and Chapter 37 of Ref.~\cite{akemann2011}). Questions about generic distributions and the natural generation of random quantum states have been a focus of attention~\cite{Hall:1998mh, 2004JPhA...37.8457S}. The answers to those questions are still debated as there are several measures of the set of quantum states and each has its benefits and flaws; for instance, two of those are based on the Hilbert-Schmidt metric and the Bures metric~\cite{Bures1969, Hall:1998mh}. Those measures define some kind of ``uniform distribution'' on the set of all quantum states and, actually, generate random matrix ensembles that have been studied to some extent~\cite{Hall:1998mh, 2001JPhA...34.7111Z, 2004JPhA...37.8457S, 2003JPhA...3610083S, 2010JPhA...43e5302O, 2016CMaPh.342..151F, wei2021quantum}. In this tutorial, we encounter one of the aforementioned ensembles, namely, the one related to the Hilbert-Schmidt metric, which naturally arises from a group action so that the states are Haar distributed according to this group action.
	
\subsection{Typicality and entanglement}
	
An important question that one can ask, which relates to the latest observations made in the context of random matrix ensembles, is what are the entanglement properties of typical pure quantum states. This was the earliest question to be addressed. Following work by Lubkin~\cite{lubkin1978entropy} and Lloyd and Pagels~\cite{lloyd1988complexity}, Page~\cite{page1993average} obtained a closed analytical formula for the average entanglement entropy (over all pure quantum states) as a function of the system and subsystem Hilbert space dimensions. This formula was rigorously proven later in Refs.~\cite{foong1994proof, sanchez1995simple, Sen:1996ph}. In lattice systems in which the dimension of the Hilbert space per site is finite, one can show that Page's formula results in a ``volume-law'' behavior, \ie the entanglement entropy scales linearly in the volume $V_A$ of the subsystem, $S_A\propto V_A$ (for a large system of volume $V$ and a subsystem with $V_A<V/2$). The prefactor of this volume law is the same that was later found in studies of highly excited eigenstates of nonintegrable Hamiltonians and, separately, within random matrix theory. Deviations from Page's result have been discussed in the context of highly excited eigenstates of number-preserving Hamiltonians away from half-filling~\cite{vidmar2017entanglement2, garrisson_grover_18}. The entanglement entropy of pure quantum states with particle-number conservation was also studied in Refs.~\cite{nakagawa_watanabe_18, huang_19}.
	
\begin{figure*}
	\centering
	\includegraphics[width=\linewidth]{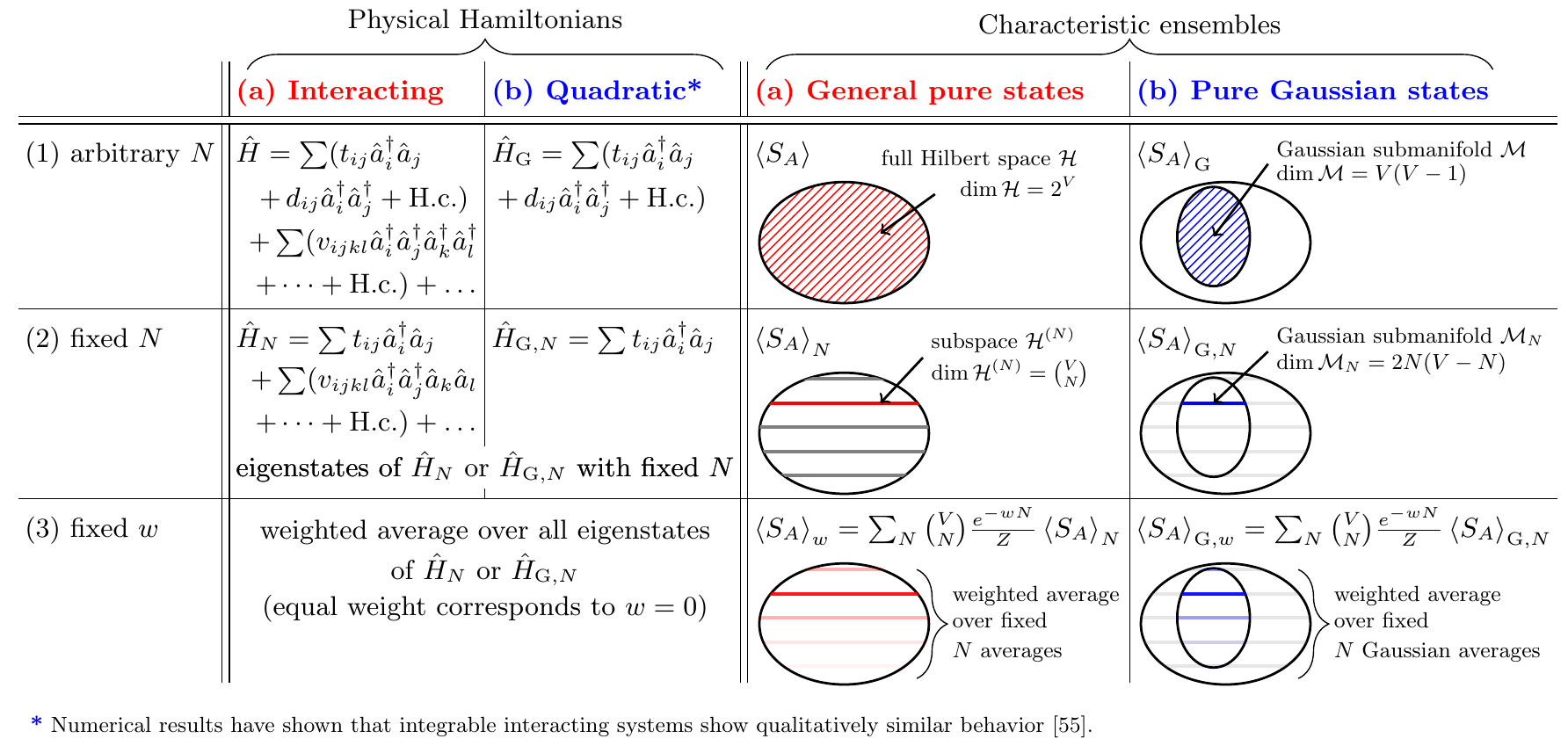}
	\caption{Physical Hamiltonians versus characteristic ensembles. This tutorial relates the typical entanglement entropy of eigenstates of physical Hamiltonians (left) to the typical entanglement entropy computed analytically for six characteristic ensembles constructed from random matrix theory (right). On the left we show four representative physical Hamiltonians, (a) generic interacting and (b) quadratic, (1) without and (2) with number conservation. On the right we show the six characteristic ensembles, which are obtained considering three sets of eigenstates [(1) arbitrary $N$, (2) fixed $N$, and (3) weighted average] for (a) general pure states and (b) pure Gaussian states.}
	\label{fig:ensembles}
\end{figure*}
	
In many physically relevant situations, constraints such as particle-number conservation are present and, as a result, the Hilbert space of the system does not factor into the tensor product of Hilbert spaces of subsystems. Other notable examples are gauge theories that have a Gauss constraint, and quantum gravitational systems where the Hamiltonian itself is a quantum constraint~\cite{rovelli2004quantum}. When the constraint is additive over subsystems, \ie of the form $C_A+C_B=0$, one can resolve the Hilbert space as a direct sum over a tensor product of simultaneous eigenspaces of $C_A$ and $C_B$ that solve the constraint. An example of this phenomenon is the previously mentioned systems with a fixed total number of particles, where $C_A$ is the number of particles in the subsystem. For this class of systems, a formula for the typical entanglement entropy of pure constrained states (and its variance) was obtained recently by one of us (E.B.) in collaboration with P.~Don\`a~\cite{bianchi2019typical}.
	
Another important class of pure states are fermionic Gaussian states. They are of special interest because, as mentioned before, the many-body eigenstates of quadratic Hamiltonians are Gaussian states. In the quantum computing community those states are known as matchgate states, and they are used to implement classical computations on a quantum computer~\cite{valiant2002quantum}. The average entanglement entropy over translationally invariant Gaussian states was studied by four of us (E.B., L.H., M.R., and L.V.) in Refs.~\cite{vidmar2017entanglement, vidmar2018volume, hackl2019average}, where tight bounds were derived. The average over all fermionic Gaussian states was studied later by three of us (E.B., L.H., and M.K.) in Ref.~\cite{bianchi2021page}, and a closed-form expression was derived as a function of the total ($V$) and the subsystem ($V_A$) volumes. To leading order in $V_A$, that expression agreed with the one previously derived in Ref.~\cite{lydzba2020entanglement} for the average entropy over all eigenstates of particle-number-preserving random quadratic Hamiltonians that, in turn, was shown to agree with the average over random quadratic Hamiltonians without particle-number conservation in Ref.~\cite{lydzba2021entanglement}. Entanglement entropies of Gaussian states with particle-number conservation were also studied in Refs.~\cite{liu_chen_18, pengfei_chunxiao_20} in the context of SYK models. We connect all these results throughout this tutorial.
	
\subsection{Outline}
	
We provide a detailed understanding of the behavior of the entanglement entropy of typical pure quantum states in (a) the entire Hilbert space (Sec.~\ref{sec:general}) and (b) the subset of Gaussian states (Sec.~\ref{sec:gaussian}). Both for pure quantum states in the entire Hilbert space, and within the subset of Gaussian states, we consider case (1) in which the number of particles is not fixed separately from cases (2) in which it is fixed and (3) in which we take a weighted sum over all fixed particle-number sectors. Overall, we thus consider \emph{six} characteristic ensembles, which we can compare with the respective typical eigenstate entanglement entropy found in physical Hamiltonians, as illustrated in Fig.~\ref{fig:ensembles}. For general states without fixed particle number, the results for the average over all quantum states are well known~\cite{page1993average}, while for general Gaussian states, the results were recently obtained by three of us (E.B., L.H., and M.K.)~\cite{bianchi2021page}. For quantum states with a fixed number of particles, all the results discussed are derived here (the derivations are explained in detail in the appendices). We identify and explain the qualitative changes that occur in subleading terms when one fixes the number of particles in the quantum states. We explore the behavior of the average entanglement entropy as a function of the system volume $V$, the subsystem volume $V_A$, and the total number of particles $N$. We show that general pure states and Gaussian states share common properties for $f=V_A/V\ll V$, but become increasingly distinct as $f\rightarrow\tfrac{1}{2}$.
	
In Sec.~\ref{sec:RMT}, we examine the relation between typical entanglement entropies in the Hilbert space and typical entanglement entropies generated by random matrices. In Sec.~\ref{sec:relphysham}, we examine the relation between typical entanglement entropies in the Hilbert space and typical entanglement entropies of eigenstates of specific model Hamiltonians. While we mostly have fermionic lattice systems (with and without spin) in mind, our results apply to any fermionic system with a well-defined bipartition, hard-core bosons, and spin-$\frac{1}{2}$ systems. In the outlook, we discuss how the same methods can be used to study regular (soft-core) bosons, large spins, and systems of distinguishable particles.
	
It was recently conjectured that the entanglement entropy of typical eigenstates of quantum many-body Hamiltonians can be used as a diagnostic of quantum chaos and integrability~\cite{leblond_mallayya_19}. This conjecture was motivated by the finding that the leading volume-law term of the entanglement entropy in typical eigenstates of an integrable model that is not mappable onto a noninteracting one behaves qualitatively (and quantitatively) like in typical eigenstates of translationally invariant quadratic Hamiltonians, and in stark contrast to the behavior in typical eigenstates of quantum-chaotic Hamiltonians. With that conjecture in mind, we expect that the analytical expressions derived in the sections to follow can be used as benchmarks for numerical results obtained for physical Hamiltonians (and analytical results obtained using different techniques). The entanglement entropy of typical eigenstates of physical Hamiltonians is complementary to diagnostics of integrability and quantum chaos currently in use, such as the previously mentioned Wigner surmise that is based on the statistical properties of the eigenenergies.

\section{GENERAL PURE STATES}\label{sec:general}
	
We consider the general setting of a system with $V$ fermionic modes (potentially arranged in a lattice of arbitrary dimension) and a bipartition of the system into a subsystem $A$ (with $V_A$ fermionic modes) and its complement $B$ (with $V-V_A$ fermionic modes). Hence, the Hilbert space has the structure $\mathcal{H} = \mathcal{H}_A \otimes \mathcal{H}_B$ and dimension $\dim\mathcal{H}=d_A d_B$, where $d_A=\dim\mathcal{H}_A$ and $d_B=\dim\mathcal{H}_B$.
	
This setup can be used in different contexts. For instance, on a $D$-dimensional lattice with $L$ sites per space direction and $N_{\rm int}$ fermionic modes per site (representing internal degrees of freedom, such as spin), the total number of fermionic modes is
\begin{align}
	V=L^D N_{\rm int}.\label{eq:Vdef}
\end{align}
For each mode, we can denote the creation (annihilation) operators as $\hat{f}_i^\dagger$ $(\hat{f}^{}_i)$, where $1\leq i\leq V$. The bipartition with respect to a subsystem of $V_A\leq V$ modes then yields the Hilbert space dimensions
\begin{equation}
	d_A=2^{V_A}\,,\qquad d_B=2^{V-V_A}\,.
\end{equation}
When we fix the subsystem fraction $f=V_A/V$, our results hold in full generality for arbitrary space dimensions $D$ and fermions with an arbitrary spin. They also hold for hard-core bosons and spin-$\frac{1}{2}$ systems (which have two states per lattice site)---but not for (soft-core) bosons for which the average entanglement entropy diverges due to the infinite local dimension of the Hilbert space. 
	
We note that, in the Introduction, we referred to $V$, $V_A$, and $V-V_A$ as volumes. We continue to use that terminology in the rest of the tutorial. Readers should keep in mind that $V$, $V_A$, and $V-V_A$ quantify the number of fermionic modes; the volume (or, similarly, the number of sites) being just one of the possible ways in which this is achieved.
	
\subsection{Arbitrary number of particles}\label{sec:generalpage}
	
A natural approach to gain an understanding of the entanglement entropy of pure quantum states is to consider randomly chosen vector states $\ket{\psi}\in\mathcal{H}$. Since the Hilbert space is finite dimensional, the set of all states describes a hypersphere that is compact. Thus, the uniform distribution on this set provides a natural unbiased probability distribution over pure states~\cite{page1993average}. The corresponding density operator $\ket{\psi}\bra{\psi}$ is also called Haar random as it is equal to the induced Haar measure on the coset ${\rm U}(d_A d_B)/{\rm U}(d_A d_B-1)$ with ${\rm U}(d)$ the $d$-dimensional unitary group. This measure is the unique one that is normalized and invariant under unitary transformations. In practice, we can construct such a state by first fixing a reference state $\ket{\psi_0}$ and then defining $\ket{\psi}=U\ket{\psi_0}$, where $U: \mathcal{H}\to\mathcal{H}$ is a random unitary transformation drawn from the Haar measure of unitary matrices.
	
For every such $\ket{\psi}$, the bipartite entanglement entropy with respect to the tensor product decomposition $\mathcal{H} = \mathcal{H}_A\otimes\mathcal{H}_B$ is, once again, defined by
\begin{align}
	S_A(\ket{\psi})=-\mathrm{Tr}_{\mathcal{H}_A} (\hat\rho_A\ln\hat\rho_A),\quad\text{with}\quad \hat\rho_A=\mathrm{Tr}_{\mathcal{H}_B}\ket{\psi}\bra{\psi}\,,
\end{align}
where $\mathrm{Tr}_{\mathcal{H}_B}$ refers to the partial trace over the sub-Hilbert space $\mathcal{H}_B$. We are interested in the average and the variance of $S_A$ with respect to the statistical ensemble. In Page's setting, the ensemble is given by Haar-random density operators $\ket{\psi}\bra{\psi}$. We can write those quantities as
\begin{align}
	\braket{S_A}&=\int S_A(U\ket{\psi_0}) d\mu(U)\,,\\
	(\Delta S_A)^2&=\int [S_A(U\ket{\psi_0})-\braket{S_A}]^2 d\mu(U)\,,
\end{align}
where $d\mu(U)$ represents the normalized Haar measure over the unitary group $\mathrm{U}(d_Ad_B)$. The average and variance can be computed from the joint probability distribution of the eigenvalues $\lambda_i$ of $\hat\rho_A$, which were derived in Ref.~\cite{lloyd1988complexity}, as $S_A(\ket{\psi}) = -\sum_{j=1}^{d_A}\lambda_j\, \ln(\lambda_j)$.
	
In the case of Page's setting, where one draws a random uniformly distributed state $\ket{\psi}\in\mathcal{H}$, the average entanglement entropy (and any other statistical quantities) will depend only on the dimensions $d_A$ and $d_B$. This shows that the ensuing statements are independent of any chosen particle statistics (bosons, fermions) and also not restricted to the special case of qubit-based (two-state per-site) systems in the context of which they are usually used. They apply much more generally.
	
\subsubsection{Statistical ensemble of states}
	
Let us outline the derivation of Page's result in Eq.~\eqref{Page}. The derivation allows us to draw some relations to random matrix theory, and we use some of the ideas behind it later when studying fermionic Gaussian states.
	
Let us consider a general state vector in the tensor space $\mathcal{H}=\mathcal{H}_A\otimes\mathcal{H}_B$. Such a vector can always be decomposed into some factorizing orthonormal basis $\ket{a}\otimes\ket{b}\in\mathcal{H}$:
\begin{eqnarray}
	\ket{\psi}=\sum_{a=1}^{d_A}\sum_{b=1}^{d_B} w_{ab}\ket{a}\otimes\ket{b}
\end{eqnarray}
with coefficients $w_{ab}\in\mathbb{C}$. As physical state vectors are normalized, the coefficients satisfy
\begin{eqnarray}\label{normalisation-coefficients}
	\sum_{a=1}^{d_A}\sum_{b=1}^{d_B} |w_{ab}|^2=1.
\end{eqnarray}
These coefficients and their distribution contain all the information about the random pure state and, hence, about the entanglement entropy. Once we take the partial trace over the subsystem $B$, the density operator in system $A$ is explicitly given by 
\begin{eqnarray}
	\hat\rho_A=\sum_{a_1,a_2=1}^{d_A}\sum_{b=1}^{d_B} w_{a_1b}w_{a_2b}^*\ket{a_1}\bra{a_2}.
\end{eqnarray}
The coefficients $w_{ab}$ can be seen as the entries of a $d_A\times d_B$ complex matrix $W$. Thence, we can identify the density operator $\hat\rho_A=WW^\dagger$. The density operator for subsystem $B$, when tracing over subsystem $A$, is given by $\hat\rho_B = \Tr_{\mathcal{H}_A} \ket{\psi}\bra{\psi}=W^\dagger W$. This allows us to explicitly see the duality between the two subsystems, and what changes when switching from $A$ to $B$. In general, the dimensions $d_A$ and $d_B$ are not equal. Thus, one of the two density operators always has zero eigenvalues but otherwise comprises the very same eigenvalues with the same multiplicity as the other density operator.
	
Equation~\eqref{normalisation-coefficients} implies that the matrix $W$ satisfies $\Tr WW^\dagger=1$. Page's setting of uniformly distributed states means that $W$ is distributed uniformly on the unit sphere described by this normalization condition. Such a matrix is a random matrix and the ensemble is known as the fixed trace ensemble~\cite{mehta2004}. In quantum information theory, it has been used in several studies, \eg such as those in Refs.~\cite{2001JPhA...34.7111Z, 2004JPhA...37.8457S, wei2021quantum, 2010JPhA...43.5303C, 2009arXiv0912.3999L, 2011JSP...142..403N}.
	
With this knowledge at hand, let us compute the entanglement entropy, which can be expressed in terms of the eigenvalues of $WW^\dagger$, \ie it holds that
\begin{align}
	\begin{split}
		S_A(\ket{\psi})&=-\mathrm{Tr}[WW^\dagger\ln(WW^\dagger)]\\
		&=-\mathrm{Tr}[ W^\dagger W\ln(W^\dagger W)]=S_B(\ket{\psi}).
	\end{split}
\end{align}
It is the spectral decomposition theorem that ensures the symmetry of the entanglement entropy between the two subsystems. This symmetry always holds.
	
To compute the ensemble average of $S_A(\ket{\psi})$, one implements the normalization condition in terms of a Dirac delta function, which can be written as a Fourier-Laplace transform of the complex Wishart-Laguerre ensemble~\cite{mehta2004, Forrester_2010},
\begin{align}
	\begin{split}
		\braket{S_A}&=-\frac{\int_{\mathbb{C}^{d_A\times d_B}} d[W] S_A(\ket{\psi}) \delta(1-\Tr WW^\dagger)}{\int_{\mathbb{C}^{d_A\times d_B}} d[W] \delta(1-\Tr WW^\dagger)}\\
		&=-\frac{\int_{\mathbb{C}^{d_A\times d_B}} d[W]\int_{-\infty}^\infty dt S_A(\ket{\psi}) e^{(1+\ii t)(1-\Tr WW^\dagger)}}{\int_{\mathbb{C}^{d_A\times d_B}} d[W]\int_{-\infty}^\infty dt e^{(1+\ii t)(1-\Tr WW^\dagger)}},
		\end{split}
\end{align}
where $d[W]$ is the product of all differentials of all matrix elements of $W$ and $W^\dagger$. The shift of $\ii t$ to $1+\ii t$ is important since it allows us to rescale $WW^\dagger\to WW^\dagger/(1+\ii t)$, which describes a rotation of the integration contours in the complex plane where the integrand is absolutely integrable. Before rescaling, we use a standard trick to rewrite the entanglement entropy removing the logarithm:
\begin{equation}
	S_A(\ket{\psi})=-\partial_\epsilon\Tr(WW^\dagger)^{\epsilon}|_{\epsilon=1}.
\end{equation}
We also use this relation when computing the average entanglement entropy of fermionic Gaussian states. After the aforementioned rescaling of the random matrix $WW^\dagger$, the entanglement entropy reads
\begin{align}\label{page.average.integrals}
	\begin{split}
		\braket{S_A}&=-\partial_\epsilon\biggl[\frac{\int_{-\infty}^\infty dt(1+\ii t)^{-d_Ad_B-\epsilon}e^{(1+\ii t)}}{\int_{-\infty}^\infty dt(1+\ii t)^{-d_Ad_B}e^{(1+\ii t)}}\\
		&\quad\times\frac{\int_{\mathbb{C}^{d_A\times d_B}} d[W]\Tr(WW^\dagger)^\epsilon e^{-\Tr WW^\dagger}}{\int_{\mathbb{C}^{d_A\times d_B}} d[W] e^{-\Tr WW^\dagger}}\biggl]_{\epsilon=1}.
		\end{split}
\end{align}
The first factor can be computed using standard techniques in complex analysis, yielding
\begin{align}
	\begin{split}
		\frac{\int_{-\infty}^\infty dt(1+\ii t)^{-d_Ad_B-\epsilon}e^{(1+\ii t)}}{\int_{-\infty}^\infty dt(1+\ii t)^{-d_Ad_B}e^{(1+\ii t)}}=\frac{\Gamma[d_Ad_B]}{\Gamma[d_Ad_B+\epsilon]},
	\end{split}
\end{align}
with $\Gamma[x]$ the gamma function. Once we apply the derivative in $\epsilon$, we see that the first term of Page's result~\eqref{Page} follows from the fixed trace condition. This was also observed and exploited in the original works in which Eq.~\eqref{Page} was computed~\cite{page1993average, sanchez1995simple}.
	
The remaining integral over $W$ is an average over the complex Wishart-Laguerre ensemble~\cite{mehta2004, Forrester_2010}, which is one of the three classical random matrix ensembles that also include the Gaussian~\cite{mehta2004, Forrester_2010,akemann2011} and the Jacobi~\cite{2000JPhA...33.2045Z, Forrester_2010} ensembles. Interestingly, it is the Jacobi ensemble that we encounter when studying fermionic Gaussian states later on.
	
In the final step, one uses the eigenvalues $x_1,\ldots,x_{d_A}\geq0$ of $WW^\dagger$ and expresses the average in terms of the level density of the Laguerre ensemble. In this step, one needs to decide whether $d_A\leq d_B$ or $d_A\geq d_B$, as the density is not analytic at $d_A=d_B$. This reflects the fact that either $\hat\rho_A$ or $\hat\rho_B$ has exact zero eigenvalues, and it is the source of the case distinction in Page's result~\eqref{Page}. When assuming that $d_A\leq d_B$, the level density is equal to~\cite{Forrester_2010}
\begin{align}
	\begin{split}
		R_{1,\rm Lag}(x)&=\frac{d_A!}{(d_B-1)!}x^{d_B-d_A}e^{-x}[L_{d_A-1}^{(d_B-d_A+1)}(x)\\
		&\quad\times L_{d_A-1}^{(d_B-d_A)}(x)-L_{d_A-2}^{(d_B-d_A+1)}(x)L_{d_A}^{(d_B-d_A)}(x)],
	\end{split}
\end{align}
in terms of the generalized Laguerre polynomials $L_n^{(\alpha)}(x)$. Using the series representation of $L_{a}^{(d_B-d_A+1)}(x)$ in Eq.~(18.5.12) of Ref.~\cite{NIST:DLMF}, and the Rodrigues formula for $L_{b}^{(d_B-d_A)}(x)$ in Eq.~(18.5.5) of Ref.~\cite{NIST:DLMF}, one can show for $a\leq b$ that
\begin{align}
	\begin{split}
		&\int_0^\infty x^{\epsilon+\alpha}e^{-x} L_{a}^{(d_B-d_A+1)}(x)L_{b}^{(d_B-d_A)}(x)dx\\
		&=\sum_{l=0}^{a}\frac{\Gamma[a+\alpha+2]\Gamma[\epsilon+l+1]\Gamma[\epsilon+\alpha+l+1]}{\Gamma[l+\alpha+2]\Gamma[\epsilon+l-n+1](a-l)!\,l!\,b!}(-1)^{l+b},
	\end{split}
\end{align}	
which is different from the formula used in Ref.~\cite{sanchez1995simple}. We use this approach as it parallels our computation for Gaussian states. Putting everything together in Eq.~\eqref{page.average.integrals}, using the identity
\begin{align}
	\begin{split}
		\frac{\int d[W]\Tr(WW^\dagger)^\epsilon e^{-\Tr WW^\dagger}}{\int d[W] e^{-\Tr WW^\dagger}}=d_A\int_0^\infty x^\epsilon R_{1,\rm Lag}(x)dx,
	\end{split}
\end{align}
we arrive at Page's result in Eq.~\eqref{Page}. As we have mentioned before, the symmetry in $d_A$ and $d_B$ reflecting the symmetry in the two subsystems $A\leftrightarrow B$ has to be implemented by hand. The random matrix approach underscores this loss of analyticity when going over to the generic nonzero eigenvalues of $WW^\dagger$ and selecting the smaller of the two dimensions.
	
\subsubsection{Average and variance}
	
The average entanglement entropy of a uniformly distributed pure state in $\mathcal{H}$ restricted to subsystem $A$ is given by the Page formula~\cite{page1993average}
\begin{equation}\label{Page}
	\hspace{-2mm}\braket{S_A}=
	\begin{cases}
		\Psi(d_A d_B\!+\!1)\!-\!\Psi(d_B\!+\!1)\!-\!\frac{d_A-1}{2d_B}& d_A\leq d_B  \\[0.5em]
		\Psi(d_A d_B\!+\!1)\!-\!\Psi(d_A\!+\!1)\!-\!\frac{d_B-1}{2d_A} & d_A> d_B
	\end{cases}
\end{equation}
where $\Psi(x)=\Gamma'(x)/\Gamma(x)$ is the digamma function. In the thermodynamic limit $V\to \infty$ when $V_A,V-V_A\to\infty$ also so that the subsystem fraction
\begin{equation}
	f=\frac{V_A}{V}
\end{equation}
is fixed, Page's formula~\eqref{Page} reduces to
\begin{equation}
	\braket{S_A}\!=\!
	f\,V\ln 2-2^{-|1-2f|V-1}+O(2^{-V})\,,
	\label{eq:Page-therm}
\end{equation}
where we will be careful to consistently use Landau's ``big $O$'' and ``little $o$'' notation in this manuscript, such that
\begin{align}
	f(V)&=O(V^n) & \Longleftrightarrow && \lim_{V\to\infty}\frac{f(V)}{V^n}&=c\neq 0\,,\\
	& & \text{and}&&\nonumber \\
	f(V)&=o(V^n) & \Longleftrightarrow &&\lim_{V\to\infty}\frac{f(V)}{V^n}&=0\,.
\end{align}
	
\begin{figure*}[!t]
	\centering  
	\includegraphics[width=\linewidth]{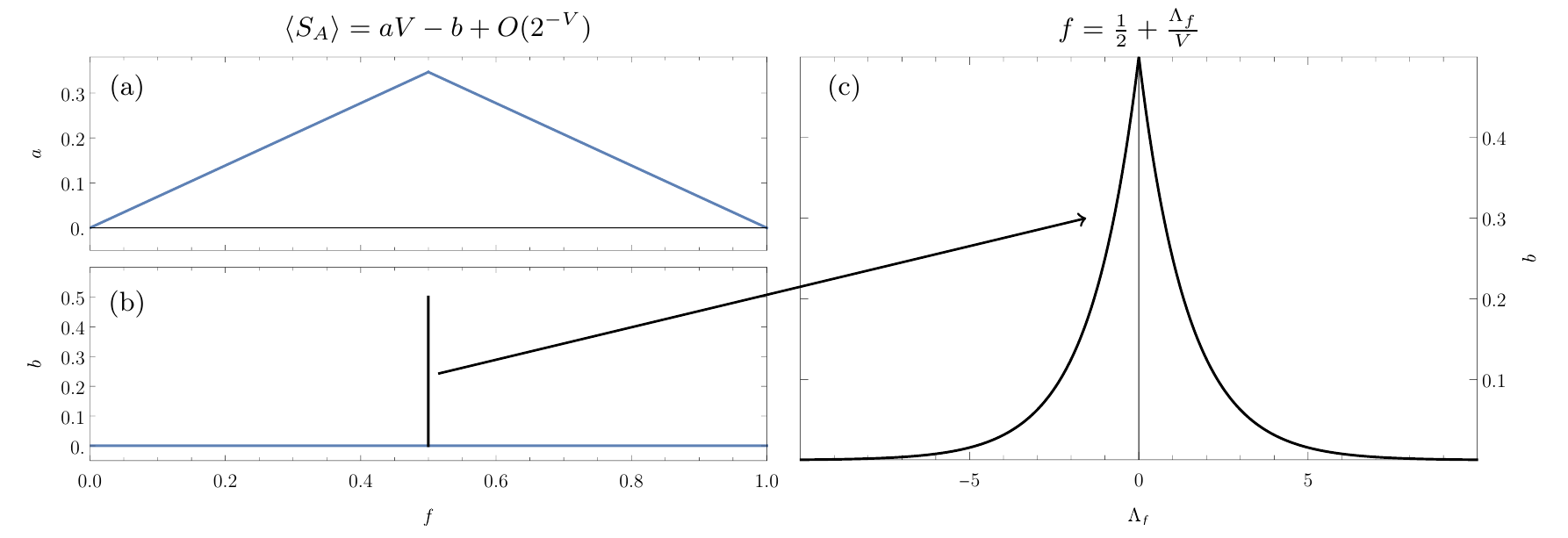}
	\caption{The average entanglement entropy $\braket{S_A}=a V-b+o(1)$ as a function of the subsystem fraction $f=V_A/V$ for large $V$. (a) Leading-order behavior, also known as the Page curve. (b) The constant correction, which is given by a Kronecker delta $-\frac{1}{2}\delta_{f,\frac{1}{2}}$. This Kronecker delta is resolved in (c) by carrying out a double scaling limit $V\to\infty$ with $f=\frac{V_A}{V}=\frac{1}{2}+\frac{\Lambda_f}{V}$.}
	\label{fig:Page-discon}
\end{figure*}

The first term in Eq.~\eqref{eq:Page-therm} is a volume law: the average entanglement entropy scales as the minimum between the volumes $V_A=f V$ and $V_B=(1-f)V$. For $f\neq \frac{1}{2}$, the second term is an exponentially small correction. In fact, at fixed $f$ and in the limit $V\to\infty$, the second term $-2^{-|1-2f|V-1}$ becomes $-\frac{1}{2}\delta_{f,\frac{1}{2}}$. We can also resolve precisely how this Kronecker delta arises in the neighborhood of $f=\frac{1}{2}$. As it may be difficult to reach exactly $f=\frac{1}{2}$ in physical experiments, the more precise statement is that we see the correction whenever $f=\frac{1}{2}+O(1/V)$. Formally, we can thus resolve the correction term exactly as $-2^{-|\Lambda_f|-1}$ for $f=\frac{1}{2}+\Lambda_f/V$, as visualized in Fig.~\ref{fig:Page-discon}.
	
We find similar Kronecker delta contributions $\delta_{f,\frac{1}{2}}$ in subsequent sections where we discuss the typical entropy at fixed particle number and in the setting of Gaussian states. These terms highlight nonanalyticities in the entanglement entropy that can be resolved by double scaling limits. Those ``critical points'' occur at symmetry points and along axes. In the present case, this has happened with the dimensions $d_A$ and $d_B$ reflecting whether the density operator $\hat\rho_A=WW^\dagger$ or $\hat\rho_B=W^\dagger W$ contains generic zero eigenvalues. 
	
The variance of the entanglement entropy of a random pure state is given by the exact formula (for $d_A\leq d_B$) \cite{vivo_pato_16,wei2017proof,bianchi2019typical}
\begin{align}
	(\Delta S_A)^2=&\; \textstyle \frac{d_A+d_B}{d_A d_B+1}\Psi'(d_B+1)-\Psi'(d_A d_B+1)\nonumber\\[.5em]
	&\textstyle-\frac{(d_A-1)(d_A+2d_B-1)}{4d_B^2(d_A d_B+1)}\,,
	\label{eq:PageDeltaS}
\end{align}
where $\Psi'(x)=\frac{d\Psi(x)}{dx}=\frac{d^2[\ln{\Gamma(x)}]}{dx^2}$ is the first derivative of the digamma function. It can be derived using similar techniques as those outlined above for the average. In particular, the fixed trace condition can be separated as before via the trick of the Fourier-Laplace transform, such that one is left with an average over the complex Wishart-Laguerre ensemble. The derivation is tedious and lengthy because one has to deal with double sums, which can be computed as described in Appendix~\ref{app:Gaussfixednumber}.\footnote{Our computation of the variance for Gaussian states at fixed particle number presented in Appendix~\ref{app:Gaussfixednumber} shows how to deal with the double sums, and can also be used in the general setting. Basically, one needs to replace the Jacobi polynomials and their corresponding weight by the Laguerre polynomials and the weight function $x^{d_B-d_A}e^{-x}$.}
	
In the thermodynamic limit discussed above, Eq.~\eqref{eq:PageDeltaS} reduces to
	\begin{equation}
		(\Delta S_A)^2=
		\big(\tfrac{1}{2}-\tfrac{1}{4}\delta_{f,\frac{1}{2}}\big)\;2^{-(1+|1-2f|) V}\,+\,o(2^{-(1+|1-2f|) V}). 
		\label{eq:variance-page}
	\end{equation}
This shows that the variance is exponentially small in $V$. As a result, in the thermodynamic limit the entanglement entropy of a typical state is given by Eq.~\eqref{eq:Page-therm} \cite{bianchi2019typical}.
	
Anew, one could resolve the variance at the critical point $f=\frac{1}{2}$ via a double scaling limit $f=\frac{1}{2}+\Lambda_f/V$. This yields $(\Delta S_A)^2=2^{-V}2^{-2|\Lambda_f|-1}(1-2^{-2|\Lambda_f|-1})$.
	
\subsection{Fixed number of particles}\label{sec:page-fixedN}
	
Let us go over to a Hilbert space $\mathcal{H}^{(N)}$ with a fixed number of particles, but still carrying over the idea to draw states uniformly from the sphere in this Hilbert space. We further assume that there is a notion of a bipartition into subsystem $A$ and $B$, such that one can specify for each particle if it is in subsystem $A$ or $B$. Such a decomposition is not a simple tensor product anymore, but it is a direct sum of tensor products
\begin{align}\label{eq:Hspace-decomposition}
	\mathcal{H}^{(N)}=\bigoplus^{N}_{N_A=0}\Big(\mathcal{H}_A^{(N_A)}\otimes\mathcal{H}_B^{(N-N_A)}\Big)\,.
\end{align}
The direct sum is over the occupation number in $A$ (which labels the center of the subalgebra). Each summand represents those states where $N_A$ particles are in subsystem $A$ and $N-N_A$ particles are in subsystem $B$ (assuming indistinguishable particles).
	
When $N_A$ is larger than dimension $V_A$ of subsystem $A$, or $N-N_A$ is larger than $V-V_A$, we consider the tensor product $\mathcal{H}_A^{(N_A)}\otimes\mathcal{H}_B^{(N-N_A)}$ as the empty set and, thence, nonexistent. This is the case as, due to Pauli's exclusion principle, we cannot put more fermions in the system than there are quantum states. We also adapt this understanding for the following discussion where direct sums, ordinary sums, and products are reduced to the components that are actually present.
	
\subsubsection{Statistical ensemble of states}
	
Let us consider fermionic creation $\hat{f}_i^\dagger$ and annihilation $\hat{f}^{}_i$ operators, which satisfy the anticommutation relations $\{\hat{f}^{}_i,\hat{f}_j^\dagger\}=\delta_{ij}$, $\{\hat{f}_i,\hat{f}_j\}=0$ with $i,j=1,\ldots,V$. The corresponding number operators are
\begin{equation}
	\hat{N}=\sum_{i=1}^V\hat{f}_i^\dagger \hat{f}^{}_i\,,\quad \hat{N}_A=\sum_{i=1}^{V_A}\hat{f}_i^\dagger \hat{f}^{}_i\,,\quad \hat{N}_B=\sum_{i=V_A+1}^{V}\hat{f}_i^\dagger \hat{f}^{}_i\,,
	\label{eq:Hilbert-sum}
\end{equation}
where one can see that
\begin{equation}
	\hat{N}=\hat{N}_A+\hat{N}_B\,.
\end{equation}
The Hilbert space of the system can be decomposed as a direct sum of Hilbert spaces at fixed eigenvalue $N$ of $\hat{N}$,
\begin{equation}
	\mathcal{H}=\bigotimes_{i=1}^V\mathcal{H}_i\;=\;\bigoplus_{N=0}^{V}\,\mathcal{H}^{(N)},
\end{equation}
with $\mathcal{H}^{(N)}$ given by Eq.~\eqref{eq:Hspace-decomposition}. The dimension of each $N$-particle sector is
\begin{equation}
	d_N=\dim\mathcal{H}^{(N)}\,=\,\frac{V!}{N!\,(V-N)!}\,.
	\label{eq:dN}
\end{equation}
It is immediate to check that $\dim \mathcal{H}=\sum_{N=0}^V d_N=2^V$. Similarly, one can use the number operators $\hat{N}_A$ and $\hat{N}_B$ to decompose the Hilbert spaces $\mathcal{H}_A$ and $\mathcal{H}_B$ into sectors
\begin{equation}
	\mathcal{H}_A=\;\bigoplus_{N_A=0}^{V_A}\,\mathcal{H}_A^{(N_A)}\,,\qquad \mathcal{H}_B=\;\bigoplus_{N_B=0}^{V-V_A}\,\mathcal{H}_B^{(N_B)}\,.
\end{equation}
Let us stress once again that, while $\mathcal{H}$ is a tensor product over $A$ and $B$,
\begin{align}
	\mathcal{H}=\left(\bigotimes_{i=1}^{V_A}\mathcal{H}_i\right)\otimes\left(\bigotimes_{i=V_A+1}^{V}\mathcal{H}_i\right)\;=\;\mathcal{H}_A\otimes \mathcal{H}_B\,,
\end{align}
the sector at fixed number $N\leq V_A$ is not a tensor product. It is the direct sum of tensor products from Eq.~\eqref{eq:Hspace-decomposition}. The corresponding dimensions of the subsystems are
\begin{align}
	\begin{split}\label{eq:dAdB}
		& d_A(N_A)=\dim\mathcal{H}_A^{(N_A)}\,=\,\frac{V_A!}{N_A!\,(V_A-N_A)!}\,,\\
		& d_B(N_B)=\dim\mathcal{H}_B^{(N_B)}\,=\,\frac{(V-V_A)!}{N_B!\,((V-V_A)-N_B)!}\,.
	\end{split}
\end{align}
One can check that the dimensions add up correctly,
\begin{equation}
	\sum_{N_A=0}^N d_A(N_A)\, d_B(N-N_A)=\frac{V!}{N!(V-N)!}=d_N\,.\label{eq:normalization-varrho}
\end{equation}
The formula for $d_A$, and equivalently that of $d_B$, follows from a simple counting argument of how many choices there are to place $N_A$ indistinguishable particles on $V_A$ modes. Let us underline that it does not matter what we label particles and what holes. Note that $d_A(N_A)$ or $d_B(N-N_A)$ will vanish for $N_A$ outside of the interval $[\max(0,N+V_A-V),\min(N,V_A)]$, but we will not truncate the sum, as we will soon turn it into a Gaussian integral.
	
From these dimensions we can readily read off two exact symmetries:
	
\noindent (i) It does not matter whether one considers subsystem $A$ or $B$. One can exchange $(d_A(N_A),V_A,N_A) \leftrightarrow (d_B(N-N_A),V-V_A,N-N_A)$. This allows us to restrict the discussion to $V_A\leq V/2$. However, the dimensions of the two Hilbert spaces are exchanged, which (as we will show) yields nonanalytic points along $V_A=V/2$ due to the two branches of Page curve~\eqref{Page}.
	
\noindent (ii) Additionally, there is a particle-hole symmetry since it does not matter whether one counts particles or holes. Actually, the ``particles'' do not necessarily need to represent particles but they can be, for instance, up spins while the ``holes'' are down spins (having in mind spin-$\frac{1}{2}$ systems). Any binary structure with fermion statistics (meaning Pauli principle) can be described in this setting.  Mathematically, the particle-hole symmetry is reflected in the exchange $(N,N_A)\leftrightarrow(V-N,V_A-N_A)$. We note that in this case the dimensions are not exchanged so one does not switch branches in Page curve~\eqref{Page}. Therefore, the symmetry points at $N=V/2$ will be analytic, as we will also show. This symmetry allows us to restrict $N\leq V/2$.
	
\noindent In summary, we only need to study the behavior in the quadrant $(V_A,N)\in(0,\frac{V}{2}]^2$. The remaining quadrants are obtained by symmetry.
	
\begin{figure}[t!]
	\centering
	\includegraphics[width=\linewidth]{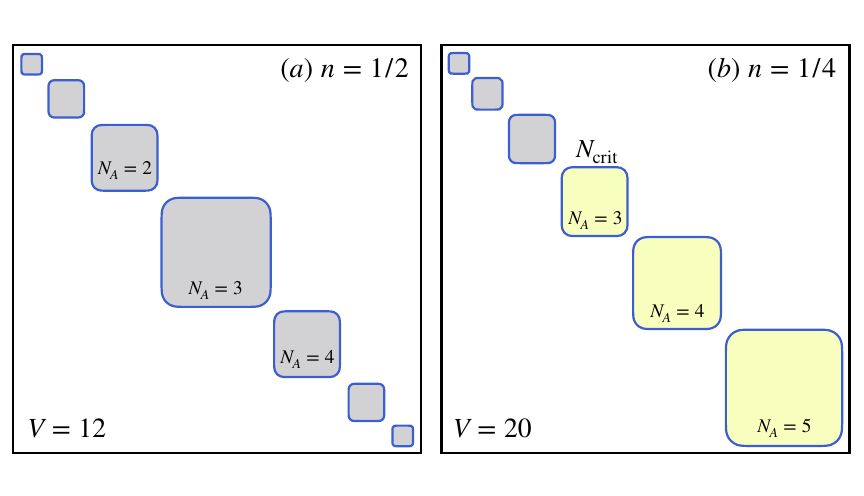}
	\caption{Sketch of the block dimensions of the reduced density matrix $\hat\rho_A$ of subsystem $A$ at the subsystem fraction $f=\frac{1}{2}$. (a) Case $V=12$ at half filling $n=\frac{1}{2}$, for which $V_A = 6$. The number of particles ranges from $N_A=0$ to $N_A=6$, with $N_A=3$ representing the largest block. (b) Case $V=20$ at quarter filling $n=\frac{1}{4}$, for which $V_A = 10$. The number of particles ranges from $N_A=0$ to $N_A=5$, with $N_A=5$ representing the largest block. The blocks with $N_A \geq N_{\rm crit} = 3$ are larger than the corresponding blocks in subsystem $B$ (not shown in the figure).}
	\label{fig:RDMsketch}
\end{figure}

Like in the setting in which we do not fix the particle number, we can relate the problem to random matrix theory. Here, we briefly recall the most important ingredients from Ref.~\cite{bianchi2019typical}. A state $\ket{\psi}\in\mathcal{H}^{(N)}$ can be again written in a basis. We choose the orthonormal basis vectors $\ket{a,N_A} \otimes \ket{b,N-N_A} \in \mathcal{H}_A^{(N_A)} \otimes \mathcal{H}_B^{(N-N_A)}$ so that the state vector has the expansion
\begin{equation}
	\ket{\psi}=\bigoplus_{N_A=0}^N \sum_{a=1}^{d_A}\sum_{b=1}^{d_B} \tilde{w}_{ab}^{(N_A)}\ket{a,N_A}\otimes\ket{b,N-N_A}
\end{equation}
with the abbreviations $d_A=d_A(N_A)$ and $d_B=d_B(N-N_A)$. The normalization is then reflected by the triple sum
\begin{equation}\label{norm.Page.fixed}
	\sum_{N_A=0}^N\sum_{a=1}^{d_A}\sum_{b=1}^{d_B}|\tilde{w}_{ab}^{(N_A)}|^2=1.
\end{equation}
The direct sum over $N_A$ is important as it tells us that the density operator $\hat\rho_A=\Tr_{\mathcal{H}_B}\ket{\psi}\bra{\psi}$ has a block diagonal form, namely,
\begin{equation}
	\hat\rho_A=\bigoplus_{N_A=0}^N \sum_{a_1,a_2=1}^{d_A} \sum_{b=0}^{d_B}\tilde{w}_{a_1b}^{(N_A)}(\tilde{w}_{a_2b}^{(N_A)})^*\ket{b,N_A}\bra{a_2,N_A}.
\end{equation}
Again, we can understand the coefficients $\tilde{w}_{ab}^{(N_A)}\in\mathbb{C}$ as the entries of a $d_A\times d_B$ matrix $\tilde{W}_{N_A}$. The point is that those matrices are coupled by condition~\eqref{norm.Page.fixed}. In Ref.~\cite{bianchi2019typical} those matrices were decoupled by understanding their squared Hilbert-Schmidt norms as probability weights, \ie defining
\begin{equation}
	p_{N_A}=\sum_{a=1}^{d_A}\sum_{b=1}^{d_B}|\tilde{w}_{ab}^{(N_A)}|^2\in[0,1]
\end{equation}
such that $\tilde{W}_{N_A}=\sqrt{p_{N_A}}\,W_{N_A}$. This notation allows one to identify the density operator of subsystem $A$ with the block diagonal matrix $\hat\rho_A = \mathrm{diag} (p_0W_0W_0^\dagger, \ldots, p_N W_NW_N^\dagger)$, as illustrated in Fig.~\ref{fig:RDMsketch}.
	
Thus, the entanglement entropy becomes the sum
\begin{align}
	S_A(\ket{\psi})=\sum_{N_A=0}^N&\Big[p_{N_A}\Tr(W_{N_A}W_{N_A}^\dagger\ln[W_{N_A}W_{N_A}^\dagger])\nonumber\\
	&+p_{N_A}\ln(p_{N_A})\Big].
	\label{page.ententro.fixed}
\end{align}
Anew, the symmetry between the two subsystems is reflected by the spectral decomposition theorem since it holds that $\hat\rho_B = \Tr_{\mathcal{H}_A}\ket{\psi}\bra{\psi} = \mathrm{diag}(p_0W_0^\dagger W_0,\ldots,p_NW_N^\dagger W_N)$.
	
Since the norms are encoded in the probability weights $p_{N_A}$, each matrix $W_{N_A}W_{N_A}^\dagger$ independently describes a fixed trace ensemble, \ie $\Tr W_{N_A}W_{N_A}^\dagger=1$. Thus, it can be dealt with in the same way as in Page's case, in particular each of those can be traced back to a complex Wishart-Laguerre ensemble of matrix dimension $d_A\times d_B$. The probability weights $p_{N_A}\in[0,1]$ are also drawn randomly via the joint probability distribution~\cite{bianchi2019typical}
\begin{equation}
	\frac{\delta\left(1-\sum_{N_A=0}^{N}p_{N_A}\right)\prod_{N_A=0}^Np_{N_A}^{d_Ad_B-1}dp_{N_A}}{\int\delta\left(1-\sum_{N_A=0}^{N}p_{N_A}\right)\prod_{N_A=0}^Np_{N_A}^{d_Ad_B-1}dp_{N_A}}.
\end{equation}
The Dirac delta function enforces condition~\eqref{norm.Page.fixed}, while the factors $p_{N_A}^{d_Ad_B-1}$ are the Jacobians for the polar decomposition of the vectors in $\mathcal{H}_A^{(N_A)} \otimes \mathcal{H}_B^{(N-N_A)}$ into their squared norm $p_{N_A}$ and the direction, which is encoded in $W_{N_A}$. The normalization of the distribution of $p_{N_A}$ was computed in Ref.~\cite{bianchi2019typical} and can be deduced by inductively tracing the integrals over $p_{N_A}$ back to Euler's beta integrals in Eq.~(5.12.1) of Ref.~\cite{NIST:DLMF}.
	
\subsubsection{Average and variance}
	
\begin{figure*}[t!]
	\centering  
	\includegraphics[width=\linewidth]{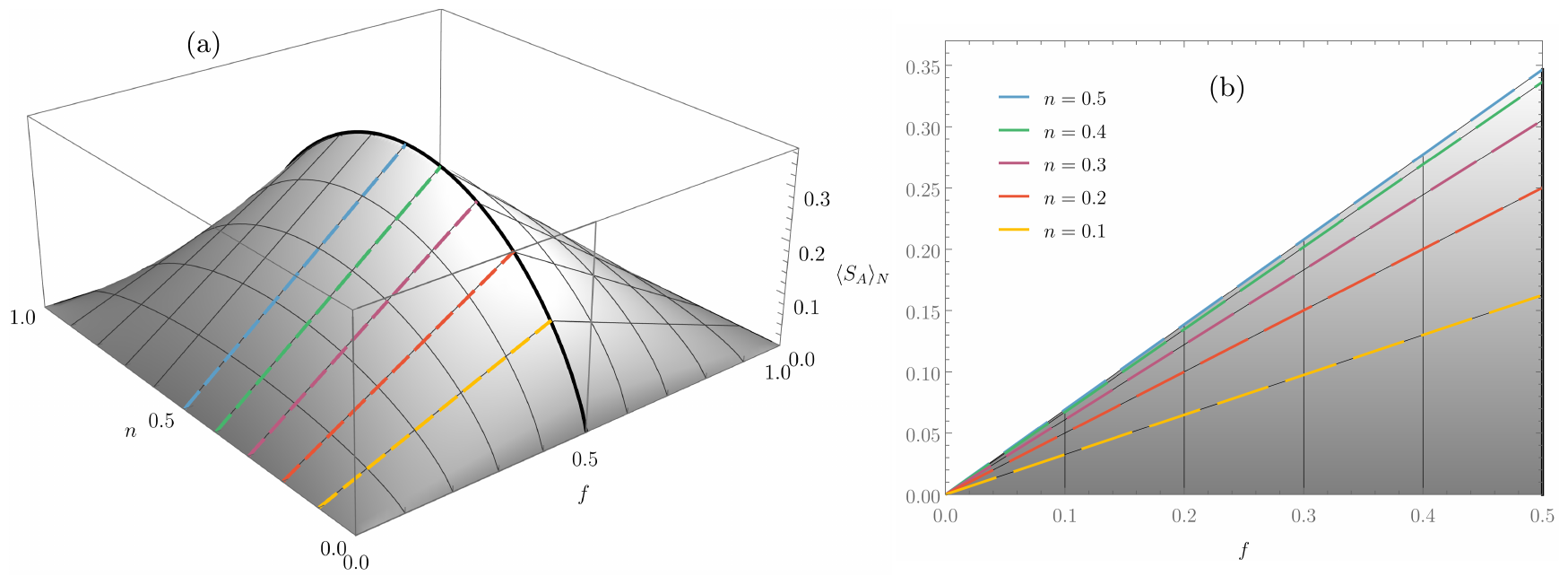}
	\caption{The leading entanglement entropy $s_A(f,n) = \lim_{V\to\infty} \braket{S_A}_N/V$ from Eq.~\eqref{eq:leading-general} [see Eq.~\eqref{eq:sA-useful}]. For $n=\frac{1}{2}$, $s_A(f,n)$ coincides with Page's result (maximal entanglement). (a) Three-dimensional plot as a function of the subsystem fraction $f=V_A/V$ and the filling ratio $n=N/V$. One can see the mirror symmetries $V_A\to V-V_A$ and $N\to V-N$. (b) Results at fixed $n$ plotted as functions of $f$. The colored lines agree in both plots so that the right one can be seen as sections of the left one along the colored lines.}
	\label{fig:Page}
\end{figure*}
	
With these definitions and discussions, we are now ready to state the main result in Eq.~(23) of Ref.~\cite{bianchi2019typical}: the average entanglement entropy in system $A$ of a uniformly distributed random state in $\mathcal{H}^{(N)}$ is given by
\begin{align}\label{eq:Scenter}
	\begin{split}
		\hspace{-2mm}\braket{S_A}_N\!&=\!\!\!\!\sum^{\min(N,V_A)}_{N_A=0}\! \frac{d_Ad_B}{d_N}\big(\braket{S_A}\!+\!\Psi(d_N\!+\!1)\!-\!\Psi(d_Ad_B\!+\!1)\big),
	\end{split}
\end{align}
where $d_A=d_A(N_A)$ and $d_B=d_B(N-N_A)$ depend on $N_A$ according to Eq.~\eqref{eq:dAdB} and $\braket{S_A}$ refers to Page's result~\eqref{Page} for given $d_A$ and $d_B$. Equation~\eqref{eq:Scenter} follows from the average over $W_{N_A}W_{N_A}^\dagger$ in Eq.~\eqref{page.ententro.fixed}, which are independent fixed trace random matrices. The prefactor $d_Ad_B/d_N$, as well as the additional digamma functions, follow from Euler's beta integral in Eq.~(5.12.1) of Ref.~\cite{NIST:DLMF}. In particular, we have used
\begin{align}
	\begin{split}
		\langle p_{N_A}^\epsilon\rangle&=\frac{\int_0^1 p_{N_A}^{\epsilon+d_Ad_B-1}(1-p_{N_A})^{d_N-d_Ad_B-1}dp_{N_A}}{\int_0^1 p_{N_A}^{d_Ad_B-1}(1-p_{N_A})^{d_N-d_Ad_B-1}dp_{N_A}}\\
		&=\frac{\Gamma[\epsilon+d_Ad_B]\Gamma[d_N]}{\Gamma[d_Ad_B]\Gamma[\epsilon+d_N]}
	\end{split}
\end{align}
for any $\epsilon>-d_Ad_B$. The average on the right-hand side can be obtained by rescaling $p_j\to (1-p_{N_A})p_j$ for any $j\neq N_A$, which decouples the average over $p_{N_A}$ with the remaining probability weights $p_j$.
	
We can write Eq.~\eqref{eq:Scenter} as
\begin{equation}
	\braket{S_A}_N=\sum^N_{N_A=0}\varrho_{N_A}\varphi_{N_A},
\end{equation}
by introducing the quantities
\begin{align}
	\begin{split}\label{eq:varphi}
		&\varrho_{N_A}=\frac{d_Ad_B}{d_N}\,,\\
		&\varphi_{N_A}\!=\!
		\begin{cases}
			\Psi(d_N\!+\!1)\!-\!\Psi(d_B\!+\!1)\!-\!\frac{d_A-1}{2d_B}&\quad d_A\leq d_B  \\[0.5em]
			\Psi(d_N\!+\!1)\!-\!\Psi(d_A\!+\!1)\!-\!\frac{d_B-1}{2d_A} &\quad d_A> d_B
		\end{cases}\\
		&=\scriptsize\Psi(d_N\!+\!1)\!-\!\Psi(\max(d_A,d_B)\!+\!1)\!-\!\min\left(\tfrac{d_A-1}{2d_B},\tfrac{d_B-1}{2d_A}\right).
	\end{split}
\end{align}
The function $\varrho_{N_A}$ can be understood as a probability distribution of having $N_A$ particles in $A$, with the normalization $\sum_{N_A}\varrho(N_A)=1$ following from Eq.~\eqref{eq:normalization-varrho}. The function $\varphi_{N_A}$, when understood as a continuous function, has a kink at $N_{\mathrm{crit}}$, which refers to the largest integer such that $d_A(N_\mathrm{crit})\leq d_B(N-N_{\mathrm{crit}})$. There is only one situation in which $N_{\rm crit}$ is not well defined, namely, when $V_A=N=V/2$ or, equivalently, when $f=n=\frac{1}{2}$ with $f=V_A/V$ and $n=N/V$. Then it always holds that $d_A(N_A)=d_B(N-N_A)$ for all $N_A=0,\ldots,N$. In this case, we do not need an $N_{\rm crit}$ as the terms in both sums are the same.
	
We are unable to evaluate this sum exactly, but we can expand $\braket{S_A}_N$ in powers of $V$ and approximate the sum by an integral
\begin{align}\label{eq:average-int}
	\hspace{-2mm}\braket{S_A}_N\!=\!\!\sum^N_{N_A=0}\!\!\varrho_{N_A}\varphi_{N_A}\!=\!\int^{\infty}_{-\infty} \!\!\!\!\!\!\!\varrho(n_A)\varphi(n_A)dn_A\!+\!o(1),
\end{align}
where $\varrho(n_A)$ is the saddle point approximation of $V\varrho_{n_AV}=Vd_Ad_B/d_N$, which represents the probability distribution for the intensive variable $n_A=N_A/V$. This is enough for computing the leading orders without double scaling. We find the normal distribution
\begin{align}\label{Gauss.approx.Page}
	\varrho(n_A)=\frac{1}{\sigma \sqrt{2\pi}}\exp\left[-\frac{1}{2}\left(\frac{n_A-\bar{n}_A}{\sigma}\right)^2\right]+o(1)
\end{align}
with mean $\bar{n}_A=fn$ and variance $\sigma^2=f(1-f)n(1-n)/V$.
	
\begin{figure*}
	\centering  
	\includegraphics[width=\linewidth]{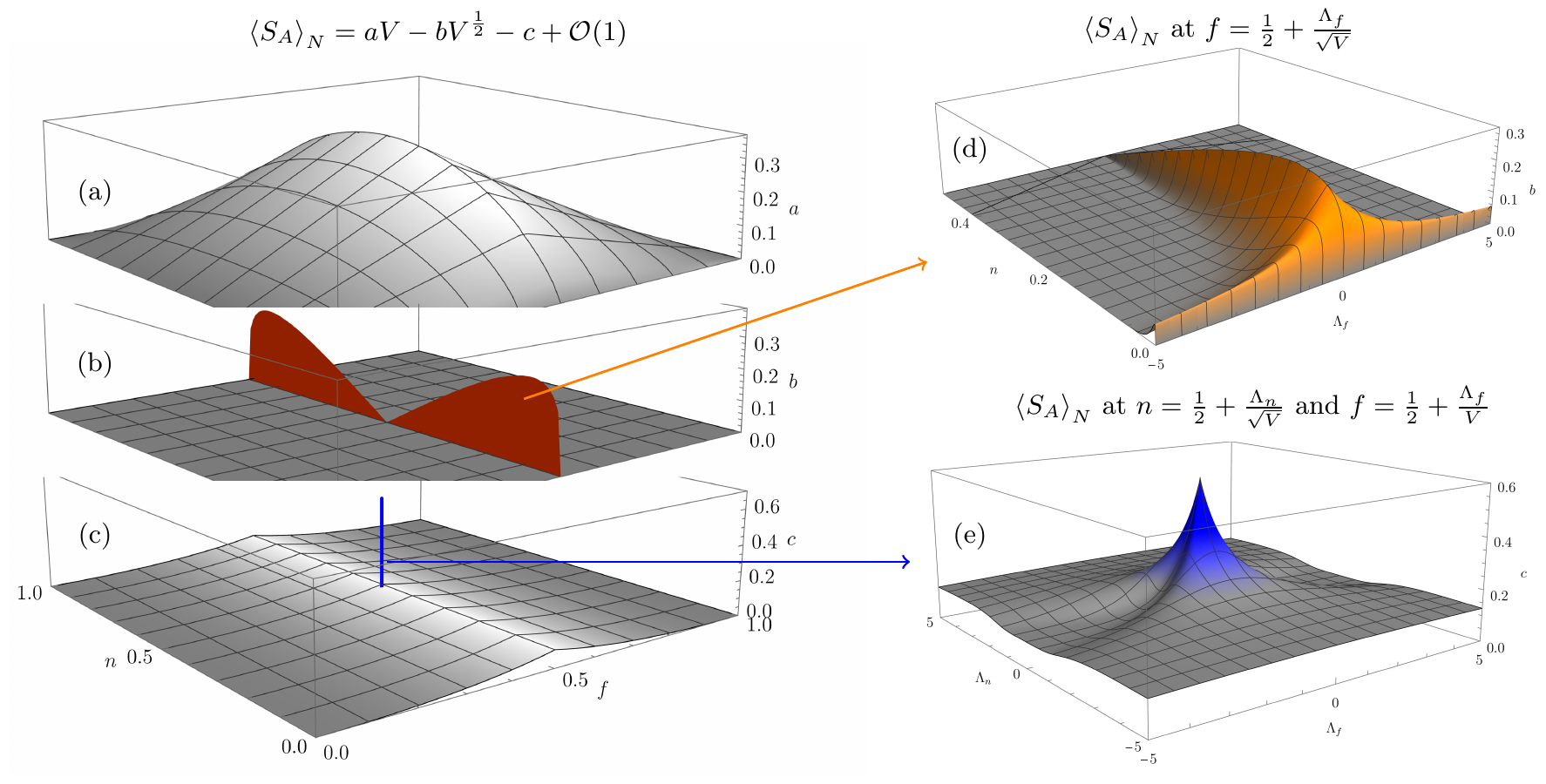}
	\caption{The entanglement entropy $\braket{S_A}_N$ from Eq.~\eqref{eq:leading-general} as viewed from the contributions of the first three terms in the expansion in $V$. (a)--(c) Three-dimensional plots as functions of the subsystem fraction $f=V_A/V$ and the filling ratio $n=N/V$. (d) Resolving the expansion coefficient $b$ for $f=\frac{1}{2}+\frac{\Lambda_f}{\sqrt{V}}$ around $f=\frac{1}{2}$, as given by Eq.~\eqref{eq:squareroot-full-Gaussian-half-system}, approaching zero for large $|\Lambda_f|$. (e) Resolving the expansion coefficient $c$ for $n=\frac{1}{2}+\frac{\Lambda_{n}}{\sqrt{V}}$ and $f=\frac{1}{2}+\frac{\Lambda_f}{V}$ around $f=n=\frac{1}{2}$, as given by Eq.~\eqref{eq:constant-full-Gaussian-half}, approaching $\frac{2\ln2-1}{4}$ for large $|\Lambda_f|$ or $|\Lambda_{n}|$. We underline that the subleading contributions are multiplied by a minus sign.}
	\label{fig:general-N-visual}
\end{figure*}
	
In Appendix~\ref{app:Ncrit}, we carefully analyze the difference $\delta n_{\mathrm{crit}}=n_{\mathrm{crit}}-\bar{n}_A$ for $n_{\mathrm{crit}}=N_{\mathrm{crit}}/V$ and find that, for fixed $f<\frac{1}{2}$, one always has $\delta n_{\mathrm{crit}}=O(1)$ and $\delta n_{\mathrm{crit}}>0$. Thus, for $f\neq\frac{1}{2}$, the center of the Gaussian $\bar{n}_A$ is sufficiently separated from $n_{\mathrm{crit}}$. This allows us to disregard the second sum in Eq.~\eqref{eq:Scenter} as it is exponentially suppressed. In the case that $f>\frac{1}{2}$, we can disregard the first sum because of the symmetry between the two subsystems $A$ and $B$.
	
To find the observable $\varphi(n_A)$ from Eq.~\eqref{eq:varphi}, we use Stirling's approximation
\begin{align}\label{approx.Digamma}
	\Psi[d_N\!+\!1]\!-\!\Psi[\max(d_A,d_B)\!+\!1]&=\ln\min\left(\tfrac{d_N}{d_B},\tfrac{d_N}{d_A}\right)\!+\!o(1).
\end{align}
Moreover, it holds for $V\gg1$ and fixed $f\in(0,1)$
\begin{align}
	\min\left(\tfrac{d_A-1}{d_B},\tfrac{d_B-1}{d_A}\right)&=\delta_{f,\frac{1}{2}}\delta_{n,\frac{1}{2}}+o(1).
\end{align}
The Kronecker-delta is, in fact, a ``relic'' of a double scaling limit, see Figs.~\ref{fig:Page-discon}(b) and~\ref{fig:Page-discon}(c) for a similar result in the context of Page's setting without fixed particle number. It can be resolved assuming that $f$ is close to $1/2$ but not exactly at $1/2$, see Appendix~\ref{app:general-average}. When collecting all terms up to order $O(1)$, we obtain
\begin{align}\label{eq:psi}
	\begin{split}
		\varphi(n_A)&=[n_A\ln(n_A)-f\ln(f)-n\ln[(1-n)/n]\\
		&\quad-\ln(1-n)+(f-n_A)\ln(f-n_A)]V\\
		&\quad+ \frac{1}{2}\ln\left[\frac{n_A (f-n_A)}{f(1-n)n}\right]-\frac{1}{2}\delta_{f,\frac{1}{2}}\delta_{n,\frac{1}{2}}+o(1)\,,
	\end{split}
\end{align}
for $n_A\geq n_{\rm crit}$. For $n_A\leq n_{\rm crit}$, we need to apply the symmetries $n_A\to n-n_A$ and $f\to 1-f$ in expansion~\eqref{eq:psi}.
	
In the limit $V\to\infty$, Gaussian~\eqref{Gauss.approx.Page} narrows because the standard deviation scales like $\sigma\sim1/\sqrt{V}$. We can, therefore, expand $\varphi(n_A)$ in powers of $(n_A-\bar{n}_A)$ around the mean $\bar{n}_A$. In order to find the average up to a constant order, it suffices to expand up to the quadratic order and then calculate integral~\eqref{eq:average-int}. Only for $f=\frac{1}{2}$, we have $\delta n_{\mathrm{crit}}=o(1)$, so that we need to take into account the nonanalyticity in $\varphi(n_A)$ introduced by the symmetry when exchanging the two subsystems. In this case, we integrate two different Taylor expansions for $n_A\leq n/2$ and $n_A\geq n/2$, which will introduce a term of order $\sqrt{V}$, as discussed below.
	
Combining these results, we arrive at the main result of this subsection,
\begin{align}\label{eq:leading-general}
	\langle S_A\rangle_{N}&=[(n-1)\ln(1-n)-n\ln(n)]\, f\,V\nonumber\\
	&-\sqrt{\frac{n(1-n)}{2\pi}}\left|\ln\left(\frac{1-n}{n}\right)\right|\delta_{f,\frac{1}{2}}\sqrt{V}\nonumber\\
	&+\frac{f+\ln(1-f)}{2}-\frac{1}{2}\delta_{f,\frac{1}{2}}\delta_{n,\frac{1}{2}}+o(1),
\end{align}
valid for $f\leq \frac{1}{2}$. The leading, volume-law, term in Eq.~\eqref{eq:leading-general} is the same as that obtained in Refs.~\cite{garrisson_grover_18, vidmar2017entanglement2} using random matrix theory, and the same as in Ref.~\cite{bianchi2019typical} [see Eq.~(27)], where it is interpreted as the typical entanglement entropy in the (highly degenerate) eigenspace of a Hamiltonian of the form $\hat{H}=\hat{N}=\hat{N}_A+\hat{N}_B$. The subleading $\sqrt{V}$ term was first discussed in Ref.~\cite{vidmar2017entanglement2}, specifically, it coincides with the bound for such a term computed at $f=\frac{1}{2}$~\cite{vidmar2017entanglement2}. It is remarkable that, for $n\neq\frac{1}{2}$, the constant term is nothing but that obtained in Ref.~\cite{vidmar2017entanglement2} within a ``mean field'' calculation, while at $n=f=\frac{1}{2}$ the extra $-\frac{1}{2}$ correction was found in Ref.~\cite{vidmar2017entanglement2} numerically, both for random states as well as for eigenstates of a nonintegrable Hamiltonian. We had all the ingredients to guess the general form in Eq.~\eqref{eq:leading-general}. Its actual derivation with all the details fills several pages, and can be found in Appendix~\ref{app:general-average}. A visualization of the leading term in Eq.~\eqref{eq:leading-general} can be found in Fig.~\ref{fig:Page}.
	
An important question concerns the resolution of the Kronecker deltas in Eq.~\eqref{eq:leading-general}, which indicate nontrivial scaling limits. The Kronecker deltas are only obtained along the critical line $f=\frac{1}{2}$, which contains a multicritical point at $n=\frac{1}{2}$ when $V\to\infty$. One needs to take the resolution into account because experiments are carried out in finite systems in which $f$ and $n$ can only be fixed within some experimental resolution. Consequently, it is important to understand within which margin of error one needs to choose $f$ and $n$ to observe the corresponding terms. This question can be answered by analyzing the limit $V\to\infty$ in the double scaling $f=\frac{1}{2}+V^{-\alpha} \Lambda_f$ and/or $n=\frac{1}{2}+V^{-\beta}\Lambda_{n}$. We find that the $\sqrt{V}$ correction in Eq.~\eqref{eq:leading-general} (for fixed $n$) becomes visible for $\alpha=\frac{1}{2}$, \ie whenever the difference between $f$ and $\frac{1}{2}$ is of order $1/\sqrt{V}$ or smaller. The constant correction requires a more detailed analysis as it depends on the relative scaling of both $f$ and $n$ around $f=n=\frac{1}{2}$. Subtle cancelations have to be taken into account as not all sources of corrections, such as $N_{\rm crit}$, approximation~\eqref{approx.Digamma}, or the rewriting of the sum as an integral, are equally important; see Appendix~\ref{app:general-average}. The visualization of the terms in Eq.~\eqref{eq:leading-general} that include Kronecker deltas, as well as their scaling, is presented in Fig.~\ref{fig:general-N-visual}.
	
The variance $(\Delta S_A)^2_{N}={\langle S_A^2\rangle}_{N}-{\langle S_A\rangle}^2_{N}$ of the entanglement entropy of pure quantum states in $\mathcal{H}^{(N)}$ can be found using the result in Eq.~(24) of Ref.~\cite{bianchi2019typical}. When expressed as a sum over the number of particles $N_A$, it takes the form
\begin{equation}
	(\Delta S_A)^2_{N}=\frac{1}{d_N+1}\Big[\!\!\sum_{N_A=0}^{N}\!\!\varrho\;(\varphi^2_{N_A}\!+\!\chi_{N_A}\big)\!-\!\big(\!\!\sum_{N_A=0}^{N}\!\!\varrho_{N_A}\;\varphi_{N_A}\big)^2\Big],
	\label{eq:DSA2N}
\end{equation}
where $\varrho_{N_A}$ and $\varphi_{N_A}$ are given in Eq.~\eqref{eq:varphi} and $\chi_{N_A}$ is defined as
\begin{align}
	\chi\!=\!\!
	&\begin{cases}
		\scriptstyle\!\! (d_A\!+d_B)\Psi'\!(d_B+1)-(d_N\!+1)\Psi'\!(d_N+1)-\frac{(d_A\!-\!1)(d_A\!+2d_B\!-1)}{4d_B^2},
		&\!\!\!\!\scriptstyle\!\! d_A\leq d_B,  \\[0.8em]
		\scriptstyle\!\! (d_A\!+d_B)\Psi'\!(d_A+1)-(d_N\!+1)\Psi'\!(d_N+1)-\frac{(d_B\!-\!1)(d_B\!+2d_A\!-1)}{4d_A^2},
		&\!\!\!\!\scriptstyle\!\! d_A> d_B.
	\end{cases}
\end{align}
As earlier, $d_N$, $d_A(N_A)$, $d_B(N-N_A)$ are understood as functions of the particle number and are given by Eqs.~\eqref{eq:dN} and~\eqref{eq:dAdB}. In the thermodynamic limit $V\to\infty$, at fixed subsystem fraction $f=V_A/V$ and fixed particle density $n=N/V$, the variance is exponentially small and its asymptotic scaling can be obtained via the saddle point methods of Appendix \ref{app:general-average}. In particular, we have
\begin{align}
	&\sum_{N_A=0}^{N}\!\!\varrho_{N_A}\;\varphi^2_{N_A}\,-\,\Big(\!\!\sum_{N_A=0}^{N}\!\!\varrho_{N_A}\;\varphi_{N_A}\Big)^2 =\\
	&\quad=\!\int^{\infty}_{-\infty} \!\!\!\!\!\!\!\varrho(n_A)\varphi^2(n_A)dn_A-\Big(\int^{\infty}_{-\infty} \!\!\!\!\!\!\!\varrho(n_A)\varphi(n_A)dn_A\Big)^2\!+\!o(1)\nonumber\\[.5em]
	&\quad=\big[f(1\!-\!f)-\frac{1}{2\pi}\delta_{f,\frac{1}{2}}\big]\big(\!\ln \frac{n}{1\!-n}\big)^2 \,n(1\!-\!n)\, V+o(V)\nonumber,
\end{align}
and
\begin{equation}
	\sum_{N_A=0}^{N}\!\!\varrho_{N_A}\chi_{N_A}=\frac{1}{4}\delta_{f,\frac{1}{2}}\delta_{n,\frac{1}{2}}+o(1)\,,
\end{equation}
where we have used the fact that, for large dimensions, $d_A\gg 1$ and $d_B\gg 1$, $\chi$ scales as
\begin{align}
	\chi_{N_A}=
	\begin{cases}
		\frac{d_A}{2d_B}+O(1/d_B^2)\,, & d_A< d_B  \\
		\frac{1}{4} +o(1)\,, & d_A= d_B  \\
		\frac{d_B}{2d_A}+O(1/d_A^2)\,, & d_A> d_B \,.
	\end{cases}
	\label{eq:chi-asympt}
\end{align}
Therefore, the term in brackets in Eq.~\eqref{eq:DSA2N} is of order $V$, while the denominator $d_N+1$ is exponentially large. Using the Stirling approximation for $d_N$ in Eq.~\eqref{eq:DSA2N}, we find that
\begin{equation}
	(\Delta S_A)^2_{N}= \alpha\, V^{\frac{3}{2}}\ee^{-\beta V}+ o(\ee^{-\beta V}),
	\label{eq:DeltaS-N}
\end{equation}
with
\begin{align}
	\alpha=&\,\scriptstyle\sqrt{2\pi} \big[f(1-f)-\frac{1}{2\pi}\delta_{f,\frac{1}{2}}\big]\left(\ln\! \frac{n}{1-n}\right)^2\,[n(1\!-\!n)]^{\frac{3}{2}}\, +\,o(1)\nonumber\\[.5em]
	\beta=&-n\ln n-(1-n)\ln(1-n)\,.
\end{align}
This means that the average entanglement entropy in Eq.~\eqref{eq:leading-general} is also the typical entanglement entropy of pure quantum states with $N$ fermions, namely, the overwhelming majority of pure quantum states with $N$ fermions have the entanglement entropy in Eq.~\eqref{eq:leading-general}.
	
\subsubsection{Weighted average and variance}\label{sec:general-mu}
	
Having computed the average entanglement entropy of pure states with $N$ particles, next we can compute the average over the entire Hilbert space. A subtlety is that the system is in either of the Hilbert spaces $\mathcal{H}_N$, but we do not know in which one. Therefore, while the distribution of the pure states with a fixed particle number is given quantum mechanically, meaning uniformly distributed over a unit sphere, we additionally have a classical probability for the particle number $N$. 
	
With this in mind, we can average over $\braket{S_A}_N$ within each sector with $N$ particles weighted by its Hilbert space dimension $d_N$ from Eq.~\eqref{eq:dN}. More generally, we can introduce a weight parameter $w$ and a probability $P_N$ of finding $N$ particles:
\begin{align}
	P_N=\frac{1}{Z}d_N \ee^{-w N}.
	\label{eq:PN-binomial}
\end{align}
Here $Z=\sum_{N=0}^V d_N \ee^{-w N}=(1+\ee^{-w})^V$ normalizes the distribution. The average filling fraction $\bar{n}$ can be expressed in terms of the weight parameter $w$ as
\begin{align}
	\bar{n}=\sum_{N=0}^V P_N \frac{N}{V}\;=\;\frac{1}{1+\ee^w}\,
	\label{eq:nbar}
\end{align}
with half-filling $\bar{n}=\frac{1}{2}$ corresponding to equiweighted sectors, \ie $w=0$. The variance of the filling fraction,
\begin{align}
	(\Delta n)^2=\sum_{N=0}^V P_N\; (\frac{N}{V}-\bar{n})^2\;=\;\frac{\bar{n}(1-\bar{n})}{V}\,
	\label{eq:Dn}
\end{align}
can be obtained easily by noting that $P_N$ is a binomial distribution.
	
We calculate the average entanglement entropy at fixed weight parameter $w$,
\begin{align}
	\braket{S_A}_w=\sum_{N=0}^V P_N \braket{S_A}_N\,,
	\label{eq:Sw-def}
\end{align}
up to constant order in $V$ by expanding $\braket{S_A}_N$ around $\bar{n}$ and then using the known variance $(\Delta n)^2$. Since $\braket{S_A}_N$ is analytic as a function of $N$ (for $f<\frac{1}{2}$) and does not have any discontinuities in its derivatives, it suffices to expand its leading order (linear in $V$) around $\bar{n}$ as
\begin{align}
	\begin{split}\label{eq:sA-useful}
		s_A(f,n)&=[(n-1)\ln(1-n)-n\ln{n}]f\\
		&=[(\bar{n}-1)\ln(1-\bar{n})-\bar{n}\ln{\bar{n}}]f\\
		&\quad+f\ln[(1-\bar{n})/\bar{n}]-\frac{f(n-\bar{n})^2}{2(1-\bar{n})\bar{n}}\\
		&\quad+o(n-\bar{n})^3,
	\end{split}
\end{align}
and calculate its expectation value with respect to the binomial distribution. Using $\braket{(n-\bar{n})^2}=\sigma^2$, we find the constant correction $-\frac{f}{2}$, which cancels the identical term in Eq.~\eqref{eq:leading-general}. Terms of order $V^{1/2}$ and $V^0$ can be directly evaluated at $n=\bar{n}$, where the binomial distribution is centered, because its finite width on those terms will only contribute corrections of subleading order $O(1)$. Hence, the resulting average is equal to
\begin{align}\label{eq:Page-weighted}
	\begin{split}
		\braket{S_A}_{w}&=\left[(\bar{n}-1)\ln(1-\bar{n})-\bar{n}\ln(\bar{n})\right] fV\\
		&\quad-\sqrt{\frac{\bar{n}(1-\bar{n})}{2\pi}}\left|\ln\left(\frac{1-\bar{n}}{\bar{n}}\right)\right|\delta_{f,\frac{1}{2}}\sqrt{V}\\
		&\quad+\frac{\ln(1-f)}{2}-\frac{2}{\pi}\,\delta_{f,\frac{1}{2}}\delta_{\bar{n},\frac{1}{2}}+o(1)\,,
	\end{split}
\end{align}
where $\bar{n}=1/(1+e^{w})$ was computed in Eq.~\eqref{eq:nbar}. A pedagogical derivation of Eq.~\eqref{eq:Page-weighted} can be found in Appendix~\ref{app:general-weighted}. Interestingly, Eq.~\eqref{eq:Page-weighted} can be summarized by the simple relation $\braket{S_A}_{w} = \braket{S_A}_{N=\bar{N}} - \frac{f}{2}+o(1)$ except at $f=\bar{n}=\frac{1}{2}$, where the Kronecker delta from Eq.~\eqref{eq:leading-general} leads to additional integrals, as explained in Appendix~\ref{app:general-weighted}.
	
For $w=0$ with $\bar{n}=\frac{1}{2}$, Eq.~\eqref{eq:Page-weighted} describes the average entanglement entropy of uniformly weighted eigenstates of the number operator (with respect to the Haar measure). This average was computed in Ref.~\cite{huang_19} as $\braket{S_A}_{w=0}=f V \ln{2}+\frac{\ln(1-f)}{2}-\frac{2}{\pi}\delta_{f,1/2}$, which coincides with Eq.~\eqref{eq:Page-weighted} for $\bar{n}=\frac{1}{2}$.
	
Similarly, one can compute the variance of the weighted entanglement entropy
\begin{align}
	\begin{split}
		(\Delta S_A)^2_w&=\sum_{N=0}^V P_N\, \langle S_A^2\rangle_N-\big(\sum_{N=0}^V P_N\, \langle S_A\rangle_N\big)^2\\
		&=\bar{n}(1-\bar{n})\big(\ln \frac{\bar{n}}{1-\bar{n}}\big)^2 f V+o(V)\,.
		\label{eq:DeltaS-w}
	\end{split}
\end{align}
Note that, while the variance $(\Delta S)^2_N$ at a fixed number of particles is exponentially small at large $V$, the weighted variance $(\Delta S)^2_w$ scales linearly in $V$ because of the $O(V^{-1})$ variance $(\Delta n)^2$ in the filling fraction. For $f\neq0$ and $\bar{n}\neq 0$, the leading-order term only vanishes at $\bar{n}=\frac{1}{2}$. However, we always have $\lim_{V\to\infty}(\Delta S_A)_w/\braket{S_A}_w=0$, \ie the \emph{relative standard deviation} vanishes in the thermodynamic limit, so that the average entanglement entropy $\braket{S_A}_w$ and the \emph{typical} eigenstate entanglement entropy always coincide.

\section{PURE FERMIONIC GAUSSIAN STATES} \label{sec:gaussian}
	
In this section, we define fermionic Gaussian states and calculate the average and variance of the entanglement entropy for this family of states. Following Ref.~\cite{bianchi2021page}, we do this first for pure fermionic Gaussian states, for which the number of particles is not fixed. Next, we derive new results for fermionic Gaussian states with a fixed number of particles. In both cases we mimic the idea of a uniformly distributed state. This works because in both cases there is a natural action of a compact group and the set is given by a single orbit of this group action. Thus, one can choose the unique Haar measure to generate an ensemble of fermionic Gaussian states.
	
It may be natural to ask whether the same analysis could also be carried out for bosonic Gaussian states. Unfortunately, the answer is in the negative. The ensemble of bosonic Gaussian states is noncompact with unbounded entanglement entropy since the corresponding invariance group is a noncompact one. So any group invariant average would diverge. Moreover, the only bosonic Gaussian state that has a fixed particle number is the vacuum with zero particles and zero entanglement. To circumvent the problem, one could fix the \emph{average} number of particles. Then, the corresponding manifold would be again compact and one can average over all those Gaussian states (in a similar spirit as in Refs.~\cite{serafini2007canonical, fukuda2019typical}), but the resulting analysis would be rather different from our approach here. It may be possible to use a duality between bosonic and fermionic entanglement entropy of Gaussian states~\cite{jonsson2021entanglement} for this, but we will not carry out this analysis here.
	
\subsection{Definition of fermionic Gaussian states}
	
Instead of starting with pure fermionic Gaussian states, it is easier to begin with mixed Gaussian states because the pure ones can be understood as limits of this definition. We choose a Majorana basis $\{\gamma_j\}_{j=1,\ldots,2V}$ in the $2^V$-dimensional Hilbert space $\mathcal{H}$ since the corresponding ensemble is easier to describe. This Majorana basis satisfies the anticommutation relation $\{\gamma_j,\gamma_k\}=\delta_{jk}$, meaning that they create a Clifford algebra and can be chosen to be Hermitian, $\gamma_j^\dagger=\gamma_j$. Moreover, it holds that $\Tr\left(\prod_{l=1}^m\gamma_{j_l}\right)=0$ with $j_{l}\in\{1,\ldots,V\}$ and any positive integer $m$ whenever there is a $\gamma_j$ that does not appear in this product with an even order. Otherwise, it holds that $\Tr\left(\prod_{l=1}^m\gamma_{j_l}\right)=\pm2^{V-m/2}$, which is up to a factor $2^{-m/2}$ the dimension of the representation of the Clifford algebra as well as the dimension of the Hilbert space $\mathcal{H}$.
	
A Gaussian state is then any density operator of the form
\begin{equation}
	\hat\rho(\gamma)=\frac{\exp(-\sum_{j,k=1}^{2V}q_{jk}\gamma_j\gamma_k)}{\Tr \exp(-\sum_{j,k=1}^{2V}q_{jk}\gamma_j\gamma_k)}=\frac{\exp(-\gamma^\dagger Q\gamma)}{\Tr \exp(-\gamma^\dagger Q\gamma)}
\end{equation}
with the Majorana operator-valued column vector $\gamma = (\gamma_1, \ldots, \gamma_{2V})^\dagger$. This form gives the Gaussian states their name. The Hermiticity of $\hat\rho(\gamma)$ implies that the coefficient matrix $Q=\{q_{jk}\}_{j,k=1,\ldots,2V_A}$ needs to be Hermitian, while the anticommutation relations of the Majorana basis allows us to set the real symmetric part to zero. Indeed, due to
\begin{equation}
	\begin{split}
		\sum_{j,k=1}^{2V}q_{jk}\gamma_j\gamma_k&=\sum_{j=1}^{2V}q_{jj}+\sum_{1\leq j<k\leq 2V}q_{jk}(\gamma_j\gamma_k-\gamma_k\gamma_j)\\
		&=\sum_{j=1}^{2V}q_{jj}+\sum_{1\leq j<k\leq 2V}(q_{jk}-q_{kj})\gamma_j\gamma_k,
	\end{split}
\end{equation}
we see that the diagonal part of $Q$ only yields a constant while the symmetric one drops out. Thence, the coefficient matrix is a $2V\times 2V$ imaginary antisymmetric matrix $Q=-Q^*=-Q^T$. Such a matrix can be block diagonalized by an orthogonal matrix $M\in{\mathrm{O}}(2V)$. In particular, it holds that $Q=M^T{\rm diag}(\lambda_1\tau_2,\ldots,\lambda_V\tau_2)M$, with the second Pauli matrix $\tau_2$ and $\lambda_j\geq0$. Introducing $\eta=(\eta_1,\ldots,\eta_{2V})^\dagger=M\gamma$, whose entries create another Majorana basis, the expression simplifies because of $\gamma^\dagger Q\gamma=-2\ii\sum_{j=1}^V\lambda_j\eta_{2j-1}\eta_{2j}$. We can readily compute the exponent
\begin{equation}
	\exp[-\gamma^\dagger Q\gamma]=\prod_{j=1}^V(\cosh(\lambda_j)+2\ii \sinh(\lambda_j)\eta_{2j-1}\eta_{2j}),
\end{equation}
and the normalization
\begin{equation}
	\Tr \exp[-\gamma^\dagger Q\gamma]=2^{V}\prod_{j=1}^V\cosh(\lambda_j).
\end{equation}
Summarizing, any Gaussian state has the compact form
\begin{equation}
	\hat\rho(\gamma)=  2^{-V}\prod_{j=1}^V\left[1+2\ii \tanh(\lambda_j)\eta_{2j-1}\eta_{2j}\right],
\end{equation}
where the $\lambda_j$ are the singular values and the $\eta_j$ are the Majorana basis in the corresponding eigenbasis of the matrix $Q$.
	
Gaussian states satisfy the Wick-Iserlis theorem, meaning that all moments can be expressed in terms of the first and second moments. Since the first moments vanish for fermions, all the information of a fermionic Gaussian state is encoded in the covariance matrix. When subtracting the identity and multiplying by the imaginary unit, we obtain the symplectic form
\begin{equation}
	\begin{split}
		&-\ii\tilde{J}=\Tr_{\mathcal{H}}[\hat\rho(\gamma)(\gamma\gamma^\dagger-\frac{1}{2}\id_{2V})]\\
		&=M^T\Tr_{\mathcal{H}}\left[\prod_{j=1}^V(\frac{1}{2}+\ii \tanh(\lambda_j)\eta_{2j-1}\eta_{2j})(\eta\eta^\dagger-\id_{2V})\right]M.
	\end{split}
\end{equation}
We have emphasized that the trace is only over the Hilbert space $\mathcal{H}$ and not over the indices of the Majorana basis, which explains why we could take the orthogonal matrix $M$ out the trace. The shift by half of the identity matrix $\tfrac{1}{2}\id_{2V}$ only subtracts the diagonal terms $\gamma_j^2=\tfrac{1}{2}$, which do not contain any information. The symmetries of the Majorana basis tell us that the symplectic form $\tilde{J}$ is real antisymmetric. In a straightforward computation one can show that
\begin{equation}
	\begin{split}
		& \Tr_{\mathcal{H}}\left[\prod_{j=1}^V(\frac{1}{2}+\ii \tanh(\lambda_j)\eta_{2j-1}\eta_{2j})(\eta\eta^\dagger-\frac{1}{2}\id_{2V})\right]\\
		&={\rm diag}\left[\tanh(\lambda_1)\tau_2,\ldots,\tanh(\lambda_V)\tau_2\right].
	\end{split}
\end{equation}
Therefore, the eigenvalues of $\tilde{J}$ are equal to $\pm\ii x_j=\pm\ii\tanh(\lambda_j)$, and the link between $Q$ and $\tilde{J}$ is given by the bijective relation
\begin{equation}
	\tilde{J}=\ii\tanh(Q).
\end{equation}
As we can go back and forth between these two matrices, $\hat\rho(\gamma)$ is fully determined by $\tilde{J}$ so that it is suitable to adopt the notation $\hat\rho(\tilde{J})$. For instance, we can express the von Neumann entropy in terms of $\tilde{J}$ because of
\begin{equation}
	\begin{split}
		&-\Tr[\hat\rho(\tilde{J})\ln\hat\rho(\tilde{J})]\\
		&=\Tr\left[\prod_{j=1}^V(\frac{1}{2}+\ii x_j\eta_{2j-1}\eta_{2j})\,\sum_{k=1}^V\ln\left(\frac{1}{2}+\ii x_j\eta_{2k-1}\eta_{2k}\right)\right]\\
		&=\sum_{k=1}^Vs(x_k)=\Tr s(\ii\tilde{J})
	\end{split}
\end{equation}
with~\cite{Peschel2003,hackl2018aspects,hackl2020bosonic}
\begin{align}
	s(x)=-\left(\frac{1+x}{2}\right)\ln\left(\frac{1+x}{2}\right)-\left(\frac{1-x}{2}\right)\ln\left(\frac{1-x}{2}\right).\label{eq:Gaussian-s}
\end{align}
	
With the help of the von Neumann entropy it is straightforward to identify the pure fermionic Gaussian states. Those are given when all eigenvalues are equal to $x_j=\pm1$. Indeed, a density operator of the form $\hat\rho(\tilde{J})=  2^{-V} \prod_{j=1}^V (1\pm2\ii\eta_{2j-1}\eta_{2j})$ satisfies the necessary and sufficient condition for pure states (in combination with positive semidefiniteness and the normalized trace), \ie
\begin{equation}
	\begin{split}
		\hat\rho^2(\tilde{J})&=2^{-2V}\prod_{j=1}^V(1\pm2\ii\eta_{2j-1}\eta_{2j})^2\\
		&=2^{-V}\prod_{j=1}^V(1\pm2\ii\eta_{2j-1}\eta_{2j})=\hat\rho(\tilde{J}).
	\end{split}
\end{equation}
The corresponding normalized state vector of $\hat\rho(\tilde{J})=\ket{\tilde{J}}\bra{\tilde{J}}$ is denoted by $\ket{\tilde{J}}$ and it is only determined up to a complex phase. The set of all real antisymmetric matrices $\tilde{J}$ with eigenvalues $\pm \ii$ is described by $\{\tilde{J}=M^TJ_0M|\,M\in{\rm O}(2V)\ {\rm and}\ J_0=\ii\tau_2\otimes\id_V\}$. This gives a natural parametrization of pure fermionic Gaussian states, which will be our starting point in Sec.~\ref{sec:gaussian-arbitraryN}.
	
\subsection{Arbitrary number of particles}\label{sec:gaussian-arbitraryN}
We first focus on the family of pure fermionic Gaussian states in which one does not fix the particle number, \ie we include Gaussian states that consist of a superposition of states with different total particle number.
	
\subsubsection{Statistical ensemble of states}
	
As we have seen in the previews subsection, all pure fermionic Gaussian states can be described by their symplectic form $\tilde{J} = \ii \braket{\tilde{J}| \gamma\gamma^\dagger-\id_{2V} |\tilde{J}}=M^TJ_0M$, with an orthogonal matrix $M\in{\rm O}(2V)$ and the symplectic unit $J_0=\ii\tau_2\otimes\id_V$, which is essentially a canonical embedding of the second Pauli matrix $\tau_2$ in the $2V$-dimensional space and defines a complex structure, see Refs.~\cite{hackl2018aspects, hackl2020bosonic, bianchi2021page}. One can render the relation between pure states and real antisymmetric matrices $\tilde{J}$ with eigenvalues $\pm \ii$, \ie $\tilde{J}^2=-\id_{2V}$, to a one-to-one correspondence when dividing all orthogonal matrices out of ${\rm O}(2V)$ that commute with $J_0$. These matrices build a subgroup that is the real representation of the unitary group ${\rm U}(V)$ (the direct sum of the fundamental and antifundamental representations). Thence, the set of all pure fermionic Gaussian states can be identified with the coset ${\rm O}(2V)/{\rm U}(V)$, which is $V(V-1)$ dimensional.
	
There is a natural ${\rm O}(2V)$ group action on the pure states by $\tilde{J}\to M^T\tilde{J}M$ that corresponds to the change of an orthonormal Majorana basis. Therefore, adopting Page's idea of a uniform distribution that is given by a group action, we assume that the ensemble of random pure fermionic Gaussian states is created by the normalized Haar distribution on ${\rm O}(2V)/{\rm U}(V)$. Practically, this can be realized by drawing a Haar-distributed orthogonal matrix $M\in{\rm O}(2V)$, and considering the pure state corresponding to the real antisymmetric matrix $\tilde{J}=M^TJ_0M$; see Ref.~\cite{bianchi2021page}.
	
When restricting a pure fermionic Gaussian state $\hat\rho(\tilde{J})$ to a subsystem $A$ with a $(d_A=2^{V_A})$-dimensional Hilbert space $\mathcal{H}_A$, one obtains a mixed Gaussian state. The corresponding symplectic form can be obtained by a projection of the matrix $\tilde{J}$. Without loss of generality, we assume that $\tilde{\gamma}=(\gamma_1,\ldots,\gamma_{2V_A})^\dagger$ is an orthonormal Majorana basis only acting nontrivially on $\mathcal{H}_A$ but has a trivial action on the other Hilbert space $\mathcal{H}_B$. Defining $\Pi_A$ as the projection of a $2V$ vector onto the first $2V_A$ components, it holds that $\tilde{\gamma}=\Pi_A\gamma$ and the new symplectic form corresponding to the state $\hat\rho_A(\tilde{J})=\Tr_{\mathcal{H}_B}\hat\rho(\tilde{J})$ is
\begin{equation}
	\begin{split}
		\tilde{J}_A&=\Tr_{\mathcal{H}_A}[\hat\rho_A(\tilde{J})(\tilde{\gamma}\tilde{\gamma}^\dagger-\frac{1}{2}\id_{2V_A})]\\
		&=\Pi_A\Tr_{\mathcal{H}}[\hat\rho(\tilde{J})(\gamma\gamma^\dagger-\frac{1}{2}\id_{2V})]\Pi_A^T=\Pi_A\tilde{J}\Pi_A^T.
	\end{split}
\end{equation}
Hence, the new covariance matrix is only an orthogonal projection of the old one onto its upper left $2V_A\times 2V_A$ block. Surely, it can be any diagonal block or even a more complicated embedding of this $2V_A\times 2V_A$ matrix $\tilde{J}_A$ in $\tilde{J}$. However, the group invariant generation of the pure fermionic Gaussian states tells us that all these embeddings are equivalent. Physically, this means that all these subsystems of $\mathcal{H}=\mathcal{H}_A\otimes \mathcal{H}_B$ are essentially the same once the dimension $d_A$ is fixed. We have already seen this picture in Page's setting.
	
In Ref.~\cite{bianchi2021page}, it was shown that the random matrix $\tilde{J}_A=\Pi_AM^TJ_0M\Pi_A^T$ with a Haar distributed $M\in{\rm O}(2V)$ has a joint probability density of its eigenvalues $\ii{\rm diag} (x_1\tau_2, \ldots, x_{V_A}\tau_2)$ of the form
\begin{align}\label{jacobi-arbitrary}
	P(x)=\mathcal{N}\prod_{j<k}(x_j-x_k)^2\prod^{V_A}_{l=1}(1-x_l^2)^{V-2V_A},
\end{align}
with $\mathcal{N}$ the normalization. Here, we already see that it is crucial to assume that $V_A\leq V/2$; otherwise, we need to consider the density operator $\hat\rho_B(\tilde{J})=\Tr_{\mathcal{H}_A}\hat\rho(\tilde{J})$. Indeed, it is again the same symmetry between subsystems $A$ and $B$ that is still true here, and the breakdown of analyticity is due to some eigenvalues, either of $\hat\rho_A(\tilde{J})$ or $\hat\rho_B(\tilde{J})$, being exactly zero. Those eigenvalues are related to the eigenvalues of $\tilde{J}_A$ or, equivalently, $\tilde{J}_B$ that are exactly $\pm\ii$.
	
The main idea that enters the computation of the joint probability distribution~\eqref{jacobi-arbitrary} is Proposition~A.2 of Ref.~\cite{kieburg2019multiplicative}, which shows what the eigenvalues of a corank-$2$ projection of a real antisymmetric matrix are. Then, one needs only repetitively apply this proposition, leading to the distribution above.
	
One important ingredient in the computation is that all $k$-point correlation functions can be expressed in terms of a single kernel function $K(x_l,x_j)$ as\footnote{Note that Ref.~\cite{bianchi2021page} uses a different normalization constant, where the level density is $\rho(x)=R_1(x)$. In contrast, we use the normalization $\int_0^1 R_1(x)dx=V_A$, which is standard in random matrix theory, and the level density is $\rho(x)=R_1(x)/V_A$.}
\begin{equation}
	\begin{split}\label{k-point}
		R_k(x_1,\ldots,x_k)&=\frac{V_A!}{(V_A-k)!}\int_{-1}^1 dx_{k+1}\cdots\int_{-1}^1 dx_{V_A}P(x)\\
		&=\det[K(x_l,x_j)]_{l,j=1,\ldots, k}\,,
	\end{split}
\end{equation}
In the mathematical branch of random matrix theory, this structure is known as a determinantal point process~\cite{1998NuPhB.536..704B}. The average and variance of the entanglement entropy can then be traced back to an integral of the one-point function $R_1(x)$ and the two-point correlation function $R_2(x_1,x_2)$, respectively. In general, higher cumulants of the entanglement entropy are averages of specific $k$-point correlation functions.
	
Distribution~\eqref{jacobi-arbitrary} is shared with the unitary Jacobi ensemble~\cite{2000JPhA...33.2045Z, Forrester_2010}, which is a truncation of a Haar-distributed unitary matrix. Despite the fact that the eigenvalue statistics is the same with a unitary Jacobi ensemble, the eigenvector statistics is different. This can be readily seen in that the eigenvectors of the unitary Jacobi ensemble must be ${\rm U}(V_A)$ invariant, while in the present case they are only ${\rm O}(2V_A)$ invariant.
	
\subsubsection{Average and variance}
	
Our goal is again to compute the mean and variance of the entanglement entropy of subsystem $A$,
\begin{align}
	\braket{S_A}_{\mathrm{G}}&=\int S_A(\ket{MJ_0M^{-1}})d\mu(M)\,,\\
	(\Delta S_A)^2_{\mathrm{G}}&=\int(S_A(\ket{MJ_0M^{-1}})-\braket{S_A}_{\mathrm{G}})^2 d\mu(M)\,,
\end{align}
where $S_A(\ket{\tilde{J}})=\Tr s(\ii\tilde{J}_A)$ [see Eq.~\eqref{eq:Gaussian-s}] and $d\mu(M)$ now represents the Haar measure over the orthogonal group. Computing the average entanglement entropy over all Gaussian states was the main technical achievement of Ref.~\cite{bianchi2021page}. It was facilitated by recent results in random matrix theory~\cite{kieburg2019multiplicative}, from which one can deduce what the joint probability distribution of the singular values $\tilde{J}_A$ are. For this, we make repetitive use of Proposition A.2 of Ref.~\cite{kieburg2019multiplicative} by always projecting away two rows of matrix $J$; first to $[J]_{N-1}$, then $[J]_{N-2}$, until we arrive at $[J]_{N_A}$. This yields for $x=(x_1,\dots,x_{N_A})$ the distribution
\begin{align}
	\textstyle  P(x)=\frac{\left(\det X\right)^2}{N_A!}\Big(\prod^{N_A-1}_{j=0}c_j^{-1}(1-x_{j+1}^2)^{\Delta}\Big)\,,\label{eq:probability2}
\end{align}
where we have the $N_A\times N_A$ matrix $X$ and $c_j$ given by
\begin{align}
	X_{ij}&=p_{j-1}(x_i)=\mathcal{P}^{(\Delta,\Delta)}_{2j-2}(x_i)\,,\\
	c_{j}&=\frac{2^{2\Delta}\,[(2j+\Delta)!]^2}{(2j)!\,(2j+2\Delta)!\,(4j+2\Delta+1)}\,,
\end{align}
with $\Delta=N_B-N_A\geq0$. The $k$-point correlation functions from Eq.~\eqref{k-point} are fully encoded by $K(x_a,x_b)$, which is given by the $k\times k$ matrix (with $a,b=1,\dots,k$) given by~\cite{Forrester_2010}
\begin{align}
	\begin{split}
		\hspace{-3mm}K(x,y)=\!\!\sum^{N_A-1}_{j=0}\!\psi_j(x)\psi_j(y),\,\,\,
		\psi_j(x)=\frac{(1-x^2)^{\Delta/2}}{\sqrt{c_j}} p_j(x),
	\end{split}\label{eq:Kmatrix}
\end{align}
with $\int_0^1 \psi_j(x)\psi_k(x)dx=\delta_{jk}$. The average entanglement entropy can be written in terms of the one-point function,
\begin{align}
	\begin{split}
		\textstyle \hspace{-2mm} \braket{S_A}_\mathrm{G}&=\int_0^1R_1(x)s(x)dx\\
		&=(V\!-\!\frac{1}{2})\Psi(2V)+(\frac{1}{2}\!+\!V_A\!-\!V)\Psi(2V\!-\!2V_A)\\[.5em]
		&\quad +(\frac{1}{4}\!-\!V_A)\Psi(V)-\frac{1}{4}\Psi(V\!-\!V_A)-V_A\,,
	\end{split}\label{eq:Gaussian-average}
\end{align}
where the details can be found in Appendix~\ref{app:average-Gaussian-general}. Anew, the symmetry $V_A\leftrightarrow V-V_A$ is not reflected in this result and needs to be introduced by hand. The origin of this breaking in the analytical result is as in Page's setting, one of the two density operators $\hat\rho_A$ and $\hat\rho_B$ has an exact number of zero modes. The result corresponds to this system without these generic zero modes. This selection is a nonanalytical step. Indeed, there is a nonanalytic kink at $V_A=V/2$ (first derivative jumps there). However, it is difficult to see when plotting the result even for moderately small $V$ (say of the order $10$). The reason is that this kink vanishes like $1/V^2$, so that it is only of the order of $1\%$ when $V=10$.
	
For $f=V_A/V\leq \frac{1}{2}$, the thermodynamic limit reads
\begin{align}
	\begin{split}
		\braket{S_A}_\mathrm{G}&= V\,\big((\ln{2}\!-\!1)f+\!(f\!-\!1)\ln(1\!-\!f)\big)\\[.5em]
		&\quad +\frac{1}{2}f+\frac{1}{4}\ln{(1-f)}\,+\,O(1/V)\,,
	\end{split}
	\label{eq:thermodynamic-limit}
\end{align}
whose leading-order term was found earlier in Ref.~\cite{lydzba2020entanglement} to give the average eigenstate entanglement entropy of number-preserving random quadratic Hamiltonians. This match is not a coincidence, as discussed in Sec.~\ref{sec:Gaussian-mu}.
	
Interestingly, result~\eqref{eq:thermodynamic-limit} glued to its reflection $f\to1-f$ at $f=\frac{1}{2}$ is 2 times differentiable at $f=\frac{1}{2}$. Thus, the nonanalyticity is hardly visible. Starting with the third derivative one can actually see the breaking of analyticity. We note that, in contrast to the case of general pure states considered by Page, in Eq.~\eqref{eq:thermodynamic-limit} there is no Kronecker delta contribution at $f=\frac{1}{2}$.
	
The variance $(\Delta S_A)^2$ can be computed from the matrix representation of the entanglement entropy function $s(x)$ with respect to the function $\psi_i(x)$:
\begin{align}
	s_{ij}=\int^1_{-1} s(x)\psi_i(x)\psi_j(x)\,dx\,.
\end{align}
In full analogy to the calculation in Ref.~\cite{bianchi2021page}, we find that
\begin{eqnarray}
	&&(\Delta S_A)^2_{\mathrm{G}}\nonumber\\&& =\int_{-1}^1 s^2(x) K(x,x)dx-\int_{-1}^1s(x_1)s(x_2) K^2(x_1,x_2)d^2x\nonumber\\&& =\int_{-1}^1 s(x_1)s(x_2) K(x_1,x_2)[\delta(x_1-x_2)-K(x_1,x_2)]d^2x\nonumber\\ &&=\int_{-1}^1 s(x_1)s(x_2) \left[\sum^{V_A-1}_{i=0}\psi_i(x_1)\psi_i(x_2)\right]\nonumber\\&&\quad\quad\times \left[ \sum^{\infty}_{j=V_A} \psi_j(x_1)\psi_j(x_2) \right] d^2x =\sum^{V_A-1}_{i=0} \sum^\infty_{j=V_A} s^2_{ij}\,.\ \label{eq:variance-sum}
\end{eqnarray}
The variance of the entanglement entropy for a pure fermionic Gaussian state was computed to leading order in Ref.~\cite{bianchi2021page} and is given in
\begin{align}
	(\Delta S_A)^2_{\mathrm{G}}=\frac{f+f^2+\ln(1-f)}{2}+o(1)\,.\label{eq:Gaussian-variance}
\end{align}
We present a systematic derivation in Appendix~\ref{app:variance-Gaussian-general} based on certain integrals of Jacobi polynomials. Using the techniques of Refs.~\cite{wei2021quantum, com-witte}, we expect that the variance can be calculated in a closed compact form for fixed $V$ as an expression in terms of digamma functions, analogous to Eq.~\eqref{eq:Gaussian-average}. In fact, Huang and Wei~\cite{huang2021second} conjectured such an analytical formula.
	
\subsection{Fixed number of particles}\label{sec:gaussian-fixedN}

We consider fermionic Gaussian states with a fixed particle number $N$, \ie the intersection of the family of fermionic Gaussian states and the Hilbert space $\mathcal{H}^{(N)}$.
	
For this, it is useful to change from the basis of Majorana operators $\gamma_i$ to the basis of creation and annihilation operators. This basis $\{\hat{f}_j = \frac{1}{\sqrt{2}}(\gamma_{2j-1}+\ii\gamma_{2j}), \hat{f}_j^\dagger = \frac{1}{\sqrt{2}}(\gamma_{2j-1}-\ii\gamma_{2j})\}_{j=1,\ldots,V}$ will be helpful when studying Gaussian states with fixed particle numbers. Those are given by three $\pi/4$ rotations of the form
\begin{equation}\label{gammafrel}
	(\hat{f}_1,\ldots,\hat{f}_V,\hat{f}_1^\dagger,\ldots,\hat{f}_V^\dagger)=(\gamma_1,\ldots,\gamma_{2V})T^\dagger=\gamma^\dagger T^\dagger
\end{equation}
with
\begin{equation}\label{Tdef}
	T=e^{\ii\pi/4}\exp\left[-\ii\frac{\pi}{4}\tau_3\otimes\id_{V}\right]\exp\left[-\ii\frac{\pi}{4}\tau_1\otimes\id_{V}\right],
\end{equation}
and $\tau_1$ and $\tau_3$ being the first and third Pauli matrices. Hence, the symplectic form becomes a complex structure in this basis that is given by
\begin{align}
	\begin{split}
		J&=\ii T\Tr_{\mathcal{H}}[\hat\rho(\gamma)(\gamma\gamma^\dagger-\tfrac{1}{2}\id_{2V})] T^\dagger\\
		&=\ii\begin{pmatrix}
			\braket{\tilde{J}|\hat{f}_i\hat{f}_j^\dagger-\hat{f}_j^\dagger\hat{f}_i|\tilde{J}} & \braket{\tilde{J}|\hat{f}_i^\dagger\hat{f}_j^\dagger-\hat{f}_j^\dagger\hat{f}_i^\dagger|\tilde{J}}\\
			\braket{\tilde{J}|\hat{f}_i\hat{f}_j-\hat{f}_j\hat{f}_i|\tilde{J}} & \braket{\tilde{J}|\hat{f}_i^\dagger\hat{f}_j-\hat{f}_j\hat{f}_i^\dagger|\tilde{J}}
		\end{pmatrix}_{i,j=1,\ldots,V}\,,
	\end{split}\label{eq:J-Gaussian}
\end{align}
where we have used the anticommutation relation $\{\hat{f}_k,\hat{f}_l^\dagger\} = \delta_{kl}$ and $\{\hat{f}_k,\hat{f}_l\} = \{\hat{f}_k^\dagger,\hat{f}_l^\dagger\} = 0$. The transformation of $\tilde{J}\to J$ is a unitary transformation.
	
For fermionic Gaussian states with a fixed number of particles, the off-diagonal blocks in Eq.~\eqref{eq:J-Gaussian} vanish because those contain expectation values of operators that change the particle number. Thus, we are left with the eigenvalues of the two $V$-by-$V$ matrices $F_{ij} = -\ii\braket{J|\hat{f}_i\hat{f}_j^\dagger - \hat{f}_j^\dagger\hat{f}_i|J}$ and $G_{ij} = -\ii\braket{J|\hat{f}_i^\dagger\hat{f}_j - \hat{f}_j\hat{f}_i^\dagger|J}$. These two matrices are intimately related via $F=-G^\intercal$ due to the anticommutation relations. Actually, it is also a direct consequence of the anti-Hermiticity and the $\tau_1\otimes\id_V$ antisymmetry of $J=-J^\dagger=-(\tau_1\otimes\id_V)J^T(\tau_1\otimes\id_V)$. Therefore, when $\ii x_j$ is an eigenvalue of the anti-Hermitian $V\times V$ matrix $F$, then $-\ii x_j$ has to be an eigenvalue of $G$. There are no additional symmetries of $F$ and $G$, meaning that they can be arbitrary Hermitian matrices. Only their singular values are bounded to be inside the interval $[0,1]$ because this is already the case for the complex structure $J$ that is inherited from the positive semidefiniteness of state $\hat\rho$.
	
Using the canonical commutation relations, it holds that
\begin{align}\label{eq:relation}
	F_{ij}=2\ii\braket{J|\hat{f}_j^\dagger\hat{f}^{}_i|J}-\ii\delta_{ij}=-G_{ji}.
\end{align}
This equation relates $F$ and $G$ to the one-body reduced density matrix that is defined as $C_{ij}=\braket{J|\hat{f}_j^\dagger \hat{f}^{}_i|J}$. Indeed, the matrix $C$ is Hermitian,
\begin{equation}
	\{C^\dagger\}_{ij}= C_{ji}^*=(\braket{J|\hat{f}_i^\dagger \hat{f}^{}_j|J})^* =\braket{J|\hat{f}_j^\dagger \hat{f}^{}_i|J}=C_{ji}, 
\end{equation}
and positive semidefinite,
\begin{equation}
	v^\dagger C v=\sum_{i,j=1}^VC_{ij}v_i^*v_j=\left\|\sum_{j=1}^Vv_j\hat{f}_j\ket{J}\right\|^2\geq0.
\end{equation}
Moreover, its trace is fixed, $\Tr C= \braket{J|\sum_{j=1}^V\hat{f}_j^\dagger \hat{f}^{}_j|J}=\braket{J|\hat{N}|J}=N$, in an eigenspace of the total number operator $\hat{N}$. Hence, after a proper normalization one can interpret $C$ as a density matrix.
	
\subsubsection{Statistical ensemble of states}
	
We have seen that, for a pure fermionic Gaussian state, the eigenvalues of $J$ must be $\pm\ii$ or invariantly written  $J^2=-\id_{2V}$. Therefore, the eigenvalues of sub-blocks $F$ and $G$ are also $\pm \ii$ when we assume a fixed particle number; see Ref.~\cite{hackl2020bosonic}. Any basis $\{\hat{f}'_j\}_{j=1,\ldots, V}$, which admits the same canonical anticommutation relations and spanning the same space of creation operators as the original basis $\{\hat{f}_j\}_{j=1,\ldots, V}$, can be chosen in the construction of $F$ and $G$. The set of these bases of creation operators is given by the action of the unitary group ${\rm U}(V)$, \ie $(\hat{f}'{}^\dagger_1,\ldots,\hat{f}'{}^\dagger_V)=({\hat{f}}^\dagger_1,\ldots,{\hat{f}}^\dagger_V)U$ with $U\in{\rm U}(V)$. As each basis is bijectively related to a pure state $\hat\rho=\ket{J}\bra{J}$, the set of all pure fermionic Gaussian states with a fixed particle number $N$ can be generated by $F=U^\dagger F_0U$ and $G= U^T G_0 U^*$ where $F_0$ and $G_0$ are diagonal matrices with $\pm \ii$ on their diagonal. One can bring the number of eigenvalues with $+\ii$ and $-\ii$ in connection with the fixed number of particles $N$ when tracing the matrix $F$ yielding
\begin{eqnarray}
	\Tr F=2\ii\braket{J|\sum_{j=1}^V\hat{f}_j^\dagger\hat{f}^{}_j|J}-\ii V=-\ii(V-2N).
\end{eqnarray}
Thus, $F_0$ can be chosen as $F_0=\ii{\rm diag}(\id_{N},-\id_{V-N})$, and, equivalently $G_0=\ii{\rm diag}(-\id_{N},\id_{V-N})$. Similarly, we can write the one-body reduced density matrix $C=U^\dagger C_0 U$ with $C_N={\rm diag}(\id_N,0,\ldots,0)$ that comprises $V-N$ zeros, and where the subscript highlights the number of particles.
	
The group action of ${\rm U}(V)$ on pure fermionic Gaussian states with a fixed particle number again suggests the notion of a uniform distribution. Hence, we generate the state ensemble by parameterizing the associated complex structure as $J=\ii{\rm diag}(U^\dagger[2C_N -\id_V]U,U^T[ \id_V-2C_N]U^*)$ with a Haar distributed $U\in{\rm U}(V)$.
	
When we consider a subsystem consisting of the first $V_A$ sites, we need to restrict $J$ to the $2V_A\times2V_A$ matrix $J_A$, in which both $F$ and $G$ are restricted to the left upper $V_A\times V_A$ blocks. This choice, as before, results in no loss of generality since the Haar-distributed matrix $U$ covers any other kind of orthogonal projection. That the restriction to a sub-block is indeed directly related to the restriction of a subsystem follows along the same lines as in the case without a fixed particle number. One needs to compute the covariance matrix that is given by the annihilation and creation operators that only act nontrivially in the Hilbert space $\mathcal{H}_A$, which is again equivalent with a projection $J_A=\mathrm{diag}(F_A,G_A)$ and, thus, $F_A=\hat{\Pi}_AF\hat{\Pi}_A^T$ and $G_A=\hat{\Pi}_A G\hat{\Pi}_A^T$, where $\hat{\Pi}_A$ projects onto the first $V_A$ rows. 
	
Instead of using the spectrum $\pm \ii x$ of $J_A$, it is sometimes convenient to use the eigenvalues $y$ of the $V_A\times V_A$ restricted one-body reduced density matrix $C_A=\hat{\Pi}_AC\hat{\Pi}_A^T$. It still holds that $J_A=\ii{\rm diag}(2C_A-\id_{V_A},\id_{V_A}-2C_A)$. This implies that, for the entanglement entropy, we have $\Tr s(\ii J)=2\Tr s(2C_A-\id_{V_A})$ based on Eq.~\eqref{eq:relation}. In terms of eigenvalues this reads $s(x)=s(2y-1)$. The entanglement entropy per volume $s(2y-1)$ vanishes for $y=0$ and $y=1$, due to $s(\pm1)=0$. Therefore, the entanglement entropy $S_A$ is invariant under changing the number of eigenvalues $0$ or $1$ in $C_A$.
	
The generation of $C_A$ is then given by a Haar-random unitary $V\times V$ matrix $U$ and the matrix product
\begin{align}
	C_A=[U]_{V_A\times N}[U]_{N\times V_A}^\dagger\,,
\end{align}
where $[U]_{V_A\times N}$ is the $V_A\times N$ upper left sub-block of the matrix $U$. The matrix $[U]_{V_A\times N}$ is also known as the truncated unitary ensemble or simply the unitary Jacobi ensemble in random matrix theory~\cite{2000JPhA...33.2045Z, Forrester_2010}. It appears in several contexts such as quantum transport~\cite{1997RvMP...69..731B} and quantum scattering~\cite{MELLO1985254}, as it can be seen as a sub-block of an $S$-matrix.
	
\begin{figure}[!t]
	\centering
	\includegraphics[width=\linewidth]{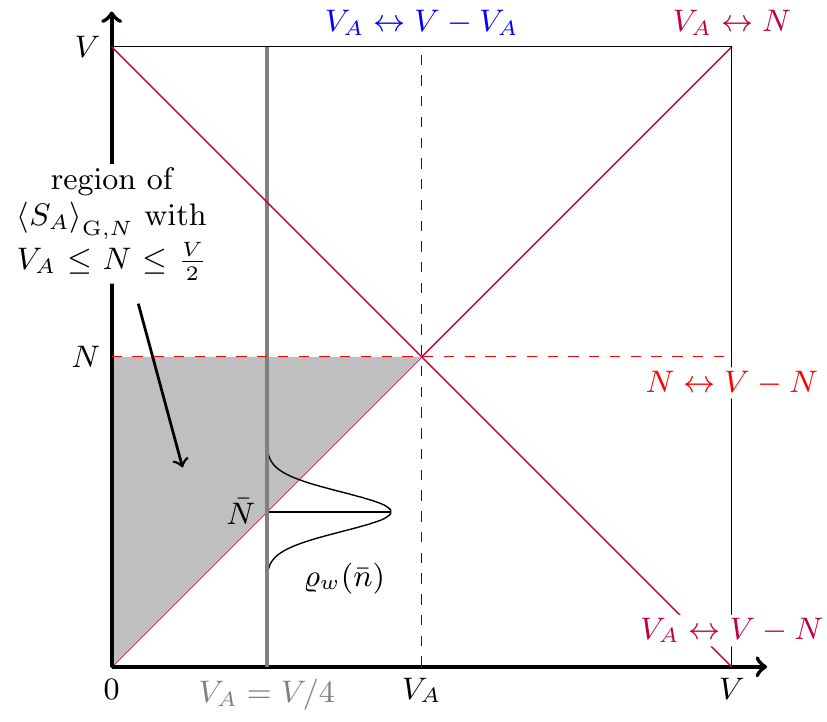}
	\caption{Illustration of the symmetries of the average entanglement entropy $\braket{S_A}_{\mathrm{G},N}$ as a function of $V_A$ and $N$. When we compute the average $\braket{S_A}_{\mathrm{G},w}$ as a function of $(w,V_A,V)$, we are effectively integrating against the density function $\varrho_w(N)$ at fixed $V_A$, which is approximately a Gaussian (plus corrections) centered at the expectation value $\bar{N}=V e^{-w}/(1+e^{-w})$. Transitions are characterized by an enhanced correction of order $1/\sqrt{V}$ to $\braket{S_A}_{\mathrm{G},w}$, and occur whenever $\bar{N}=V_A$, as this is the point where the increasingly narrow Gaussian is integrated against two different analytic functions on either side of its peak (due to the particle-subsystem symmetry). The function $\braket{S_A}_{\mathrm{G},N}$ is analytical across the subsystem and particle-hole symmetries (dashed lines), but \emph{not} across the particle-subsystem symmetries (full lines).}
	\label{fig:phase-transitions}
\end{figure}
	
Let us summarize the symmetries of the above setting.
	
\noindent(i) The particle-hole symmetry, which is given by $N\leftrightarrow V-N$, is reflected when replacing $C=U^\dagger C_NU \to \id_V-C = U^\dagger(\id_V-C_N)U$. Exploiting the symmetry $s(x)=s(-x)$, it holds that
\begin{equation}
	\Tr s(\hat{\Pi}_A[2C-\id_{V}]\hat{\Pi}_A)=\Tr s(\hat{\Pi}_A[2(\id_V-C)-\id_{V}]\hat{\Pi}_A)
\end{equation}
which underlines this symmetry.
	
\noindent(ii) There is again a symmetry between subsystems $A$ and $B$. Anew, it is not immediate as the selection is always given by the smallest of the two complementary diagonal blocks [one of size $V_A\times V_A$ and  of size $(V-V_A)\times (V-V_A)$] of $C$. These are anew given by having a density matrix without zero modes. Thus, we actually expect to put this symmetry in by hand as before.
	
\noindent(iii)  Surprisingly, this manual implementation of the exchange of subsystems is not really needed for fermionic Gaussian states as there is another, more subtle symmetry that relates to the number of exact eigenvalues at $+\ii$ of the symplectic form $\tilde{J}_A$.  This is manifested in an additional particle-subsystem symmetry $V_A\leftrightarrow N$. Its mathematical origin is that the spectra of $[U]_{V_A\times N}[U^\dagger]_{N\times V_A}$ and $[U^\dagger]_{N\times V_A}[U]_{V_A\times N}$ only differ by the number of zero eigenvalues, which correspond to exact eigenvalues $\pm\ii$ of $\tilde{J}_A$. Physically, this means that there are fermionic modes in the eigenbasis of $\tilde{J}_A$ that only act on $\mathcal{H}_A$ and do not act on the sub-Hilbert space $\mathcal{H}_B$. The particle-subsystem symmetry also needs to be introduced by hand as the calculation requires that $C_A$ has no generic zero eigenvalues. Therefore, we can expect a breaking of analyticity at the symmetry axes $N=V_A$ and $N=V-V_A$ because of the particle-hole symmetry $N\leftrightarrow V-N$. One consequence of the particle-subsystem symmetry is that the symmetry axis defined by $V_A=V/2$ must have the same analyticity properties as the symmetry axis $N=V/2$. This is the reason why the implementation of the symmetry between the two subsystems is not needed.
	
\begin{figure*}[!t]
	\centering  
	\includegraphics[width=\linewidth]{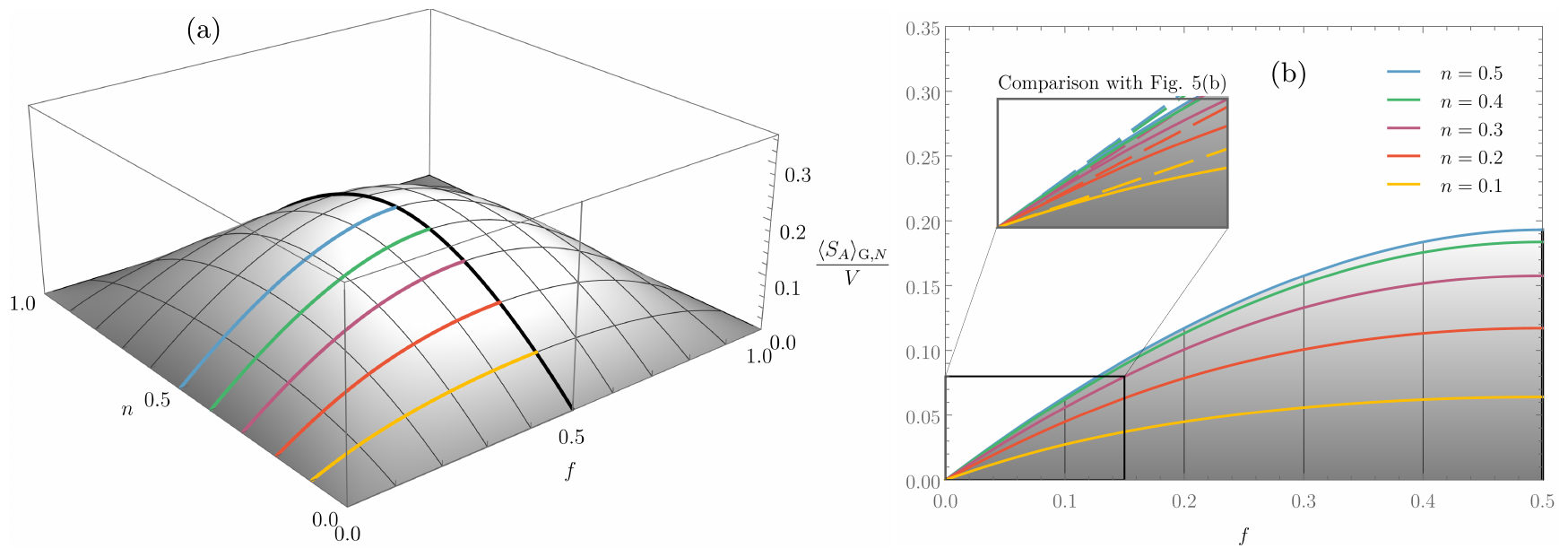}
	\caption{The leading order of the entanglement entropy $s^{\mathrm{G}}_A(f,n)=\lim_{V\to\infty}\braket{S_A}_{\mathrm{G},N}/V$ from Eq.~\eqref{eq:Gaussian-expansion} [see also Eq.~\eqref{eq:sAG}]. One can see the mirror symmetries $V_A\to V-V_A$, $N\to V-N$, and $V_A\to N$. For $n=\frac{1}{2}$, $s^{\mathrm{G}}_A(f,n)$ coincides with the formula derived in Refs.~\cite{lydzba2020entanglement, bianchi2021page}. (a) Three-dimensional plot as a function of the subsystem fraction $f=V_A/V$ and the filling ratio $n=N/V$. One can see the mirror symmetries $V_A\to V-V_A$ and $N\to V-N$. (b) Results at fixed $n$ plotted as functions of $f$. The inset shows that $s_A^{\mathrm{G}}(f,n)\sim s_A(f,n)$ with $s_A(f,n)=\lim_{V\to\infty}\braket{S_A}_{N}/V$  [meaning Eq.~\eqref{eq:sA-useful} in Page's setting] as $f\to 0$. As before, the colored curves are the same in the left and right plot.}
	\label{fig:Page-Gaussian}
\end{figure*}
	
The three symmetries create the overall symmetry group $\mathbb{Z}_2\times \mathbb{Z}_2\times \mathbb{Z}_2\simeq \mathbb{Z}_2\otimes \mathbb{Z}_4$, which can be visualized by the respective mirror axes. The latter product $\mathbb{Z}_2\otimes \mathbb{Z}_4$ reflects the fact that there is a finite rotation group $\mathbb{Z}_4$ and a point reflection group $\mathbb{Z}_2$, which commute.  When we compute $\braket{S_A}_{\mathrm{G},N}$ in Eq.~\eqref{eq:SA-full} for $V_A\leq N\leq N/2$, we only compute it on one eighth of the available parameter space. Using the above symmetries, one can easily deduce $\braket{S_A}_{\mathrm{G},N}$ for any other values of $V_A$ and $N$. We illustrate the symmetries and the respective transitions in Fig.~\ref{fig:phase-transitions}.
	
The eigenvalue distribution of $\ii F$ is given by the unitary Jacobi ensemble, as discussed in Refs.~\cite{Forrester_2010, bernard2021entanglement}. This distribution has the form
\begin{align}
	P(x)=\mathcal{N}\prod_{j<k}(x_j-x_k)^2\prod^{V_A}_{i=1}(1+x_i)^{N-V_A}(1-x_i)^{V-V_A-N}\,.
\end{align}
We can rewrite this probability distribution as
\begin{align}
	\textstyle  P(x)=\frac{\left(\det X\right)^2}{V_A!}\left[\prod^{V_A-1}_{j=0}c_j^{-1}(1-x_{j+1})^{\alpha}(1+x_{j+1})^{\beta}\right]\,,\label{eq:probability}
\end{align}
where we have the $V_A\times V_A$ matrix $X$ and $c_j$ given by
\begin{align}
	X_{ij}&=p_{j-1}(x_i)=\mathcal{P}^{(\alpha,\beta)}_{j-1}(x_i)\,,\\
	c_{j}&=\frac{2^{\alpha+\beta+1}}{2j+\alpha+\beta+1}\frac{(j+\alpha)!(j+\beta)!}{(j+\alpha+\beta)!j!}\,,
\end{align}
with
\begin{align}
	\alpha&=V-N\geq0,\\
	\beta&=V-N-V_A\geq 0\,,
\end{align}
and $\mathcal{P}^{(\alpha,\beta)}_n(z)$ the Jacobi polynomials. We can define the function
\begin{align}
	\psi_j(x)=\frac{1}{\sqrt{c_j}}\mathcal{P}^{(\alpha,\beta)}_j(x)\,,\label{eq:def-psi}
\end{align}
which allows us to express the level density and the two-point kernel as
\begin{align}
	\begin{split}
		R_1(x)&=\sum^{V_A-1}_{i=0}\psi^2_i(x),\label{eq:G-level-dens}
	\end{split}\\
	K(x,y)&=\sum^{V_A-1}_{i=0}\psi_i(x)\psi_i(y)\,.
\end{align}
Equation~\eqref{k-point} underlines that the kernel is a centerpiece in the general spectral statistics of determinantal point processes, as it is here.
	
The above analytical preparations are our starting point for the computations that are performed in Appendix~\ref{app:Gaussfixednumber} and whose results are summarized in Sec.~\ref{sec:aveGaussian}.
	
\subsubsection{Average and variance} \label{sec:aveGaussian}

\begin{figure*}[!t]
	\centering  
	\includegraphics[width=\linewidth]{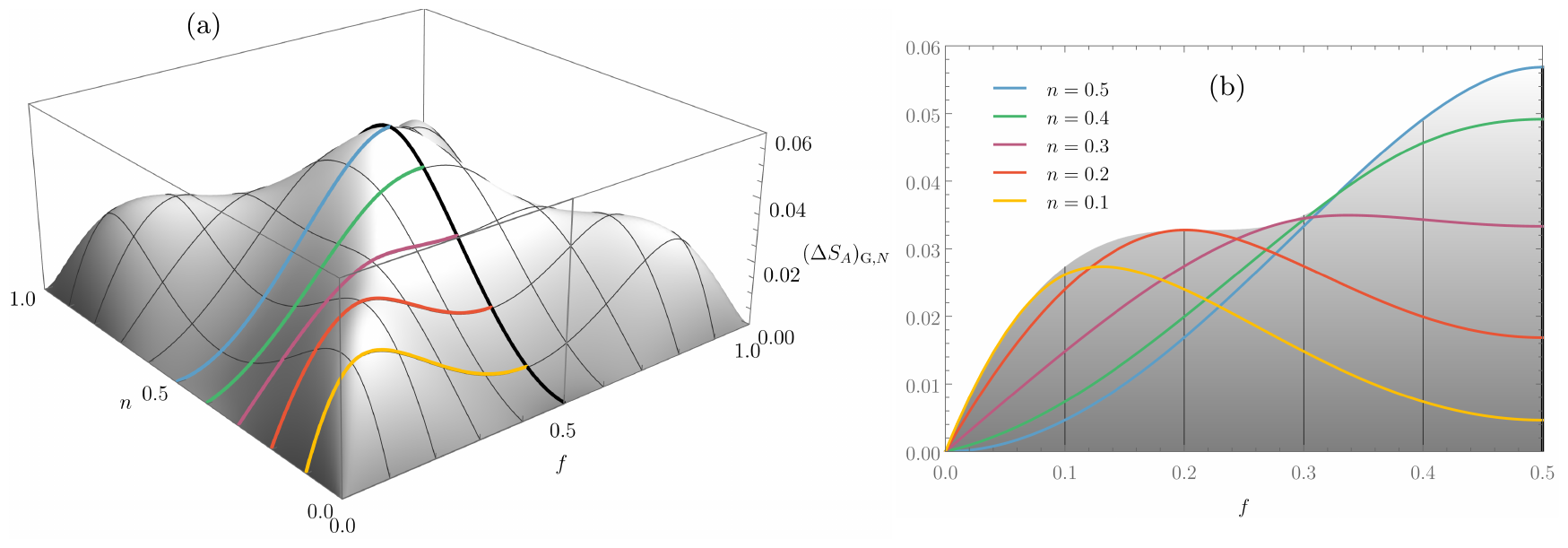}
	\caption{The leading order of the standard deviation $(\Delta S_A)_{\mathrm{G},N}$ of the entanglement entropy $S_A$, from Eq.~\eqref{eq:variance-Gaussian}. (a) Three-dimensional plot as a function of the subsystem fraction $f=V_A/V$ and the filling ratio $n=N/V$. (b) Results at fixed $n$ plotted as functions of $f$. The colored curves in the right plot are the sections with the same color in the left plot.}
	\label{fig:variance}
\end{figure*}
	
The average entanglement entropy over all pure fermionic Gaussian states with total particle number $N$ and a subsystem volume $V_A$ out of a total volume $V$ (with $V_A\leq N\leq V/2$) is
\begin{align}
	\braket{S_A}_{\mathrm{G},N}=V_A\int^1_{-1}R_1(x)\, s(x)\, dx\,,
\end{align}
with $s(x)$ from Eq.~\eqref{eq:Gaussian-s} and $R_1(x)$ is the one point function. The evaluation of this integral is explained in detail in Appendix~\ref{app:average}. We obtain
\begin{align}
	\begin{split}\label{eq:SA-full}
		\hspace{-2mm}\braket{S_A}_{\mathrm{G},N}&=1-\frac{V_A}{V}(1\!+\!V)-\frac{N V_A}{V}\Psi(N)+V\Psi(V)\\
		&\quad+\!\frac{V_A(N-V)}{V}\Psi(V\!-\!N)\!+\!(V_A\!-\!V)\Psi(V\!-\!V_A\!+\!1)
	\end{split}
\end{align}
for $V_A\leq N\leq V/2$, where $\Psi(x)=\Gamma'(x)/\Gamma(x)$ is the digamma function. All other values of $N$ and $V_A$ can be computed by using the fact that the entanglement entropy is symmetric under $N\to V-N$, $V_A\to V-V_A$, and $N\leftrightarrow V_A$. Let us emphasize that Eq.~\eqref{eq:SA-full} is already symmetric under $N\to V-N$. This will play an important role when identifying $\sqrt{V}$ contributions for averages at fixed weight parameter $w$.
	
If we define $n=N/V$ and $f=V_A/V$, we can expand this formula in $V$ to find the thermodynamic limit
\begin{align}
\braket{S_A}_{\mathrm{G},N}&\!=\!\left((f\!\!-\!\!1)\ln(1\!\!-\!\!f)\!\!+\!\!f\left[(n\!\!-\!\!1)\ln(1\!\!-\!\!n)\!\!-\!\!n\ln{n}\!\!-\!\!1\right]\right)V\nonumber\\
&\quad+\frac{f[1-f+n(1-n)]}{12(1-f)(1-n)n}\frac{1}{V}+O(V^{-3})\,,\label{eq:Gaussian-expansion}
\end{align}
where we assume that $f\leq n\leq \frac{1}{2}$. One can use the symmetries discussed in Fig.~\ref{fig:phase-transitions} to find $\braket{S_A}_{\mathrm{G},N}$ for the other parameters. We note that the leading orders for $f\leq \frac{1}{2}$ and $n=\frac{1}{2}$ read
\begin{align}\label{eq:sA-half}
\begin{split}
    \braket{S_A}_{\mathrm{G},N=V/2}&=[(\ln{2}-1)f+(f-1)\ln(1-f)]V\\
    &\quad +\frac{(5/4-f)f}{3(1-f)}\frac{1}{V}+O(V^{-3}).
\end{split}
\end{align}
Remarkably, the leading-order in $V$ is the same as that found in Ref.~\cite{lydzba2020entanglement} for the average over eigenstates of number-conserving random Hamiltonians, and in Ref.~\cite{bianchi2021page} for the ensemble of all fermionic Gaussian states. Why this is no coincidence will become apparent in Sec.~\ref{sec:Gaussian-mu}. In Fig.~\ref{fig:Page-Gaussian}, we visualize our analytical results for the leading order of $\braket{S_A}_{\mathrm{G},N}$.
	
We compute the variance $(\Delta S_A)_{\mathrm{G},N}$ using the same strategy as Eq.~\eqref{eq:Gaussian-variance} based on $s_{ij}$ from Eq.~\eqref{eq:sij-value}, where $\psi_i(x)$ now comes from Eq.~\eqref{eq:def-psi}. In Appendix~\ref{app:variance}, we study the asymptotics of $s_{ij}$ around the leading contribution $s_{V_A-1,V_A}$ in the limit $V\to\infty$ at a fixed subsystem fraction $f=V_A/V$ and particle number $n=N/V$. We find that
\begin{align}
	(\Delta S_A)^2_{\mathrm{G},N}&=\ln(1\!-\!f)\!+\!f\!+\!f^2\!+\!f^2(2n\!-\!1)\ln\left(\tfrac{1}{n}\!-\!1\right)\nonumber\\
	&\quad+f(f-1)(n-1)n\ln^2\left(\tfrac{1}{n}-1\right)+o(1),\label{eq:variance-Gaussian}
\end{align}
for $0<f\leq n\leq \tfrac{1}{2}$. We visualize this result in Fig.~\ref{fig:variance}. At $n=\frac{1}{2}$, the expression above simplifies to $\lim_{V\to\infty} (\Delta S_A)^2_{\mathrm{G},N} (f,\tfrac{1}{2})=f+f^2+\ln(1-f)$, which is exactly twice the variance~\eqref{eq:Gaussian-variance} found for the entanglement entropy of \emph{all} fermionic Gaussian states~\cite{bianchi2021page}.\\
	
\subsubsection{Weighted average and variance} \label{sec:Gaussian-mu}
	
Following the definitions in Sec.~\ref{sec:general-mu}, we can define the weighted average entanglement entropy of Gaussian states
\begin{align}
	\braket{S_A}_{\mathrm{G},w}=\sum^V_{N=0}P_N\braket{S_A}_{\mathrm{G},N},\label{eq:sum-chem-gauss}
\end{align}
where $P_N$ is a function of $w$ as defined in Eq.~\eqref{eq:PN-binomial}, and produces an average filling ratio $\bar{n}=\bar{N}/V=1/(1+\ee^w)$ [see Eq.~\eqref{eq:nbar}]. In the thermodynamic limit, $V\to \infty$, we can approximate the binomial distribution by a continuous probability distribution $\varrho_w(n)$, which approaches a Gaussian plus corrections (see Appendix~\ref{app:weighted}) that becomes increasingly peaked at $n=\bar{n}$.
	
\begin{figure*}
	\centering  
	\includegraphics[width=\linewidth]{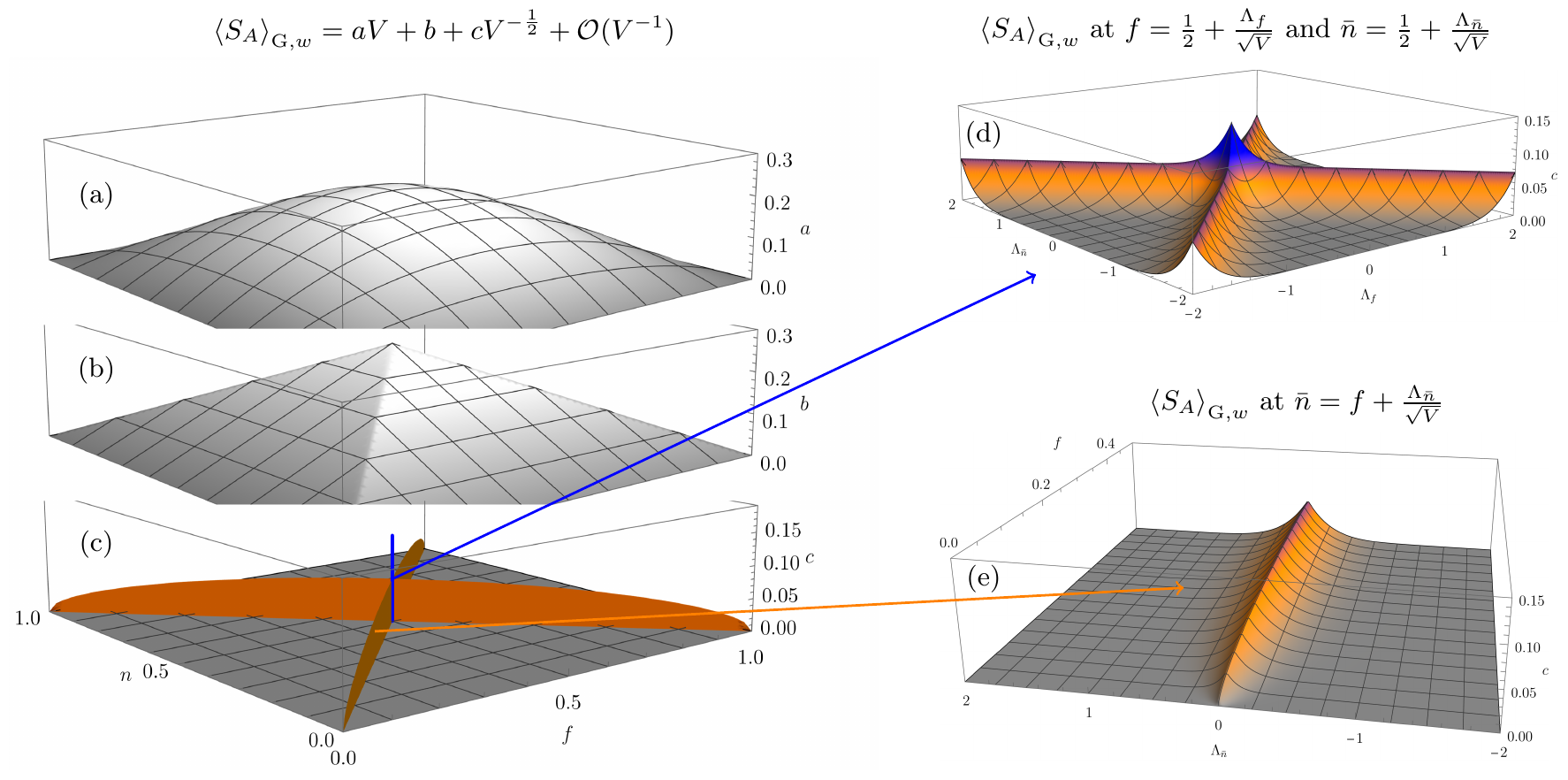}
	\caption{The subsystem entanglement entropy $\braket{S_A}_{\mathrm{G},w}$ from Eq.~\eqref{eq:SAmu-asymp} as viewed from the contributions of the first three terms in the expansion in $V$. (a)--(c) Three-dimensional plots as functions of the subsystem fraction $f=V_A/V$ and the filling ratio $n=N/V$. (d) Resolving $b$ for $\bar{n} = \frac{1}{2} + \frac{\Lambda_{\bar{n}}}{\sqrt{V}}$ and $f = \frac{1}{2} + \frac{\Lambda_f}{\sqrt{V}}$ around $f=\bar{n}=\frac{1}{2}$, as given by Eq.~\eqref{eq:center-gaussian-weighted}. (e) Resolving discontinuities of $c$ for $\bar{n}=f+\frac{\Lambda_{\bar{n}}}{\sqrt{V}}$ around $f=\bar{n}<\frac{1}{2}$ given by Eq.~\eqref{eq:chemGaussiangeneral}.}
	\label{fig:Gaussian-mu-visual}
\end{figure*}
	
When we average over all Gaussian states with a fixed number of particles, a natural weight is given when $w=0$, \ie we weigh each particle-number sector by its Hilbert space dimension $\Omega_N$. If we would draw a random eigenstate of a random quadratic number-conserving Hamiltonian (see Sec.~\ref{sec:rig-res-rmt}), the resulting \emph{eigenstate entanglement entropy} will thus correspond to $w=0$. The leading-order average for eigenstates of such Hamiltonians was derived in Ref.~\cite{lydzba2020entanglement}, and was later shown numerically~\cite{lydzba2021entanglement} and analytically~\cite{bianchi2021page} to coincide with the leading-order average over eigenstates of random quadratic Hamiltonians without number conservation or, equivalently, over all fermionic Gaussian states. The present calculation explains these coincidences by showing explicitly how the average at $w=0$ corresponds to the peak of the binomial distribution at $n=\bar{n}=\tfrac{1}{2}$, so at leading order $\braket{S_A}_{\mathrm{G},w=0}=\braket{S_A}_{\mathrm{G}}+O(1)$.

\begin{figure*}[!t]
	\centering  
	\includegraphics[width=\linewidth]{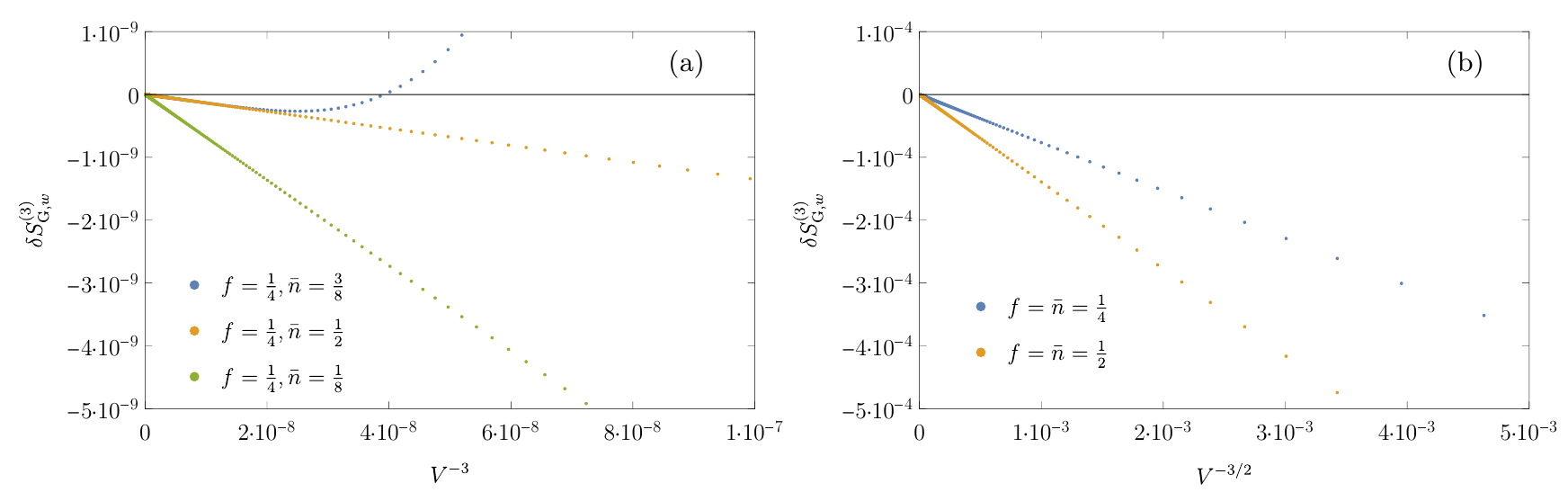}
	\caption{Asymptotics of $\braket{S_A}_{\mathrm{G},w}$. We compare the exact values of the average entanglement entropy $\braket{S_A}_{\mathrm{G},w}$ computed numerically evaluating the sum in Eq.~\eqref{eq:sum-chem-gauss} using Eq.~\eqref{eq:Gaussian-expansion} for $V\leq 2000$, with the asymptotic results, Eq.~\eqref{eq:expansion}, \ie we show the difference $\delta S^{(3)}_{\mathrm{G},w}=\braket{S_A}_{\mathrm{G},w}-\braket{S_A}^{(3)}_{\mathrm{G},w}$, where $\braket{S_A}^{(3)}_{\mathrm{G},w}$ corresponds to expansion~\eqref{eq:expansion} up to order $1/V$. In particular, one can see that the next order is $1/V^{3/2}$ if $f=\bar{n}$, and $1/V^3$ otherwise.}
	\label{fig:asymptotics}
\end{figure*}
	
We calculate the binomial average over $N$ analytically up to order $1/V$ in Appendix~\ref{app:weighted}. The resulting leading-order behavior, as a function of $\bar{n}$ and $f$, is given by
\begin{widetext}
	\begin{align}\label{eq:expansion}
		\hspace{-3mm}\braket{S_A}_{\mathrm{G},w}&=\begin{cases}
			[(f-1)\ln(1\!-\!f)\!+\!f((\bar{n}\!-\!1)\ln(1\!-\!\bar{n})\!-\!1-\!\bar{n}\ln{\bar{n}})]V
			-\frac{f}{2}
			+\frac{(f-2)f}{12(f-1)}\frac{1}{V}
			+O(1/V^3),
			& f<\bar{n}\leq\frac{1}{2},\\
			[(f\!-\!1)\bar{n}\ln(1\!-\!f)\!-\!\bar{n}(1\!+\!f\ln{f})\!+\!(\bar{n}\!-\!1)\ln(1\!-\!\bar{n})]V
			-\frac{\bar{n}}{2}
			+\frac{\bar{n}(1-f+f^2)}{12f(1-f)}\frac{1}{V}+O(1/V^3),
			& \bar{n}<f\leq\frac{1}{2},\\
			[(f^2-1)\ln(1-f)-f(1+f\ln{f})]V
			-\frac{f}{2}
			+\sqrt{\frac{(1-f)f}{18\pi}}\frac{1}{\sqrt{V}}
			+\frac{1+f}{24(1-f)}\frac{1}{V}
			+O(1/V^{3/2}),
			& f=\bar{n}<\frac{1}{2},\\
			[\ln{2}-\frac{1}{2}]V
			-\frac{1}{4}
			+\frac{1}{3\sqrt{2\pi}}\frac{1}{\sqrt{V}}
			+\frac{1}{8}\frac{1}{V}
			+O(1/V^{3/2}),
			& f=\bar{n}=\frac{1}{2},\\
		\end{cases}
	\end{align}
	for $f,\bar{n}\leq\frac{1}{2}$. To $O(1/\sqrt{V})$, the above results can be recast in the following compact equation
	\begin{align}\label{eq:SAmu-asymp}
		\hspace{-3mm}\begin{array}{l}
			\braket{S_A}_{\mathrm{G},w}=[(\mu_--1)\ln(1-\mu_-)+\mu_-((\mu_+-\!1)\ln(1-\mu_+)-1-\mu_+\ln{\mu_+})]V\\
			\quad\qquad\qquad-\frac{\mu_-}{2}+\,\delta_{f,\bar{n}}\left(\sqrt{\frac{f(1-f)}{18\pi}}+\delta_{f,\frac{1}{2}}\frac{1}{6\sqrt{2\pi}}\right)\frac{1}{\sqrt{V}}+O(\tfrac{1}{V}),
		\end{array}\qquad\text{with}\qquad\begin{array}{l}
			\mu_+=\max(f,\bar{n})\\[1mm]
			\mu_-=\min(f,\bar{n})
		\end{array},
	\end{align}
\end{widetext}
Equation~\eqref{eq:SAmu-asymp} is visualized in Figs.~\ref{fig:Gaussian-mu-visual}(a)--\ref{fig:Gaussian-mu-visual}(c). Note that $\braket{S_A}_{\mathrm{G},w}$ satisfies the particle-subsystem symmetry $\bar{n}\leftrightarrow f$ only up to $\frac{1}{\sqrt{V}}$, which is not surprising considering that this symmetry is only exact for averages at \emph{fixed} $N$. We compared our analytical results with those of numerical calculations. In Fig.~\ref{fig:asymptotics}, we show some of the finite-size scaling analyses that we carried out.
	
We note that the asymptotic behavior of the average entanglement entropy $\braket{S_A}_{\mathrm{G},w}(f)$ is characterized by a nonanalytic behavior along the symmetry axes $f=\bar{n}$ and $f=1-\bar{n}$. This gives rise to distinct corrections in the thermodynamic limit. Most importantly, we have an enhancement of order $1/\sqrt{V}$ whenever $f=\bar{n}$ and, in particular, at $f=\bar{n}=\frac{1}{2}$. These regimes are due to the various symmetries of the average entanglement entropy $\braket{S_A}_{\mathrm{G},N}$ as a function of $V_A$ and $N$. In particular, there is the aforementioned particle-subsystem symmetry, which states that $\braket{S_A}_{\mathrm{G},N}$ is invariant under interchanging $N\leftrightarrow V_A$. This symmetry has to be put in by hand and results in Kronecker deltas as in Page's setting, although there the Kronecker deltas resulted from a different symmetry. In Appendix~\ref{app:res-Gaussian-fixed-w}, we resolve the Kronecker deltas in Eq.~\eqref{eq:SAmu-asymp} by studying the asymptotics for either $\bar{n}=f+\frac{\Lambda_f}{\sqrt{V}}$ or $\bar{n}=\frac{1}{2}+\frac{\Lambda_f}{\sqrt{V}}$ and $f=\frac{1}{2}+\frac{\Lambda_f}{\sqrt{V}}$. The resulting expressions~\eqref{eq:center-gaussian-weighted} and~\eqref{eq:chemGaussiangeneral} are visualized in Figs.~\ref{fig:Gaussian-mu-visual}(d) and~\ref{fig:Gaussian-mu-visual}(e).
	
When computing the average entanglement entropy $\braket{S_A}_{\mathrm{G},w}$ from Eq.~\eqref{eq:fequaln} for weight parameter $w$ and subsystem size $V_A\leq V/2$, one expects transitions to occur whenever the expectation value $\bar{N}=V\bar{n}=V e^{-w}/(1+e^{-w})$ crosses a symmetry axis, where there is a discontinuity in the third derivative of $\braket{S_A}_{\mathrm{G},N}$. Interestingly, we only have a transition at $\bar{n}=f$ (\ie $\bar{N}=V_A$) and $\bar{n}=f=\frac{1}{2}$, but not at $\bar{n}=\frac{1}{2}$ (\ie $\bar{N}=V/2$) due to the fact that Eq.~\eqref{eq:SA-full} is symmetric in $N\leftrightarrow V-N$, so that our function $\braket{S_A}_{\mathrm{G},N}$ is analytic across the particle-hole symmetry. The reason for this is that $\braket{S_A}_{\mathrm{G},N}$ has continuous derivatives up to order three at $\bar{n}=\frac{1}{2}$, so that we would not see a transition if we only expand up to order $1/V$. However, we expect that the fifth-order term of $s^{\mathrm{G}}_A(f,n)=\lim_{V\to\infty}\braket{S_A}_{\mathrm{G},N}/V$ [see also Eq.~\eqref{eq:sAG}] expanded in powers of $(\bar{n}-n)$ will contribute a square root enhancement of order $V^{-3/2}$ for $\bar{n}=\frac{1}{2}$.
	
Finally, let us also comment about the leading order of the variance at fixed weight parameter $w$. The leading-order contribution is due to the variance $(\Delta n)^2$ in the number of particles, Eq.~\eqref{eq:Dn}. As a result, while the variance at fixed particle number is $O(1)$, Eq.~\eqref{eq:variance-Gaussian}, the variance at fixed $w$ scales linearly with the volume
\begin{align}
	&(\Delta S_A)^2_{\mathrm{G},w}=\label{eq:variance-Gw}\\
	&\begin{cases}
		\scriptstyle\bar{n}(1-\bar{n})\big[\ln(\tfrac{\bar{n}}{1-\bar{n}})\big]^2\,f^2\,V+o(V),& f\leq \bar{n}\\[.5em]
		\scriptstyle \bar{n}(1-\bar{n})\big[(1-f)\ln(1-f)+f \ln f+\ln(1-\bar{n})\big]^2\,\,V+o(V), & f>\bar{n}
	\end{cases}\nonumber
\end{align}
with $f\leq \frac{1}{2}$. Note that, at $w=0$ (corresponding to $\bar{n}=\frac{1}{2}$), the leading-order $O(V)$ term vanishes. In general, we have $\lim_{V\to \infty} (\Delta S_A)_{\mathrm{G},w}/\braket{S_A}_{\mathrm{G},w}=0$, which shows that in the thermodynamic limit the average \eqref{eq:expansion} also gives the typical value of the entanglement entropy.

\section{EXACT RELATION TO RANDOM HAMILTONIANS}\label{sec:RMT}
	
So far, we have focused on ensembles of quantum states and computed statistical properties of the entanglement entropy with respect to the following six ensembles: (1a) random states, (2a) random states with fixed total particle number, (3a) weighted averages over random states with fixed total particle number, (1b) random fermionic Gaussian states, (2b) random fermionic Gaussian states with fixed total particle number, and (3b) weighted averages over random fermionic Gaussian states with fixed total particle number. In this section, we shift the focus from ensembles of quantum states to random Hamiltonians, their eigenstates, and their dynamics.
	
\subsection{Random many-body Hamiltonians}\label{sec:rig-res-rmt}
	
Ensembles (1a), (2a), and (3a) can be realized using eigenstates (even only ground states) of random Hamiltonians that are traditional random matrices. The ensuing Hamiltonians give an \emph{exact} correspondence to Page's setting, \ie the averages and variances will agree at all orders (meaning even at finite $V$) when the respective random Hamiltonian satisfies the properties discussed next.
	
We first consider case (1a), for which the number of particles is not fixed. The state vector in this case explores the entire sphere of the Hilbert space $\mathcal{H}$. Thus, any random Hamiltonian that creates a Haar-distributed random state vector is suitable. For instance, let us study the random-matrix Hamiltonian
\begin{equation}
	\hat{H}_\text{1a}=\sum^{2^V}_{\kappa,\lambda=1}C_{\kappa\lambda}\ket{v_\kappa}\bra{v_\lambda},
\end{equation}
where $\ket{v_\lambda}$ is an orthonormal basis of the Hilbert space and $C_{\kappa\lambda}$ is a Haar-distributed random matrix. To get Haar-distributed eigenvectors, the diagonalization $C=U^\dagger E U$ must involve random matrices $U$ drawn from the Haar measure of ${\rm U}(2^V)$, while the distribution of the eigenvalues appearing in the diagonal matrix $E$ can be arbitrary. A simple, and one of the most common examples of such a distribution for $C$ is given by the GUE~\cite{mehta2004, Forrester_2010, akemann2011},
\begin{equation}\label{GUE-dist}
	\begin{split}
		P(\hat{H}_\text{1a})=&2^{-2^{V-1}}\pi^{-2^{2V-1}}\exp\left[-\frac{1}{2}\sum_{\kappa,\lambda=1}^{2^V}|C_{\kappa\lambda}|^2\right]\\
		=&2^{-2^{V-1}}\pi^{-2^{2V-1}}e^{-\Tr\hat{H}_\text{1a}^2/2}.
	\end{split}
\end{equation}
	
To relate the Hamiltonian $\hat{H}_\text{1a}$ to many-body Hamiltonians, we rewrite it into a polynomial in fermionic creation and annihilation operators
\begin{align}\label{H1a}
	\hat{H}_\text{1a}\;\;&=\sum_{l=0}^{2V}\sum_{j_1,\ldots,j_l=1}^{2V}c^{(l)}_{j_1\ldots j_l}\,\hat{\xi}_{j_1}\cdots\hat{\xi}_{j_l},
\end{align}
with $\{\hat{\xi}_j\}_{j=1,\ldots,2V} = (\hat{f}_1, \dots, \hat{f}_V, \hat{f}_1^\dagger, \dots, \hat{f}_V^\dagger)$. The coefficients $c^{(l)}_{j_1,\ldots,j_{l}}$ satisfy symmetries that reflect the anticommutation relations, $\{\hat{f}_k,\hat{f}_l\} = \{\hat{f}_k^\dagger,\hat{f}_l^\dagger\}=0$ and $\{\hat{f}_k,\hat{f}_l^\dagger\} = \delta_{kl}$, the Hermiticity of $\hat{H}_\text{1a}$, and the fact that in each sum over $c^{(l)}_{j_1,\ldots,j_{l}}$ there are exactly $l$ operators involved that cannot be reduced to a smaller order of a many-body interaction. Exploiting the unitary matrix $T$ in Eq.~\eqref{Tdef}, in particular going into a Majorana basis, shows that $\tilde{c}^{(l)}_{k_1,\ldots,k_{l}}=\sum_{j_1,\ldots,j_l=1}^{2V}c^{(l)}_{j_1,\ldots,j_{l}}\prod_{a=1}^lT_{j_a k_a}$ is totally skew symmetric in the indices and is real when $l(l-1)/2$ is even and imaginary when $l(l-1)/2$ is odd.
	
The statistical distribution of the coefficients $c^{(l)}_{j_1,\ldots,j_{l}}$ is determined by the distribution of matrix $C_{\mu\nu}$. The best way to see this is to go into the Majorana basis $\gamma_1,\ldots,\gamma_{2V}$ via relation~\eqref{gammafrel}. Then, one needs to take into account the normalization $\gamma_j^2=\tfrac{1}{2}\id_{2^V}$ to determine this distribution, which leads to
\begin{equation}
	\begin{split}
		P(\hat{H}_\text{1a})&=\prod_{l=1}^{2V}\prod_{1\leq j_1<\ldots<j_{l}\leq l}\sqrt{\frac{2^{V-l-1}l!}{\pi}}\\
		&\quad\times \exp\left[-2^{V-l-1}\,l!|\tilde{c}^{(l)}_{j_1,\ldots,j_{l}}|^2\right].
	\end{split}
\end{equation}
To derive this result, one needs to use the fact that the trace of a product of $\gamma_j$ is only nonvanishing when each $\gamma_j$ appears with an even number in this product.
	
The statistical properties of the entanglement entropy in eigenstates of the random Hamiltonian $\hat{H}_\text{1a}$ are described exactly by the results of Sec.~\ref{sec:generalpage} since the Hamiltonian is invariant under the conjugations of the unitary group ${\rm U}(2^V)$. Hence, its eigenstates are uniformly distributed over the unit sphere in $\mathcal{H}$. We emphasize that this Hamiltonian is not parity preserving such that the superselection rule (either only even or odd powers in $\xi_j$) does not apply.
	
SYK models~\cite{1993PhRvL..70.3339S, 2016PhRvD..94j6002M} are related to this construction. For the $q$-body SYK (or in short SYK$q$), one sets $c_{j_1,\ldots,j_{l}}^{(l)}=0$ for all $l\neq q$, and chooses the Gaussian distribution for $c_{j_1,\ldots,j_{q}}^{(q)}$, as we have done here. For a fixed $q$ these models have more symmetries than we have by adding up all $q$. Thus, they reflect all kinds of random matrix symmetry classes (actually one can find all ten classes of the Altland-Zirnbauer classification~\cite{Zirnbauer:1996zz, 1997PhRvB..55.1142A}) and follow the Bott periodicity~\cite{Bott} in $V$ and $q$; see Refs.~\cite{Kanazawa:2017dpd, Garcia21}. When mixing SYK$q$ with different $q$'s, it is likely that in the large-$V$ limit one ends up in the same class as our random Hamiltonian $\hat{H}_\text{1a}$. However, for a fixed $q$, it might happen that the subleading orders differ from our results.
	
We turn now to case (2a), in which we need to implement number conservation in the random Hamiltonian. Based on the direct sum decomposition~\eqref{eq:Hilbert-sum} of the Hilbert space, we define an orthonormal basis $\ket{v_\mu^{(N)}}$ of the $N$-particle Hilbert space $\mathcal{H}^{(N)}$ of dimension $d_N$ given in Eq.~\eqref{eq:dN}. To parallel (2a), we consider a random-matrix Hamiltonian in this particle sector given by
\begin{equation}
	\hat{H}_\text{2a}^{(N)}=\sum^{d_N}_{\kappa,\lambda=1}\tilde{C}_{\kappa\lambda}^{(N)}\ket{v_\kappa^{(N)}}\bra{v_\lambda^{(N)}}\,,
\end{equation}
where the Hermitian matrix $\tilde{C}^{(N)} = \{\tilde{C}_{\kappa\lambda}^{(N)}\}_{\kappa,\lambda = 1,\ldots,d_N}$ is assumed to be a ${\rm U}(d_N)$ group invariant random matrix, \ie it can be diagonalized, $\tilde{C}^{(N)}=U^\dagger EU$, via a Haar-distributed unitary matrix $U\in{\rm U}(d_N)$. Anew, the GUE is a simple example of such a distribution [in particular, Eq.~\eqref{GUE-dist} with $2^V$ replaced by $d_N$], but the class is more general and does not constrain the diagonal matrix $E$ that comprises the energies.
	
As before, we want to express the Hamiltonian $\hat{H}^{(N)}_\text{2a}$ in terms of a polynomial in fermionic creation and annihilation operators. This cannot be done without adding an orthogonal projection onto the Hilbert space with $N$ particles. Instead of $\hat{H}^{(N)}_\text{2a}$, we consider the direct sum
\begin{equation}
	\hat{H}_\text{2a/3a}=\bigoplus_{N=0}^V\hat{H}^{(N)}_\text{2a},
\end{equation}
which is now expressible in terms of $(\hat{f}_1, \dots, \hat{f}_V, \hat{f}_1^\dagger, \dots, \hat{f}_V^\dagger)$. While an arbitrary random Hamiltonian $\hat{H}^{(N)}_\text{2a}$ has $d_N^2$ degrees of freedom, the random Hamiltonian $\hat{H}_\text{2a/3a}$ has $\sum_{N=0}^Vd_N^2=\binom{2V}{V}$ degrees of freedom. The subscript already indicates that this random Hamiltonian also describes case (3a), where one averages over states with different particle numbers.
	
Another subtle point is that particle-number conservation does not allow any odd powers of these operators nor does it allow an unbalanced number of creation and annihilation operators. Thus, the general form of $\hat{H}_\text{2a/3a}$ is given by
\begin{align}\label{H2a3a}
	\hat{H}_\text{2a/3a}= \sum_{l=0}^{V} \sum_{j_1,\ldots,j_l,k_1,\ldots,k_l=1}^V\hspace*{-0.4cm}c^{(l)}_{j_1\ldots j_l,k_1\ldots k_l}\hat{f}^\dagger_{j_1}\cdots\hat{f}_{j_l}^\dagger\hat{f}^{}_{k_l}\cdots\hat{f}_{k_1},
\end{align}
which, by construction, commutes with the particle-number operator. The coefficients $c^{(l)}_{j_1\ldots j_l,k_1\ldots k_l}$ are totally skew symmetric in the first $l$ indices as well as skew symmetric in the last $l$ ones. To ensure the Hermiticity of $\hat{H}_\text{2a/3a}$, it holds that the $\binom{V}{l}\times\binom{V}{l}$ matrix $C^{(l)}$ created by $c^{(l)}_{j_1\ldots j_l,k_1\ldots k_l}$ for a fixed $l$ is Hermitian. It is simple to check that the degrees of freedom given by all $c^{(l)}_{j_1\ldots j_l,k_1\ldots k_l}$ for all $l$ add up to the formerly obtained $\binom{2V}{V}$.
	
The statistical distribution of the coefficients $c^{(l)}_{j_1\ldots j_l,k_1\ldots k_l}$ is determined by the distribution of the matrices $\tilde{C}^{(N)}$ for all $N$. Since the coefficients $c^{(l)}_{j_1\ldots j_l,k_1\ldots k_l}$ are linear combinations of matrix elements of all $\tilde{C}^{(0)},\ldots,\tilde{C}^{(V)}$, their distributions will in general be coupled despite the fact that one can choose the distributions of $\tilde{C}^{(N)}$ to be completely independent and not identical.
	
As mentioned before, the simplest choice are GUEs in $\tilde{C}^{(N)}$ with the same variance, which yields
\begin{equation}\label{H2a}
	P(\hat{H}_\text{2a/3a})=\frac{\sqrt{\det\Sigma}}{2^{2^{V-1}}\pi^{\ell^2/2}}\exp\left[-\frac{1}{2}\Tr \hat{H}_\text{2a/3a}^2\right],
\end{equation}
where $\ell=\binom{2V}{V}$. The matrix $\Sigma$ is $\ell\times\ell$ dimensional, with matrix entries $\Sigma_{j_1\ldots j_lj'_1\ldots j'_l,k_1\ldots k_mk'_1\ldots k'_m}$ equal to $(l!)^2(m!)^2\Tr[ \hat{f}_{j'_1}^\dagger\cdots\hat{f}_{j'_l}^\dagger \hat{f}_{j_l}\cdots \hat{f}_{j_1}\hat{f}_{k_1}^\dagger\cdots\hat{f}_{k_m}^\dagger \hat{f}_{k'_m}\cdots \hat{f}_{k'_1}]$ with $l,m=0,\ldots,V$ and $1\leq j_1<\ldots<j_l\leq V$,  $1\leq j'_1<\ldots<j'_l\leq V$, $1\leq k_1<\ldots<k_m\leq V$ as well as $1\leq k'_1<\ldots<k'_m\leq V$. It is very sparse because the trace is only nonvanishing when each $\hat{f}_j$ appears as often as its Hermitian conjugate $\hat{f}_j^\dagger$, and it might happen that there are operators that annihilate each other such as $\hat{f}_j^2=(\hat{f}_j^\dagger)^2=0$. Thence, $\Sigma$ only contains entries equal to $\pm 2^{L}(l!)^2(m!)^2$ with an $L\in\{0,\ldots,V\}$. The factor $1/\pi^{\ell^2/2}$ reflects the number of total degrees of freedom while $1/2^{2^{V-1}}$ results from the counting of how many real coefficients exist, namely, $\sum_{l=0}^V\binom{V}{l}=2^V$. The simplest nontrivial example is obtained for $V=2$ where $\Sigma$ is equal to
\begin{equation}
	\Sigma=\left(\begin{array}{cccccc} 4 & 2 & 0 & 0 & 2 & 4 \\ 2 & 2 & 0 & 0 & 1 & 4 \\ 0 & 0 & 0 & 1 & 0 & 0 \\ 0 & 0 & 1 & 0 & 0 & 0 \\ 2 & 1 & 0 & 0 & 2 & 4 \\ 4 & 4 & 0 & 0 & 4 & 4 \end{array}\right)
\end{equation}
with the ordering of the coefficients $(c^{(0)}, c_{1,1}^{(1)}, c_{1,2}^{(1)}, c_{2,1}^{(1)}, c_{2,2}^{(1)}, c_{12,12}^{(2)})$. This example underscores how intertwined the correlations of the coefficients $c^{(l)}_{j_1\ldots j_l,k_1\ldots k_l}$ can become if one wants to write the exact realization of cases (2a) and (3a) in terms of annihilation and creation operators. Thence, from a practical point of view, it is simpler to generate these random Hamiltonians in terms of the independent GUE generated coefficients $\tilde{C}_{\kappa\lambda}^{(N)}$.
	
When focusing on a certain particle-number sector, the statistical properties of the entanglement entropy in eigenstates of $\hat{H}_\text{2a/3a}$ are described exactly by the results of Sec.~\ref{sec:page-fixedN}, meaning case (2a). When considering all sectors, we have case (3a). For distribution~\eqref{H2a}, picking any eigenstate is equally likely, this yields the weight $\binom{V}{N}$ to find a state in the sector $\mathcal{H}^{(N)}$, which corresponds to a weighted average with $w=0$. Implementing a weighted average with $w>0$ is also possible, but must be largely done by hand, \ie we would organize the eigenstates of a random Hamiltonian based on their particle number and then choose one at random using the statistical weight encoded by $w$.
	
Many-body interacting Hamiltonians studied in nuclear physics~\cite{monfrench1975, FRENCH1970449, BOHIGAS1971261, BOHIGAS1971383, PhysRev.120.1698, FRENCH19715} are related to these kinds of Hamiltonians. They, as well as the SYK models, are called embedded random matrices~\cite{RevModPhys.53.385, Guhr1998, Benet:2000cy, Kota2001, Kota2014}. For instance, for a $q$-body Hamiltonian, we set $c^{(l)}_{j_1\ldots j_l,k_1\ldots k_l}=0$ for all $l\neq q$ and choose the above Gaussian distribution for $c^{(l)}_{j_1\ldots j_l,k_1\ldots k_l}$. As for case (1a) and SYK$q$ for a fixed $q$, the many-body Hamiltonian may satisfy additional global symmetries so that subleading terms may deviate from our results. However, we expect that a mixture of $q$-body interactions should speed up the convergence to the leading-order result in the thermodynamic limit $V\to\infty$.
	
\subsection{Random quadratic Hamiltonians}
	
Case (1b) for random pure fermionic Gaussian states is obtained from $\hat{H}_\text{1a}$ by setting all coefficients $c^{(l)}_{i_1, \ldots, i_{l}} = 0$ whenever $l\neq2$ in Eq.~\eqref{H1a}; the resulting random quadratic Hamiltonian reads
\begin{align}
	\hat{H}_\text{1b}&=\sum^{2V}_{i,j=1}c^{(2)}_{ij}\,\hat{\xi}_i\hat{\xi}_j\,,
\end{align}
with coefficients $c^{(2)}_{ij}$, drawn from a probability distribution that depends only on matrix invariants of $TC_{(2)}T^T$ with $C_{(2)}=\{c^{(2)}_{ij}\}_{i,j=1,\ldots,2V}$, such as traces $\Tr (TC_{(2)}T^T)^{2k}$. Then the invariance under ${\rm O}(2V)$ is guaranteed, which is needed for the uniformly distributed pure fermionic Gaussian states that are the eigenvectors of this Hamiltonian. The Gaussian choice as the distribution of the coefficients $c^{(2)}_{ij}$ is equal to
\begin{equation}
	\begin{split}
		P(\hat{H}_\text{1b})=&\prod_{1\leq j_1<j_2\leq 2V}\sqrt{\frac{2^{V}}{\pi}}e^{-2^{V}|\tilde{c}^{(2)}_{j_1,j_2}|^2},
	\end{split}
\end{equation}
when starting from the distribution of $\hat{H}_\text{1a}$. Since $C_{(2)}$ is unitarily equivalent to the real antisymmetric $2V\times 2V$ matrix $2TCT^T$, which can be seen in the Majorana basis, the Gaussian ensemble is also known as the Gaussian real antisymmetric ensemble~\cite{mehta2004} or GAOE in Ref.~\cite{Kieburg:2017rrk} (it also has other acronyms such as BdG-S in Ref.~\cite{Garcia21} or class BD in Refs.~\cite{Zirnbauer:1996zz, 1997PhRvB..55.1142A} as it is not as canonical as the GUE). We would like to emphasize that the real antisymmetry of $TCT^T$ is the reason why there is no minus sign in the exponent of the distribution, \ie $\Tr(T C_{(2)}T^T)^2<0$. When going over from the coefficients $\tilde{c}_{jk}^{(2)}$ in the Majorana basis to $c^{(2)}_{jk}$ in the creation-annihilation basis, we need to employ $TT^T=\tau_1\otimes\id_V/2$ and properly normalize the distribution (multiplying it by the Jacobian), which gives rise to a factor $2^{-V(V-1)}$. Then, the distribution is
\begin{equation}
	\begin{split}
		P(\hat{H}_\text{1b})=&\frac{2^{V(V-1)(V-3)/2}}{\pi^{V(V-1)/2}}\exp[2^{V-3}\Tr( C_{(2)}[\tau_1\otimes\id_V])^2].
	\end{split}
\end{equation}
Certainly, the factors of $2$ can be absorbed when choosing a general standard deviation for the Gaussian ensemble.
	
The eigenvectors of $\hat{H}_\text{1b}$ are Haar distributed with respect to the group $\mathrm{O}(2V)$, as discussed in Ref.~\cite{hackl2020bosonic}, though they are slightly rotated by the unitary matrix $\sqrt{2}T$; see Eq.~\eqref{Tdef}. The eigenstates of this Hamiltonian are fermionic Gaussian states, and the statistical properties of the entanglement entropy in these eigenstates are described exactly by the results of Sec.~\ref{sec:gaussian-arbitraryN}.
	
For cases (2b) and (3b), we can repeat the fixed-particle-number steps for random quadratic Hamiltonians, specifically, we set $c^{(l)}_{j_1\ldots j_l,k_1\ldots k_l}=0$ for all $l\neq1$ in Eq.~\eqref{H2a3a}. The most general quadratic fermionic Hamiltonian that commutes with the total number operator $\hat{N}$ defined in Eq.~\eqref{eq:Hilbert-sum} can be written as
\begin{align}\label{H2a2b}
	\hat{H}_\text{2b/3b}= \sum^{V}_{i,j=1}\tilde{c}^{(2)}_{i,j}\hat{f}^\dagger_i\hat{f}^{}_j
\end{align}
with the coefficients $\tilde{c}^{(2)}_{i,j}$ drawn from an ensemble that is invariant under the conjugate action of ${\rm U}(V)$; in particular the Hermitian random matrix $\tilde{C}=\{\tilde{c}^{(2)}_{i,j}\}_{i,j=1,\ldots, V}$ and $U\tilde{C}U^\dagger$ with an arbitrary fixed $U\in{\rm U}(V)$ are equally distributed. The simplest example for such a distribution is a $V\times V$ GUE. That one agrees with a quadratic particle-number conserving SYK model in its Dirac fermion formulation (the Dirac SYK2 model for short), which is a free random Hamiltonian. The volume-law coefficient $s_A^{\mathrm{G}}(f,n)=\lim_{V\to\infty}\braket{S_A}_{\mathrm{G},N}/V$ [see Eq.~\eqref{eq:sA-half}] was first conjectured in the context of quadratic Hamiltonians whose single-particle eigenstates can be well approximated by eigenstates of random matrices~\cite{lydzba2020entanglement}.
	
In general, all eigenstates of nondegenerate quadratic random Hamiltonians of this ${\rm U}(V)$ invariance are Gaussian states. Note that degenerate eigenspaces will contain superpositions of Gaussian states, which are not Gaussian themselves. For most random Hamiltonians, the subset of degenerate Hamiltonians is a set of measure zero, so it can be ignored.
	
\begin{figure}[!t]
	\centering
	\includegraphics[width=\linewidth]{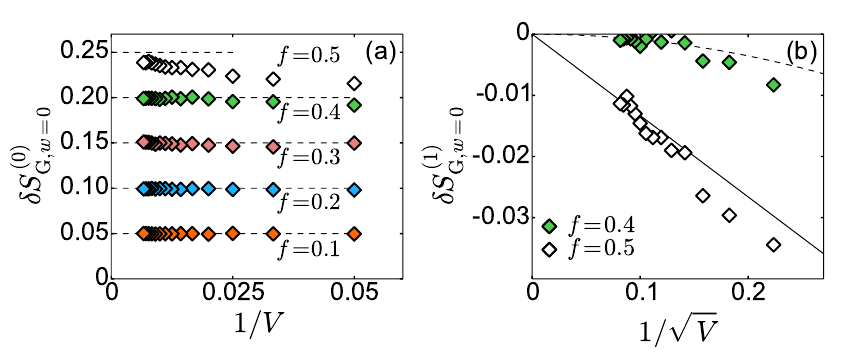}
	\caption{Finite-size scaling of the average entanglement entropy difference for random quadratic fermionic Hamiltonians. (a) Plot of $\delta S_{{\rm G},w=0}^{(0)} = \langle S_A\rangle_{{\rm G},w=0}^{(0)}- \bar S$, where $\langle S_A\rangle_{{\rm G},w=0}^{(0)}$ is the volume-law term in Eq.~\eqref{eq:expansion} and $\bar S$ denotes the numerical results for eigenstates of the particle-number conserving SYK2 model at $\bar n = \frac{1}{2}$, versus $1/V$ at subsystem fractions (from bottom to top) $f=0.1,\,0.2,\,0.3,\,0.4,\,0.5$. Horizontal lines are constants $f/2$. (b) Plot of $\delta S_{{\rm G},w=0}^{(1)} = \langle S_A\rangle_{{\rm G},w=0}^{(1)}- \bar S$ versus $1/\sqrt{V}$ at subsystem fractions $f=0.4$ and 0.5, where $\langle S_A\rangle_{{\rm G},w=0}^{(1)} = \langle S_A\rangle_{{\rm G},w=0}^{(0)} - f/2$. The lines are the second subleading terms from Eq.~\eqref{eq:expansion}. Specifically, the solid line (corresponding to $f=0.5$) is the function $-1/(3\sqrt{2\pi}) (1/\sqrt{V})$, and the dashed line (corresponding to $f=0.4$) is the function $-f(f-2)/[12(f-1)] (1/V)$. The numerical results for $\bar S$ are from Ref.~\cite{lydzba2020entanglement}.} \label{fig:S_syk2_scaling}
\end{figure}
	
Anew, we have treated cases (2b) and (3b) with a single Hamiltonian. For the former case, one needs to restrict the Hamiltonian eigenstates to the sector with $N$ particles, while for the latter case, one needs to compute the weighted average over all eigenstates. For the average over all sectors, we again pick an arbitrary eigenstate of the Hamiltonian $H_{\text{2b/3b}}$. A random many-body eigenstate of this Hamiltonian will be in the sector with $N$ particles with probability $2^{-V}\binom{V}{N}$, so a \emph{typical} eigenstate will have the entanglement entropy $\braket{S_A}_{\mathrm{G},w=0}$ found in Eq.~\eqref{eq:expansion}. If one restricts the analysis to eigenstates with $N$ particles, then the statistical properties of the entanglement entropy are described exactly by the results of Sec.~\ref{sec:aveGaussian}.
	
In Fig.~\ref{fig:S_syk2_scaling}, we show numerical results for the average entanglement entropy of the particle-number conserving SYK2 model~\eqref{H2a2b}. The numerical results in Fig.~\ref{fig:S_syk2_scaling}(a) behave as expected from the analytical predictions for the volume-law contribution~\eqref{eq:expansion} as a function of $V$, and the ones in Fig.~\ref{fig:S_syk2_scaling}(b) behave as expected from the analytical predictions for the subleading terms given in Eq.~\eqref{eq:expansion}. In the simulations, the average is carried out over all sectors with a fixed particle number, where the weight is given by the dimension of the sector. This corresponds to $w=0$ or, equivalently, $\bar n = \frac{1}{2}$.
	
\subsection{Dynamical averages}
	
Another important application of our results concerns the study of quantum many-body stochastic dynamics~\cite{Onorati:2016met}. We consider a time-dependent random Hamiltonian (with or without particle conservation) for which the time derivative $\dot{C}_{\mu\nu}$ or $\dot{c}^{(2)}_{ij}$ of the coefficient matrices are delta correlated in time, \ie for which we have
\begin{align}
	\big{\langle} \dot{C}_{\mu\nu}(t)\,\dot{C}_{\tau\sigma}(t')\big{\rangle}\;\;&\propto \gamma\, \delta(t-t')\\[.5em]
	\big{\langle} \dot{c}^{(2)}_{ij}(t)\,\dot{c}^{(2)}_{kl}(t')\big{\rangle}\;\;&\propto \gamma\, \delta(t-t')\,.
\end{align}
If we evolve an initial quantum state (Gaussian for a quadratic Hamiltonian and with fixed particle number when it is a particle-conserving Hamiltonian), the time evolution leads to an \emph{ergodic} exploration of the respective space of states considered in (1a), (1b), (2a), and (2b), provided that the strength $\gamma$ of the fluctuations is sufficiently large in the thermodynamic limit. Indeed, it is known from the Brownian motion on the unitary group ${\rm U}(d)$ (see Ref.~\cite{2020arXiv201211993K} for the random matrix version and Ref.~\cite{2013arXiv1312.7390L} for the corresponding eigenvalues) that there is a phase transition at a critical value $t_{\rm crit}=t_{\rm crit}(d)\propto d$ when the matrix size $d$ goes to infinity. In Ref.~\cite{2013arXiv1312.7390L} it was found that the level density of such a unitary random matrix does not have the entire complex unit circle as a support. We presume that this has a direct consequence for the states, too, since they have been constructed via the natural group action on the states.
	
This implies that the time evolution will only uniformly sample the full ensemble of states with respect to the Haar measure when the time is sufficiently large compared to the underlying group dimension, which is ${\rm U}(2^V)$ or ${\bigoplus}_{N=0}^V {\rm U}(d_N)$ for the Page setting and ${\rm O}(2V)$ or ${\rm U}(V)$ for pure fermionic Gaussian states without and with particle-number conservation, respectively. Then, the asymptotic time average of the entanglement entropy will coincide with the respective averages computed in the previous sections. Moreover, we also expect that the standard deviation gives a good approximation of the expected fluctuations about this average over time.
	
Again, this analysis applies to both general and quadratic Hamiltonians. The latter corresponds to the quantum simple symmetric exclusion process introduced in Ref.~\cite{bauer2017stochastic, bauer2019equilibrium}, for which the statistical properties of the Renyi entropies were studied in Ref.~\cite{bernard2021entanglement}.

\section{RELATION TO PHYSICAL HAMILTONIANs}\label{sec:relphysham}
	
The goal of this section is to contrast the results for the entanglement entropies from Secs.~\ref{sec:general} and~\ref{sec:gaussian} to those in eigenstates of physical Hamiltonians on a lattice. For the latter, we mostly have in mind local Hamiltonians with short-range hoppings and interactions (involving only a few neighboring lattice sites). Results for interacting Hamiltonians are discussed in Sec.~\ref{sec:qchaoticinteracting}, while results for quadratic (noninteracting) Hamiltonians are discussed in Secs.~\ref{sec:qchaoticquadratic} and~\ref{sec:translational}.
	
\subsection{Quantum-chaotic interacting model} \label{sec:qchaoticinteracting}
	
\begin{figure*}[!t]
	\centering
	\includegraphics[width=\linewidth]{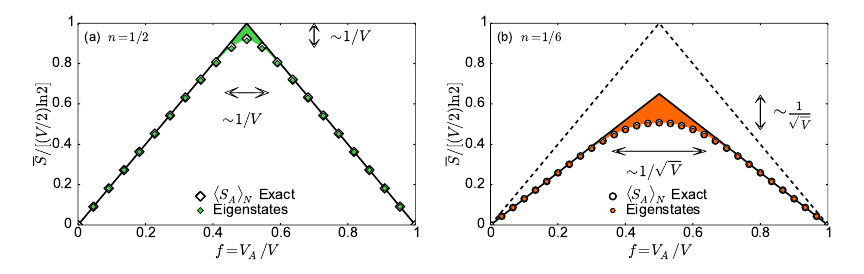}
	\caption{Average entanglement entropy density $\bar S/[(V/2)\ln 2]$ (filled symbols) of the eigenstates of the quantum-chaotic interacting Hamiltonian in Eq.~(\ref{def_H_chaotic}), which is particle-number conserving, as a function of the subsystem fraction $f=V_A/V$. The average is carried out over the central 20$\%$ of the eigenstates. Open symbols (overlapping with the filled ones) depict the corresponding exact result for general pure states, given by $\langle S_A\rangle_N$ from Eq.~(\ref{eq:Scenter}) for the same filling and system size, while the lines are the thermodynamic limit results from Eq.~(\ref{eq:leading-general}). The particle filling $n$ and the number of lattice sites $V$ are: (a) $n=\frac{1}{2}$ and $V=22$, and (b) $n=\frac{1}{6}$ and $V=30$.}
	\label{fig:S_f_chaotic}
\end{figure*}
	
We focus on a model of interacting hard-core bosons in a one-dimensional (1D) lattice with $V$ sites, as described by the Hamiltonian
\begin{eqnarray} \label{def_H_chaotic}
	\hat H_{\rm HCB} & = & - t_1 \sum_{l=1}^{V} (\hat b_{l+1}^\dagger \hat b^{}_l + \hat b_{l}^\dagger \hat b^{}_{l+1}) - t_2 \sum_{l=1}^{V} (\hat b_{l+2}^\dagger \hat b^{}_l + \hat b_{l}^\dagger \hat b^{}_{l+2}) \nonumber \\ &  & + V_1 \sum_{l=1}^V  \hat n_l \hat n_{l+1} + V_2 \sum_{l=1}^V \hat n_l \hat n_{l+2} \, ,
\end{eqnarray}
where $\hat b_l^\dagger$ ($\hat b^{}_l$) creates (annihilates) a boson at site $l$ and $\hat n_l = \hat b_l^\dagger \hat b^{}_l$ is the site occupation operator. The operators $\hat{b}^\dagger_l$ and $\hat b_l$ satisfy the commutation relations $[\hat{b}_j,\hat{b}_k] = [\hat{b}_j^\dagger,\hat{b}_k^\dagger]=0$ and $[\hat{b}_j,\hat{b}_k^\dagger] = \delta_{jk}$, supplemented by a hard-core constraint $(\hat b_l)^2 = (\hat b_l^\dagger)^2 = 0$ on physical states, which tells us that in physical states there can be at most one boson in a lattice site. 
	
Implementing this constraint is subtle; see Refs.~\cite{rey_satija_06a, 2011PhRvA..83b3611H}. We cannot assume the relation $(\hat b_l)^2 = (\hat b_l^\dagger)^2 = 0$ in an operator way as otherwise one finds that the algebra is zero. One needs to interpret this constraint as follows: $\braket{\tilde{n}|(\hat b_l)^2|\tilde{m}}=\braket{\tilde{n}|(\hat b_l^\dagger)^2|\tilde{m}}=0$ for physical states $\ket{\tilde{n}}$ and $\ket{\tilde{m}}$, which are only given by the occupation numbers $\tilde{n}=(n_1,\ldots,n_V),\tilde{m}=(m_1,\ldots,m_V)\in\{0,1\}^V$. The crucial point is that virtual states are allowed to have more than one boson on a site. For instance, for the expectation value in a single site ($V=1$), it holds that
\begin{eqnarray}
	\braket{0|\hat{b}^2(\hat{b}^\dagger)^2|0}=\braket{0|2\hat{b}\hat{b}^\dagger|0}=2,
\end{eqnarray}
where we have exploited $[\hat{b},(\hat{b}^\dagger)^2]=2\hat{b}^\dagger$ and $\hat{b}\ket{0}=0$. If one wants to replace the creation and annihilation operators by the regular spin operators in ${\rm sl}_{\mathbb{R}}(2)$, one first needs to normal order the operators via the standard bosonic commutation relations, meaning that the creation operators $\hat{b}_j^\dagger$ need to be placed to the left and the annihilation operators $\hat{b}_j$ to the right, and then replace them by spin operators. Mathematically, this means that each operator has to be dealt with in the infinite-dimensional Fock space, but at the end this space is projected onto the subspace with zero or one boson per site. This recipe naturally follows from how expectation values need to be calculated to describe the results of measurements in bosonic systems with very large on-site interactions~\cite{rey_satija_06a, 2011PhRvA..83b3611H}.
	
Model~\eqref{def_H_chaotic} contains both nearest-neighbor and the next-nearest-neighbor hoppings and interactions, it is translationally invariant, and conserves the total number of hard-core bosons. When written in terms of $\frac{1}{2}$ spins, Hamiltonian~\eqref{def_H_chaotic} is known as the (extended) spin-$\frac{1}{2}$ XXZ chain. It has been of much interest to the condensed matter and mathematical physics communities, and been used to describe the behavior of solid-state materials~\cite{cazalilla_citro_review_11}.
	
Hamiltonian~\eqref{def_H_chaotic} is integrable when $t_2=V_2=0$, independently of the values of $t_1$ and $V_1$. Then, it can be solved exactly using a Bethe ansatz~\cite{cazalilla_citro_review_11}. When $t_2\neq0$ and/or $V_2\neq0$ (for $t_1\neq0$ and $V_1\neq0$), Hamiltonian~\eqref{def_H_chaotic} is nonintegrable. In the thermodynamic limit, one expects such a nonintegrable Hamiltonian to exhibit {\it many-body} quantum chaos, namely, one expects the distribution of the spacings of nearest-neighbor levels of the many-body energy spectrum to be described by the Wigner-Dyson statistics of random matrix theory~\cite{santos_rigol_10a, santos_rigol_10b, leblond_sels_21, d2016quantum}. This is why in the nonintegrable regime we refer to Hamiltonian~\eqref{def_H_chaotic} as a quantum-chaotic interacting Hamiltonian.
	
In finite (and relatively small, tens of sites) systems, i.e., those that can be solved using full exact diagonalization (as we do here), there is a crossover regime between integrability and quantum chaos as the magnitude of $t_2$ and/or $V_2$ depart from zero (for $t_1\sim V_1\neq0$)~\cite{rigol_09a, rigol_09b, santos_rigol_10a, santos_rigol_10b, leblond_sels_21}. In such a crossover regime, for small values of $t_2$ and/or $V_2$ relative to $t_1\sim V_1$, the statistical properties of the many-body energy spectrum cannot be described by classical random matrix ensembles. When $t_1$, $t_2$, $V_1$, and $V_2$ are all similar in magnitude, full exact diagonalization calculations in Refs.~\cite{santos_rigol_10a, leblond_sels_21} showed that Hamiltonian~\eqref{def_H_chaotic} exhibits many-body quantum chaos. To be in this quantum chaotic regime, in the calculations reported here we set $t_1=t_2 = 1$ and $V_1=V_2=1.1$.
	
\begin{figure}[!t]
	\centering
	\includegraphics[width=\linewidth]{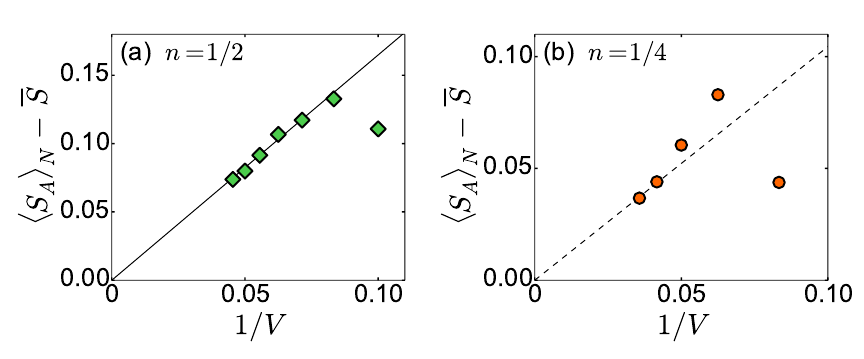}
	\caption{Finite-size scaling of the average entanglement entropy difference $\langle S_A\rangle_N- \bar S$ versus $1/V$, at the subsystem fraction $f=\frac{1}{2}$. $\langle S_A\rangle_N$ is given by Eq.~(\ref{eq:Scenter}) and $\bar S$ denotes the numerical results for the eigenstates of the quantum-chaotic interacting Hamiltonian in Eq.~(\ref{def_H_chaotic}), averaged over the central 20$\%$ of the energy spectrum. The particle fillings $n$ are: (a) $n=\frac{1}{2}$, and (b) $n=\frac{1}{4}$. The solid line in (a) shows a fit of a function $a_1/V$ to the numerical results  at $V\geq 14$, with $a_1 = 1.65$. The dashed line in (b) depicts $1/V$ behavior and is plotted as a guide to the eye.}
	\label{fig:S_f_chaotic_scaling}
\end{figure}

The entanglement entropy of eigenstates of $\hat H_{\rm HCB}$ in Eq.~(\ref{def_H_chaotic}) was calculated using full exact diagonalization after resolving all the symmetries of the model. The average entanglement entropy $\bar S$ was computed over the central 20\% of the energy eigenstates (from all symmetry sectors; see Ref.~\cite{vidmar2017entanglement2} for details). Figure~\ref{fig:S_f_chaotic} shows the behavior of the average entanglement entropy density $\bar S/[(V/2)\ln 2]$ as a function of the subsystem fraction $f$. Two remarkable features of those numerical results are as follows. (i) They are nearly identical for the Hamiltonian eigenstates (filled symbols) and the result from Eq.~(\ref{eq:Scenter}). The small differences between them are quantified in Fig.~\ref{fig:S_f_chaotic_scaling}. (ii) The deviations of $\bar S$ from the maximal entanglement entropy (shaded region) due to subleading terms may be substantial in finite systems, and they depend strongly on the particle filling $n$ [compare the results for $n=\frac{1}{2}$ in Fig.~\ref{fig:S_f_chaotic}(a) and for $n\frac{1}{6}$ in Fig.~\ref{fig:S_f_chaotic}(b)]. 
	
Figure~\ref{fig:S_f_chaotic} also mirrors the finite-size effects observed in the asymptotic expansion~\eqref{eq:leading-general}, illustrated in Fig.~\ref{fig:general-N-visual}. The vertical (double arrow) difference at $f=\frac{1}{2}$ refers to the next-to-leading-order finite-size correction of the entanglement entropy density, which is $O(1/V)$ for $n=\frac{1}{2}$ and $O(1/\sqrt{V})$ for $n\neq\frac{1}{2}$. The horizontal (double arrow) difference refers to the spread of the Kronecker delta $\delta_{f,1/2}$ for finite systems, which is $O(1/V)$ for $n=\frac{1}{2}$ and $O(1/\sqrt{V})$ for $n\neq\frac{1}{2}$. We resolved them in Figs.~\ref{fig:general-N-visual}(e) and~\ref{fig:general-N-visual}(d), respectively.
	
Next, we resolve to which degree the average entanglement entropy $\bar S$ of Hamiltonian eigenstates agrees with the analytical predictions from Sec.~\ref{sec:general}, in particular with $\langle S_A\rangle_N$ in Eq.~(\ref{eq:Scenter}). In Fig.~\ref{fig:S_f_chaotic_scaling}, we plot the finite-size scaling of $\langle S_A\rangle_N - \bar S$. In all cases under consideration, the differences appear to vanish in the thermodynamic limit. This suggests that not only the volume-law contribution of Eq.~(\ref{eq:Scenter}), but also subleading terms including the $O(1)$ terms correctly predict the eigenstate entanglement entropies of quantum-chaotic interacting Hamiltonians. The differences between the latter and Eq.~(\ref{eq:Scenter}) appear to scale algebraically as $1/V^\zeta$, with $\zeta = O(1)$. The numerical results suggest that $\zeta=1$ at $n=\frac{1}{2}$ [see the solid line in Fig.~\ref{fig:S_f_chaotic_scaling}(a)], while they are not conclusive at $n=\frac{1}{4}$ [the dashed line in Fig.~\ref{fig:S_f_chaotic_scaling}(b) depicts $1/V$ behavior and is plotted as a guide to the eye]. We note that, in Fig.~\ref{fig:S_f_chaotic_scaling}, $\langle S_A\rangle_N - \bar S > 0$, implying that the asymptotic entanglement entropy is approached from below as the system size increases.
	
\subsection{Quantum-chaotic quadratic model} \label{sec:qchaoticquadratic}
	
Next, we focus on a quadratic model, namely, a model whose Hamiltonian is bilinear in fermionic creation and annihilation operators. We explore how well the results for fermionic Gaussian states from Sec.~\ref{sec:gaussian} predict the behavior of the entanglement entropy in eigenstates of a particle-number-conserving quadratic model that exhibits {\it single-particle} quantum chaos. By single-particle quantum chaos we mean that the statistical properties of the single-particle energy spectrum are described by the Wigner-Dyson statistics of random matrix theory. Hence, we refer to this model as a quantum-chaotic quadratic model~\cite{lydzba2021entanglement}. This is to be contrasted to the model in Sec.~\ref{sec:qchaoticinteracting}, which exhibits {\it many-body} quantum chaos, and to which we referred to as a quantum-chaotic interacting model.
	
\begin{figure}[!t]
	\centering
	\includegraphics[width=\linewidth]{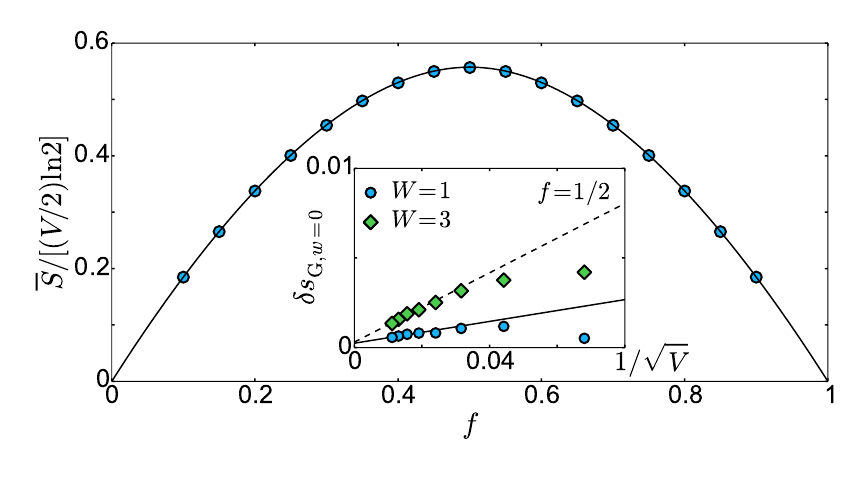}
	\caption{Average entanglement entropy density $\bar S/[(V/2)\ln 2]$ in the 3D Anderson model~(\ref{def_H_Anderson}) at $\bar n=\frac{1}{2}$. Main panel: plot of $\bar S/[(V/2)\ln 2]$ versus $f$ at disorder strength $W=1$, in a cubic lattice with $V=8000$ sites (symbols). The results are obtained averaging over 100 randomly selected many-body eigenstates and 10 Hamiltonian realizations. The solid line is the corresponding thermodynamic limit result for fermionic Gaussian states given by $\langle S_A \rangle_{{\rm G},w=0}$ in Eq.~(\ref{eq:expansion}). Inset: plot of $\delta s_{{\rm G},w=0} = (\langle S_A \rangle_{{\rm G},w=0} - \bar S)/[(V/2) \ln 2]$ versus $1/\sqrt{V}$ at $f=\frac{1}{2}$, for $W=1$ and 3, where $\langle S_A \rangle_{{\rm G},w=0}$ corresponds to the fermionic Gaussian states [Eq.~(\ref{eq:sum-chem-gauss})] at $w=0$ and the same $V$ as $\bar S$. The results for $\bar S$ are obtained averaging over $10^2$ to $10^4$ randomly selected many-body eigenstates and over 5 to 500 Hamiltonian realizations. Lines are linear fits $a_0 + a_1/\sqrt{V}$ to the results for $V \geq 2000$. We get $a_0 = 2.4 \times 10^{-4}$ and $a_1 = 0.03$ for $W=1$ (solid line), and $a_0 = 3.0 \times 10^{-4}$ and $a_1 = 0.10$ for $W=3$ (dashed line). The numerical results for $\bar S$ are from Ref.~\cite{lydzba2021entanglement}.} \label{fig:S_Anderson_scaling}
\end{figure}
	
A well-known quadratic model that exhibits single-particle quantum chaos is the 3D Anderson model below the localization transition. The Hamiltonian of this model reads
\begin{equation} \label{def_H_Anderson}
	\hat H_{\rm And} = -t \sum_{\langle i,j\rangle} (\hat f_i^\dagger \hat f^{}_j + \hat f_j^\dagger \hat f^{}_i) + \frac{W}{2}\sum_i \varepsilon_i \hat n_i \, , 
\end{equation}
where the first sum runs over nearest-neighbors sites on a cubic lattice. The operator $\hat f_j^\dagger$ ($\hat f^{}_j$) creates (annihilates) a spinless fermion at site $j$, and $\hat n_j = \hat f_j^\dagger \hat f^{}_j$ is the site occupation operator. The operators $\hat f_j^\dagger$ and $\hat f^{}_j$ satisfy the standard anticommutation relations $\{\hat{f}_l,\hat{f}_k\} = \{\hat{f}_l^\dagger,\hat{f}_k^\dagger\} = 0$ and $\{\hat{f}_l,\hat{f}_k^\dagger\} = \delta_{lk}$. The single-site occupation energies $\varepsilon_i \in [-1,1]$ are independently and identically distributed random numbers drawn from a box distribution. The 3D Anderson model exhibits a delocalization-localization transition at the critical disorder $W_c \approx 16.5$ (see, \eg Refs.~\cite{kramer_mackinnon_93, markos_06, evers_mirlin_08, suntajs_prosen_21} for reviews). Our focus here is on disorder strengths well below this transition, $W \ll W_c$. We stress that, when referring to single-particle quantum chaos in the context of the 3D Anderson model~\eqref{def_H_Anderson}, we have in mind the fixed Hilbert space $\mathcal{H}_1$ as the model of a single particle.
	
Even though it has been known for decades that the single-particle spectral properties of the 3D Anderson model in the delocalized regime are well described by the Wigner-Dyson statistics~\cite{altshuler_shklovskii_86, altshuler_zharekeshev_88, shklovskii_shapiro_93}, the entanglement entropy of energy eigenstates was studied only recently~\cite{lydzba2021entanglement}. The latter study showed that the volume-law contribution of typical many-body eigenstates is accurately described by the volume-law term of the asymptotic expression in Eq.~(\ref{eq:thermodynamic-limit}) for $n=\frac{1}{2}$, which is the same as that in Eq.~(\ref{eq:expansion}) for $\bar n=\frac{1}{2}$. This result suggests that the leading (volume-law) term in the eigenstate entanglement entropy of the 3D Anderson model deep in the delocalized regime is universal. In the main panel of Fig.~\ref{fig:S_Anderson_scaling}, we plot the average eigenstate entanglement entropy density $\bar S/[(V/2)\ln 2]$ of randomly selected eigenstates as a function of the subsystem fraction $f$. The results show remarkable agreement with the corresponding thermodynamic limit expression for the weighted average entanglement entropy over fermionic Gaussian states $\langle S_A\rangle_{{\rm G},w=0}$ in Eq.~(\ref{eq:expansion}).
	
In spite of the latter agreement, we note that the average entanglement entropy over fermionic Gaussian states does not describe the first subleading term of the average entanglement entropy in the 3D Anderson model. As shown in the inset of Fig.~\ref{fig:S_Anderson_scaling}, the first subleading term in the latter model scales $\propto \sqrt{V}$ at $f=\frac{1}{2}$. No such term appears in $\langle S_A\rangle_{{\rm G},w=0}$ in Eq.~(\ref{eq:expansion}). The fact that, for the 3D Anderson model, the subleading $O(\sqrt{V})$ term is not described by Eq.~(\ref{eq:expansion}) is in stark contrast to what we found in Sec.~\ref{sec:qchaoticinteracting} for a quantum-chaotic {\it interacting} model. In the latter case, subleading terms that are $O(1)$ or greater in the physical model are properly described by the average $\langle S_A\rangle_N$ in Eq.~(\ref{eq:Scenter}). Hence, the origin of the $O(\sqrt{V})$ contribution to the entanglement entropy of eigenstates in the 3D Anderson model remains an open question. Such a contribution is not present in our analytical calculations of the averages over Gaussian states.
	
\subsection{Translationally invariant noninteracting fermions} \label{sec:translational}
	
Next, we consider a paradigmatic quadratic model that does not exhibit quantum chaos at the single-particle level. Namely, translationally invariant noninteracting fermions, for which the Hamiltonian is a sum of hopping terms over nearest-neighbor sites [the first term in Eq.~\eqref{def_H_Anderson}]. For simplicity, we focus on the 1D case
\begin{equation} \label{def_H_Tinvariant}
	\hat H_\text{T}^\text{1D} = - \sum_{i=1}^{V} \left( \hat f_i^\dagger \hat f^{}_{i+1} + \hat f_{i+1}^\dagger \hat f^{}_{i} \right) ,
\end{equation}
with periodic boundary conditions, $\hat f^{}_{V+1} \equiv \hat f^{}_1$. The single-particle eigenenergies of the model in Eq.~(\ref{def_H_Tinvariant}) are given by the well-known expression $\epsilon_n = -2\cos(2\pi n/V)$ with $n = 0, 1, ..., V-1$, which makes apparent that the statistical properties of the single-particle spectrum are not described by the Wigner-Dyson statistics.
	
\begin{figure}[!t]
	\centering
	\includegraphics[width=\linewidth]{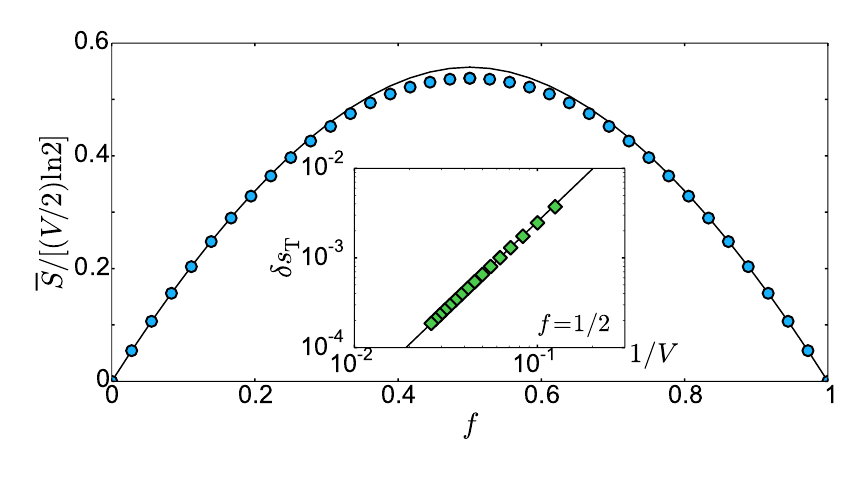}
	\caption{Average entanglement entropy density $\bar S/[(V/2) \ln 2]$ of translationally invariant noninteracting fermions in a one-dimensional lattice, described by the Hamiltonian in Eq.~(\ref{def_H_Tinvariant}). Main panel: plot of $\bar S/[(V/2) \ln 2]$ versus $f$ in the lattice with $V=36$ sites. The results are obtained by averaging over all $2^V$ many-body eigenstates. The solid line is the corresponding thermodynamic limit result for fermionic Gaussian states given by $\langle S_A \rangle_{{\rm G},w=0}$ in Eq.~\eqref{eq:expansion}. Inset: plot of $\delta s_{\rm T} = (\bar S_{\rm T} - \bar S)/([V/2] \ln 2)$ versus $1/V$ at $f=\frac{1}{2}$, where $\bar S_{\rm T}/([V/2] \ln 2) = 0.5378$. The solid line shows the function $a/V^\zeta$, with $a = 0.23$ and $\zeta=1.96$. The numerical results for $\bar S$ are from Ref.~\cite{vidmar2017entanglement}.}\label{fig:S_Tinvariant_scaling}
\end{figure}

The average eigenstate entanglement entropy of the model in Eq.~(\ref{def_H_Tinvariant}) was studied in Ref.~\cite{vidmar2017entanglement} (before the universal predictions for the quantum-chaotic quadratic models and the fermionic Gaussian states were derived). The numerical calculations in Ref.~\cite{vidmar2017entanglement} were carried out by averaging the entanglement entropy over the full set of $2^V$ many-body eigenstates. Remarkably, the numerical results were found to converge rapidly to the thermodynamic limit result, as shown for the case of $f=\frac{1}{2}$ in the inset of Fig.~\ref{fig:S_Tinvariant_scaling}. Thanks to that scaling, we find the volume-law coefficient $s^\infty_\text{T}$ of the average entanglement entropy $\bar S_\text{T} = s^\infty_\text{T} V_A \ln2$ at $f=\frac{1}{2}$ to high numerical accuracy, $s^\infty_\text{T} = 0.5378(1)$, which is consistent with the result reported in Ref.~\cite{vidmar2017entanglement}. This is to be contrasted to the volume-law coefficient $s^\infty_{{\rm G},w=0}$ of fermionic Gaussian states $\langle S_A\rangle_{{\rm G},w=0} = s^\infty_{{\rm G},w=0} V_A \ln2$ from Eq.~(\ref{eq:expansion}), which yields $s^\infty_{{\rm G},w=0} = 0.5573$. We then see that $s^\infty_\text{T}$ and $s^\infty_{{\rm G},w=0}$ are close but different. The full curve for $S_\text{T}$ as a function of $f$, for $V=36$, is shown in Fig.~\ref{fig:S_Tinvariant_scaling} together with the full curve for $\langle S_A\rangle_{{\rm G},w=0}$ from Eq.~(\ref{eq:expansion}). They are clearly different and, given the abovementioned fast convergence of the numerical results with $V$, we expect the differences to remain in the thermodynamic limit. The exact analytical form of the $\bar S_\text{T}(f)$ curve for translationally invariant free fermions remains elusive, but tight bounds have already been calculated~\cite{hackl2019average}.
	
We conclude by noting that, for the translationally invariant quantum-chaotic interacting model studied in Sec.~\ref{sec:qchaoticinteracting}, the average eigenstate entanglement entropy is accurately described by the corresponding entanglement entropy of general pure states. The role of Hamiltonian symmetries in the average entanglement entropy of energy eigenstates in quantum-chaotic interacting and quantum-chaotic quadratic models remains an important question to be explored in future studies.

\begin{table*}[!t]
	\renewcommand{\arraystretch}{1.7}
	\hspace*{-0.6cm}\begin{center}\begin{tabular}{l||ll|ll}
			& \textbf{(a) General pure states} & & \multicolumn{2}{l}{\textbf{(b) Pure fermionic Gaussian states}} \\
			\hline
			\hline
			\multirow{2}{*}{\shortstack[l]{(1) no\\ particle\\ number}} & $\braket{S_A}=a V\!-\!b+O(2^{-V})$ \& \textbf{exact} & $\rightarrow$ \eqref{eq:Page-therm}, Fig.~\ref{fig:Page-discon}, \cite{page1993average} & $\braket{S_A}_{\mathrm{G}}=a V\!+\!b\!+\!O(\frac{1}{V})$ \& \textbf{exact} & $\rightarrow$ \eqref{eq:Gaussian-average}, Fig.~\ref{fig:Page-Gaussian}, \cite{bianchi2021page}\\
			& $(\Delta S_A)^2=\alpha e^{-\beta V}+o(e^{-\beta V})$ & $\rightarrow$ \eqref{eq:variance-page}, \cite{vivo_pato_16} & $(\Delta S_A)^2_{\mathrm{G}}=a+o(1)$ & $\rightarrow$ \eqref{eq:Gaussian-variance}, \cite{bianchi2021page}\\
			\hline
			\multirow{2}{*}{\shortstack[l]{(2) fixed\\particle\\ number}} & $\braket{S_A}_N=a V\!-\!b\sqrt{V}\!-\!c\!+\!o(1)$ & $\rightarrow$ \eqref{eq:leading-general}, Fig.~\ref{fig:general-N-visual} & $\braket{S_A}_{\mathrm{G},N}=aV\!-\!\frac{b}{V}\!+\!O(\frac{1}{V^2})$ \& \textbf{exact} & $\rightarrow$ \eqref{eq:thermodynamic-limit}\\
			& $(\Delta S_A)^2_N=\alpha\, V^{\frac{3}{2}}\ee^{-\beta V}$ & $\rightarrow$ \eqref{eq:DeltaS-N} & $(\Delta S_A)^2_{\mathrm{G},N}=a\!+\!o(1)$ & $\rightarrow$ \eqref{eq:variance-Gaussian}, Fig.~\ref{fig:variance}\\
			\hline
			\multirow{2}{*}{\shortstack[l]{(3) fixed\\ weight}} & $\braket{S_A}_w=aV\!+\!b\!+\!c\sqrt{V}\!+\!o(1)$ & $\rightarrow$ \eqref{eq:Page-weighted} & $\braket{S_A}_{\mathrm{G},w}=\!a V\!+\!b\!+\!\tfrac{c}{\sqrt{V}}\!+\!\tfrac{d}{V}\!+\!o(\tfrac{1}{V})$ & $\rightarrow$ \eqref{eq:expansion}, Fig.~\ref{fig:Gaussian-mu-visual}\\
			& $(\Delta S_A)^2_w= a V+o(V)$ & $\rightarrow$ \eqref{eq:DeltaS-w} & $(\Delta S_A)^2_{\mathrm{G},w}=a V+o(V)$ & $\rightarrow$ \eqref{eq:variance-Gw}
	\end{tabular}\end{center}
	\caption{Overview of the results discussed in this tutorial. We list the main results, indicate up to which order in $V$ we derived the respective expressions (and if there exists an exact formula), and where the respective formulas can be found (equations, figures, references). Most results for fixed particle number are new, but if special cases or the leading order term were already known before, we cite the relevant works after the equation in the main text.}
	\label{tab:results}
\end{table*}
	
\section{SUMMARY AND OUTLOOK}
	
In this section, we briefly summarize the key results discussed in this tutorial, and give an outlook of where we envision the methods introduced to be applicable. We also mention some open questions in the context of the entanglement entropy of typical pure states.
	
\subsection{Summary}
	
We provided a pedagogical introduction to the current understanding of the behavior of the entanglement entropy of pure quantum states. We derived analytical expressions for the average entanglement entropy of general and Gaussian states, and considered states with and without a fixed number of particles. A comprehensive summary of the results discussed can be found in Table~\ref{tab:results}, where we contrast results for: (1) arbitrary particle number, (2) fixed particle number $N$ and (3) fixed weight parameter $w$ for both (a) general pure states and (b) Gaussian states. This yields the six state ensembles (1a) through (3b).
	
For both Gaussian and general pure states, the leading-order behavior $\braket{S_A}$ at half-filling $N=V/2$ coincides with the full average without fixing the total particle number, while the next-to-leading-order terms differ. For general pure states, we confirmed an additional contribution proportional to $\sqrt{V}$ at $f=\frac{1}{2}$ in Eq.~\eqref{eq:leading-general}, previously found in Ref.~\cite{vidmar2017entanglement2}. For Gaussian states, we derived the exact formula, which does not contain such a term and has a next-to-leading-order term of order $1/V$ [Eq.~\eqref{eq:Gaussian-expansion}]. However, we did find a contribution of order $1/\sqrt{V}$ in the asymptotic average $\braket{S_A}_{\mathrm{G},w}$ at fixed $w$ with $f=\bar{n}$, \ie whenever the subsystem fraction $f$ equals the average filling ratio $\bar{n}=\braket{N/V}=1/(1+e^{w})$.
	
We traced back these contributions to the nonanalytic behavior of the average entanglement entropy as a function of the subsystem fraction $f$ and the filling ratio $n$. In the case of Gaussian states, we identified the additional particle-subsystem symmetry $n\leftrightarrow f$, which is responsible for the $1/\sqrt{V}$ term. From a mathematical perspective, the origin of the $\sqrt{V}$ term in $\braket{S_A}_N$ is therefore the same as that of the $1/\sqrt{V}$ term in $\braket{S_A}_{\mathrm{G},w}$, namely, both calculations involve the average of a nonanalytic function with respect to an approximately Gaussian statistical distribution. Square root powers of $V$ appear exactly when the mean of the Gaussian lies in a neighborhood of the nonanalyticity, \ie there is a jump in one of the function's derivatives.
	
Finally, we connected the results obtained for the average entanglement entropy in the six ensembles of states mentioned before to the average entanglement entropy in eigenstates of specific random matrices and of physical Hamiltonians. Maybe the most surprising result in the context of quantum-chaotic interacting Hamiltonians is that not only does the leading term in the average agree with the corresponding ensemble average, but also subleading terms that are $O(1)$ or larger in the volume, \eg $O(\sqrt{V})$. Why this is so is a question that deserves to be further explored. Equally intriguing is to understand why the same is not true in the case of quantum-chaotic quadratic Hamiltonians.
	
\subsection{Outlook}
	
Looking forward, an important question is how general are the methods and results discussed here. We focused on fermionic systems, for which we can compare general pure states with Gaussian pure states, and unveiled the effect of fixing the total particle number. Our results for general pure states apply equally to hard-core bosons and spin-$\frac{1}{2}$ systems. In the latter, the total magnetization plays the role that the total particle number plays in fermionic and hard-core boson models.
	
\subsubsection{Typical eigenstate entanglement entropy as a diagnostic of quantum chaos and integrability}
	
As mentioned in the Introduction, a novel picture that the recent numerical studies such as those discussed in Sec.~\ref{sec:relphysham} have started to consolidate is that typical many-body eigenstates of quantum-chaotic interacting Hamiltonians have similar entanglement properties as typical pure states in the Hilbert space. In parallel, typical many-body eigenstates of quantum-chaotic quadratic Hamiltonians have similar entanglement properties as typical Gaussian pure states. We quantified how similar they are by showing that typical eigenstates of a specific quantum-chaotic interacting Hamiltonian exhibit $O(1)$ and greater terms in the entanglement entropy that are the same than in typical pure states in the Hilbert space. For typical many-body eigenstates of quantum-chaotic quadratic Hamiltonians, we showed that the $O(V_A)$ term is the same as in typical Gaussian pure states. These statements (for $V_A=fV\leq V/2$) are true independently of whether one deals with states in which the number of particles is fixed or not.
	
In the context of Hamiltonians that do not exhibit many-body quantum chaos, namely, in which the many-body level spacing distributions are not described by the Wigner surmise~\cite{d2016quantum}, we showed that typical many-body energy eigenstates of translationally invariant noninteracting fermions exhibit an $O(V_A)$ term that behaves qualitatively similar (but is not equal) to that obtained for typical Gaussian pure states, namely, the prefactor of such a term is a function of the subsystem fraction $f$. The same behavior was found in Ref.~\cite{leblond_mallayya_19} for the typical entanglement entropy of many-body eigenstates of the integrable spin-$\frac{1}{2}$ XXZ chain. This is fundamentally different from what happens in typical many-body eigenstates of quantum-chaotic interacting Hamiltonians, in which the prefactor is maximal (it depends only on the filling $n$) as in typical pure states.
	
Hence, as conjectured in Ref.~\cite{leblond_mallayya_19}, the entanglement entropy of typical many-body energy eigenstates can be used to distinguish models that exhibit many-body quantum chaos (whose level spacing distributions are described by the Wigner surmise, and are expected to thermalize when taken far from equilibrium~\cite{d2016quantum}) from those that do not. This is a welcome addition to the toolbox for identifying quantum chaos as it relies on the properties of the eigenstates as opposed to the properties of the eigenenergies. Other entanglement-based diagnostics of quantum chaos and integrability have been proposed in recent years, among them are the operator entanglement growth~\cite{prosen07, alba19, alba21}; the diagonal entropy~\cite{santos_11, rigol_16},  the mutual information scrambling~\cite{alba19}, and entanglement revivals~\cite{modak20} after quantum quenches; the tripartite operator mutual information~\cite{hosur16, ryu21}; and the entanglement negativity between two subsystems in a tripartition of many-body energy eigenstates~\cite{grover20}.
	
It is important to emphasize that an advantage of using the entanglement properties of energy eigenstates, instead of the properties of the eigenenergies, is that one does not need to resolve all the symmetries of the model nor does one need to do an unfolding of the spectrum, which are of paramount importance to identify quantum chaos using the eigenenergies as discussed in Sec.~\ref{sec:localspec}. In addition, in comparison to some of the entanglement diagnostics that were mentioned above, one does not need to study dynamics. Further works are needed on interacting integrable models to establish whether the leading term of the entanglement entropy of typical many-body energy eigenstates is universal or not, and to understand the nature of the subleading terms. So far, results are available only for the integrable spin-$\frac{1}{2}$ XXZ chain~\cite{leblond_mallayya_19}.
	
\subsubsection{Beyond qubit-based systems}
	
The analytical tools introduced and explained in this tutorial can be used beyond the fermionic systems we studied (and beyond the spin-$\frac{1}{2}$ and hard-core boson systems we mentioned), and facilitate the study of bosonic systems with a fixed particle number. To be concrete, a bosonic subsystem with $V_A$ out of $V$ bosonic modes and total particle number $N$ can be treated analogously to Eq.~\eqref{eq:Scenter}, but with dimensions respecting the bosonic commutation statistics, \ie
	\begin{align}\label{eq:boson-dim}
		d_A(N_A)&=\frac{(N_A+V_A-1)!}{N_A!(V_A-1)!}\,,\\
		d_B(N-N_A)&=\frac{(N-N_A+V-V_A-1)}{(N-N_A)!(V-V_A-1)!}\,,\\
		d_N&=\frac{(N+V-1)!}{N!(V-1)!}\,,
	\end{align}
which follows from the combinatorics of sampling with replacement without caring about the order, \eg for $d_A$, we ask how many ways there are to distribute $N_A$ indistinguishable particles over $V_A$ sites (where each site can hold arbitrarily many particles). Anew, it holds that ${\sum}_{N_A=0}^N d_A(N_A)d_B(N-N_A)=d_N$.
	
Following Page's approach, we again choose an arbitrary uniformly distributed random vector state in the Hilbert space $\mathcal{H}_N$. Thus, the invariance of the state under the unitary group ${\rm U}(d_N)$, now with a different dimension $d_N$, still applies. Therefore, we can follow the same strategy as in Sec.~\ref{sec:page-fixedN}, in particular we can exploit Eq.~\eqref{eq:Scenter} with dimensions~\eqref{eq:boson-dim}. This yields, in the thermodynamic limit with fixed $f\in(0,\frac{1}{2})$ and $n\in(0,\infty)$,\footnote{We evaluate Eq.~\eqref{eq:average-int}, where $\varrho(n_A)$ and $\varphi(n_A)$ slightly change from expanding Eq.~\eqref{eq:boson-dim} via a saddle point approximation. This yields the normal distribution $\varrho(n_A)$, with mean $\bar{n}_A=fn$ and variance $\sigma^2=(1-f)f(1+n)n/V$, and $\varphi(n_A)$ in Eq.~\eqref{eq:psi} becomes
\begin{align*}
	\begin{split}
		\varphi(n_A)&=[n_A\ln(n_A)+f\ln(f)+n\ln[(1+n)/n]\\
		&\quad+\ln(1+n)-(f+n_A)\ln(f+n_A)]V\\
		&\quad+ \tfrac{1}{2}\ln\left(\tfrac{n_A (f+n_A)}{f(1+n)n}\right)-\tfrac{1}{2}\delta_{f,\tfrac{1}{2}}\delta_{n_A,n/2}+o(1)
	\end{split}
\end{align*}
for $n_A\geq n_{\rm crit}$ with $n_{\rm crit}=N_{\rm crit}$ again given by $d_A(N_{\rm crit})=d_B(N-N_{\rm crit})$. For $n_A\leq n_{\rm crit}$ one needs to apply the symmetry $(n_A,f)\leftrightarrow(n-n_A,1-f)$. The summand at $N_A=N/2$ reflected by the term $\delta_{n_A,n/2}$ has to be taken as it is and is not integrated. Nevertheless, one can check numerically that it yields a term of order $1/\sqrt{V}$ and is thus subleading in Eq.~\eqref{eq:bosonic-results}.}
\begin{align}
	\begin{split}\label{eq:bosonic-results}
		\braket{S_A}_{\mathrm{bos},N}&=fV[n\ln(1+n^{-1})+\ln(1+n)]\\
		&\quad+\sqrt{V}\sqrt{\frac{n+n^2}{8\pi}}\ln(1+n^{-1})\,\delta_{f,\frac{1}{2}}\\
		&\quad+\frac{f+\ln(1-f)}{2}+o(1),
	\end{split}\\
	\braket{S_A}_{\mathrm{bos},w}&=\braket{S_A}_{\mathrm{bos},N=\bar{n}V}-\frac{f}{2}+o(1)\,,
\end{align}
where the weighted average is only meaningful for $w>0$, for which $\bar{n}=1/(e^w-1)$. Note that there is no particle-hole symmetry for bosons, and that $n=N/V$ can be arbitrarily large.
	
Other natural generalizations are spin-$s$ systems with $s>\frac{1}{2}$ and systems consisting of distinguishable particles. These cases can also be studied using the methods discussed in this tutorial, after carrying out the respective combinatorics of the Hilbert space dimensions $d_A$ and $d_B$. Also, systems with global symmetries such as time-reversal invariance or chirality can be considered, which have an impact on the respective symmetry group so that the Hilbert space is not invariant anymore under ${\rm U}(d_N)$ but only under ${\rm O}(d_N)$ or ${\rm U}(d_{N_1})\times{\rm U}(d_{N_2})$. The leading terms are expected to be the same, as the respective random matrix ensembles share the same level densities. Deviations are expected to occur in subleading terms.
	
\subsubsection{Other ensembles and entanglement measures}
	
We focused on ensembles of states, general and Gaussian pure states for arbitrary and fixed particle numbers, which mirror the entanglement properties of typical (``infinite-temperature'') eigenstates of physical lattice models. It is also possible to construct ensembles of pure states in which one fixes the energy, which mirror the entanglement properties of ``finite-temperature'' eigenstates of physical lattice models. Steps in this direction have already been taken using different tools; see, \eg Refs.~\cite{Deutsch_2010, nakagawa_watanabe_18, Fujita:2018wtr, lu_grover_19, murthy_19, bianchi2019typical}. In the context the scaling of the eigenstate entanglement entropy at different energy densities (``temperatures''), let us also emphasize that all the average entanglement entropies computed in this tutorial exhibited a leading volume-law term, namely, the leading term in the average entropies scales with the number of modes $V$ and is thus agnostic to the individual shape or area of the subsystem. In contrast, as discussed in the Introduction, it is well known that low-energy states of many physical systems of interest exhibit a leading area law term. An important open question is whether one can define ensembles of pure states that exhibit leading terms in the entanglement entropy that are area law.
	
Instead of considering the von Neumann entanglement entropy, one can also consider other quantities that are defined with respect to the invariant spectrum of the reduced density operator $\hat\rho_A=\mathrm{Tr}_{\mathcal{H}_B}\ket{\psi}\bra{\psi}$ of a pure state $\ket{\psi}$. Such quantities include the well-known Renyi entropies $S^{(n)}_A(\ket{\psi})$, and the eigenstate capacity~\cite{de2019aspects}. We focused on the von Neumann entropy, as it is arguably the most prominent measure of bipartite entanglement. Nonetheless, we expect that our findings can also be extended to the aforementioned quantities; see, \eg Refs.~\cite{liu_chen_18, pengfei_chunxiao_20, lydzba2021entanglement, ulcakar_vidmar_22} for studies of Renyi entropies and Refs.~\cite{bhattacharjee2021eigenstate, huang2021second} for studies of the eigenstate capacity.
	
It would also be interesting to explore multipartite entanglement measures for different ensembles of pure states. This will likely require new techniques, and it is not clear what the most suitable measure is. The latter question is the subject of ongoing research.
	
\section*{Acknowledgments}
We would like to thank Pietro Don\`a, Peter Forrester, Patrycja  \L yd\.{z}ba, Lorenzo Piroli, and Nicholas Witte for inspiring discussions. E.B.~acknowledges support from the National Science Foundation, Grant No.~PHY-1806428, and from the John Templeton Foundation via the ID 61466 grant, as part of the “Quantum Information Structure of Spacetime (QISS)” project (\hyperlink{http://www.qiss.fr}{qiss.fr}). L.H.~gratefully acknowledges support from the Alexander von Humboldt Foundation. M.K.~acknowledges support from the Australian Research Council (ARC) under grant No.~DP210102887. M.R.~acknowledges support from the National Science Foundation under Grant No.~2012145. L.V.~acknowledges support from the Slovenian Research Agency (ARRS), Research core fundings Grants No.~P1-0044 and No.~J1-1696. L.H.~and M.K.~are also grateful to the MATRIX Institute in Creswick for hosting the online research programme and workshop ``Structured Random Matrices Downunder'' (26 July–13 August 2021).
	
\onecolumngrid
\appendix

\section{GENERAL PURE STATES WITH A FIXED NUMBER OF PARTICLES}\label{app:general-average}
	
\subsection{Average entanglement entropy}
	
In the sum~\eqref{eq:Scenter} we have different contributions. Since our goal is to asymptotically expand the sum as a whole up to $O(1)$, which includes three different orders, namely $V$, $\sqrt{V}$ and $1$, we need to approach it systematically. The first step is to identify the various contributions of such an expansion. Let us list them as follows.

\begin{enumerate}

	\item  The difference of the digamma functions in Eq.~\eqref{eq:Scenter} can be always asymptotically expanded because the dimensions $d_N$, $d_B(N-N_A)$, and $d_A(N_A)$ grow with $V$. The only exceptions are when either the volumes of subsystem $A$ are $V_A=0,V$ or the occupation numbers are $N=0,V$. These situations correspond to the trivial cases that we do not split the system into two subsystems or that all sites are empty or occupied, which we always exclude. When assuming $n=N/V$ and $f=V_A/V$ being of order one and having a finite distance from $1$, the dimensions grow exponentially in $V$.
		
	The dimension do not grow or only grow polynomially in $V$ when $N_A\approx0,V_A$ for $d_A(N_A)$ or $N-N_A=0,V-V_A$ for $d_B(N-N_A)$. However, the prefactor $\varrho_{N_A}=d_A(N_A)d_B(N-N_A)/d_N$ will be exponentially small for these cases as we will see below.
		
	Thus, we can expand the difference of the digamma functions as follows
	\begin{equation}\label{diagamma-approx}
		\begin{split}
			\Psi[d_N+1]-\Psi[d_B(N-N_A)+1]&=\ln\left(\frac{d_N}{d_B(N-N_A)}\right)+O(e^{-\gamma_1V}),\\
			\Psi[d_N+1]-\Psi[d_A(N_A)+1]&=\ln\left(\frac{d_N}{d_A(N_A)}\right)+O(e^{-\gamma_2V}),
		\end{split}
	\end{equation}
	where $\gamma_1,\gamma_2>0$ are two non-vanishing rates that only state how fast the exponential correction of this approximation vanishes. These logarithms of the ratios are expanded in Appendix~\ref{app:log-asymp}.
		
	\item   A second contribution results from $N_{\rm crit}$, which determines the fraction of the contribution of the two sums. It is defined by the largest positive integer such that it still holds $d_A(N_A)\leq d_B(N-N_A)$. We need a $1/V$ expansion of this integer as slight deviations may yield order one terms. We derive such an expansion in Appendix~\ref{app:Ncrit}.
	
	\item   Once we know the $1/V$ expansion of $N_{\rm crit}$, we can combine it with the approximation~\eqref{diagamma-approx} and consider the first part of the sum which is
	\begin{equation}\label{sigma1}
		\begin{split}
			\Sigma_1&=\sum_{N_A=0}^{N_{\rm crit}}\frac{d_A(N_A)d_B(N-N_A)}{d_N}(\Psi[d_N+1]-\Psi[d_B(N-N_A)+1])\\
			&\quad+\sum_{N_A=N_{\rm crit}+1}^{N}\frac{d_A(N_A)d_B(N-N_A)}{d_N}(\Psi[d_N+1]-\Psi[d_A(N_A)+1])\\
			&=\sum_{N_A=0}^{N_{\rm crit}}\frac{d_A(N_A)d_B(N-N_A)}{d_N}\ln\left(\frac{d_N}{d_B(N-N_A)}\right)+\sum_{N_A=N_{\rm crit}+1}^{N}\frac{d_A(N_A)d_B(N-N_A)}{d_N}\ln\left(\frac{d_N}{d_A(N_A)}\right)+O(e^{-\tilde\gamma V}),
		\end{split}
	\end{equation}
	with a fixed $\tilde{\gamma}>0$. We note that we can extend the second sum to the upper terminal $N$ as the binomial factor is equal to $0$ whenever $N_A>V_A$.
		
	What is still needed is to expand the dimensions. The factor $\varrho_{N_A}=d_A(N_A)d_B(N-N_A)/d_N$ is a crucial ingredient for this purpose, as already pointed out. For large $V$, it will have approximately a Gaussian shape in $N_A$ with a center $\bar{n}_AV$ and a width $\sigma_A$. The problematic point is that the two sums create a kink at $N_{\rm crit}$. Thence, one needs to make a case discussion when $|N_{\rm crit}-\bar{n}_AV|$ is maximally of the order $\sigma V$ or not. We have already pointed out above Eq.~\eqref{approx.Digamma} that $\bar{n}_A=nf$ and $\sigma\sqrt{f(1-f)n(1-n)}/\sqrt{V}$. The logarithm of the ratio of the dimensions will be Taylor expanded in $(N_A/V- nf)$ and will be of order $1/\sqrt{V}$. Therefore, we only need to understand the asymptotic expansion of the moments
	\begin{equation}\label{moments.a}
		\begin{split}
			\mathcal{M}_j&= c_-\sum_{N_A=0}^{N_{\rm crit}}\frac{d_A(N_A)d_B(N-N_A)}{d_N}\left(\frac{N_A}{V}-nf\right)^j+c_+\sum_{N_A=N_{\rm crit}+1}^{N}\frac{d_A(N_A)d_B(N-N_A)}{d_N}\left(\frac{N_A}{V}-nf\right)^j,
		\end{split}
	\end{equation}
	where $c_-$ and $c_+$ are some $N_A$-independent coefficients, which in principle depend on the order $j$ of the moment, though.
		
	\begin{figure*}
		\centering  
		\includegraphics[width=.7\linewidth]{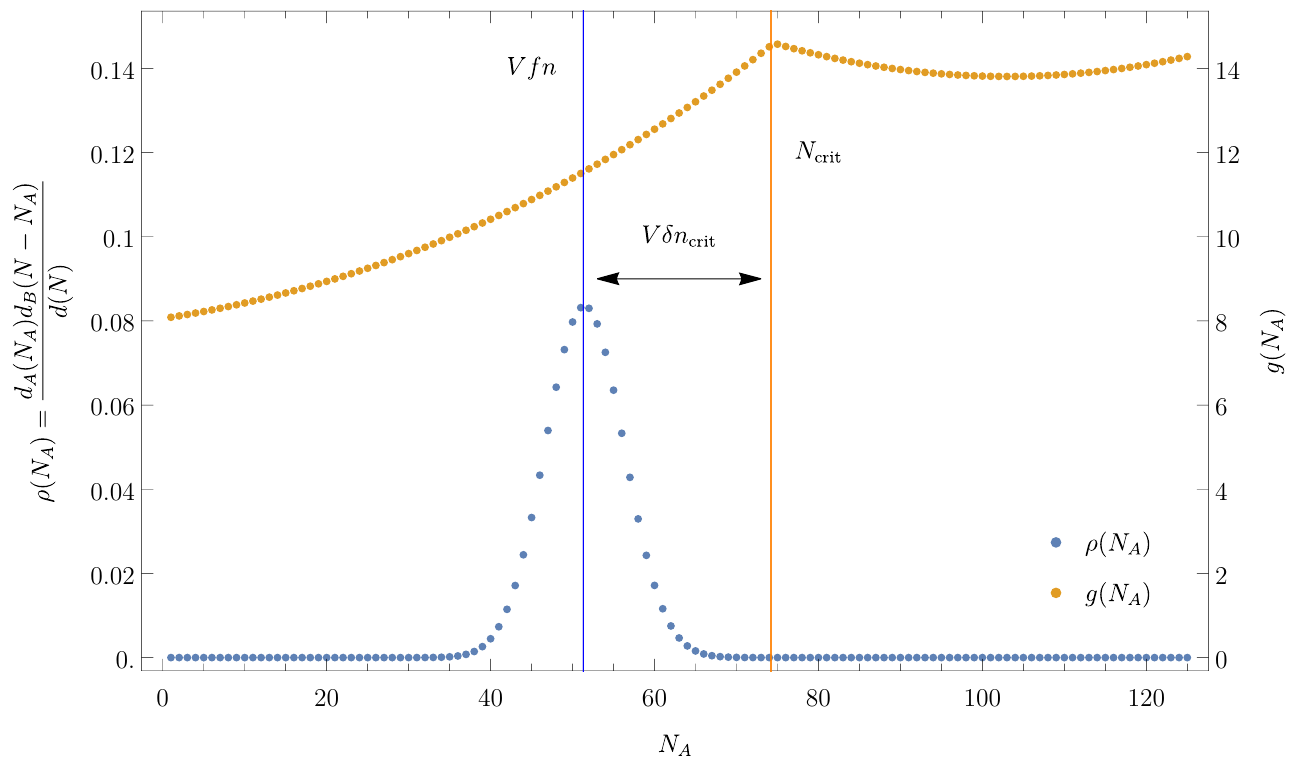}
		\caption{Illustration of the contributions of the sum~\eqref{eq:Scenter} for $V=500$, $f=\frac{1}{2}$ and $n=\frac{1}{4}-\frac{4}{\sqrt{L}}$, where we plot the density $\varrho_{N_A}=d_A(N_A)d_B(N-N_A)/d_N$ (blue dots) and the approximate observable $g(N_A)$ (yellow dots), which is given by $\ln{d_N/d_A(N_A)}$ for $d_A(N_A)\leq d_B(N-N_A)$ and $\ln(d_N/d_B(N-N_A))$ otherwise. Note that we use different scales for $\varrho_{N_A}$ and $g(N_A)$. As reference points we have added the center of the Gaussian approximation of $\varrho_{N_A}$ (blue vertical line) and the point $N_{\rm crit}$ of the kink of the summands (yellow vertical line).}
		\label{fig:N-sum}
	\end{figure*}
		
	We show in Appendix~\ref{app:Ncrit} that $\bar{n}_AV$ is never larger than $N_{\rm crit}$ so that the maximum of $\varrho_{N_A}$ is always in the first sum while the second sum only contains an exponential tail, see Fig.~\ref{fig:N-sum}. Thus, it is suitable to consider the splitting
	\begin{equation}\label{moments.b}
		\begin{split}
			\mathcal{M}_j&= c_-\sum_{N_A=0}^{N}\frac{d_A(N_A)d_B(N-N_A)}{d_N}\left(\frac{N_A}{V}-nf\right)^j+(c_+-c_-)\Delta \mathcal{M}_j,\\
			\Delta \mathcal{M}_j&=\sum_{N_A=N_{\rm crit}+1}^{N}\frac{d_A(N_A)d_B(N-N_A)}{d_N}\left(\frac{N_A}{V}-nf\right)^j.
		\end{split}
	\end{equation}
	The advantage is that the first sum with $N_A$ running from $0$ to $N$ can be seen as a contribution that is always present. Only when $|N_{\rm crit}-\bar{n}_AV|$ is at most of order $\sigma V=\sqrt{f(1-f)n(1-n)V}$ the second sum $\Delta \mathcal{M}_j$ will be of importance. We carry out the expansion of $M_j$ in Appendix~\ref{app:binom}.
		
	\item   The remaining part of the sum~\eqref{eq:Scenter} is
	\begin{equation}
		\begin{split}
			\Sigma_2&=\sum_{N_A=0}^{N_{\rm crit}}\frac{d_A(N_A)d_B(N-N_A)}{d_N}\frac{d_A(N_A)-1}{2d_B(N-N_A)}+\sum_{N_A=N_{\rm crit}+1}^{N}\frac{d_A(N_A)d_B(N-N_A)}{d_N}\frac{d_B(N-N_A)-1}{2d_A(N_A)}.
		\end{split}
	\end{equation}
	As mentioned, the dimensions $d_A(N_A)$ and $d_B(N-N_A)$ usually grow exponentially in $V$, and where they do not grow like that they will be suppressed by the exponential tails of $\varrho_{N_A}$. Thus, we can omit the terms $1/[2d_B(N-N_A)]$ and $1/[2d_A(N_A)]$ resulting in an exponentially suppressed correction. Secondly, the two sums can be combined into the form
	\begin{equation}\label{dimension.average.a}
		\begin{split}
			\Sigma_2&=\frac{1}{2}\sum_{N_A=0}^{\min(N,V_A)}\frac{d_A(N_A)d_B(N-N_A)}{d_N}\min\{d_A(N_A)/d_B(N-N_A),d_B(N-N_A)/d_A(N_A)\}+O(e^{-\tilde\gamma V})
		\end{split}
	\end{equation}
	with another fixed $\tilde{\gamma}>0$.
		
	One may ask why we do not deal with this sum in the same way as we have done for the difference of the digamma functions. First of all, we cannot extend the upper terminal of the sum to $N$ as the ratio $d_A(N_A)/d_B(N-N_A)$ can cancel the zero of the weight $\varrho_{N_A}$. Secondly, the ratio $d_A(N_A)/d_B(N-N_A)$ can exponentially grow as well as shrink in $V$ which, in principle, can shift the maximum of the weight $\varrho_{N_A}$. We will analyze this behavior in Appendix~\ref{app:dimratio}.
		
\end{enumerate}
	
Once all these contributions have been analyzed and expanded in $1/\sqrt{V}$, we combine the intermediate results in Appendix~\ref{app:result.page}.
	
\subsubsection{Asymptotic expansion of the logarithm of the ratio of dimensions}\label{app:log-asymp}
	
As we have seen in the approximation~\eqref{diagamma-approx} of the differences of the digamma functions it is suitable to expand the logarithm of the ratio
\begin{equation}\label{log.dimension}
	\ln\left(\frac{d_N}{d_A(N_A)}\right)=\ln\left(\frac{V!N_A!(V_A-N_A)!}{N!(V-N)!V_A!}\right)=\ln\left(\frac{V!(n_AV)!([f-n_A]V)!}{(nV)!([1-n]V)!(fV)!}\right).
\end{equation}
The approximation of the other logarithm follows from the symmetry relation $(N_A,V_A)\to(N-N_A,V-V_A)$ or, equivalently, $(n_A,f)\to(n-n_A,1-f)$.
	
We carry out the series expansion of Eq.~\eqref{log.dimension} in two steps. First, we take the asymptotic expansion in $V$. For this purpose, we take into account that $f,\,1-f,\,n,\,1-n>0$ are of order one so that one can use of the Stirling formula~\eqref{Stirling.appr} for four of the six factorials, namely $V!/(nV)!([1-n]V)!(fV)!$. For the other two terms $(n_AV)!([f-n_A]V)!$, we exploit the knowledge that the maximal contribution of the binomial distribution is given at $\bar{n}_A=n f$.  Hence, we have that also the argument in the remaining factorials grows linearly with $V$. Therefore, we exploit the Stirling approximation
\begin{equation}\label{Stirling.appr}
	(kV)!=\sqrt{2\pi kV}(kV+1)^{kV} e^{-kV-1}[1+O(V^{-1})]=\sqrt{2\pi kV}(kV)^{kV} e^{-kV}[1+O(V^{-1})]'
\end{equation}
with $k$ being of order one, for any of the six factorials. Collecting these approximations, the large $V$ expansion yields
\begin{equation}\label{log.da.a}
	\begin{split}
		\ln\left(\frac{d_N}{d_A(N_A)}\right)&=V\biggl[n_A\ln(n_A)+(f-n_A)\ln(f-n_A)-n\ln(n)-(1-n)\ln(1-n)-f\ln(f)\biggl]\\
		&\quad+\frac{1}{2}\ln\left(\frac{n_A(f-n_A)}{n(1-n)f}\right)+O(V^{-1}).
	\end{split}
\end{equation}
In the second step, we replace $n_A=n f+\delta n_A$ with $\delta n_A$ of order $1/\sqrt{V}$ to obtain
\begin{equation}\label{log.da}
	\begin{split}
		\ln\left(\frac{d_N}{d_A(N_A)}\right)&=V(f-1)\biggl[n\ln(n)+(1-n)\ln(1-n)\biggl]+V\delta n_A\ln\left(\frac{n}{1-n}\right)+\frac{V\delta n_A^2}{2n(1-n) f}+\frac{1}{2}\ln(f)+O(V^{-1/2}).
	\end{split}
\end{equation}
	
As aforementioned, the asymptotic for the second logarithm $\ln\left(d_N/d_B(N-N_A)\right)$ can also be obtained using Eq.~\eqref{log.da} when exploiting the symmetry $(n_A,f)\leftrightarrow(n-n_A,1-f)$. We can indeed apply this symmetry to Eq.~\eqref{log.da} as also the weight $\varrho_{N_A}$ shares this symmetry. Here, we note that the maximum of the weight $\bar{n}_A=n f$ indeed implies $n-\bar{n}_A=n(1-f)$, reflecting the symmetry. Thence, we have
\begin{equation}\label{term1}
	\begin{split}
		\ln\left(\frac{d_N}{d_B(N-N_A)}\right)&=-Vf\biggl[n\ln(n)+(1-n)\ln(1-n)\biggl]+V\delta n_A\ln\left(\frac{1-n}{n}\right)+\frac{V\delta n_A^2}{2n(1-n) (1-f)}\\
		&\quad+\frac{1}{2}\ln(1-f)+O(V^{-1/2}).
	\end{split}
\end{equation}
	
For computational purposes in the ensuing section, we need to take the difference of Eqs.~\eqref{log.da} and~\eqref{term1}, recall the third step of the outline of computing the asymptotic expansion. Especially when $1-2f$ is much smaller than one, this is important because the expectation value of the two terms tells us how many terms we need to take into account in the computation. The difference is equal to
\begin{equation}\label{log.asym}
	\begin{split}
		\ln\left(\frac{d_B(N-N_A)}{d_A(N_A)}\right)&=
		-V(1-2f)\biggl[n\ln(n)+(1-n)\ln(1-n)\biggl]+2V\delta n_A\ln\left(\frac{n}{1-n}\right)+o(1)
	\end{split}
\end{equation}
where we have set $1-2f\ll1$. Hence, this approximation is only true in this case. For the other case that $f$ is not close enough to $1/2$, we do not need the respective expression as we will see.
	
\subsubsection{Computation of \texorpdfstring{$N_{\rm crit}$}{critical N}}\label{app:Ncrit}
	
As already mentioned, $N_{\rm crit}$ is given by the condition $d_A(N_A)/d_B(N-N_A)=1$. This ratio can be approximated via Stirling's formula~\eqref{Stirling.appr},
\begin{equation}\label{ratio-start}
	\begin{split}
		\frac{d_A(N_A)}{d_B(N-N_A)}&=\frac{V_A!(N-N_A)!(V-V_A-N+N_A)!}{N_A!(V_A-N_A)!(V-V_A)!}=\frac{(fV)!([n-n_A]V)!([1-f-n+n_A]V)!}{(n_AV)!([f-n_A]V)!([1-f]V)!}\\
		&=\sqrt{\frac{f(n-n_A)(1-f-n+n_A)}{n_A(f-n_A)(1-f)}}\left[\frac{f^f(n-n_A)^{n-n_A}(1-f-n+n_A)^{1-f-n+n_A}}{n_A^{n_A}(f-n_A)^{f-n_A}(1-f)^{1-f}}\right]^V[1+O(V^{-1})].
	\end{split}
\end{equation}
As in the previous subsection, we need to go even further and take into account that $n_A$ is concentrated at $\bar{n}_A=n f$ and the small deviation $\delta n_A=n_A-\bar{n}_A$ is of order $1/\sqrt{V}$. Thus, the Taylor expansion in $\delta n_A$ about this point leads to
\begin{equation}\label{ratio-dim}
	\begin{split}
		\frac{d_A(N_A)}{d_B(N-N_A)}&=\sqrt{\frac{1-f}{f}}\left[n^{n}(1-n)^{1-n}\right]^{V(1-2f)}\exp\left[2V\delta n_A\ln\left(\frac{1-n}{n}\right)+\frac{V(2f-1)\delta n_A^2}{2nf(1-f)(1-n)}\right]\biggl[1+O(V^{-1/2})\biggl].
	\end{split}
\end{equation}
We would like to recall that we choose $0<f\leq1/2$ due to the symmetries of the setting. Thence, the leading prefactor shows that the ratio vanishes exponentially when $1-2f$ is larger than order $1/V$ because of $1/2\leq n^n(1-n)^{(1-n)}<1$ for any $0<n\leq 1/2$. There is, however, a transition regime like in the Page setting without particle number conservation, namely, when $1-2f\propto 1/V$. Then the two dimensions become comparable. From the equation above, we can also read off that only for $f\approx1/2$ the maximum $\bar{n}_AV=n fV$ or equivalently $\delta n_A\approx 0$ of the weight $\varrho_{N_A}=d_A(N_A)d_B(N-N_A)/d_N$ is close to $N_{\rm crit}$. Although it does not say, yet, when $N_{\rm crit}$ is in a distance of order $\sqrt{V}$ away from  the maximum $\bar{n}_AV=n fV$. Indeed when $\delta n_A$ and $1-2f$ are of order $1/\sqrt{V}$, there is a possibility that the two growing terms cancel each other so that it is not necessary that $1-2f$ needs to be of order $1/V$ as $1/\sqrt{V}$ is already suitable. This is exactly what happens in a particular regime as we will see below.
	
One additional comment, when $1-2f>0$ is much larger than $1/V$, we have $d_A(N_A)\ll d_B(N-N_A)$. Thus, the maximum $\bar{n}_AV$ lies in the first part of the sum~\eqref{sigma1}. Thus, in the case that the terminal $N_{\rm crit}$ is in the exponentially suppressed part of the tail of $\varrho_{N_A}$, we can concentrate only on the first part of the sum~\eqref{sigma1} and disregard the second part.
	
Thence, we only need to understand what $N_{\rm crit}-nfV$ is when $1-2f\ll1$ in the limit $V\to\infty$ for the average of Eq.~\eqref{term1}. It is crucial to know when to cut the sum of $N_A$ in the lower terminal and exploit the other branch of the Page curve for the remaining sum. A subtle point is that we need to find $\delta n_{\rm crit}=N_{\rm crit}/V-nf$ up to order $1/V$, since $N_{\rm crit}=V(nf+\delta n_{\rm crit})$ is multiplied by $V$ and the spacing between the step size of the summation index is $1$. Therefore, we have to refine the expansion in Eq.~\eqref{ratio-dim}.
	
The Taylor expansion of the Stirling approximation one order higher is given by Eq.(5.11.1) of Ref.~\cite{NIST:DLMF}
\begin{equation}
	\begin{split}
		\ln[(kV)!]&=\frac{1}{2}\ln[2\pi kV]+kV\ln(kV)-kV-\frac{1}{12kV}+O(V^{-2}).
	\end{split}
\end{equation}
We exploit this expansion in Eq.~\eqref{ratio-start} and use the notation $N_{\rm crit}=nfV+\delta n_{\rm crit}V$. As $N_{\rm crit}$ is defined by the condition $d_A(N_{\rm crit})/d_B(N-N_{\rm crit})=1$, we take its logarithm and evaluate the resulting equation up to order $1/V$,
\begin{equation}
	\begin{split}
		0&=\ln\left(\frac{d_A(N_{\rm crit})}{d_B(N-N_{\rm crit})}\right)\\
		&=\frac{1}{2}\ln\left(\frac{f[n(1-f)-\delta n_{\rm crit}][(1-f)(1-n)+\delta n_{\rm crit}]}{[nf+\delta n_{\rm crit}][f(1-n)-\delta n_{\rm crit}](1-f)}\right)\\
		&\quad+V\ln\left[\frac{f^f[n(1-f)-\delta n_{\rm crit}]^{n(1-f)-\delta n_{\rm crit}}[(1-f)(1-n)+\delta n_{\rm crit}]^{(1-f)(1-n)+\delta n_{\rm crit}}}{(nf+\delta n_{\rm crit})^{nf+\delta n_{\rm crit}}[f(1-n)-\delta n_{\rm crit}]^{f(1-n)-\delta n_{\rm crit}}(1-f)^{1-f}}\right]\\
		&\quad-\frac{1}{12V}\left(\frac{1}{f}+\frac{1}{n(1-f)-\delta n_{\rm crit}}+\frac{1}{1-f(1-n)+\delta n_{\rm crit}}-\frac{1}{nf+\delta n_{\rm crit}}-\frac{1}{f(1-n)-\delta n_{\rm crit}}-\frac{1}{1-f}\right)+O(V^{-2}).
	\end{split}
\end{equation}
The very last term before the correction is at least of order $(1-2f)/V\ll1/V$ when $\delta n_{\rm crit}$ is of order $1/\sqrt{V}$ because of
\begin{equation}
	\begin{split}
		&\quad\frac{1}{12V}\left(\frac{1}{f}+\frac{1}{n(1-f)-\delta n_{\rm crit}}+\frac{1}{1-f(1-n)+\delta n_{\rm crit}}-\frac{1}{nf+\delta n_{\rm crit}}-\frac{1}{f(1-n)-\delta n_{\rm crit}}-\frac{1}{1-f}\right)\\
		&=\frac{1-2f}{12Vf(1-f)}\left(1-\frac{1}{n(1-n)}\right)+O(V^{-{3/2}})=O\left(\frac{1-2f}{V}\right)=o(V^{-1}),
	\end{split}
\end{equation}
and $1-2f\ll1$. Thus, we can neglect it.
	
The other terms still need to be Taylor expanded in $\delta n_{\rm crit}$ as it is of order $1/\sqrt{V}$ or smaller. Expanding this expression in $\delta n_{\rm crit}$ up to order $1/V$, which is equivalent with the fourth order in $\delta n_{\rm crit}$ as we multiply one term with $V$, we find
\begin{equation}
	\begin{split}
		0&=\frac{1}{2}\ln\left(\frac{1-f}{f}\right)+\frac{(2n-1)\delta n_{\rm crit}}{2n(1-n)f(1-f)}+\frac{(1-2f)\delta n_{\rm crit}^2}{4f^2(1-f)^2}\left(\frac{1}{n^2}+\frac{1}{(1-n)^2}\right)\\
		&\quad +V(1-2f)\ln\left[n^n(1-n)^{1-n}\right]+2V\delta n_{\rm crit}\ln\left[\frac{1-n}{n}\right]+\frac{V(2f-1)\delta n_{\rm crit}^2}{2nf(1-f)(1-n)}\\
		&\quad+\frac{V(1-2n)\delta n_{\rm crit}^3}{6n^2(1-n)^2}\left(\frac{1}{(1-f)^2}+\frac{1}{f^2}\right)+\frac{V(2f-1)(f^2-f+1)\delta n_{\rm crit}^4}{12f^3(1-f)^3}\left(\frac{1}{n^3}+\frac{1}{(1-n)^3}\right)+o(V^{-1})\\
		&=\frac{1}{2}\ln\left(\frac{1-f}{f}\right)+\frac{(2n-1)\delta n_{\rm crit}}{2n(1-n)f(1-f)}+V(1-2f)\ln\left[n^n(1-n)^{1-n}\right]+2V\delta n_{\rm crit}\ln\left[\frac{1-n}{n}\right]\\
		&\quad+\frac{V(2f-1)\delta n_{\rm crit}^2}{2nf(1-f)(1-n)}+\frac{V(1-2n)\delta n_{\rm crit}^3}{6n^2(1-n)^2}\left(\frac{1}{(1-f)^2}+\frac{1}{f^2}\right)+o(V^{-1}).
	\end{split}
\end{equation}
In the second equality, we have omitted all terms that are of lower order than $1/V$ making use of the fact that $1-2f\ll1$.
	
The equation above can be recast into an implicit expression for $\delta n_{\rm crit}$. One first puts all $\delta n_{\rm crit}$-dependent terms on one side, then pulls out the common factor $\delta n_{\rm crit}$ and finally divides the entire expression by the prefactor of $\delta n_{\rm crit}$ yielding
\begin{equation}
	\begin{split}
		\delta n_{\rm crit}&=-\left[\frac{(2n-1)}{2n(1-n)f(1-f)}+2V\ln\left[\frac{1-n}{n}\right]+\frac{V(2f-1)\delta n_{\rm crit}}{2nf(1-f)(1-n)}+\frac{V(1-2n)\delta n_{\rm crit}^2}{6n^2(1-n)^2}\left(\frac{1}{(1-f)^2}+\frac{1}{f^2}\right)\right]^{-1}\\
		&\times\left[\frac{\ln[(1-f)/f]}{2}+V(1-2f)\ln\left[n^n(1-n)^{1-n}\right]\right]+O(V^{-3/2}).
	\end{split}
\end{equation}
	
In the next step, we need to find out the leading contribution of the denominator. We pull out the factor $\ln[(1-n)/n]$ in the denominator.  Then it is immediate that there are only two possibilities since the ratio $(1-2n)/\ln[(1-n)/n]$ is of order one whenever $n,1-n>0$ are of order one. Either $(1-2f)\delta n_{\rm crit}/(1-2n)$ is of order one or larger or it is much smaller than $1$. In the former case, $n$ needs to be very close to $1/2$ as $(1-2f)\delta n_{\rm crit}=o(V^{-1/2})$. This means $V(1-2f)\ln[n^n(1-n)^{1-n}]\approx-2\ln(2)V(1-2f)$ is the dominant term in the numerator. The latter, however, implies that $\delta n_{\rm crit}$ and $V(1-2f)/[V(1-2f)\delta n_{\rm crit}]=1/\delta n_{\rm crit}$ are of the same order. Hence, $\delta n_{\rm crit}$ must be of order one which is in contradiction that it is of at most of order $1/\sqrt{V}$. In conclusion, the ratio $(1-2f)\delta n_{\rm crit}/(1-2n)$ is never of order one or larger, and the denominator is always dominated by $2V\ln\left[(1-n)/n\right]$ about which we need to expand.
	
The question is how far we need to go with this expansion. Here, we have to discuss different regimes.
\begin{enumerate}
	\item The leading order term of $\delta n_{\rm crit}$, which is given in Eq.~\eqref{deltancrit}, tells us that for $1-2n\ll\sqrt{V}(1-2f)$ the terminal $V\delta n_{\rm crit}$ is of an order larger than $\sqrt{V}$, meaning the maximum of the weight $\varrho_{N_A}=d_A(N_A)d_B(N-N_A)/d_N$ plus the standard deviation is out of reach of $N_{\rm crit}$. Thus, we can disregard the second sum in Eq.~\eqref{sigma1} and can extend the first sum to the upper terminal $N$.
	\item When $V(1-2f)/(1-2n)$ ranges between order $\sqrt{V}$ and one, then it holds
	\begin{equation}
		\begin{split}\label{deltancrit}
			\delta n_{\rm crit}&=-\frac{(1-2f)\ln\left[n^n(1-n)^{1-n}\right]}{2\ln\left[(1-n)/n\right]}+o(V^{-1}).
		\end{split}
	\end{equation}
	This means $N_{\rm crit}$ is the integer smaller than or equal to $nfV-\frac{V(1-2f)\ln\left[n^n(1-n)^{1-n}\right]}{2\ln\left[(1-n)/n\right]}$.
	\item In the case $(1-2n)\gg V(1-2f)$, $\delta n_{\rm crit}$ is so close to $0$ that we can set $N_{\rm crit}$ as the integer smaller than or equal to $nfV$. This case can also be seen as a particular limit of the second case.
\end{enumerate}
Note that for all three cases we have $1-2f\ll 1$. However, case 1 shows that for $1-2f\gg1/\sqrt{V}\gtrsim (1-2n)/\sqrt{V}$ we will never find a contribution from $N_{\rm crit}$ as it will be always exponentially suppressed by the weight $\varrho_{N_A}$. We recall that the cases $1-2f\gg1/\sqrt{V}$ are always suppressed due to the exponentially small ratio $d_A(N_A)/d_B(N-N_A)$ regardless what $n$ is.  Thus, the earliest occurrence of order $\sqrt{V}$ and order one terms in the entanglement entropy will be when $f$ enters the regime about $f\approx1/2$ at a distance of order $1/\sqrt{V}$.
	
Let us highlight that this approximation of $N_{\rm crit}$ only works for the average over Eq.~\eqref{term1}. For the average over the ratio $\min\{d_A(N_A)/d_B(N-N_A),d_B(N-N_A)/d_A(N_A)\}$, we need to be more careful as they have additional exponential terms in $\delta n_A$ which may shift the maximum of the summands in the average over $\varrho_{N_A}$.
	
\subsubsection{Moments of \texorpdfstring{$\varrho_{N_A}$}{rho}} \label{app:binom}
	
To compute the entanglement entropy, we need the moments of the weight $\varrho_{N_A}=d_A(N_A)d_B(N-N_A)/d_N$ either for the full sum with terminals at $0$ and $N$ or for the one with a cutoff due to $N_{\rm crit}$. In general, we need to compute $\mathcal{M}_j$ defined in Eq.~\eqref{moments.b}. It can be rewritten in terms of a derivative acting on a sum, \ie
\begin{equation}\label{moments}
	\begin{split}
		\mathcal{M}_j&=(\partial_\lambda-nf)^j\left[c_-\sum_{N_A=0}^{N}\frac{V_A!(V-V_A)!N!(V-N)!}{V!N_A!(V_A-N_A)!(N-N_A)!(V-V_A-N+N_A)!}e^{N_A\lambda/V}\right.\\
		&\left.+(c_+-c_-)\sum_{N_A=N_{\rm crit}+1}^{N}\frac{V_A!(V-V_A)!N!(V-N)!}{V!N_A!(V_A-N_A)!(N-N_A)!(V-V_A-N+N_A)!}e^{N_A\lambda/V}\right]_{\lambda=0},
	\end{split}
\end{equation}
with $j\in\mathbb{N}$. The two coefficients  $c_\pm$ grow maximally with $V$, and $\Delta c=c_+-c_-$ grows at most like $\sqrt{V}$ for $j=0$ and like $V$ for $j=1$, cf., Eqs.~\eqref{term1} and~\eqref{log.asym}. The second sum without $\Delta c$ has been denote by $\Delta\mathcal{M}_j$, see~\eqref{moments.b} while the first sum will be labeled by
\begin{equation}\label{moments.far}
	\begin{split}
		\widetilde{\mathcal{M}}_j&=(\partial_\lambda-nf)^j\sum_{N_A=0}^{N}\frac{V_A!(V-V_A)!N!(V-N)!}{V!N_A!(V_A-N_A)!(N-N_A)!(V-V_A-N+N_A)!}e^{N_A\lambda/V}\biggl|_{\lambda=0}.
	\end{split}
\end{equation}
	
\paragraph{Evaluation of the first sum~\eqref{moments.far}.}
	
As we have seen the maximum of $\varrho_{N_A}$ is close to $\bar{n}_A=nfV$, which always lies below $N_{\rm crit}$. It can be approximated by a Gaussian with variance $\sqrt{V}$. Hence, when $N_{\rm crit}-nfV$ is larger than order $\sqrt{V}$ only the first sum contributes, and we can approximate Eq.~\eqref{moments} as $\mathcal{M}_j=\widetilde{\mathcal{M}}_j+o(1)$. The error is exponentially small due to the Gaussian tail. Otherwise the sum $\Delta\mathcal{M}_j$ will contribute, too. Yet, the ensuing computation still applies for the first sum~\eqref{moments.far} as it only takes into account that the upper terminal in the sum is $N$.
	
To compute Eq.~\eqref{moments.far}, we make use of the contour integral
\begin{equation}
	\frac{(V-N)!}{(V_A-N_A)!(V-V_A-N+N_A)!}=\oint_{|z|=1}\frac{dz}{2\pi i z}\frac{(1+z)^{V-N}}{z^{V_A-N_A}},
\end{equation}
which is based on the binomial sum $(1+z)^{V-N}=\sum_{l=0}^{V-N}\binom{V-N}{l}z^l$. Then, we can carry out the sum exactly and find
\begin{equation}
	\begin{split}
		\widetilde{\mathcal{M}}_j&= c_-\frac{V_A!(V-V_A)!}{V!}\left.(\partial_\lambda-nf)^j\oint_{|z|=1}\frac{dz}{2\pi i z}\frac{(1+z)^{V-N}(1+ze^{\lambda/V})^N}{z^{V_A}}\right|_{\lambda=0}.
	\end{split}
\end{equation}
	
Comparing with Eq.~\eqref{log.da}, we only need the first and second centered moment. We note that due to the proper normalization of $\varrho_{N_A}$ it holds $\widetilde{\mathcal{M}}_0=1$. The first centered moment,
\begin{equation}
	\begin{split}
		\widetilde{\mathcal{M}}_1&= \frac{V_A!(V-V_A)!}{V!}\oint_{|z|=1}\frac{dz}{2\pi i z}(nz-nf(1+z))\frac{(1+z)^{V-1}}{z^{V_A}}+o(1)\\
		&=\frac{V_A!(V-V_A)!}{V!}\left[n(1-f)\frac{(V-1)!}{(V_A-1)!(V-V_A)!}-nf\frac{(V-1)!}{V_A!(V-V_A-1)!}\right]\\
		&=\frac{n(1-f)V_A}{V}-nf\frac{(V-V_A)}{V}=0,
	\end{split}
\end{equation}
vanishes exactly without any approximation; note that $V_A/V=f$. In the same manner we compute the second centered moment, where we have to go up to order $1/V$ as we multiply by $V$, see Eq.~\eqref{term1},
\begin{equation}
	\begin{split}
		\widetilde{\mathcal{M}}_2&= \frac{V_A!(V-V_A)!}{V!}\oint_{|z|=1}\frac{dz}{2\pi i z}\left[n^2(1-f)^2z^2+\left(\frac{n}{V}-2n^2f(1-f)\right)z+n^2f^2\right]\frac{(1+z)^{V-2}}{z^{V_A}}\\
		&= \frac{V_A!(V-V_A)!}{V!}\left[n^2(1-f)^2\frac{(V-2)!}{(V_A-2)!(V-V_A)!}+\left(\frac{n}{V}-2n^2f(1-f)\right)\frac{(V-2)!}{(V_A-1)!(V-V_A-1)!}\right.\\
		&\qquad\left.+n^2f^2\frac{(V-2)!}{V_A!(V-V_A-2)!}\right]\\
		&= n^2(1-f)^2f\left(f+\frac{f-1}{V}\right)+\left(\frac{n}{V}-2n^2f(1-f)\right)f(1-f)\left(1+\frac{1}{V}\right)+n^2f^2(1-f)\left(1-f-\frac{f}{V}\right)+O(V^{-2})\\
		&= \frac{n(1-n)f(1-f)}{V}+O(V^{-2}).
	\end{split}
\end{equation}
The third equality follows by Taylor expanding in $1/V$, which amounts to an error term starting with the order $1/V^2$. Actually, the leading orders of the two centered moments are not very surprising as they are the mean and the variance of $\varrho_{N_A}$. This calculation only checks that there are no additional corrections that may become relevant.
	
Returning to the approximation $\mathcal{M}_j=\widetilde{\mathcal{M}}_j+o(1)$ for $N_{\rm crit}-nfV\gg\sqrt{V}$, the first moment $\mathcal{M}_1$ is exponentially suppressed as it is only given by the second sum $\Delta \mathcal{M}_1$ that lies in the exponentially small tail of  $\varrho_{N_A}$. In contrast, the correction to the leading order in the second centered moment $\mathcal{M}_2$  is of order $1/V$ when $c_-$ is of order $V$, which is indeed the case, see Eq.~\eqref{term1}.
	
Finally, we combine the two moments~$\widetilde{\mathcal{M}}_1$ and $\widetilde{\mathcal{M}}_2$ with the expansion~\eqref{term1} to compute the sum
\begin{equation}\label{log.d.db.average}
	\sum_{N_A=0}^N\frac{d_A(N_A)d_B(N-N_A)}{d_N} \ln\left(\frac{d_N}{d_B(N-N_A)}\right)=-Vf\biggl[n\ln(n)+(1-n)\ln(1-n)\biggl]+\frac{f}{2}+\frac{\ln(1-f)}{2}+O\left(\frac{1}{\sqrt{V}}\right).
\end{equation}
This is our first intermediate result, which will be needed when combining it to the full entanglement entropy.
	
\paragraph{The case $N_{\rm crit}-nfV\leq O(\sqrt{V})$ and the term $\Delta\mathcal{M}_j$.}
	
When $N_{\rm crit}$ is in the vicinity of $\bar{n}_AV=nfV=nV/2+n(2f-1)V/2$ (namely of order $\sqrt{V}$) we need to take into account that both sums, $\widetilde{\mathcal{M}}_j$ and $\Delta\mathcal{M}_j$, contribute. Recall that $\delta f=1-2f$ must be of order $1/\sqrt{V}$  or smaller and $\delta n= 1-2n$ can be not less than the order $\sqrt{V}(1-2f)$ to find this regime. This implies that the coefficient for the constant term ($j=0$) of Eq.~\eqref{log.asym} will be at most of order $\sqrt{V}$. Thus, we need to expand the sum for this case up to order $1/\sqrt{V}$. For $j=1$, we still need to expand up to order $1/V$, though this expansion does not have a term of order one as we will see.
	
We consider the sum
\begin{equation}\label{moments2}
	\Delta\mathcal{M}_j= \biggl(\partial_\lambda+\frac{N_{\rm crit}+1}{V}-nf\biggl)^j\left[\sum_{N_A=N_{\rm crit}+1}^N\frac{V_A!(V-V_A)!N!(V-N)!}{V!N_A!(V_A-N_A)!(N-N_A)!(V-V_A-N+N_A)!}e^{(N_A-N_{\rm crit}-1)\lambda/V}\right]_{\lambda=0},
\end{equation}
for $j=0,1$, where we have pushed a factor $\exp[-(N_{\rm crit}+1)\lambda/V]$ into the derivative as it will be beneficial for the ensuing calculation. The summands can be extended by $(N-N_{\rm crit}-1)!N_{\rm crit}!(N_A-N_{\rm crit}-1)!$ in the numerator and the denominator to exploit the identity
\begin{equation}
	\frac{N_{\rm crit}!(N_A-N_{\rm crit}-1)!(V-N)!}{N_A!(V_A-N_A)!(V-V_A-N+N_A)!}= \int_0^1dx\oint_{|z|=1}\frac{dz}{2\pi iz}\frac{(1+z)^{V-N}}{z^{V-V_A-N+N_A}}x^{N_A-N_{\rm crit}-1}(1-x)^{N_{\rm crit}}.
\end{equation}
The integrand over the auxiliary variable $x$ is also known as the beta-distribution. Let us underline that it always holds $N_{\rm crit}< N_A\leq N$ for $\Delta\mathcal{M}_j$. The sum can be anew carried out exactly, and we find for the zeroth moment
\begin{equation}
	\begin{split}
		\Delta\mathcal{M}_0=&\frac{V_A!(V-V_A)!N!}{N_{\rm crit}!(N-N_{\rm crit}-1)!V!} \int_0^1dx\oint_{|z|=1}\frac{dz}{2\pi iz}\frac{(1+z)^{V-N}}{z^{V-V_A}}(1-x)^{N_{\rm crit}}(x+z)^{N-N_{\rm crit}-1},
	\end{split}
\end{equation}
and for the first one
\begin{equation}
	\begin{split}
		\Delta\mathcal{M}_1&=\frac{V_A!(V-V_A)!N!}{N_{\rm crit}!(N-N_{\rm crit}-1)!V!}\int_0^1 dx\oint_{|z|=1}\frac{dz}{2\pi iz}\left[\frac{N-N_{\rm crit}-1}{V}x+\left(\frac{N_{\rm crit}+1}{V}-nf\right)(x+z)\right]\\
		&\quad\times\frac{(1+z)^{V-N}}{z^{V-V_A}}(1-x)^{N_{\rm crit}}(x+z)^{N-N_{\rm crit}-2}.
	\end{split}
\end{equation}
	
This time the integral is too involved to compute the results directly in a closed form. We use a saddle point approximation. The derivatives in $x$ and $z$ of the logarithm of the integrand yield in leading order the saddle point equations
\begin{equation}
	\frac{N-N_{\rm crit}}{x_0+z_0}-\frac{N_{\rm crit}}{1-x_0}=0\quad {\rm and}\quad \frac{V-N}{1+z_0}-\frac{V-V_A}{z_0}+\frac{N-N_{\rm crit}}{x_0+z_0}=0.
\end{equation}
Since $N_{\rm crit}=nfV+\Delta N_{\rm crit}$ with $\Delta N_{\rm crit}=O(\sqrt{V})$, and the standard deviation about the saddle points is of order $1/\sqrt{V}$, we can replace $N_{\rm crit}$ by $nfV$ in the saddle point equations without changing the integrals. The same holds for the deviation of $f$ from $1/2$ as $1-2f$ is at most of order $1/\sqrt{V}$ in the present case. This yields the approximate saddle point solutions
\begin{equation}
	x_0=0 \quad{\rm and}\quad z_0=1.
\end{equation}
Thus, we expand $x=\delta x/\sqrt{V}$ and $z=1+i\delta z/\sqrt{V}$ with $\delta x\in\mathbb{R}_+$ and $\delta z\in\mathbb{R}$.
	
For the first moment this yields a drastic simplification because of the factor 
\begin{equation}
	\frac{N-N_{\rm crit}-1}{V}x+\biggl(\frac{N_{\rm crit}+1}{V}-nf\biggl)(x+z)=\frac{n}{2\sqrt{V}}\delta x+\frac{\Delta N_{\rm crit}}{V}+\frac{1}{2V}\left(2+\sqrt{V}(1-2f)n\delta x+2i\frac{\Delta N_{\rm crit}}{\sqrt{V}}\delta z\right)+O(V^{-3/2}),
\end{equation}
which already starts with order $1/\sqrt{V}$. Therefore, we only need to go to order $1/\sqrt{V}$ in all other factors in $\Delta\mathcal{M}_1$ like in $\Delta\mathcal{M}_0$.
	
The single terms we need to expand are [note, $1-2f=O(1/\sqrt{V})$ or smaller]
\begin{equation}
	(1+z)^{V-N}=2^{(1-n)V}\left(1+i\frac{1}{2\sqrt{V}}\delta z\right)^{(1-n)V}
	=2^{(1-n)V}\exp\left[i\frac{\sqrt{V}(1-n)}{2}\delta z+\frac{1-n}{8}\delta z^2\right]\left[1-i\frac{1-n}{24\sqrt{V}}\delta z^3+O(V^{-1})\right],
\end{equation}
\begin{equation}
	\begin{split}
		z^{-V+V_A-1}=&\left(1+i\frac{\delta z}{\sqrt{V}}\right)^{-(1+1-2f)V/2-1}\\
		=&\exp\left[-i\frac{\sqrt{V}}{2}\delta z-\frac{1}{4}\left(2i\sqrt{V}(1-2f)\delta z+\delta z^2\right)\right]\left[1-i\frac{\delta z}{\sqrt{V}}-\frac{1-2f}{4}\delta z^2+i\frac{\delta z^3}{6}+O(V^{-1})\right],
	\end{split}
\end{equation}
\begin{equation}
	\begin{split}
		(1-x)^{N_{\rm crit}}=\left(1-\frac{\delta x}{\sqrt{V}}\right)^{n(1-1+2f)V/2+\Delta N_{\rm crit}}
		=&\exp\left[-\frac{n\sqrt{V}}{2}\delta x-\left(\frac{\Delta N_{\rm crit}}{\sqrt{V}}-\frac{\sqrt{V}(1-2f)n}{2}\right)\delta x-\frac{n}{4}\delta x^2\right]\\
		&\times\left[1-\frac{1}{2\sqrt{V}}\left(\frac{\Delta N_{\rm crit}}{\sqrt{V}}-\frac{\sqrt{V}(1-2f)n}{2}\right)\delta x^2-\frac{n}{6\sqrt{V}}\delta x^3+O(V^{-1})\right],
	\end{split}
\end{equation}
and
\begin{equation}
	\begin{split}
		(x+z)^{N-N_{\rm crit}-j}=&\left(1+\frac{\delta x+i\delta z}{\sqrt{V}}\right)^{n(1+1-2f)V/2-\Delta N_{\rm crit}-j}\\
		=&\exp\left[\left(\frac{n\sqrt{V}}{2}-\frac{\Delta N_{\rm crit}}{\sqrt{V}}+\frac{\sqrt{V}(1-2f)n}{2}\right)\left(\delta x+i \delta z\right)-\frac{n}{4}(\delta x+i\delta z)^2\right]\\
		&\times\biggl[1-\frac{j}{\sqrt{V}}(\delta x+i\delta z)+\left(\frac{\Delta N_{\rm crit}}{2V}-\frac{(1-2f)n}{4}\right)(\delta x+i\delta z)^2+\frac{n}{6\sqrt{V}}(\delta x+i\delta z)^3+O(V^{-1})\biggl],
	\end{split}
\end{equation}
with $j=1,2$. Additionally, we expand the factorial prefactors using Stirling's formula
\begin{equation}
	\begin{split}
		&\frac{V_A!(V-V_A)!N!}{N_{\rm crit}!(N-N_{\rm crit}-1)!V!}=\frac{2^{(n-1)V}(1-1+2f)^{(1-1+2f)V/2}}{[1-1+2f+2\Delta N_{\rm crit}/(nV)]^{n(1-1+2f) V/2+\Delta N_{\rm crit}}}\\
		&\qquad\qquad\times\frac{(1+1-2f)^{(1+1-2f)V/2}}{[1+1-2f-2\Delta N_{\rm crit}/(nV)]^{n(1+1-2f)V/2-\Delta N_{\rm crit}}}\left(1+1-2f- \frac{2\Delta N_{\rm crit}}{nV}\right)\frac{V\sqrt{n}}{2}\left[1+O(V^{-1})\right] \\
		&=2^{(n-1)V}\frac{V\sqrt{n}}{2}\exp\left[\frac{V}{2}(1-2f)^2-\frac{(V(1-2f)n-2\Delta N_{\rm crit})^2}{2nV}\right]\left[1+(1-2f)-\frac{2\Delta N_{\rm crit}}{V}+O(V^{-1})\right]. 
	\end{split}
\end{equation}
	
Collecting all our previous results, and integrating over $\delta x\in\mathbb{R}_+$ and $\delta z\in\mathbb{R}$, we find
\begin{equation}
	\begin{split}
		\Delta\mathcal{M}_0=&\frac{1}{2}{\rm erfc}\left[\sqrt{\frac{2}{n(1-n)V}}\Delta N_{\rm crit}\right]-\frac{1}{\sqrt{2\pi n(1-n)V}}\exp\left[-\frac{2\Delta N_{\rm crit}^2}{n(1-n)V}\right]+O(V^{-1}),
	\end{split}
\end{equation}
and
\begin{equation}
	\begin{split}
		\Delta\mathcal{M}_1=&\frac{(1-n)n+2\Delta N_{\rm crit}/V}{\sqrt{8\pi(1-n)n V}}\exp\left[-\frac{2\Delta N_{\rm crit}^2}{n(1-n)V}\right]+O(V^{-{3/2}}).
	\end{split}
\end{equation}
These two equations are our second ingredient. Especially when combining those with Eq.~\eqref{log.asym}, we find
\begin{equation}\label{log.db.da.average}
	\begin{split}
		&\sum_{N_A=N_{\rm crit}+1}^N\frac{d_A(N_A)d_B(N-N_A)}{d_N} \ln\left(\frac{d_B(N-N_A)}{d_A(N_A)}\right)\\
		=&-V(1-2f)\biggl[n\ln(n)+(1-n)\ln(1-n)\biggl]\left(\frac{1}{2}{\rm erfc}\left[\sqrt{\frac{2}{n(1-n)V}}\Delta N_{\rm crit}\right]-\frac{1}{\sqrt{2\pi n(1-n)V}}\exp\left[-\frac{2\Delta N_{\rm crit}^2}{n(1-n)V}\right]\right)\\
		&+\sqrt{V}\ln\left(\frac{n}{1-n}\right)\frac{(1-n)n+2\Delta N_{\rm crit}/V}{\sqrt{2\pi(1-n)n }}\exp\left[-\frac{2\Delta N_{\rm crit}^2}{n(1-n)V}\right]+o(1),
	\end{split}
\end{equation}
which is our second intermediate result. As one can readily check, this term starts at most with order $\sqrt{V}$ as $1-2f$ is at most of order $1/\sqrt{V}$ in the present case. Additionally it will contribute order one terms.
	
\subsubsection{Average over dimension ratios}\label{app:dimratio}
	
We now turn to evaluate the sum~\eqref{dimension.average.a}. We need to expand the average
\begin{equation}\label{sum-factorials-Page}
	\begin{split}
		&\sum_{N_A=0}^{\min(N,V_A)}\frac{d_A(N_A)d_B(N-N_A)}{d_N}\min\{d_A(N_A)/d_B(N-N_A),d_B(N-N_A)/d_A(N_A)\}\\
		&=\sum_{N_A=0}^{N_{\rm crit}}\frac{d_A^2(N_A)}{d_N}+\sum_{N_A=N_{\rm crit}+1}^{\min(N,V_A)}\frac{d_B^2(N-N_A)}{d_N}.
	\end{split}
\end{equation}
For that purpose, it is paramount to take the correct bounds of the summing index as either $d_B(N-N_A)$ or $d_A(N_A)$ are canceled in the weight $\varrho_{N_A} = d_A(N_A)d_B(N-N_A)/d_N$ by the observable $\min\{d_A(N_A)/d_B(N-N_A), d_B(N-N_A)/d_A(N_A)\}$. Note that, in Eq.~\eqref{moments}, the weight $\varrho_{N_A}$ allowed us to extend the sum from $N_A=0$ to $N_A=N$ instead of $N_A=\min(N,V_A)$. The weight $\varrho_{N_A}$ implemented the correct terminal.
	
We need to understand where $d_A(N_A)$ and $d_B(N-N_A)$ become maximal. This is indeed given for $\bar{N}_A^{(1)}=V_A/2=fV/2$ for $d_A(N_A)$ and $\bar{N}_A^{(2)}=N-(V-V_A)/2=(2n-1+f)V/2$ for $d_B(N-N_A)$ because both are binomial weights whose maximum is always at the center of the distribution.
	
To see which of the two dimensions is largest, we take their ratio and expand in large $V$
\begin{equation}
	\frac{d_A(\bar{N}_A^{(1)})}{d_B(N-\bar{N}_A^{(2)})}=\frac{V_A!}{(V-V_A)!}\left(\frac{[(V-V_A)/2]!}{(V_A/2)!}\right)^2=\frac{(fV)!}{[(1-f)V]!}\left(\frac{[(1-f)V/2]!}{(fV/2)!}\right)^2
	=\sqrt{\frac{1-f}{f}}2^{(2f-1)V}\left[1+O(V^{-1})\right].
\end{equation}
Hence, for suitably large $V$ it is always $d_A(\bar{N}_A^{(1)})\leq d_B(N-\bar{N}_A^{(2)})$ because $f\leq1/2$. Note that we compare the maximums of the two dimensions at different $N_A$ and not at the same one. However, the ratio of the two dimensions at the same $N_A=\bar{N}_A^{(2)}$ will be always smaller one, \ie $d_A(\bar{N}_A^{(2)})\leq d_A(\bar{N}_A^{(1)})\leq d_B(N-\bar{N}_A^{(2)})$ (the first inequality follows from the fact that the maximum of $d_A(N_A)$ is achieved at $\bar{N}_A^{(1)}$). Hence, it is clear that $\bar{N}_A^{(2)}\leq N_{\rm crit}$. In particular, the maximum of the $d_B^2(N-N_A)/d_N$ cannot be achieved at $\bar{N}_A^{(2)}$ but is only given at the lower terminal $N_A=N_{\rm crit}+1$. Since $d_B^2(N-N_A)/d_N<\varrho_{N_A}$ for $N_A\geq N_{\rm crit}$ the second sum can only contribute when $N_{\rm crit}$ has a distance of order $\sqrt{V}$ to the maximum $\bar{n}_AV=nf V$ of $\varrho_{N_A}$. As we have seen this is only the case when $1-2f$ is of order $1/\sqrt{V}$ or less and $1-2n\ll \sqrt{V}(1-2f)$.
	
Let us first consider the case $1-2f\gg V^{-1/2}$ as it is still unclear whether the sum~\eqref{sum-factorials-Page} yields  anything of order one in this case. When $N_A\leq N_{\rm crit}$ we have for the first part of the sum
\begin{equation}\label{inequality-dimensions}
	\frac{d_A^2(N_A)}{d_N}<\frac{d_A(N_A)d_B(N-N_A)}{d_N}=\varrho_{N_A}.
\end{equation}
This means that $d_A^2(N_A)/d_N$ is exponentially suppressed when $N_A$ has a distance larger than the order $\sqrt{V}$ to the maximum $\bar{n}_AV=nf V$ of $\varrho_{N_A}$. The maximum $\bar{N}_A^{(1)}=f V/2$ of $d_A(N_A)$ is evidently far away when $n$ is not close to $1/2$. When expanding $d_A^2(N_A)/d_N$ about $N_A=nf V+\delta n_A V$ with $\delta n_A$ of order $1/\sqrt{V}$ the expansion up to order  one is
\begin{eqnarray}\label{inequality-dimensions.b}
	\frac{d_A^2(N_A)}{d_N}=\frac{1}{\sqrt{2\pi Vn(1-n)f^2}}\left(\frac{1-n}{n}\right)^{2V\delta n_A}\left((1-n)^{1-n}n^n\right)^{V(1-2f)}\exp\left[-\frac{V\delta n_A^2}{n(1-n)f}\right][1+O(V^{-1/2})].
\end{eqnarray}
Since $f,n\in(0,1/2]$, the leading term in $V$ exponentially suppresses all summands if $1-2f\gg V^{-1/2}$.  Similarly, we have an exponential suppression of $d_B^2(N-N_A)/d_N<d_A(N_A)d_B(N-N_A)/d_N=\varrho_{N_A}$ for any $N_A>N_{\rm crit}$ because then $\bar{n}_AV=nfV$ is certainly further away than a distance $\sqrt{V}$ from $N_A$. Therefore, the second part~\eqref{dimension.average.a} of the entanglement entropy does not contribute for $1-2f\gg V^{-1/2}$ in the thermodynamic limit.
	
When $1-2f=O(V^{-1/2})$ the leading term in Eq.~\eqref{inequality-dimensions.b} slightly shifts the maximum of the summands as we have seen it for $N_{\rm crit}$. Actually $N_{\rm crit}$, which is $(nf+\delta n_{\rm crit})V$ with Eq.~\eqref{deltancrit}  in this scaling, will be the maximum  since $(1-n)/n\geq1$ for $n\leq 1/2$, such that whenever $N_{\rm crit}/V- n_A=\delta n_{\rm crit}-\delta n_A$ is larger than order $1/V$ the ratio $d_A^2(N_A)/d_N$ will be again exponentially suppressed. Also for $d_B^2(N-N_A)/d_N$ the maximum lies at $N_{\rm crit}$ when $N_A>N_{\rm crit}$, as we have seen that the maximum $\bar{N}_A^{(2)}<N_{\rm crit}$ and $d_B(N-N_A)$ is monotonously decreasing for $N_A>\bar{N}_A^{(2)}$. Its expansion about $N_A=(nf+\delta n_A)V$ is equal to
\begin{equation}\label{inequality-dimensions.c}
	\frac{d_B^2(N_A)}{d_N}=\frac{1}{\sqrt{2\pi Vn(1-n)(1-f)^2}}\left(\frac{n}{1-n}\right)^{2V\delta n_A}\left((1-n)^{1-n}n^n\right)^{V(2f-1)}\exp\left[-\frac{V\delta n_A^2}{n(1-n)(1-f)}\right][1+O(V^{-1/2})].
\end{equation}
	
Let us try to simplify both expansions~\eqref{inequality-dimensions.b} and~\eqref{inequality-dimensions.c} further. We know that they are only valid when $N_{\rm crit}$ has a distance of the order $\sqrt{V}$ to $\bar{n}_AV=nfV$. In Sec.~\ref{app:Ncrit}, we have seen that this is only the case when $\sqrt{V}(1-2f)$ is maximally of the order $1-2n$. Hence $1-2f$ must be maximally of order $1/\sqrt{V}$ in which we can expand. However, there are two cases to discuss in the scaling of the variable $n$ depending on the exact scaling of $1-2f$.
	
When $1-2f=O(V^{-1/2})$, we have already mentioned that $N_A$ takes its maximum at $N_{\rm crit}=\lfloor nf+\delta n_{\rm crit})V\rfloor$ which is the largest integer that is smaller than or equal to $(nf+\delta n_{\rm crit})V=(nf-(1/2-f)\ln[n^n(1-n)^{1-n}]/\ln[(1-n)/n])V+o(1)$, see Eq.~\eqref{deltancrit}. This holds true for both sums~\eqref{sum-factorials-Page}. One can readily see from the Eqs.~\eqref{inequality-dimensions.b} and~\eqref{inequality-dimensions.c} that only summands where $|N_A-N_{\rm crit}|$ is of order one are of the same order as the maximum while the other terms are suppressed. Thus, we substitute $N_A=N_{\rm crit}-j=nfV+\Delta N_{\rm crit}-j$ for $N_A\leq N_{\rm crit}$ and $N_A=N_{\rm crit}+1+j=nfV+\Delta N_{\rm crit}+1+j$ for $N_A>N_{\rm crit}$ with  $j=0,1,2,\ldots$ and
\begin{align}
	\Delta N_{\mathrm{crit}}=N_{\rm crit}-nfV=\left\lfloor\delta n_{\rm crit}V\right\rfloor+O(1)=\left\lfloor-\frac{V(1-2f)\ln\left[n^n(1-n)^{1-n}\right]}{2\ln\left[(1-n)/n\right]}\right\rfloor+O(1),
\end{align}
where the error is only a number in $[0,1)$. Substituting the expansions~\eqref{inequality-dimensions.b} and~\eqref{inequality-dimensions.c} into Eq.~\eqref{sum-factorials-Page}, we find
\begin{equation}\label{sum-factorials-Page.case1}
	\begin{split}
		&\sum_{N_A=0}^{\min(N,V_A)}\frac{d_A(N_A)d_B(N-N_A)}{d_N}\min\{d_A(N_A)/d_B(N-N_A),d_B(N-N_A)/d_A(N_A)\}\\
		=&\sum_{j=0}^{\infty}\frac{1}{\sqrt{2\pi Vn(1-n)f^2}}\left(\frac{1-n}{n}\right)^{2\Delta N_{\rm crit}-2\delta n_{\rm crit}V-2j}\exp\left[-\frac{(\Delta N_{\rm crit}-j)^2}{n(1-n)fV}\right]\\
		&+\sum_{j=0}^{\infty}\frac{1}{\sqrt{2\pi Vn(1-n)(1-f)^2}}\left(\frac{n}{1-n}\right)^{2\Delta N_{\rm crit}-2\delta n_{\rm crit}V+2+2j}\exp\left[-\frac{(\Delta N_{\rm crit}+1+j)^2}{n(1-n)(1-f)V}\right]+O(V^{-1}),
	\end{split}
\end{equation}
since in the first sum it is $\delta n_a V=\Delta N_{\rm crit}-j$ and in the second sum it is $\delta n_AV=\Delta N_{\rm crit}+1+j$. We have extended the sum to $\infty$ because we only add exponentially suppressed terms to it, and we have used the expression of $\delta n_{\rm crit}$ from Eq.~\eqref{deltancrit}. The $j$-dependence in the Gaussian part of Eq.~\eqref{sum-factorials-Page.case1} can be dropped since $j/\sqrt{V}$ is of order $1/\sqrt{V}$; note that $\Delta N_{\rm crit}=O(\sqrt{V})$. Using the fact that the geometric series converge because $1-2n\in(0,1)$ is of order one and all the summands are of order $1/\sqrt{V}$ due to the prefactor, it becomes immediate that the sum~\eqref{dimension.average.a} still vanishes; although it vanishes now like $1/\sqrt{V}$.
	
A similar argument holds when $1-2n\gg 1/\sqrt{V}$ is of order one even if $(1-2f)\sqrt{V}$ is maximally of order $1-2n$ and, hence, rapidly vanishing.  In this case the variable $x=(1-2n)j$ becomes quasi-continuous, meaning $j$ will be of order $1/(1-2n)$. We need to scale $j$ with $1-2n$ since the quadratic part from the Gaussian, namely $j/\sqrt{V}=x/[\sqrt{V}(1-2n)]\ll 1$. This yields an integral over an exponential function showing that both sums in Eq.~\eqref{sum-factorials-Page} vanish like $1/[\sqrt{V}(1-2n)]$. It is certainly much weaker but it is vanishing nonetheless.
	
Therefore, only in the case $(1-2f)\sqrt{V}\ll 1-2n\ll 1$ the sum~\eqref{dimension.average.a} can contribute to the entanglement entropy of order one and above. The two expansions~\eqref{inequality-dimensions.b} and~\eqref{inequality-dimensions.c} simplify in this case to
\begin{align}\label{inequality-dimensions.d}
	\begin{split}
		\frac{d_A^2(N_A)}{d_N}=&2^{V(2f-1)}\sqrt{\frac{8}{\pi V}}\exp\left[-8V\delta n_A^2+4V(1-2n)\delta n_A\right][1+O(V^{-1/2})],\\
		\frac{d_B^2(N_A)}{d_N}=&2^{V(1-2f)}\sqrt{\frac{8}{\pi V}}\exp\left[-8V\delta n_A^2-4V(1-2n)\delta n_A\right][1+O(V^{-1/2})].
	\end{split}
\end{align}
The variable $\delta n_A$ becomes quasi continuous as it is of order $1/\sqrt{V}$ meaning the sum is replaced by an integral. Thus, the final computation of the sum~\eqref{sum-factorials-Page} is
\begin{equation}\label{sum-factorials-Page.final}
	\begin{split}
		&\hspace{-0.7cm}\sum_{N_A=0}^{\min(N,V_A)}\frac{d_A(N_A)d_B(N-N_A)}{d_N}\min\{d_A(N_A)/d_B(N-N_A),d_B(N-N_A)/d_A(N_A)\}\\
		=&2^{V(2f-1)}\int_{-\infty}^{\ln(2)(1-2f)/[4(1-2n)]} \frac{4\sqrt{V}d\delta n_A}{\sqrt{2\pi}}\exp\left[-8V\delta n_A^2+4V(1-2n)\delta n_A\right]\\
		&+2^{V(1-2f)}\int_{\ln(2)(1-2f)/[4(1-2n)]}^\infty \frac{4\sqrt{V}d\delta n_A}{\sqrt{2\pi}}\exp\left[-8V\delta n_A^2-4V(1-2n)\delta n_A\right]+o(1)\\
		=&2^{V(2f-1)-1}\exp\left[\frac{V(1-2n)^2}{2}\right]{\rm erfc}\left[\sqrt{V}\frac{(1-2n)^2-\ln(2)(1-2f)}{\sqrt{2}(1-2n)}\right]\\
		&+2^{V(1-2f)-1}\exp\left[\frac{V(1-2n)^2}{2}\right]{\rm erfc}\left[\sqrt{V}\frac{(1-2n)^2+\ln(2)(1-2f)}{\sqrt{2}(1-2n)}\right]+o(1),
	\end{split}
\end{equation}
where we made use of the complementary error function.
	
Also, in this formula one can readily check the various asymptotic limits. If either $1-2n\gg1/\sqrt{V}$ or $1-2f\gg1/V$, the term vanishes as already pointed out. For the three cases discussed in Sec.~\ref{app:Ncrit}, we have
\begin{enumerate}
	\item $(1-2n)\ll\sqrt{V}(1-2f)=O(1/\sqrt{V})$, it holds
	\begin{equation}
		\sum_{N_A=0}^{\min(N,V_A)}\frac{d_A(N_A)d_B(N-N_A)}{d_N}\min\left\{\frac{d_A(N_A)}{d_B(N-N_A)},\frac{d_B(N-N_A)}{d_A(N_A)}\right\}=2^{V(2f-1)}+o(1),
	\end{equation}
	\item $(1-2n)=O[\sqrt{V}(1-2f)]=O(1/\sqrt{V})$, we need to keep the result as it is,
	\item $O(1/\sqrt{V})=(1-2n)\gg\sqrt{V}(1-2f)$, we obtain
	\begin{equation}
		\sum_{N_A=0}^{\min(N,V_A)}\frac{d_A(N_A)d_B(N-N_A)}{d_N}\min\left\{\frac{d_A(N_A)}{d_B(N-N_A)},\frac{d_B(N-N_A)}{d_A(N_A)}\right\}=\exp\left[\frac{V(1-2n)^2}{2}\right]{\rm erfc}\left[\sqrt{V}\frac{(1-2n)}{\sqrt{2}}\right]+o(1).
	\end{equation}
\end{enumerate}
The third case covers part of the second case and the third case in Sec.~\ref{app:Ncrit}.  In the first case, the exponential decay is not always completely correct as we have seen in the discussion above. The reason is that the error term still contains algebraic decaying terms which will take over.
	
\subsubsection{Resulting formula}\label{app:result.page}
	
Collecting all intermediate results~\eqref{log.d.db.average},~\eqref{log.db.da.average} and~\eqref{sum-factorials-Page.final}, we arrive at
\begin{equation}\label{finalresult-Page}
	\begin{split}
		\braket{S_A}_N=&-Vf\biggl[n\ln(n)+(1-n)\ln(1-n)\biggl]+\frac{f}{2}+\frac{\ln(1-f)}{2}\\
		&-V(1-2f)\ln[n^n(1-n)^{1-n}]\left(\frac{1}{2}{\rm erfc}\left[\sqrt{\frac{2}{n(1-n)V}}\Delta N_{\rm crit}\right]-\frac{1}{\sqrt{2\pi n(1-n)V}}\exp\left[-\frac{2\Delta N_{\rm crit}^2}{n(1-n)V}\right]\right)\\
		&+\sqrt{V}\ln\left(\frac{n}{1-n}\right)\frac{(1-n)n+2\Delta N_{\rm crit}/V}{\sqrt{2\pi(1-n)n }}\exp\left[-\frac{2\Delta N_{\rm crit}^2}{n(1-n)V}\right]\\
		&-2^{V(2f-1)-2}\exp\left[\frac{V(1-2n)^2}{2}\right]{\rm erfc}\left[\sqrt{V}\frac{(1-2n)^2-\ln(2)(1-2f)}{\sqrt{2}(1-2n)}\right]\\
		&-2^{V(1-2f)-2}\exp\left[\frac{V(1-2n)^2}{2}\right]{\rm erfc}\left[\sqrt{V}\frac{(1-2n)^2+\ln(2)(1-2f)}{\sqrt{2}(1-2n)}\right]+o(1).
	\end{split}
\end{equation}
What remains to be discussed is the parameter
\begin{align}
	\Delta N_{\mathrm{crit}}=\left\lfloor-\frac{V(1-2f)\ln\left[n^n(1-n)^{1-n}\right]}{2\ln\left[(1-n)/n\right]}\right\rfloor+O(1).
\end{align}
Two terms in Eq.~\eqref{finalresult-Page} can be combined with the help of this expansion, \ie
\begin{equation}
	\begin{split}
		\frac{V(1-2f)\ln[n^n(1-n)^{1-n}]}{\sqrt{2\pi n(1-n)V}}\exp\left[-\frac{2\Delta N_{\rm crit}^2}{n(1-n)V}\right]+\ln\left(\frac{n}{1-n}\right)\frac{2\Delta N_{\rm crit}}{\sqrt{2\pi(1-n)n V}}\exp\left[-\frac{2\Delta N_{\rm crit}^2}{n(1-n)V}\right]=o(1)
	\end{split}
\end{equation}
implying that we can omit these two terms. This yields the simplification
\begin{equation}\label{finalresult-Page.2}
	\begin{split}
		\braket{S_A}_N=&-Vf\biggl[n\ln(n)+(1-n)\ln(1-n)\biggl]+\frac{f}{2}+\frac{\ln(1-f)}{2}\\
		&-\frac{V(1-2f)}{2}\ln[n^n(1-n)^{1-n}]{\rm erfc}\left[\sqrt{\frac{2}{n(1-n)V}}\Delta N_{\rm crit}\right]+\sqrt{\frac{(1-n)nV}{2\pi }}\ln\left(\frac{n}{1-n}\right)\exp\left[-\frac{2\Delta N_{\rm crit}^2}{n(1-n)V}\right]\\
		&-2^{V(2f-1)-2}\exp\left[\frac{V(1-2n)^2}{2}\right]{\rm erfc}\left[\sqrt{V}\frac{(1-2n)^2-\ln(2)(1-2f)}{\sqrt{2}(1-2n)}\right]\\
		&-2^{V(1-2f)-2}\exp\left[\frac{V(1-2n)^2}{2}\right]{\rm erfc}\left[\sqrt{V}\frac{(1-2n)^2+\ln(2)(1-2f)}{\sqrt{2}(1-2n)}\right]+o(1).
	\end{split}
\end{equation}
The two terms in the second line can give rise to additional order one terms from the fact that $\Delta N_{\rm crit}$ has an order one distance to the smallest integer. To see whether this is indeed the case we split $\Delta N_{\rm crit}=\Delta N_{\rm crit}^{(1)}+\Delta N_{\rm crit}^{(2)}$ with
\begin{equation}
	\Delta N_{\mathrm{crit}}^{(1)}=-\frac{V(1-2f)\ln\left[n^n(1-n)^{1-n}\right]}{2\ln\left[(1-n)/n\right]}.
\end{equation}
We need to expand
\begin{equation}
	{\rm erfc}\left[\sqrt{\frac{2}{n(1-n)V}}\Delta N_{\rm crit}\right]={\rm erfc}\left[\sqrt{\frac{2}{n(1-n)V}}\Delta N_{\rm crit}^{(1)}\right]-\sqrt{\frac{8}{\pi n(1-n)V}}\exp\left[-\frac{2(\Delta N_{\rm crit}^{(1)})^2}{n(1-n)V}\right]\Delta N_{\rm crit}^{(2)}+O(V^{-1}),
\end{equation}
and
\begin{equation}
	\exp\left[-\frac{2\Delta N_{\rm crit}^2}{n(1-n)V}\right]=\exp\left[-\frac{2(\Delta N_{\rm crit}^{(1)})^2}{n(1-n)V}\right]-\frac{4\Delta N_{\rm crit}^{(1)}}{n(1-n)V}\exp\left[-\frac{2(\Delta N_{\rm crit}^{(1)})^2}{n(1-n)V}\right]\Delta N_{\rm crit}^{(2)}+O(V^{-1}).
\end{equation}
Plugging in $\Delta N_{\rm crit}^{(1)}$, we  notice that the leading order correction due to a non-zero $\Delta N_{\rm crit}^{(2)}$ vanishes, and the next order is of order $1/\sqrt{V}$.
	
Thence, the final result for the average entanglement entropy reads
\begin{equation}\label{finalresult-Page.3}
	\begin{split}
		\braket{S_A}_N=&-Vf\biggl[n\ln(n)+(1-n)\ln(1-n)\biggl]+\frac{f}{2}+\frac{\ln(1-f)}{2}\\
		&-\frac{V(1-2f)}{2}\ln[n^n(1-n)^{1-n}]{\rm erfc}\left[\sqrt{\frac{2}{n(1-n)V}}\Delta N_{\rm crit}^{(1)}\right]+\sqrt{\frac{(1-n)nV}{2\pi }}\ln\left(\frac{n}{1-n}\right)\exp\left[-\frac{2(\Delta N_{\rm crit}^{(1)})^2}{n(1-n)V}\right]\\
		&-2^{V(2f-1)-2}\exp\left[\frac{V(1-2n)^2}{2}\right]{\rm erfc}\left[\sqrt{V}\frac{(1-2n)^2-\ln(2)(1-2f)}{\sqrt{2}(1-2n)}\right]\\
		&-2^{V(1-2f)-2}\exp\left[\frac{V(1-2n)^2}{2}\right]{\rm erfc}\left[\sqrt{V}\frac{(1-2n)^2+\ln(2)(1-2f)}{\sqrt{2}(1-2n)}\right]+o(1).
	\end{split}
\end{equation}
When $n>1/2$ and/or $f>1/2$, we need to apply the symmetries and reflect $(f,n)\leftrightarrow(1-f,n)\leftrightarrow(f,1-n)\leftrightarrow(1-f,1-f)$.
	
Note that the result in Eq.~\eqref{finalresult-Page.3} naturally holds everywhere as long as $f,n,(1-n),(1-f)$ stay of order one. Only at these boundaries there are significant deviations. When additionally $1-2f$ and/or $1-2n$ are of order one, the last four terms vanish naturally as they are exponentially suppressed then.
	
At fixed $f$ and $n$, we therefore find
\begin{align}\label{eq:SAN-app}
	\braket{S_A}_N=&-Vf\biggl[n\ln(n)+(1-n)\ln(1-n)\biggl]+\frac{f+\ln(1-f)}{2}-\sqrt{\frac{n(1-n)}{2\pi}}\left|\ln\frac{1-n}{n}\right|\delta_{f,\frac{1}{2}}\sqrt{V}-\frac{1}{2}\delta_{f,\frac{1}{2}}\delta_{n,\frac{1}{2}}+o(1),
\end{align}
as already derived in the main text. Thus, there are discontinuities at $f=\frac{1}{2}$ and $f=n=\frac{1}{2}$, which can be further resolved with the help of Eq.~\eqref{finalresult-Page.3}, as discussed next.
	
\subsubsection{Resolving the critical regimes}
	
In the main text, we found that the formula at fixed $f,n\in(0,1/2)$ comprises additional terms at the line $f=\frac{1}{2}$ and as well as at the multicritical point  $f=n=\frac{1}{2}$. With the help of Eq.~\eqref{finalresult-Page.3}, we can resolve these critical points by considering double scaling limits as follows:
\begin{align}
	&\text{zooming in at $f=n=\frac{1}{2}$:} &&\hspace{-4cm} f=\frac{1}{2}+\frac{\Lambda_f}{V}\,,\quad n=\frac{1}{2}+\frac{\Lambda_{\bar{n}}}{\sqrt{V}}\,,\label{eq:A63}\\
	&\text{zooming in at $f=\frac{1}{2}$:} &&\hspace{-4cm} f=\frac{1}{2}+\frac{\Lambda_f}{\sqrt{V}}\,,
\end{align}
where the relevant scales (\ie powers of $V$) were determined based on the discussion of the previous section. When choosing different dependencies in $V$ for the deviations $\Lambda_f$ and $\Lambda_{\bar{n}}$ either create the Kronecker deltas in Eq.~\eqref{eq:SAN-app} (higher powers in $V$) or the terms take their limit for vanishing deviations (lower powers in $V$). Actually, the multicritical point $f=n=\frac{1}{2}$ is more subtle. While the last two terms in Eq.~\eqref{finalresult-Page.3} scale like those shown in Eq.~\eqref{eq:A63}, the two terms in the second line of Eq.~\eqref{finalresult-Page.3} are still resolved when keeping $\sqrt{V}(1-2f)/(1-2n)$ of order one. We do not discuss this subtlety, in the present subsection. 
	
\paragraph{Critical line at $f=\frac{1}{2}$ and $n<\frac{1}{2}$.}
	
It is important to resolve the Kronecker delta about $f=\frac{1}{2}$ for fixed $n$. In this case, we need to expand Eq.~\eqref{finalresult-Page.3} with $f=\frac{1}{2}+\Lambda_f/\sqrt{V}$, where we need to take into account that Eq.~\eqref{finalresult-Page.3} is only valid for $\Lambda_f<0$ because we assumed $f,n<1/2$ to derive this formula. The last two terms become exponentially suppressed and can be omitted. Thus the expression simplifies to
\begin{align}
	\begin{split}
		\braket{S_A}_N&=-\frac{V}{2}\left[n\ln(n)+(1-n)\ln(1-n)\right]+\sqrt{V}\left[n\ln(n)+(1-n)\ln(1-n)\right]\,|\Lambda_f|+\frac{1}{4}-\frac{\ln(2)}{2}\\
		&\quad-\sqrt{V}\left[|\Lambda_f|\ln[n^n(1-n)^{1-n}]\mathrm{erfc}\left(-\frac{\sqrt{2}|\Lambda_f|\ln[n^n(1-n)^{1-n}]}{\sqrt{n(1-n)}|\ln[(1-n)/n]|}\right)\right.\\
		&\quad\left.+\sqrt{\frac{n(1-n)}{2\pi}}\left|\ln\left[\frac{1-n}{n}\right]\right|\exp\left(-\frac{2\Lambda_f^2\ln^2[n^n(1-n)^{1-n}]}{(1-n)n\ln^2[(1-n)/n]}\right)\right].
	\end{split}
\end{align}
The terms of the first line agree with the expansion~\eqref{eq:SAN-app} around $f=\frac{1}{2}+\frac{\Lambda_f}{\sqrt{V}}$, while the second line resolves the Kronecker delta describing the subleading term $-b\sqrt{V}$ given by
\begin{align}\label{eq:squareroot-full-Gaussian-half-system}
	\begin{split}
		b&=|\Lambda_f|\ln[n^n(1-n)^{1-n}]\mathrm{erfc}\left(-\frac{\sqrt{2}|\Lambda_f|\ln[n^n(1-n)^{1-n}]}{\sqrt{n(1-n)}|\ln[(1-n)/n]|}\right)\\
		&\quad+\sqrt{\frac{n(1-n)}{2\pi}}\left|\ln\left[\frac{1-n}{n}\right]\right|\exp\left(-\frac{2\Lambda_f^2\ln^2[n^n(1-n)^{1-n}]}{(1-n)n\ln^2[(1-n)/n]}\right).
	\end{split}
\end{align}
This term is visualized in Fig.~\ref{fig:general-N-visual}(d). Both equations above are already written in such a way that they hold for $f,n\in(0,1)$, meaning that one can also plug in values with $n>1/2$ and $\Lambda_f>0$.
	
\paragraph{Multicritical point at $f=n=\frac{1}{2}$.}
	
When expanding the final result~\eqref{finalresult-Page.3} about $f=n=\frac{1}{2}$, where we assume $f=\frac{1}{2}+\frac{\Lambda_f}{V}$ and $n=\frac{1}{2}+\frac{\Lambda_{\bar{n}}}{\sqrt{V}}$, we resolve the Kronecker delta at the point $f=n=1/2$, see Eq.~\eqref{eq:SAN-app}. Now all terms in Eq.~\eqref{finalresult-Page.3} become important and need to be taken into account
\begin{align}
	\begin{split}
		\braket{S_A}_{N}=&\frac{\ln(2)}{2}V-\Lambda_{\bar{n}}^2-\sqrt{\frac{2}{\pi}}|\Lambda_{\bar{n}}|\exp\left(-\frac{\Lambda_f^2\ln^2(2)}{2\Lambda_{\bar{n}}^2}\right)-|\Lambda_f|\ln(2)\,\mathrm{erf}\left(\frac{|\Lambda_f|\ln(2)}{\sqrt{2}|\Lambda_{\bar{n}}|}\right)\\
		&\qquad-\frac{1}{4}\left(\ln{4}-1+e^{2\Lambda_{\bar{n}}^2}\left[4^{\Lambda_f}\mathrm{erfc}\left(\frac{2\Lambda_{\bar{n}}^2+\Lambda_f\ln{2}}{\sqrt{2}|\Lambda_{\bar{n}}|}\right)+4^{-\Lambda_f}\mathrm{erfc}\left(\frac{2\Lambda_{\bar{n}}^2-\Lambda_f\ln{2}}{\sqrt{2}|\Lambda_{\bar{n}}|}\right)\right]\right)+o(1)\,.
	\end{split}
\end{align}
The constant terms of the first line are the result of expanding higher order terms in $n$ and $f$ around $n=f=\frac{1}{2}$. The second line then contains the resolution of the Kronecker delta at $f=n=1/2$ plus the offset $(\ln4-1)/4$. The offset is not the result of higher order terms which is the reason why we have included it in the negative term $-c$ which is of constant order in $V$. This constant is given by
\begin{align}\label{eq:constant-full-Gaussian-half}
	c=\frac{1}{4}\left(\ln{4}-1+e^{2\Lambda_{\bar{n}}^2}\left[4^{\Lambda_f}\mathrm{erfc}\left(\frac{2\Lambda_{\bar{n}}^2+\Lambda_f\ln{2}}{\sqrt{2}|\Lambda_{\bar{n}}|}\right)+4^{-\Lambda_f}\mathrm{erfc}\left(\frac{2\Lambda_{\bar{n}}^2-\Lambda_f\ln{2}}{\sqrt{2}|\Lambda_{\bar{n}}|}\right)\right]\right)\,.
\end{align}
We have plotted this term in Fig.~\ref{fig:general-N-visual}(e). Let us mention that only the last two terms in this constant create the Kronecker delta  at $f=n=1/2$, cf., Eq.~\eqref{eq:SAN-app}.
	
\subsection{Weighted average over sectors} \label{app:general-weighted}
	
Finally, we turn to the average over different sectors of fixed particle number $N=0,\ldots,V$, which is weighted by  $\exp[-wN]$. Explicitly this means that we need to compute
\begin{equation}
	\begin{split}
		\braket{S_A}_w=&\frac{1}{(1+e^{-w})^V}\sum^V_{N=0} \frac{V!}{N!(V-N)!}e^{-w N}\, \braket{S_A}_N\\
		=&\frac{1}{(1+e^{-w})^V}\sum^{V/2}_{N=0} \frac{V!}{N!(V-N)!}e^{-w N}\, \braket{S_A}_N+\frac{1}{(1+e^{-w})^V}\sum^{V/2}_{N=0} \frac{V!}{N!(V-N)!}e^{w (N-V)}\, \braket{S_A}_N.
	\end{split}
\end{equation}
The second identity follows from the particle-hole symmetry $\braket{S_A}_N=\braket{S_A}_{V-N}$. It is useful to write the sum in this form since Eq.~\eqref{finalresult-Page.3} is only applicable for $N=nV<V/2$. It becomes symmetric in $n\leftrightarrow1-n$ when replacing all terms in $\ln[n/(1-n)]$ by $-|\ln[(1-n)/n]|$ and in the last two terms $(1-2n)\to|1-2n|$. All the other terms are already symmetric. Due this symmetry, we may also assume that $w$ is positive meaning that the average particle number is smaller than $1/2$.
	
The average of the first four terms in Eq.~\eqref{finalresult-Page.3} do not involve an exponential behavior in $n$. Thus, it is suitable to expand $n$ about its mean, which is $\langle n\rangle=\langle N\rangle/V=1/(1+e^{w})+\delta n$ in the first sum and $\langle n\rangle=\langle N\rangle/V=1/(1+e^{-w})+\delta n$ in the second one, with $\delta n=O(1/\sqrt{V})$. The order of this deviation tells us when the second sum plays a role, namely, when $w=O(1/\sqrt{V})$. Otherwise it is exponentially suppressed due to the Gaussian tail of the large $V$ approximation of the binomial weight. Indeed, this Gaussian would also exponentially suppress the last two terms in~\eqref{finalresult-Page.3} when $w\gg1/\sqrt{V}$, as those only contribute when $1-2n$ is of order $1/\sqrt{V}$ or smaller.
	
Hence, for $w\gg1/\sqrt{V}$ the average entanglement entropy becomes
\begin{equation}
	\begin{split}
		\braket{S_A}_w=&\frac{1}{(1+e^{-w})^V}\sum^{V/2}_{N=0} \frac{V!}{N!(V-N)!}e^{-w N}\, \braket{S_A}_N+o(1)\\
		=&\frac{1}{(1+e^{-w})^V}\sum^{V/2}_{N=0} \frac{V!}{N!(V-N)!}e^{-w N}\biggl(-Vf\biggl[n\ln(n)+(1-n)\ln(1-n)\biggl]+\frac{f}{2}+\frac{\ln(1-f)}{2}-\frac{V(1-2f)}{2}\\
		&\times\ln[n^n(1-n)^{1-n}]{\rm erfc}\left[\sqrt{\frac{2}{n(1-n)V}}\Delta N_{\rm crit}^{(1)}\right]+\sqrt{\frac{(1-n)nV}{2\pi }}\ln\left(\frac{n}{1-n}\right)\exp\left[-\frac{2(\Delta N_{\rm crit}^{(1)})^2}{n(1-n)V}\right]\biggl)+o(1)\\
		=&\braket{S_A}_{N=\bar{N}}-\frac{f}{2V\langle n\rangle(1-\langle n\rangle)}\frac{1}{(1+e^{-w})^V}\sum^{V}_{N=0} \frac{V!}{N!(V-N)!}e^{-w N}(N-\langle N\rangle)^2+o(1)\\
		=&\braket{S_A}_{N=\bar{N}}-\frac{f}{2}+o(1),
	\end{split}
\end{equation}
with $\bar{N}=1/(1+e^{w})$. The $-f/2$ results from the very first term in Eq.~\eqref{finalresult-Page.3}, as it is the average of the second order term in the Taylor expansion in $(n-\langle n\rangle)$. The average in the first order terms vanish because of the centered Gaussian approximation of the binomial weight. This is also the reason why the other terms do not contribute as the second order Taylor terms will be of order $O(1/\sqrt{V})$ or smaller.
	
The calculation is more complicated for $w=O(1/\sqrt{V})$ or smaller. Then $N$ is concentrated about $V/2$ and its difference is of the order $\sqrt{V}$. The question is where we get new contributions in Eq.~\eqref{finalresult-Page.3}. Certainly, this can only happen when $1-2f$ is of order $1/V$ or smaller as otherwise the maximum is always sufficiently away from the multicritical point $f=n=1/2$. We will go step by step through the single terms.
	
The terms in the first line of Eq.~\eqref{finalresult-Page.3} are symmetric and smooth about $n=1/2$ so that the Taylor expansion and average will be exactly the same as in the case for $w\gg1/\sqrt{V}$. The only thing to consider is that $\langle N\rangle=\bar{N}=V\bar{n}$ is not exactly at $V/2$, but $1/2-\bar{n}$ is of order $1/\sqrt{V}$. Therefore, we have an expansion about $N=V/2$ yielding
\begin{equation}\label{w.average.large}
	\braket{S_A}_{N=\bar{N}}=\braket{S_A}_{N=V/2}-V\left(\bar{n}-\frac{1}{2}\right)^2+O(V^{-1}).
\end{equation}
	
The first term in the second line of Eq.~\eqref{finalresult-Page.3} is of order $V(1-2f)=O(1)$ or smaller, so that any Taylor expansion about $n=1/2$ leads to terms that are vanishing in the large $V$ limit. The second term in the second line of Eq.~\eqref{finalresult-Page.3} yields an additional integral as the leading order vanishes and the first order Taylor expansion gives a term of order one. This integral is
\begin{equation}
	\begin{split}
		&\hspace{-0.5cm}\left\langle\sqrt{\frac{(1-n)nV}{2\pi }}\left|\ln\left(\frac{n}{1-n}\right)\right|\exp\left[-\frac{2(\Delta N_{\rm crit}^{(1)})^2}{n(1-n)V}\right]\right\rangle_N\\
		=&4e^{-Vw^2/8}\int_0^\infty\frac{d\delta n}{\sqrt{2\pi}}\exp[-2\delta n^2]\cosh[\sqrt{V}w\delta n]\sqrt{\frac{2}{\pi }}\delta n\exp\left[-\frac{(\ln2)^2V^2(1-2f)^2}{8\delta n^2}\right]+O\left(\frac{1}{\sqrt{V}}\right)\\
		=&\frac{4}{\pi}e^{-Vw^2/8}\int_0^\infty d\delta n \exp\left[-2\delta n^2-\frac{(\ln2)^2V^2(1-2f)^2}{8\delta n^2}\right]\cosh[\sqrt{V}w\delta n]\delta n+O\left(\frac{1}{\sqrt{V}}\right),
	\end{split}
\end{equation}
because $\Delta N_{\rm crit}^{(1)} = \ln(2)V^{3/2}(1-2f)/(8\delta n)$ with $n=N/V = 1/2-\delta n/\sqrt{V}$. We have not found a way to further simplify this integral so we have evaluated it numerically.
	
Also, for the two last terms in Eq.~\eqref{finalresult-Page.3}, we can go over to a Gaussian integral as any correction will be of vanishing order in $V\to\infty$ as the two terms are of order one or smaller. Hence, we have
\begin{equation}
	\begin{split}
		&\frac{1}{(1+e^{-w})^V}\sum^{V/2}_{N=0} \frac{V!}{N!(V-N)!}\left(e^{-w N}+e^{w(N-V)}\right) 2^{V(2f-1)-2}\exp\left[\frac{V(1-2n)^2}{2}\right]{\rm erfc}\left[\sqrt{V}\frac{(1-2n)^2-\ln(2)(1-2f)}{\sqrt{2}(1-2n)}\right]\\
		&=2^{V(2f-1)}e^{-Vw^2/8}\int_0^\infty \frac{d\delta n}{\sqrt{2\pi}} \cosh[\sqrt{V}w\delta n]{\rm erfc}\left[\frac{4\delta n^2-\ln(2)V(1-2f)}{\sqrt{8}\delta n}\right]+O\left(\frac{1}{\sqrt{V}}\right),
	\end{split}
\end{equation}
and similarly
\begin{equation}
	\begin{split}
		&\frac{1}{(1+e^{-w})^V}\sum^{V/2}_{N=0} \frac{V!}{N!(V-N)!}\left(e^{-w N}+e^{w(N-V)}\right) 2^{V(1-2f)-2}\exp\left[\frac{V(1-2n)^2}{2}\right]{\rm erfc}\left[\sqrt{V}\frac{(1-2n)^2+\ln(2)(1-2f)}{\sqrt{2}(1-2n)}\right]\\
		&=2^{V(1-2f)}e^{-Vw^2/8}\int_0^\infty \frac{d\delta n}{\sqrt{2\pi}} \cosh[\sqrt{V}w\delta n]{\rm erfc}\left[\frac{4\delta n^2+\ln(2)V(1-2f)}{\sqrt{8}\delta n}\right]+O\left(\frac{1}{\sqrt{V}}\right).
	\end{split}
\end{equation}
We evaluate all those integrals numerically as an analytical treatment seems to be out of reach.
	
In summary, the entanglement entropy averaged over the particle number $N$ at $w=O(1/\sqrt{V})$ or smaller and for $1-2f=O(V^{-1})$ is
\begin{equation}\label{w.average.result.page}
	\begin{split}
		\langle S_A\rangle_w=&\langle S_A\rangle_{\langle N\rangle=V/2}-\frac{V(1-2\bar{n})^2}{4}-\frac{1}{4}-\frac{4}{\pi}e^{-Vw^2/8}\int_0^\infty d\delta n \exp\left[-2\delta n^2-\frac{(\ln2)^2V^2(1-2f)^2}{8\delta n^2}\right]\cosh[\sqrt{V}w\delta n]\delta n\\
		&-2^{V(2f-1)}e^{-Vw^2/8}\int_0^\infty \frac{d\delta n}{\sqrt{2\pi}} \cosh[\sqrt{V}w\delta n]{\rm erfc}\left[\frac{4\delta n^2-\ln(2)V(1-2f)}{\sqrt{8}\delta n}\right]\\
		&-2^{V(1-2f)}e^{-Vw^2/8}\int_0^\infty \frac{d\delta n}{\sqrt{2\pi}} \cosh[\sqrt{V}w\delta n]{\rm erfc}\left[\frac{4\delta n^2+\ln(2)V(1-2f)}{\sqrt{8}\delta n}\right]+O\left(\frac{1}{\sqrt{V}}\right).
	\end{split}
\end{equation}
Let us stress that
\begin{equation}
	\begin{split}
		\langle S_A\rangle_{\langle N\rangle=V/2}=&\ln(2) Vf+\frac{1}{4}-\frac{\ln(2)}{2}+\ln(2)V(1-2f)-2^{V(2f-1)-1}+O\left(\frac{1}{\sqrt{V}}\right)\\
		=&\ln(2)V(1-f)+\frac{1-\ln(4)}{4}-2^{V(2f-1)-1}+O\left(\frac{1}{\sqrt{V}}\right),
	\end{split}
\end{equation}
for $f\leq 1/2$.
	
At last we would like to consider the particular case $f=\frac{1}{2}$ and $w=0$ which corresponds to a multicritical point, too. In this case Eq.~\eqref{w.average.result.page} simplifies drastically. Indeed, the last two integrals are both equal to
\begin{align}\label{eq:c-resolved}
	\int_0^\infty \frac{d\delta n}{\sqrt{2\pi}} {\rm erfc}\left[\sqrt{2}\delta n\right]=\frac{1}{2\pi}\,.
\end{align}
Also the third, remaining integral can be carried out
\begin{equation}
	\frac{4}{\pi}\int_0^\infty d\delta n e^{-2\delta n^2}\delta n =\frac{1}{\pi}.
\end{equation}
Collecting everything we find that, at $w=O(1/\sqrt{V})$ or smaller,
\begin{equation}
	\begin{split}
		\langle S_A\rangle_w=&\langle S_A\rangle_{\langle N\rangle=V/2}-\frac{1}{4}-\frac{2}{\pi}.
	\end{split}
\end{equation}
This result can be combined with Eq.~\eqref{w.average.large} to get Eq.~\eqref{eq:Page-weighted}, when choosing $f$ and $\bar{n}$ of order one and fixed without double scaling. Then, the last term $-2/\pi$ only appears when $f=\bar{n}=1/2$ resulting in a Kronecker delta.

\section{INTEGRATION FORMULA FOR JACOBI POLYNOMIALS}
	
When considering the average and the variance of the entanglement entropy for ensembles of fermionic Gaussian states, we will routinely encounter expressions involving integrals of Jacobi polynomials given by
\begin{align}
	I^{(\alpha_1,\beta_1,\alpha_2,\beta_2)}_{kl}=\left[\frac{d}{d\epsilon}\int^1_{-1}(1-x)^{\alpha_2+\epsilon}(1+x)^{\beta_2}\mathcal{P}_k^{(\alpha_1,\beta_1)}\mathcal{P}_l^{(\alpha_2,\beta_2)}\right]_{\epsilon\to 1}\,.\label{eq:Ikl}
\end{align}
In order to evaluate this expression, we write the Jacobi polynomials using one of their two representations:
\begin{align}
	&\text{Sum representation:}&\mathcal{P}_k^{(\alpha_1,\beta_1)}(x)&=\sum^k_{m=0}\underbrace{\frac{(\alpha_1+k)!}{k!(\alpha_1+\beta_1+k)!}\binom{k}{m}\frac{(\alpha_1+\beta_1+k+m)!}{(\alpha_1+m)!}\left(-\frac{1}{2}\right)^m}_{=X^{(\alpha_1,\beta_1,k)}_m}(1-x)^m\,,\label{eq:X}\\
	&\text{Rodrigues formula:}&\mathcal{P}_l^{(\alpha_2,\beta_2)}(x)&=\underbrace{\frac{(-1)^l}{2^ll!}}_{=Y^{(l)}}\frac{1}{(1-x)^{\alpha_2}(1+x)^{\beta_2}}\partial_x^{l}[(1-x)^{\alpha_2+l}(1+x)^{\beta_2+l}]\,.\label{eq:Y}
\end{align}
Note that it is important that $(1-x)^{\alpha_2}(1+x)^{\beta_2}$ in Eq.~\eqref{eq:Ikl} matches the ones appearing in $\mathcal{P}_l^{\alpha_2,\beta_2)}(x)$. The appearing integrals can be evaluated by, first, an integration by parts and then exploiting the beta function integral, Eq.~(5.12.1) of Ref.~\cite{NIST:DLMF}, which can be identified when substituting $x=2t-1$ and $t\in[0,1]$,
\begin{align}
	\begin{split}
		T_{\alpha_2,\beta_2,m,l}(\epsilon)&=\int^{1}_{-1}(1-x)^{m+\epsilon}\partial^{l}_{x}[(1-x)^{\alpha_2+l}(1+x)^{\beta_2+l}]dx\\&=2^{\alpha_2+\beta_2+m+\epsilon+l+1}\frac{\Gamma(\alpha_2+m+\epsilon+1)\Gamma(\beta_2+l+1)\Gamma(m+\epsilon+1)}{\Gamma(\alpha_2+\beta_2+m+\epsilon+l+2)}\frac{1}{\Gamma(m-l+\epsilon+1)}\,,
	\end{split}
\end{align}
where $\Gamma(z)$ is the gamma function. Note that this function has removable singularities at integers when $m\leq l-2$ in the limit $\epsilon\to 1$. When $0\leq m\leq l-2$, we can use Euler's reflection formula, Eq.~(5.5.3) of Ref.~\cite{NIST:DLMF},
\begin{align}
	\Gamma(z)\Gamma(1-z)=\frac{\pi}{\sin(\pi z)}\quad\Leftrightarrow\quad \frac{1}{\Gamma(m-l+\epsilon+1)}=\frac{\sin{\pi(m-l+\epsilon+1)}}{\pi}\Gamma(1-(m-l+\epsilon+1))\,,
\end{align}
to remove those singularities, such that
\begin{align}
	T_{\alpha_2,\beta_2,m,l}=\begin{cases}
		2^{\alpha_2+\beta_2+m+\epsilon+l+1}\frac{\Gamma(\alpha_2+m+\epsilon+1)\Gamma(\beta_2+l+1)}{\Gamma(\alpha_2+\beta_2+m+\epsilon+l+2)\Gamma(m+\epsilon+1)}\frac{\sin{\pi(m-l+\epsilon+1)}\Gamma(l-m-\epsilon)}{\pi},& m\leq l-2,\\
		2^{\alpha_2+\beta_2+m+\epsilon+l+1}\frac{\Gamma(\alpha_2+m+\epsilon+1)\Gamma(\beta_2+l+1)}{\Gamma(\alpha_2+\beta_2+m+\epsilon+l+2)\Gamma(m+\epsilon+1)}\frac{1}{\Gamma(m-l+\epsilon+1)}, & m\geq l-1.
	\end{cases}
\end{align}
We can evaluate the derivative in $\epsilon$ at $\epsilon=1$ analytically to find
\begin{align}
	\hspace{-3mm}T'_{\alpha_2,\beta_2,m,l}(1)\!=\!\!
	\begin{cases}
		\!\frac{2^{2+m+l+\alpha_2+\beta_2}(m+1)!(l+\beta_2)!(m+\alpha_2+1)!(l-m-2)!}{(\alpha_2+\beta_2+l+m+2)!}(-1)^{m+l}, & \!\!\!\!m\!\leq\! l\!-\!2,\\
		\!\frac{2^{2+m+l+\alpha_2+\beta_2}(m+1)!(l+\beta_2)!(m+\alpha_2+1)!\left[\ln{2}+\Psi(2+m)-\Psi(2+m-l)+\Psi(2+m+\alpha_2)-\Psi(3+m+l+\alpha_2+\beta_2)\right]}{(\alpha_2+\beta_2+l+m+2)!(m-l+1)!}, & \!\!\!\!m\!\geq\! l\!-\!1
	\end{cases}\label{eq:Tcases}
\end{align}
for the respective integer cases of $m$, where the divergences have been removed accordingly. This yields
\begin{align}
	I^{(\alpha_1,\beta_1,\alpha_2,\beta_2)}_{kl}=Y^{{l}}\sum^{k}_{m=0}\underbrace{X^{(\alpha_1,\beta_1,k)}_{m}T'_{\alpha_2,\beta_2,m,l}(1)}_{F}\,.\label{eq:Ikl-sum}
\end{align}
The summand $F$ for $m\leq l-2$ can be written as
\begin{align}
		F&=(-1)^{l}2^{2+l+\alpha_2+\beta_2}\frac{k!(k+\alpha_1)!(l+\beta_2)!}{k!}\frac{(1+m)!}{m!}\frac{(l-m-2)!}{(k-m)!}\frac{(1+m+\alpha_2)!}{(m+\alpha_1)!}\frac{(k+m+\alpha_1+\beta_1)!}{(2+l+m+\alpha_2+\beta_2)!}\\
		&=(-1)^{l}2^{2+l+\alpha_2+\beta_2}(k+\alpha_1)!(l+\beta_2)!(1+m)\frac{(l-m-2)!}{(k-m)!}\frac{(1+m+\alpha_2)!}{(m+\alpha_1)!}\frac{(k+m+\alpha_1+\beta_1)!}{(2+l+m+\alpha_2+\beta_2)!}\,.
\end{align}
The difficulty lies in the evaluation of this sum, which simplifies drastically for suitable $k$ and $l$, as well as $\alpha_1+\beta_1$ and $\alpha_2+\beta_2$. While there may exist a method to evaluate Eq.~\eqref{eq:Ikl-sum} for general values, it will suffice for our purpose to compute for the specific situation we encounter.

\section{PURE FERMIONIC GAUSSIAN STATES}
	
\subsection{Average entanglement entropy}\label{app:average-Gaussian-general}

We start with the one-point function $R_1(x)$, normalized to $\int_0^1 R_1(x)dx=V_A$, which can be shown~\cite{bianchi2021page,Forrester_2010} to be given by
\begin{align}
	R_1(x)=\sum^{V_A-1}_{n=0}\psi_n^2(x)=(1-x)^{\Delta}\sum^{V_A-1}_{n=0}\frac{1}{c_n}\left[\mathcal{P}^{(\Delta,\Delta)}_{2n}(x)\right]^2\,,
\end{align}
where $\Delta=V-2V_A$, $\mathcal{P}^{(\alpha,\beta)}_n$ refers to the Jacobi polynomials, and the constant $c_n$ is given by
\begin{align}
	c_n=\frac{2^{2\Delta}[(2n+\Delta)!]^2}{(2n)!(2n+2\Delta)!(4n+2\Delta+1)}\,.
\end{align}
An important property of orthogonal polynomials is that they satisfy the Christoffel-Darboux relation, Eq.~(18.2.12) of Ref.~\cite{NIST:DLMF}, that expresses the above sum in terms of the polynomials and their first derivatives of the highest orders. This is useful for analyzing the asymptotic behavior. This Christoffel-Darboux formula reads as follows for a general set of orthogonal polynomials $p_n(x)$ with normalization $c_n$,
\begin{align}
	\sum^{V_A-1}_{n=0}\frac{p^2_n(x)}{c_n}=\frac{k_{V_A-1}}{c_{V_A-1}k_{V_A}}\left[p'_{V_A}(x)p_{V_A-1}(x)-p'_{V_A-1}(x)p_{V_A}(x)\right]\,,
\end{align}
where the coefficients $k_n$ are the leading order coefficients in $p_n(x)$. In the present case of the Jacobi polynomials, we can exploit an additional recurrence relation, Eq.~(18.9.15) of Ref.~\cite{NIST:DLMF}, and the known coefficient $k_n$ of the highest power in the Jacobi polynomials,
\begin{align}
	\frac{d}{dx}\mathcal{P}^{(\alpha,\beta)}_n(x)=\frac{\alpha+\beta+n+1}{2}\mathcal{P}^{(\alpha+1,\beta+1)}_{n-1}(x)\quad\text{and}\quad k_n=\frac{(2n+\alpha+\beta)!}{2^n(n+\alpha+\beta)!n!}\,.\label{eq:recurrence-and-kn}
\end{align}
This allows us to write $R_1(x)$ as
\begin{align}
	R_1(x)=V_A(A_1 F_1(x)-A_2 F_2(x))\,,
\end{align}
where we have introduced the abbreviation
\begin{align}
	A_1&=\frac{k_{2V_A-2}}{c_{2V_A-2}k_{2V_A}}\frac{2\Delta+2V_A+1}{2}\,, & F_1(x)&=(1-x^2)^{\Delta}\mathcal{P}^{(\Delta+1,\Delta+1)}_{2V_A-1}(x)\mathcal{P}^{(\Delta,\Delta)}_{2V_A-2}(x)\,,\\
	A_2&=\frac{k_{2V_A-2}}{c_{2V_A-2}k_{2V_A}}\frac{2\Delta+2V_A-1}{2}\,,&
	F_2(x)&=(1-x^2)^{\Delta} \mathcal{P}^{(\Delta+1,\Delta+1)}_{2V_A-3}(x)\mathcal{P}^{(\Delta,\Delta)}_{2V_A}(x)\,.
\end{align}
We make use of this simplified expression for the level density to compute the average entanglement entropy $\braket{S_A}_{{\rm G},N}=\int^1_{0}R_1(x)s(x)dx$ for Gaussian states, where the empirical entanglement entropy in terms of the eigenvalues $x$ is
\begin{align}\label{eq:s(x)-deriv2}
	s(x)=-\left(\frac{1+x}{2}\right)\ln\left(\frac{1+x}{2}\right)-\left(\frac{1-x}{2}\right)\ln\left(\frac{1-x}{2}\right)=\ln{2}-\frac{1}{2}\partial_\epsilon[(1-x)^{\epsilon}+(1+x)^\epsilon]_{\epsilon=1},
\end{align}
where $s(x)$ was introduced in Eq.~\eqref{eq:Gaussian-s}. The trick with generating the logarithm by a derivative is a standard one and it is related to the replica trick. The advantage is that we can understand the factors $(1\pm x)^{\epsilon}$ as a tractable deformation of the original weight $(1-x^2)^{\Delta}=(1-x)^{\Delta}(1+x)^{\Delta}$.
	
Due to the normalization of the one-point function $\int_0^1R_1(x)dx=V_A$ and the symmetry $s(x)=s(-x)$, the average becomes
\begin{align}
	\braket{S_A}_{{\rm G},N}=V_A\left[\ln{2}-A_1 I_1+A_2 I_2\right],\label{eq:SAinI2}
\end{align}
where we have introduced the integrals
\begin{align}
	I_1=I^{(\Delta+1,\Delta+1,\Delta,\Delta)}_{(2V_A-1),(2V_A-2)}\,,\qquad
	I_2=I^{(\Delta+1,\Delta+1,\Delta,\Delta)}_{(2V_A-3),2V_A}\,,
\end{align}
where $I^{(\alpha_1,\beta_1,\alpha_2,\beta_2)}_{kl}$ was introduced in Eq.~\eqref{eq:Ikl-sum} and reduced to a sum in Eq.~\eqref{eq:Ikl-sum}. Note that the factors of $\frac{1}{2}$ in Eq.~\eqref{eq:s(x)-deriv2} were canceled by the symmetrizing over $x\to-x$. Both sums can be performed analytically leading to the full expression given by~\eqref{eq:Gaussian-average}.
	
\subsection{Variance}\label{app:variance-Gaussian-general}

In Eq.~\eqref{eq:variance-sum}, we have seen that the variance of the entanglement entropy can be expressed in terms of the integrals 
\begin{align}
	s_{ij}=\int^1_{-1} s(x)\psi_i(x)\psi_j(x)dx=-\frac{1}{2}\frac{d}{d\epsilon} \left[\int^1_{-1}[(1+x)^\epsilon+(1-x)^\epsilon]\psi_i(x)\psi_j(x)\right]_{\epsilon=1}=-I^{(\Delta,\Delta,\Delta,\Delta)}_{ij}\,,
\end{align}
where $\psi_i(x)$ and $s(x)$ were introduced in Eqs.~\eqref{eq:Kmatrix} and~\eqref{eq:Gaussian-s}, respectively, and we used that $\psi_i(x)=\psi_i(-x)$.
	
\begin{align}
	s^2_{ij}&=\tfrac{(2 j)! (2 \Delta +4 i+1) (\Delta +j+1) (2 \Delta +2 j+1) (2 \Delta +4 j+1) (2 (\Delta +i))! \left((1+\Delta-2 \Delta ^2) i-2 (\Delta -1) i^2+(\Delta +1) (2 j+1) (\Delta +j)\right)^2}{2 (2 i)! (2 i-2 j+1)^2 (i-j)^2 (-2 i+2 j+1)^2 (2 (\Delta +j+1))! (\Delta +i+j)^2 (\Delta +i+j+1)^2 (2 \Delta +2 i+2 j+1)^2}\,\,\,\text{for}\,\,\, i<j\,,\label{eq:sij-value}
\end{align}
where Eq.~\eqref{eq:sij-value} is only valid for $i<j$, which is all we need for the sum in Eq.~\eqref{eq:variance-sum}. Despite all terms in the sum of Eq.~\eqref{eq:variance-sum} are non-zero for large $N$, it is dominated by the summand $s^2_{N_A-1,N_A}$, so that it makes sense to consider the limit
\begin{align}
	\begin{split}
		\overline{s}^2_{lk}&=\lim_{N\to\infty}s^2_{N_A-1-l,N_A+k}=\frac{(\frac{1}{f}-1)^{-2 (k +l +1)} (2 k +2 l +3-4 f (k +l +1))^2}{4 (k +l +1)^2 (2 k +2 l +1)^2 (2 k +2l +3)^2}
	\end{split}\label{eq:sbar-lk}
\end{align}
with fixed $f=V_A/V$. Summing over~\eqref{eq:sbar-lk} then yields Eq.~\eqref{eq:Gaussian-variance}.

\section{PURE FERMIONIC GAUSSIAN STATES WITH FIXED NUMBER OF PARTICLES} \label{app:Gaussfixednumber}
	
\subsection{Average entanglement entropy}\label{app:average}
	
The idea of computing the entanglement entropy for Gaussian states is based on a different random matrix average than the one in the Page setting. Since a formula for the entanglement entropy \`a la Page is not at hand, we will derive it in the present section. Let us briefly outline the strategy:
\begin{enumerate}
	\item The goal is to compute $\braket{S_A}_{\mathrm{G},N}=\int^1_{0}s(x)R_1(x)dx$ as explained in the main text.
	\item We can express $R_1(x)$ as a sum of Jacobi polynomials and $s(x)$ as derivatives in $\epsilon$ of powers $(1\pm x)^{\epsilon}$ evaluated at $\epsilon=1$. This is a crucial trick that allows us to deal with the logarithm in $s(x)$.
	\item In order to simplify the integral, we first apply the Christoffel-Darboux formula (turning the sum of $V_A$ terms into a sum over two terms) and then express one Jacobi polynomial as sum and the other as derivatives via Rodrigues formula.
	\item This gives $\braket{S_A}_{\mathrm{G},N}$ as a sum of known integrals (of which some need to be regularized) which can be eventually evaluated.
\end{enumerate}
	
As already mentioned, we start with the level density $R_1$ from Eq.~\eqref{eq:G-level-dens} which describes the eigenvalues of a truncated unitary matrix. It is given in terms of Jacobi polynomials and the associated weight $(1-x)^\alpha (1+x)^\beta$, with respect to which they form an orthonormal set. In particular, the level density is
\begin{align}
	R_1(x)=\sum^{V_A-1}_{n=0}\psi^2_n(x)=(1-x)^{\alpha}(1+x)^{\beta}\sum^{V_A-1}_{n=0}\frac{1}{c_n}\left[\mathcal{P}^{(\alpha,\beta)}_n(x)\right]^2,
\end{align}
with $\psi_n(x)$ as in Eq.~\eqref{eq:def-psi}. An important property of orthogonal polynomials is that they satisfy the Christoffel-Darboux relation, Eq.~(18.2.12) of Ref.~\cite{NIST:DLMF}, that expresses the above sum in terms of the polynomials and their first derivatives of the highest orders. This is useful for analyzing the asymptotic behavior. This Christoffel-Darboux formula reads as follows for a general set of orthogonal polynomials $p_n(x)$ with normalization $c_n$,
\begin{align}
	\sum^{V_A-1}_{n=0}\frac{p^2_n(x)}{c_n}=\frac{k_{V_A-1}}{c_{V_A-1}k_{V_A}}\left[p'_{V_A}(x)p_{V_A-1}(x)-p'_{V_A-1}(x)p_{V_A}(x)\right]\,,
\end{align}
where the coefficients $k_n$ are the leading order coefficients in $p_n(x)$. In the present case, we can again use the known relations from Eq.~\eqref{eq:recurrence-and-kn}. Thus, $R_1(x)$ is expressed as
\begin{align}
	R_1(x)=V_A[A_1F_1(x)-A_2F_2(x)]\,,
\end{align}
where we have introduced the abbreviations
\begin{align}
	A_1&=\frac{k_{V_A-1}(\alpha+\beta+V_A+1)}{2V_A c_{V_A-1}k_{V_A}}\,, & F_1(x)&=(1-x)^{\alpha}(1+x)^{\beta} \mathcal{P}^{(\alpha+1,\beta+1)}_{V_A-1}(x)\mathcal{P}^{(\alpha,\beta)}_{V_A-1}(x)\,,\\
	A_2&=\frac{k_{V_A-1}}{c_{V_A-1}k_{V_A}}\frac{\alpha+\beta+V_A}{2}\,,&
	F_2(x)&=(1-x)^{\alpha}(1+x)^{\beta} \mathcal{P}^{(\alpha+1,\beta+1)}_{V_A-2}(x)\mathcal{P}^{(\alpha,\beta)}_{V_A}(x)\,.
\end{align}
	
We make use of this simplified expression for the level density to compute the average entanglement entropy $\braket{S_A}_{{\rm G},N}=\int^1_{0}R_1(x)s(x)dx$ for Gaussian states,  where the empirical entanglement entropy in terms of the eigenvalues $x$ is
\begin{align}\label{eq:s(x)-deriv}
	s(x)=-\left(\frac{1+x}{2}\right)\ln\left(\frac{1+x}{2}\right)-\left(\frac{1-x}{2}\right)\ln\left(\frac{1-x}{2}\right)=\ln{2}-\frac{1}{2}\partial_\epsilon[(1-x)^{\epsilon}+(1+x)^\epsilon]_{\epsilon=1},
\end{align}
where $s(x)$ was introduced in Eq.~\eqref{eq:Gaussian-s}.
	
Due to the normalization, $\int_0^1R_1(x)dx=V_A$ and the symmetry $R_1(x)=R_1(-x)$ by construction, the average reduces to
\begin{align}
	\braket{S_A}_{{\rm G},N}=V_A\left[\ln{2}-\tfrac{1}{2}A_1 I_1+\tfrac{1}{2} A_2 I_2\right]\,.\label{eq:SAinI}
\end{align}
Here, we have introduced the two symmetrized integrals
\begin{align}
	I_1=I^{(\alpha+1,\beta+1,\alpha,\beta)}_{V_A-1,V_A-1}+I^{(\beta+1,\alpha+1,\beta,\alpha)}_{V_A-1,V_A-1}\,,\qquad I_2=I^{(\alpha+1,\beta+1,\alpha,\beta)}_{V_A-2,V_A}+I^{(\beta+1,\alpha+1,\beta,\alpha)}_{V_A-2,V_A}\,,
\end{align}
where $I^{(\alpha_1,\beta_1,\alpha_2,\beta_2)}_{kl}$ was introduced in Eq.~\eqref{eq:Ikl-sum} and reduced to a sum in Eq.~\eqref{eq:Ikl-sum}. Note that we used $\mathcal{P}^{(\alpha,\beta)}_n(-x)=(-x)^n\mathcal{P}^{(\beta,\alpha)}_n(x)$ to arrive at this symmetrization.
	
Only the last two terms in the sum need to be dealt with in the second case because those correspond to $m=n-1$ and $m=n$ with $n=V_A-1$. This leads to the following lengthy expression
\begin{align}
	\begin{split}
		I_1=&\frac{2^{2+\alpha+\beta}(2V_A+\alpha+\beta)(V_A+\alpha-1)!(V_A+\beta-1)!}{(V_A-1)!(V_A+\alpha+\beta+1)!}\Bigg[\sum^{V_A-3}_{m=0}\frac{(m+1)}{(V_A-m-2)(V_A-m-1)}\\
		&+(\alpha+\beta+2V_A)\big(\ln{2}-V_A+\Psi(V_A)-\Psi(1)-\Psi(\alpha+\beta+2V_A)\big)\\
		&+(V_A+\alpha)\Psi(V_A+\alpha)+(V_A+\beta)\Psi(V_A+\beta)+\alpha+\beta+3V_A\Bigg].
	\end{split}
\end{align}
The cumbersome looking sum can  be in fact carried out exactly with the help of the following identity that is based on a series representation of the digamma function, Eq.~(5.7.6) of Ref.~\cite{NIST:DLMF},
\begin{align}
	\sum^n_{m=0}\frac{(m+1)}{(Z+m)(Z+m+1)}=\frac{Z}{1+Z+n}-1+\Psi(Z+n+1)-\Psi(Z)\,.\label{eq:sum-trick}
\end{align}
Thence, we arrive at
\begin{align}
	\begin{split}\label{term1.app.B}
		\partial_\epsilon I_1&=\frac{2^{V+2-2V_A}N!(V-N-1)!\left[V_A+V\ln{2}+N\Psi(N)-V\Psi(V)+(V-N)\Psi(V-N)\right]}{N(V_A-1)!(V-V_A+1)!}.
	\end{split}
\end{align}
	
In a very similar way, we can also evaluate $I_2$, as follows
\begin{align}
	\begin{split}\label{term2.app.B}
		I_2&=\sum^{V_A-2}_{m=0}Y^{(V_A)}\left[X_m^{(\alpha+1,\beta+1,V_A-2)}T'_{\alpha,\beta,V_A,m}(1)+X_m^{(\beta+1,\alpha+1,V_A-2)}T'_{\beta,\alpha,V_A,m}(1)\right]\\
		&=2^{2+\alpha+\beta}\frac{(2V_A+\alpha+\beta)(V_A+\alpha-1)!(V_A+\beta-1)!}{V_A!(V_A+\alpha+\beta)!}\sum^{V_A-2}_{m=0} \frac{(m+1)}{(V_A+\alpha+\beta+m+1)(V_A+\alpha+\beta+m+2)}\\
		&=-\frac{2^{2+\alpha+\beta}\Gamma(V_A+\alpha)\Gamma(V_A+\beta)\left[V_A-1+(2V_A+\alpha+\beta)(\Psi(1+\alpha+\beta+V_A)-\Psi(2V_A+\alpha+\beta)\right]}{V_A!(V_A+\alpha+\beta)!}\\
		&=-\frac{2^{V+2-2V_A}(N-1)!(V-N-1)!]\left[V_A-1+T\Psi(V+1-V_A)-T\Psi(V)\right]}{V_A!(V-V_A)!}.
	\end{split}
\end{align}
In this case, we have not needed to consider different cases of Eq.~\eqref{eq:Tcases}. It always holds $m\leq n-2$ with $n=V_A-2$. In the third line, we anew employed Eq.~\eqref{eq:sum-trick} to express the sum in terms of digamma functions.
	
Finally, we combine both terms~\eqref{term1.app.B} and~\eqref{term2.app.B} according to Eq.~\eqref{eq:SAinI}. After canceling the various factorials, we eventually arrive at the expression
\begin{align}
	\braket{S_A}_{\mathrm{G},N}=1-\frac{V_A}{V}(1+V)+V\Psi(V)-\frac{V_A}{V}[(V-N)\Psi(V-N)+N\Psi(N)]+(V_A-V)\Psi(V-V_A+1), \label{limit-gauss-result-mean}
\end{align}
quoted in the main text. This average has the particle-hole symmetry $N\leftrightarrow V-N$, as can be readily seen. Let us highlight that the symmetry between the number of particles $N$ and system size $V_A$ is not visible since the formula above only holds for $V_A\leq N,V-N$. The symmetry is enforced by hand, where one needs to reflect $V_A\leftrightarrow N$ in the formula when $V_A>N$. In this way one gets also the mirror symmetry between two subsystems $A\leftrightarrow B$ reflected in $V_A\leftrightarrow V-V_A$. The latter symmetry and how it is introduced, namely by hand, is shared with Page's setting.
	
\subsection{Variance}\label{app:variance}
	
In Eq.~\eqref{eq:variance-sum}, we have seen that the variance of the entanglement entropy can be expressed in terms of the integrals 
\begin{align}
	s_{ij}=\int^1_{-1} s(x)\psi_i(x)\psi_j(x)dx=-\frac{1}{2}\frac{d}{d\epsilon} \left[\int^1_{-1}[(1+x)^\epsilon+(1-x)^\epsilon]\psi_i(x)\psi_j(x)\right]_{\epsilon=1}=-\frac{1}{2}(I^{(\alpha,\beta,\alpha,\beta)}_{ij}+(-1)^{i+j}I^{(\alpha,\beta,\alpha,\beta)}_{ij})\,,
\end{align}
which are only needed for the indices $i<V_A\leq j$. Note that the integral in the second line is of the same form as that in
Eq.~\eqref{eq:Ikl}, which was reduced to a sum in Eq.~\eqref{eq:Ikl-sum}. Thus, we can express the result as the following sum
\begin{align}
	I_{ij}=\sum^i_{m=0}Y^{(j)}\left[X_m^{(\alpha,\beta,i)}T'_{\alpha,\beta,m,j}(1)+(-1)^{i+j}X_m^{(\beta,\alpha,i)}T'_{\beta,\alpha,m,j}(1)\right]\,,
\end{align}
where $Y$, $X$ and $T$ were defined in Eqs.~\eqref{eq:X},~\eqref{eq:Y} and~\eqref{eq:Tcases}, respectively. In order to compute the sum, we need to distinguish the cases $i+1=j=V_A$ and $i+1<j$ due to the case discussion in $T$. The resulting expressions are rather unwieldy, but we can simplify them by defining $i=V_A-1-l$ and $j=V_A+k$ and then taking the thermodynamic limit $V\to\infty$ to find
\begin{align}
	\overline{s}^2_{00}&=\lim_{V\to\infty}s^2_{V_A-1,V_A}=\frac{f \left[f-2 f n-2 (f-1) (n-1)\, n \,\ln[(1- n)/n]\right]^2}{(f-1) (n-1) n}\,,\\
	\overline{s}^2_{lk}&=\lim_{V\to\infty}s^2_{V_A-1-l,V_A+k}\nonumber\\
	&=\frac{f^{1+k+l}\left(\left[f(2+k+l-2n)+(2+k+l)(n-1)\right]n^{1+k+l}-(n-1)^{1+k+l}\left[(2+k+l)n+f(k+l+2n)\right]\right)^2}{(f-1)(k+l)^2(1+k+l^2)^2(2+k+l)^2(n-fn-n^2+fn^2)^{1+k+l}}\,.
\end{align}
The sum~\eqref{eq:variance-sum} is certainly a finite sum of generalized hypergeometric function after taking this limit. What has been rather surprising for us is that it can be performed exactly yielding the relatively simple expression
\begin{align}\label{limit-gauss-result-variance}
	\lim_{V\to\infty}(\Delta S_A)^2_{\mathrm{G},N}=\sum^\infty_{l,k=0}\overline{s}^2_{kl}=\ln(1\!-\!f)\!+\!f\!+\!f^2\!+\!f^2(2n-1)\ln\left(\frac{1-n}{n}\right)\!+\!f(f\!-\!1)(n\!-\!1)\,n\ln^2\left(\frac{1-n}{n}\right)\,,
\end{align}
which is the main result of this section.
	
While our representation of $s_{ij}$ has not been suitable to perform the full sum for finite size $V$, the asymptotic result looks as if it is possible to compute the variance $(\Delta S_A)_{\mathrm{G},N}$ as an expression of digamma functions at fixed $V$, similar to $\braket{S_A}$ from Eq.~\eqref{eq:SA-full}. For this, it will likely be beneficial to find closed formulas for the inner product of different orthogonal polynomials along the lines of Ref.~\cite{wei2021quantum}. In fact, such methods were already used to find a closed expression for the variance $(\Delta S_A)_{\mathrm{G}}$ of the entanglement entropy for the ensemble of all Gaussian states~\cite{com-witte}.
	
\subsubsection{Weighted average over sectors}\label{app:weighted}
	
As in the Page setting, we can average the entanglement entropy over all particle number sectors with a binomial weight for Gaussian states, as well. Employing the same notation as in the Page case, the binomial weight
\begin{align}
	\varrho_w(N)=\binom{V}{N}p^{N}q^{V-N}=\binom{V}{N}\frac{e^{-Nw}}{(1+e^{-w})^V}\quad\text{with}\quad p=\frac{e^{-w}}{1+e^{-w}}\quad\text{and}\quad q=1-p=\frac{1}{1+e^{-w}}\,.
\end{align}
depends on the parameter $w$, which tells us how likely it is to find a system with a particular number of particles in it. Due to the symmetry we can anew assume that $w\geq0$ where $w=0$ corresponds to the symmetric probability weight $\varrho_w(N)=2^{-V}\binom{V}{N}$.
	
Let us recall that this binomial distribution is characterized by the mean $\bar{N}$, variance $\Delta^2 N$, and skewness $\nu_N$:
\begin{align}
	\bar{N}=Vp=\frac{Ve^{-w}}{1+e^{-w}}\,,\quad (\Delta N)^2=Vpq=\frac{Ve^{-w}}{(1+e^{-w})^2}\,,\quad\text{and}\quad \nu_N=\frac{q-p}{\sqrt{Vqp}}=\frac{2\sinh(w/2)}{\sqrt{V}}\,.
\end{align}
We need these quantities for our dual approach when averaging  which has been similarly applied to the average over $N_A$ in Sec.~\ref{app:binom}. As in the Page case we have to deal with kinks in the entanglement entropy averaged over all Gaussian states, which result from the fact that we have to introduce the symmetry in the particle number $N=n V$ and the subsystem size $V_A=fV$. We would like to highlight despite the difference in the physical origin of the average, for Page over $N_A$ and in the present case over $N$, the mathematical problem is very similar. The average will be split into a sum extending over the whole range of $N=0,\ldots,V$ for the part of the quantity where the maximum of the binomial weight $\varrho_w(N)$ lies in. For this quantity, we make use of the exact cumulants shown above. In the remaining parts, which are subleading as we will see, we approximate the binomial distribution by a normal distribution and the sum by an integral.
	
In the particular case $w=0$, the skewness of the binomial distribution vanishes. This will have an important impact as we will see since many terms will drop out.
	
Let us briefly outline the strategy:
\begin{enumerate}
	\item In appendix~\ref{app:Gresulting}, we assume that $f\leq 1/2$ and $w\geq0$ (thus, the average particle number $\bar{n}\leq1/2$) are fixed and do not follow a double scaling in the limit $V\to\infty$.
	\item We expand $\braket{S_A}_{\mathrm{G},N}$ up to order $1/V$, where we make the surprising observation that there is no term of order $1$.
	\item To get the correct weighted average up to order $1/V$, it is therefore sufficient to evaluate $\braket{S_A}_{\mathrm{G},N}$ at $N=\bar{N}=V\bar{n}$, where the binomial distribution is peaked and then take the binomial average for the leading order term $s^{\mathrm{G}}_A$ in $\braket{S_A}_{\mathrm{G},N}=V s^{\mathrm{G}}_A+O(V^{-1})$ into account.
	\item For this, we expand $s^{\mathrm{G}}_A$ around $\bar{n}$ up to fourth order. The non-analyticities along the symmetry axis $f=n$ have to be dealt, separately, but as the binomial distribution becomes increasingly narrow around $\bar{n}$, we will only need to take them into account if $\bar{n}=f$.
	\item Combining all terms than yields the weighted average $\braket{S_A}_{\mathrm{G},w}$ up to order $O(1/V)$ for fixed $f$ and $\bar{n}$.
	\item In appendix~\ref{app:res-Gaussian-fixed-w}, we zoom into the critical line $\bar{n}=f<\frac{1}{2}$  and the multi-critical point $\bar{n}=f=\frac{1}{2}$. The critical regime about these points is dictated by the width of the binomial distribution which is for $n=N/V$ of the order $1/\sqrt{V}$. This allows us to resolve any kinks in the expansion up to order $O(1/\sqrt{V})$.
\end{enumerate}
	
\subsubsection{Resulting formula}\label{app:Gresulting}
	
When we take the averages, we need to respect the validity of Eq.~\eqref{limit-gauss-result-mean}, which is $V_A\leq N\leq V-V_A$ or, equivalently, $f\leq n\leq 1-f$. Hence, we must take into account the kinks we introduce when we enforce the symmetries $f\leftrightarrow n\leftrightarrow 1-n$. We have the identifications
\begin{equation}
	\braket{S_A}_{{\rm G},N}\!=\!\begin{cases} 1\!-\!\tfrac{N}{V}(1\!+\!V)\!+\!V\Psi(V)\!-\!\tfrac{N}{V}[(V\!-\!V_A)\Psi(V\!-\!V_A)\!+\!V_A\Psi(V_A)]\!+\!(N\!-\!V)\Psi(V\!-\!N\!+\!1), & N\!\leq\! V_A,\\
	1\!-\!\tfrac{V_A}{V}(1\!+\!V)\!+\!V\Psi(V)\!-\!\tfrac{V_A}{V}[(V\!-\!N)\Psi(V\!-\!N)\!+\!N\Psi(N)]\!+\!(V_A\!-\!V)\Psi(V\!-\!V_A\!+\!1), & V_A\!\leq\! N\!\leq\! V\!-\!V_A,\\
	1\!-\!\frac{V\!-\!N}{V}(1\!+\!V)\!+\!V\Psi(V)\!-\!\frac{V\!-\!N}{V}[(V\!-\!V_A)\Psi(V\!-\!V_A)\!+\!V_A\Psi(V_A)]\!-\!N\Psi(N\!+\!1), & V\!-\!V_A\!\leq\!N.
	\end{cases}
\end{equation}
Hence, when denoting the mean $\bar{n}=1/(1+e^{w})$, around which we need to expand $\braket{S_A}_{{\rm G},N}$, it is necessary to distinguish the three cases $n<f$, $f<n<1-f$, and $1-f<n$, as each of them requires a different expansion. What is very beneficial is that $\braket{S_A}_{{\rm G},N}$ agrees up to order $1/V$ in the limit $V\to\infty$ so that the expansion coefficients will be the same.
	
We begin by expanding $\braket{S_A}_{{\rm G},N}$ as
\begin{align}
	\braket{S_A}_{{\rm G},N}=V\sum^{4}_{m=0}s_A^{(m)} (n-\bar{n})^m+\frac{s_{A,0}^{}}{V}+o(V^{-1})\quad\text{with}\quad
	s_{A,0}^{}=\left\{\begin{array}{cl} \frac{f(1-f)+f\bar{n}(1-\bar{n})}{12(1-f)(1-\bar{n})\bar{n}}, & f<\bar{n},\\ \frac{\bar{n}(1-\bar{n})+f\bar{n}(1-f)}{12(1-f)(1-\bar{n})f}, & f>\bar{n}. \end{array}\right.
\end{align}
Since $n-\bar{n}$ will be of order $1/\sqrt{V}$ and we multiply by $V$, this Taylor expansion corresponds to an expansion of up to order $1/V$. We list the respective expansion coefficients $s_A^{(m)}$ in Table~\ref{tab:sA-expansion} for $n\leq 1-f$, which are all of order one when we do not choose any double scaling limit in $f$ and $\bar{n}$. The coefficients for $n\geq1-f$ can be obtained when employing the symmetry $n\rightarrow 1-n$ in the case $n<f$.  The constant $s_{A,0}$ is stated separately because it is the only one which is explicitly multiplied by $1/V$ while the $V$-dependence in the other terms only enters via averaging over $(n-\bar{n})^m$.
	
\begin{table*}[t!]
	\centering
	\renewcommand{\arraystretch}{1.5}
	\begin{tabular}{c||c|c}
		$m$ & $s_A^{(m)}$ for $n\leq f\leq\frac{1}{2}$ & $s_A^{(m)}$ for $f\leq n\leq 1-f$ \\
		\hline
		$0$ & $(f\!-\!1)\bar{n}\ln(1\!-\!f)\!-\!\bar{n}(1\!+\!f\ln{f})\!+\!(\bar{n}\!-\!1)\ln(1\!-\!\bar{n})$ & $(f\!-\!1)\ln(1\!-\!f)\!+\!f[(\bar{n}\!-\!1)\ln(1\!-\!\bar{n})\!-\!1\!-\!\bar{n}\ln{\bar{n}}]$\\
		$1$ &  $(f-1)\ln(1-f)-f\ln{f}+\ln(1-\bar{n})$ & $f\ln[(1-\bar{n})/\bar{n}]$\\[1mm]
		$2$ & $\displaystyle\frac{1}{2(\bar{n}-1)}$ & $\displaystyle\frac{f}{2(\bar{n}-1)\bar{n}}$ \\[3mm]
		$3$  & $\displaystyle-\frac{1}{6(\bar{n}-1)^2}$ &  $\displaystyle\frac{f-2f\bar{n}}{6(\bar{n}-1)^2\bar{n}^2}$\\[3mm]
		$4$ & $\displaystyle\frac{1}{12(\bar{n}-1)^3}$ & $\displaystyle\frac{f}{12}\left(\frac{1}{(\bar{n}-1)^3}-\frac{1}{\bar{n}^{3}}\right)$
	\end{tabular}
	\caption{The probability distribution $\varrho_w(n)$ is peaked around $\bar{n}$, so we expand  $s^{\mathrm{G}}_A = \sum^4_{m=0}s_A^{(m)}\,(n-\bar{n})^m+O(n-\bar{n})^5$ from $\braket{S_A}_{\mathrm{G},N}=V s^{\mathrm{G}}_A+O(V^{-1})$ up to fourth order for the cases $f<\bar{n}$ and $f>\bar{n}$, showing the non-analyticity of $\braket{S_A}_{{\rm G},N}$ around $f=\bar{n}$. This non-analyticity only shows up starting with the third order. The coefficients for $n>1-f$ follow from the first column when substituting $\bar{n}\to1-\bar{n}$ and multiplying by $(-1)^m$ which is a consequence of the particle-hole symmetry $n\leftrightarrow1-n$ of the entanglement entropy.}
	\label{tab:sA-expansion}
\end{table*}
	
When computing the leading order behavior of the respective average $\braket{S_A}_{{\rm G},w}$, we need to compute the averages $\braket{(n-\bar{n})^m}$ of these powers. As long as $|f- \bar{n}|$ is larger than order $1/\sqrt{V}$ and $\bar{n}\leq1/2$, which is equivalent to $w\geq0$, the power series will be the same on both sides of the maximum, \ie for $n>\bar{n}$ and $n<\bar{n}$. Then, the kinks are not visible as they are too far away to have a non-exponential suppression. So we can use the known averages of the binomial distribution given by
\begin{align}
	\braket{(n-\bar{n})^m}=\begin{cases}
		1 & m=0\\
		0 & m=1\\
		\frac{pq}{V} & m=2\\
		\frac{pq(q-p)}{V^2} & m=3\\
		\frac{p^2q^2}{V^2}(3+\frac{1-6qp}{Vpq}) & m=4
	\end{cases}\quad\text{with}\quad p=\frac{1}{1+e^{w}}=\bar{n}\quad\text{and}\quad q=\frac{1}{1+e^{-w}}=1-\bar{n}\,.\label{eq:BINmoments}
\end{align}
Hence, in the case $|f- \bar{n}|\gg1/\sqrt{V}$ with $0<\bar{n}\leq 1/2$  the full average is given by
\begin{align}
	\braket{S_A}_{\mathrm{G},w}=Vs_A^{(0)}+\bar{n}(1-\bar{n})s_A^{(2)}+\frac{\bar{n}(1-\bar{n})(1-2\bar{n})}{V}s_A^{(3)}+3\frac{\bar{n}^2(1-\bar{n})^2}{V}s_A^{(4)}+\frac{s_{A,0}^{}}{V}+o(V^{-1})\,,\label{eq:Gmuaverage}
\end{align}
where we used that the next order term $\braket{(n-\bar{n})^5}$ would be of order $V^{-3}$ since the leading order has a vanishing asymmetry.
	
In Eq.~\eqref{eq:Gmuaverage}, only the coefficients $s_A^{(0)}$ and $s_A^{(2)}$ are important when one is interested in terms up to order $1$. Table~\ref{tab:sA-expansion} shows that the expansion coefficients up to order $2$ are continuous at $f=\bar{n}$. Thus, Eq.~\eqref{eq:Gmuaverage} still holds up to order $1$ even when $|f- \bar{n}|$ is of order $1/\sqrt{V}$ or smaller. The key subtlety of this formula lies in the third and fourth moment. The cause for this is a discontinuity  showing in the third and higher derivatives of $\braket{S_A}_{{\rm G},N}$ at $fV=V_A=N=n f$. This requires us to compute the expectation value of $(n-\bar{n})^3$ and $(n-\bar{n})^4$, separately, when $|f- \bar{n}|=O(V^{-1/2})$.
	
For this purpose, we need to approximate the average over the binomial distribution by a Gaussian integral. However, we need to take particular care beyond the Gaussian case, which is given by the Edgeworth series
\begin{align}
	\varrho_{w}(Vn)=\frac{\sqrt{V}}{\sqrt{2\pi}(\bar{n}(1-\bar{n})+O(1/V))}e^{-\frac{V}{2\bar{n}(1-\bar{n})}\left(n-\bar{n}\right)^2}\left[1+\alpha (n-\bar{n})+O((n-\bar{n})^2)\right].
\end{align}
Such a series approximates the original probability distribution while cumulants up to a particular order are chosen to be exact. In our case, the constant $\alpha$ is chosen such that we match the skewness of the binomial distribution, \ie the case $m=3$ from Eq.~\eqref{eq:BINmoments}. This leads to the requirement $\alpha=\frac{q-p}{3pq} = \frac{1-2\bar{n}}{3\bar{n}(1-\bar{n})}$. The normalization stays the same up to order $1/V$ as the Gaussian without the Edgeworth series because the first order correction drops out when integrating over $n$.
	
We can stop with the first order since $(n-\bar{n})=O(V^{-1/2})$ and the zeroth, first and second moment of the binomial distribution do not need this approximation because the coefficients are the same at $f=\bar{n}$. The fourth moment of the binomial distribution already comes with order $1/V$, cf., Eq.~\eqref{eq:Gmuaverage}, such that the Edgeworth series is not needed for this term as it gives only higher order corrections. The only term which has been treated with the Edgeworth series is the third order term $(\bar{n}-n)^3$ which a priori  starts with an order $1/V^{3/2}$. Thence, the correction via the Edgeworth series only mixes the third with the fourth moment so that the $1/V$ term is correctly attributed. 
	
When expanding $s^{\mathrm{G}}_A$ from $\braket{S_A}_{\mathrm{G},N}=V s^{\mathrm{G}}_A+O(V^{-1})$ around $\bar{n}=f$, the expansion coefficients $s_A^{(m)}$ differ at $m=3$ for $n<f$ and $n>f$. We refer to these different coefficients by $s_A^{(m-)}$ and $s_A^{(m+)}$, respectively. The resulting average is then given by
\begin{align}\label{eq:fequaln}
	\braket{S_A}_{\mathrm{G},w}&=V\sum^4_{m=0}s_A^{(m-)}\braket{(n-\bar{n})^m}+\frac{s_{A,0}^{}}{V}+V\left(s_A^{(3+)}-s_A^{(3-)}\right)\int^{\infty}_{f}\hspace{-3mm}\varrho_w(n)(n-\bar{n})^3dn+O(V^{-\frac{3}{2}})\,,
\end{align}
where we first expand everything for $n<f$ and then correct for $n>f$ at third order. Most expansion coefficients $s_A^{(m\pm)}$ can be read off Table~\ref{tab:sA-expansion} except for $\bar{n}=f=\frac{1}{2}$. In this case, there is no region with $f< n<1-f$ such that we have the expansion for $n<f=1/2$ and $n>1-f=1/2$. The coefficients for the latter can be retrieved via the $n\to 1-n$ mirror symmetry which implies $s_A^{(m+)}=(-1)^m s_A^{(m-)}$. When expanding $s_A^{\mathrm{G}}$ around $\bar{n}=f=\frac{1}{2}$, this yields $s_A^{(3+)}-s_A^{(3-)}=\frac{4}{3}$. Recall that we consider $\bar{n}\leq\frac{1}{2}$ and hold $n$ and $f$ fixed, when taking the limit $V\to\infty.$ The double scaling limits, when zooming into the critical points, will be discussed in Sec.~\ref{app:res-Gaussian-fixed-w}.
	
Whenever $f\neq n$, the error is exponentially suppressed for $V\to\infty$, which is why we only need to take this term for $f=\bar{n}$ into account. While the averages $\braket{(n-\bar{n})^m}$ from Table~\ref{tab:sA-expansion} are exact, the integrals in Eq.~\eqref{eq:fequaln} have an additional error due to the fact that we approximate the binomial sum by a continuous integral. However, at third order this will only induce a subleading error of order $V^{-\frac{5}{2}}$ (eventually of order $V^{-\frac{3}{2}}$ after multiplying by $V$).
	
In summary, we evaluate Eq.~\eqref{eq:fequaln} to find Eq.~\eqref{eq:expansion} of the main text, where we included all terms up to order $1/V$ and indicated the correct next order based on the numerical results shown in Fig.~\ref{fig:asymptotics}.
	
\subsubsection{Resolving the critical regimes}\label{app:res-Gaussian-fixed-w}
	
The average $\braket{S_A}_w$ is described by a continuous function at linear and constant order in $V$, but the term of order $1/\sqrt{V}$ only appears when $f=\bar{n}$. Based on our previous analysis, it is therefore a natural question to analyze this term to understand how this critical regime is resolved when being close to $\bar{n}=f$. The width of the approximate binomial distribution $\varrho_w(n)$ scales as $1/\sqrt{V}$, so only if $\bar{n}=f+\frac{\Lambda_{\bar{n}}}{\sqrt{V}}$, the discontinuity in the third derivative will contribute to the average. We therefore analyze this limit to resolve the term of order $1/\sqrt{V}$ around $f=\bar{n}$.
	
In the following, we only analyze the contribution towards the $1/\sqrt{V}$ term. While there is also a discontinuity at order $1/V$, the actual calculation is rather tedious and the result is quite lengthy, but can be carried out with the same techniques presented here. For our purpose, it is indeed sufficient to resolve all discontinuities up to order $1/\sqrt{V}$.
	
\paragraph{Transition at $\bar{n}=f=\frac{1}{2}$.}
	
\begin{figure*}[!t]
	\centering  
	\includegraphics[width=\linewidth]{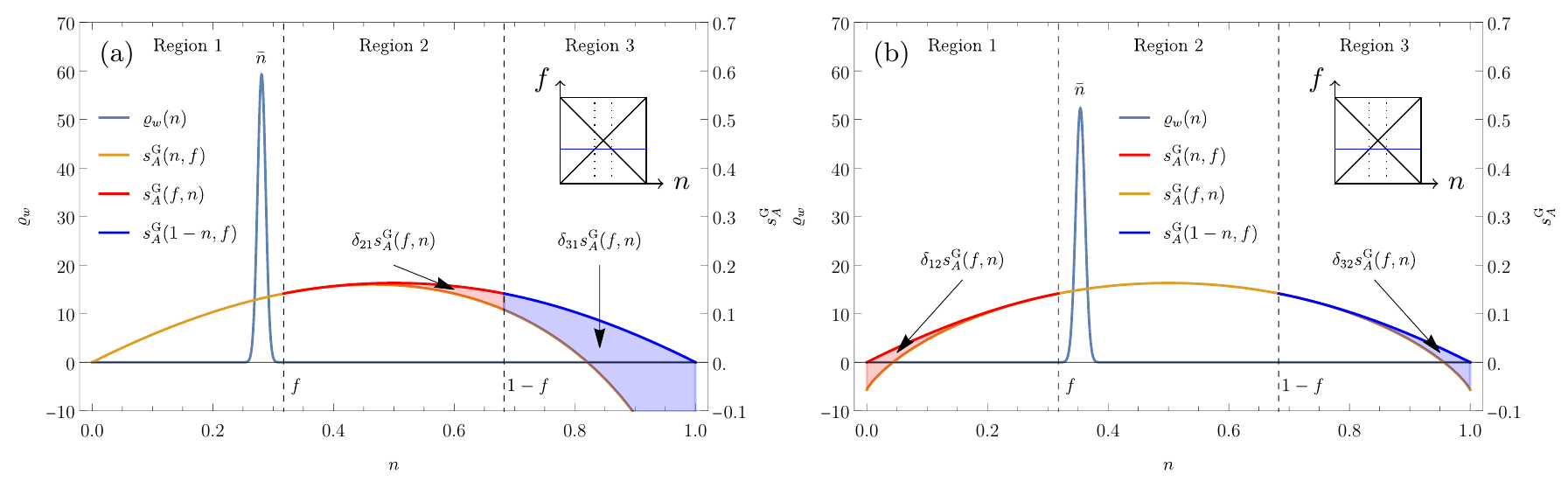}
	\caption{Illustration of integrals. We illustrate the calculation of $d\braket{S_A}_{w}$ around $\bar{n}=f=\frac{1}{2}$, where (a) refers to Case 1 (with $\Lambda_f=-1$, $\Lambda_{\bar{n}}=-1.2$, $V=30$) and (b) refers to Case 2 (with $\Lambda_f=-1$, $\Lambda_{\bar{n}}=-0.8$, $V=30$). Note that we use different scales for $\varrho_w$ and $s_A^{\mathrm{G}}$. Also the proportions are slightly off since we have used a relatively small $V$. The size of region 2 is comparable to the width of the binomial distribution when $V\gg1$.}
	\label{fig:Int-2}
\end{figure*}

Around the multicritical point $\bar{n}=f=\frac{1}{2}$, the correction of order $1/\sqrt{V}$ is due to the discontinuities of the third derivative in $s_A^{\mathrm{G}}(f,n)$ with the two cases $\bar{n}<f$ and $1-f>\bar{n}>f$ that need to be distinguished. We recall that $\braket{S_A}_{\mathrm{G},N}=V s^{\mathrm{G}}_A+O(V^{-1})$. The relevant scalings of $\bar{n}$ and $f$ are given by
\begin{align}
	\bar{n}=\frac{1}{2}+\frac{\Lambda_{\bar{n}}}{\sqrt{V}}\quad\text{and}\quad f=\frac{1}{2}+\frac{\Lambda_f}{\sqrt{V}}\,.
\end{align} 
The two kinks result in two discontinuities in the third derivative of $s_A^{\mathrm{G}}(f,n)$, namely at $n=\frac{1}{2}\pm\frac{\Lambda_f}{\sqrt{V}}$. Those imply different formulas in the entanglement in the following regions (see Fig.~\ref{fig:Int-2})
\begin{align}
	\begin{cases}
		s_A^{\mathrm{G}}(n,f) & \text{Region 1:}\,\, n\leq \frac{1}{2}-\frac{|\Lambda_f|}{\sqrt{V}}\\
		s_A^{\mathrm{G}}(f,n) & \text{Region 2:}\,\, \frac{1}{2}-\frac{|\Lambda_f|}{\sqrt{V}}< n< \frac{1}{2}+\frac{|\Lambda_f|}{\sqrt{V}}\\
		s_A^{\mathrm{G}}(1-n,f) & \text{Region 3:}\,\, \frac{1}{2}+\frac{|\Lambda_f|}{\sqrt{V}}\leq n
	\end{cases}\,,
\end{align}
where the function $s_A^{\mathrm{G}}(f,n)$ represents the leading order in $\braket{S_A}_{\mathrm{G},N} = Vs_A^{\mathrm{G}}(f,n)+O(1)$ (computed for $f\leq n\leq \frac{1}{2}$ and then analytically continued), and is given by
\begin{align}
	s^{\mathrm{G}}_A(f,n)=(f\!\!-\!\!1)\ln(1\!\!-\!\!f)\!\!+\!\!f[(n\!\!-\!\!1)\ln(1\!\!-\!\!n)\!\!-\!\!n\ln{n}\!\!-\!\!1]\label{eq:sAG}
\end{align}
in the region $f\leq n\leq \frac{1}{2}$, which is actually also valid $\frac{1}{2}\leq n\leq f$ (so that we did not have to split up region 2).
	
The main idea of the ensuing computation is to choose  the region where $\bar{n}$ lies in and then use Eq.~\eqref{eq:expansion} for this particular region. Since there also contributions from the other regions because the width of the binomial weight cover also parts therein, we need additionally compute the average over the difference between the corresponding cases in~\eqref{eq:expansion} over these regions. Let us underline, when $w\geq0$ it holds $\bar{n}\leq1/2$. Thence, it holds $\Lambda_{\bar{n}}<0$ and there are the two cases $\bar{n}<f$ and $1-f>\bar{n}>f$ to distinguish which are reflected in $\Lambda_{\bar{n}}< -|\Lambda_f|$ and $0>\Lambda_{\bar{n}}> -|\Lambda_f|$.
	
\textbf{Case 1: $\Lambda_{\bar{n}}<-|\Lambda_f|$.} As $\bar{n}<f$, the maximum of the binomial distribution lies in region 1. Hence, we rewrite the three sums that constitute the entanglement entropy into a sum where the index for $s_A^{\mathrm{G}}(n,f)$ runs over the whole range from $0$ to $V$ and then sum over the correction, which is the difference to $s_A^{\mathrm{G}}(n,f)$, in the other two sums. In particular, we write
\begin{align}
	\begin{split}
		\braket{S_A}_{\mathrm{G},w}&=V\sum_{N=0}^{V_A}\varrho_w(N)s_A^{\mathrm{G}}(n,f)+V\sum_{N=V_A+1}^{V-V_A-1}\varrho_w(N)s_A^{\mathrm{G}}(f,n)+V\sum_{N=V-V_A}^V\varrho_w(N)s_A^{\mathrm{G}}(1-n,f)+O(V^{-1})\\
		&=V\sum_{N=0}^{V}\varrho_w(N)s_A^{\mathrm{G}}(n,f)+V\sum_{N=V_A+1}^{V-V_A-1}\varrho_w(N)\delta_{21} s^\mathrm{G}_A(f,n)+V\sum_{N=V-V_A}^V\delta_{31} \varrho_w(N)s^\mathrm{G}_A(f,n)+O(V^{-1}).
	\end{split}
\end{align}
The differences are defined as
\begin{align}
	\delta_{21} s^\mathrm{G}_A(f,n)=s_A^{\mathrm{G}}(f,n)-s_A^{\mathrm{G}}(n,f)\,,\qquad
	\delta_{31} s^\mathrm{G}_A(f,n)=s_A^{\mathrm{G}}(1-n,f)-s_A^{\mathrm{G}}(n,f),\label{eq:difference}
\end{align}
and they are illustrated in Fig.~\ref{fig:Int-2}(a). For the first sum we can employ the first case in Eq.~\eqref{eq:expansion}. The other two sums simplify when noticing that the difference $\delta_{21} s^\mathrm{G}_A(f,n)$ and $\delta_{31} s^\mathrm{G}_A(f,n)$ have vanishing partial derivatives at $f=\bar{n}=1/2$ up to order $2$. Thus, we can replace $\varrho_w(Vn)$ by its Gaussian approximation without the Edgeworth series correction because the terms will already start with order $1/\sqrt{V}$. The total correction can then be summarized as the integral
\begin{align}
	d \braket{S_A}^{\bar{n}=f=\frac{1}{2}}_{\mathrm{G},w}=V\sum^3_{m=0}\left(\delta_{21} s_A^{(m)}\int^{1-f}_{f}\varrho_w(Vn) (n-\bar{n})^mdn+\delta_{31} s_A^{(m)}\int^{\infty}_{1-f}\varrho_w(Vn) (n-\bar{n})^mdn\right)+o(1/\sqrt{V})\,,
\end{align}
where $\delta_{ij} s_A^{(m)}$ are the respective expansion coefficients from Eq.~\eqref{eq:difference} expanded in $(n-\bar{n})^m$. We will directly combine the result with the second case.
	
\textbf{Case 2: $0>\Lambda_{\bar{n}}> -|\Lambda_f|$.} Now, the main contribution comes from region 2 where $\frac{1}{2}-\frac{\Lambda_f}{\sqrt{V}}\leq n\leq \frac{1}{2}+\frac{\Lambda_f}{\sqrt{V}}$. Therefore, we split the sums as follows
\begin{align}
	\begin{split}
		\braket{S_A}_{\mathrm{G},w}=V\sum_{N=0}^{V}\varrho_w(N)s_A^{\mathrm{G}}(f,n)+V\sum_{N=0}^{V_A}\varrho_w(N) \delta_{12} s^\mathrm{G}_A(f,n)+V\sum_{N=V-V_A}^V\delta_{31} \varrho_w(N)\delta_{32} s^\mathrm{G}_A(f,n)+O(V^{-1}).
	\end{split}
\end{align}
with
\begin{align}
	\delta_{12} s^\mathrm{G}_A(f,n)=s_A^{\mathrm{G}}(n,f)-s_A^{\mathrm{G}}(f,n)\,,\qquad
	\delta_{32} s^\mathrm{G}_A(f,n)=s_A^{\mathrm{G}}(1-n,f)-s_A^{\mathrm{G}}(f,n)\,.\label{eq:differenceD}
\end{align}
The latter are illustrated in Fig.~\ref{fig:Int-2}(b). As in case 1, we replace the first sum by the second case in Eq.~\eqref{eq:expansion}, and approximate the remaining two sums by the Gaussian integrals
\begin{align}\label{eq:center-gaussian-weighted}
	d \braket{S_A}^{\bar{n}=f=\frac{1}{2}}_{\mathrm{G},w}=V\sum^3_{m=0}\left(\delta_{12} s_A^{(m)}\int^{f}_{-\infty}\varrho_w(Vn) (n-\bar{n})^mdn+\delta_{32} s_A^{(m)}\int^{\infty}_{1-f}\varrho_w(Vn) (n-\bar{n})^mdn\right)+o(1/\sqrt{V})\,,
\end{align}
where $\delta_{ij} s_A^{(m)}$ are the respective expansion coefficients from~\eqref{eq:differenceD} expanded in $(n-\bar{n})^m$. 
	
We combine both cases and extend the result to positive and negative $\Lambda_f$ and $\Lambda_{\bar{n}}$ by applying the symmetries of the entanglement entropy resulting in the absolute values of $\Lambda_f$ and $\Lambda_{\bar{n}}$,
\begin{align}
	\begin{split}
		d\braket{S_A}^{\bar{n}=f=\frac{1}{2}}_{\mathrm{G},w}&=\frac{1}{12}\Bigg[\sqrt{\frac{2}{\pi}}\left(e^{-2(|\Lambda_f|-|\Lambda_{\bar{n}}|)^2}\left[1+2(|\Lambda_f|-|\Lambda_{\bar{n}}|)^2\right]+e^{-2(|\Lambda_f|+|\Lambda_{\bar{n}}|)^2}\left[1+2(|\Lambda_f|+|\Lambda_{\bar{n}}|)^2\right]\right)\\
		&\qquad\qquad-(|\Lambda_f|+|\Lambda_{\bar{n}}|)\left[3+4(|\Lambda_f|+|\Lambda_{\bar{n}}|)^2\right]\mathrm{erfc}(\sqrt{2}(|\Lambda_f|+|\Lambda_{\bar{n}}|))\\
		&\qquad\qquad-||\Lambda_f|-|\Lambda_{\bar{n}}||\left[3+4(|\Lambda_f|-|\Lambda_{\bar{n}}|)^2\right]\mathrm{erfc}(\sqrt{2}||\Lambda_f|-|\Lambda_{\bar{n}}||)\,\,\Bigg]\frac{1}{\sqrt{V}}+o(1/\sqrt{V})\,.
	\end{split}
\end{align}
We still need to expand Eq.~\eqref{eq:expansion} around $f=\frac{1}{2}+\frac{\Lambda_f}{\sqrt{V}}$ and $n=\frac{1}{2}+\frac{\Lambda_{\bar{n}}}{\sqrt{V}}$, where we expand the case $f<\bar{n}\leq1/2$ in Eq.~\eqref{eq:expansion} for the case 1 of the present subsection and the case $\bar{n}<f\leq \frac{1}{2}$ for the case 2 of the present subsection. Adding the correction from Eq.~\eqref{eq:center-gaussian-weighted} leads to the final result
\begin{align}
	\begin{split}
		\braket{S_A}^{n=f=\frac{1}{2}}_{\mathrm{G},w}&=\left(\ln{2}-\frac{1}{2}\right)V-\left(\frac{1}{4}+\Lambda_f^2+\Lambda_{\bar{n}}^2\right)-\frac{\max(|\Lambda_f|,|\Lambda_{\bar{n}}|)}{6}\left(3+12\min(\Lambda_f^2,\Lambda_{\bar{n}}^2)+4\max(\Lambda_f^2,\Lambda_{\bar{n}}^2)\right)\frac{1}{\sqrt{V}}\\
		&\qquad +d\braket{S_A}^{\bar{n}=f=\frac{1}{2}}_{\mathrm{G},w}+O(1/V)\,.
	\end{split}
\end{align}
This result reflects the fact that the non-analyticities only show in the $1/\sqrt{V}$ corrections and lower orders.
In Fig.~\ref{fig:Gaussian-mu-visual}(d), we show $d\braket{S_A}^{n=f=\frac{1}{2}}_{\mathrm{G},w}$ which resolves the Kronecker delta $\delta_{f,\bar{n}}\delta_{f,\frac{1}{2}}$ contained in Eq.~\eqref{eq:expansion}.
	
The computation above could have been also computed in different ways. For instance an expansion of all quantities about $f=\bar{n}=n=1/2$ would have led to the same result. This approach would not need any case discussion at the expense that the error functions involved have to be expanded as they contain terms that are proportional to $\sqrt{V}$.
	
\begin{figure*}
	\centering  
	\includegraphics[width=\linewidth]{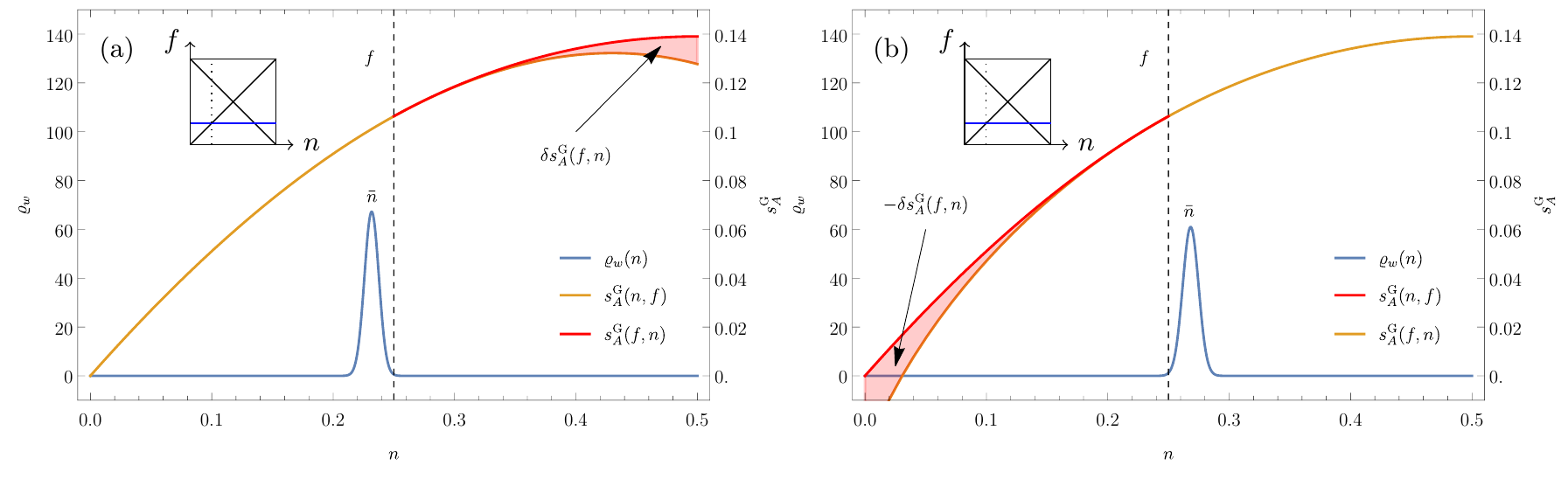}
	\caption{Illustration of integrals. We illustrate the calculation of $d\braket{S_A}_{w}$ around $f=\frac{1}{4}$, where (a) refers to Case 1 (with $\Lambda_f=-0.1$, $V=30$) and (b) refers to Case 2 (with $\Lambda_f=0.1$, $V=30$). In contrast to the previous case around $f=\bar{n}=\frac{1}{2}$, we can ignore the third region $n>1-f$, as the effect is exponentially suppressed. Note that we use different scales for $\varrho_w$ and $s_A^{\mathrm{G}}$. As in Fig.~\ref{fig:Int-2}, the proportions are a bit off due to the relatively small $V$ that has been chosen (to enhance readability). Actually, the dashed vertical line lies inside the width of the binomial distribution for $V\gg1$.}
	\label{fig:Int-1}
\end{figure*}
	
\paragraph{Transition at $\bar{n}=f<\frac{1}{2}$.}
Around $\bar{n}=f<\frac{1}{2}$ with $1-2f\gg1/\sqrt{V}$, the correction of order $1/\sqrt{V}$ is exclusively due to the discontinuity of the third derivative in $\braket{S_A}_{\mathrm{G}N}$, so we only need to analyze this case. Applying the same strategy as before, by looking for the main contribution in the sum, we need to distinguish if $\bar{n}<f$ and $\bar{n}>f$. We emphasize that contributions from region 3, where $n>1-f$, are always exponentially suppressed because this region lies at a distance that is much larger than the width of the binomial distribution.
	
\textbf{Case 1: $\bar{n}<f$.} In this case, the main contribution lies in region 1 [see Fig.~\ref{fig:Int-1}(a)] so that we express the average as follows
\begin{align}
	\begin{split}
		\braket{S_A}_{\mathrm{G},w}=V\sum_{N=0}^{V}\varrho_w(N)s_A^{\mathrm{G}}(f,n)+V\sum_{N=V_A+1}^{V}\varrho_w(N) \delta_{12} \delta s_A^{\mathrm{G}}(f,n)+O(V^{-1}),
	\end{split}
\end{align}
with $\delta s_A^{\mathrm{G}}(f,n)=s_A^{\mathrm{G}}(n,f)-s_A^{\mathrm{G}}(f,n)$ and $s_A^{\mathrm{G}}(f,n)$ is given by the first case in Eq.~\eqref{eq:expansion}. We recall that the sum over the third region is exponentially small. When expanding about $\bar{n}<f$, we need to take into account the non-analyticity that shows up with the third partial derivatives in $n$ and $f$. Like before these orders of the Taylor expansion come with the order $1/\sqrt{V}$ once we have multiplied with the prefactor $V$. Therefore, the correction is given by
\begin{align}
	d\braket{S_A}^{\bar{n}=f<\frac{1}{2}}_{\mathrm{G},w}=V\sum^3_{m=0} \delta s_A^{(m)} \int^{\infty}_{f}\varrho_w(n) (n-\bar{n})^mdn+o(1/V)\,,\label{eq:expansionGmudif}
\end{align}
where the additional error by approximating the discrete binomial sum by an integral over $\varrho_w(n)$ is subleading, as it will contribute towards the overall error of order $1/V$. The coefficients $\delta s_A^{(m)}$ are the Taylor expansion coefficients of the difference $\delta s_A^{\mathrm{G}}(f,n)$. Those are given given by
\begin{align}
	\begin{split}
		\delta s_A^\mathrm{G}(f,n)&=-\frac{(f-\bar{n})^3 [f (4 f-2 \bar{n}-3)+\bar{n}]}{12 \left[(f-1)^2 f^2\right]}+\frac{(f-\bar{n})^2 [f (7 f-4 \bar{n}-5)+2 \bar{n}]}{6 (f-1)^2 f^2}\,(n-\bar{n})\\
		&\quad-\frac{(f-\bar{n}) \left[3 f^2-2 f (\bar{n}+1)+\bar{n}\right]}{2 \left[(f-1)^2 f^2\right]}\,(n-\bar{n})^2+\frac{[f (5 f-4 \bar{n}-3)+2 \bar{n}]}{6 (f-1)^2 f^2}\,(n-\bar{n})^3+O((n-\bar{n})^4)\,.\label{eq:expansionsAGaussian}
	\end{split}
\end{align}
These coefficients are the final ingredient to evaluate Eq.~\eqref{eq:expansionGmudif}. Before we do this, we also consider the second case and then directly combine the results.
	
\textbf{Case 2: $\bar{n}>f$.} Now, the main contribution lies in region 2 [see Fig.~\ref{fig:Int-1}(b)], and we employ the splitting
\begin{align}
	\begin{split}
		\braket{S_A}_{\mathrm{G},w}=V\sum_{N=0}^{V}\varrho_w(N)s_A^{\mathrm{G}}(n,f)-V\sum_{N=0}^{V_A}\varrho_w(N) \delta_{12} \delta s_A^{\mathrm{G}}(f,n)+O(V^{-1})
	\end{split}
\end{align}
with the very same $\delta s_A^{\mathrm{G}}(f,n) = s_A^{\mathrm{G}}(n,f)-s_A^{\mathrm{G}}(f,n)$ as in case 1. Repeating the same steps as in the previous case, we find the correction term
\begin{align}
	d\braket{S_A}^{\bar{n}=f<\frac{1}{2}}_w=-V\sum^3_{m=0} \delta s_A^{(m)} \int^{f}_{-\infty}\varrho_w(n) (n-\bar{n})^mdn+o(1/\sqrt{V})\,,
\end{align}
where the overall minus sign comes from the fact that the error for $n<f$ is just opposite to the one previously calculated, \ie, the expansion coefficients $\delta s_A^{(m)}$ are still the ones listed in Eq.~\eqref{eq:expansionsAGaussian}. 
	
Using the absolute value to unify the previous results of both cases for a general $\bar{n}=f+\frac{\Lambda_{\bar{n}}}{\sqrt{V}}$, where $\Lambda_{\bar{n}}$ can be both positive or negative, we arrive at
\begin{align}
	d\braket{S_A}^{\bar{n}=f<\frac{1}{2}}_{\mathrm{G},w}=\left(\frac{e^{-\frac{\Lambda_{\bar{n}}^2}{2f(1-f)}}\left(2f^{\frac{5}{2}}-2f^{\frac{3}{2}}-f^{\frac{1}{2}}\Lambda_{\bar{n}}^2\right)}{6\sqrt{1-f}f\sqrt{2\pi}}+
	\frac{|\Lambda_{\bar{n}}|\left[3(1-f)f+\Lambda_{\bar{n}}^2\right]\mathrm{erfc}\left(\frac{|\Lambda_{\bar{n}}|}{\sqrt{2f(1-f)}}\right)}{12(1-f)f}\right)\frac{1}{\sqrt{V}}+o(1/\sqrt{V})\,.\label{eq:chemGaussiangeneral}
\end{align}
We can relate this result to our previous finding in Eq.~\eqref{eq:center-gaussian-weighted}. If we set $\Lambda_{\bar{n}}\to \Lambda_f+\Lambda_{\bar{n}}$ in Eq.~\eqref{eq:center-gaussian-weighted} and then consider the limit $\Lambda_f\to\infty$, we reproduce Eq.~\eqref{eq:chemGaussiangeneral} at $f=\frac{1}{2}$ from above. Consequently, the different limits connect at $f=\frac{1}{2}$, as expected.
	
The full asymptotic of $\braket{S_A}^{\bar{n}=f=\frac{1}{2}}_{\mathrm{G},w}$ is given by first expanding formula~\eqref{eq:expansion} in the respect regions for $f<n$ and $f>n$ (depending on the sign of $\Lambda_{\bar{n}}$) and then adding the corrections~\eqref{eq:chemGaussiangeneral}. For $n=f+\frac{\Lambda_{\bar{n}}}{\sqrt{V}}$, we eventually find
\begin{align}
	\begin{split}
		\braket{S_A}^{\bar{n}=f<\frac{1}{2}}_{\mathrm{G},w}&=[(f^2-1)\ln(1-f)-f(1+f\ln{f})]V+\Lambda_{\bar{n}} f[\ln(1-f)-\ln{f}]\sqrt{V}-\frac{1}{2}\left(\frac{\Lambda_{\bar{n}}^2}{1-f}-f\right)\\
		&\qquad+\left[\frac{\Theta(\Lambda_{\bar{n}})\Lambda_{\bar{n}}^3(1-2f)}{6f(1-f)^2}-\frac{\Theta(-\Lambda_{\bar{n}})}{6}\left(3+\frac{\Lambda_{\bar{n}}^2}{(1-f)^2}\right)\right]\frac{1}{\sqrt{V}}+d\braket{S_A}^{\bar{n}=f<\frac{1}{2}}_{\mathrm{G},w}+O(1/V)\,,
	\end{split}
\end{align}
where $\Theta$ is Heaviside step function (with $\Theta(x)=1$ for $x>0$ and $\Theta(x)=0$ otherwise). We show $d\braket{S_A}^{n=f<\frac{1}{2}}_{\mathrm{G},w}$ in Fig.~\ref{fig:Gaussian-mu-visual}(e) which resolves the Kronecker delta $\delta_{f,\bar{n}}$ contained in Eq.~\eqref{eq:expansion}.
	
Anew, one could have taken again a different approach without case discussion and an expansion about $\bar{n}=n=f$. This would have given the same result only that the Heaviside step function would have been encoded in additional error functions whose argument would have been proportional to $\sqrt{V}$.
	
\twocolumngrid
	
\bibliography{references}
	
\end{document}